\documentclass[vecphys,envcountsect]{svmult} 
\usepackage{graphicx}        
\usepackage{multicol}        
\usepackage[bottom]{footmisc}
 
\def\slash#1{\mbox{$\not \!\! #1$}} 
\def\Dslash{{\slash D}}

\newcommand{\beq}{\begin{equation}} 
\newcommand{\eeq}{\end{equation}} 
\newcommand{\beqn}{\begin{eqnarray}} 
\newcommand{\eeqn}{\end{eqnarray}} 
\newcommand{\nn}{\nonumber}

\newcommand{\half}{\frac{1}{2}}

\usepackage{amsmath,amssymb} 
\usepackage{latexsym} 
\usepackage{graphicx} 
\usepackage{bbm} 
\usepackage{bm} 
\usepackage{cite} 
 
\allowdisplaybreaks[1] 
 
\newcommand{\al}{{\alpha^{\prime}}} 
 
\newcommand{\er}{{\rm e}} 
\newcommand{\dr}{{\rm d}} 
\newcommand{\Tr}{{\rm Tr}} 
 
\newcommand{\gs}{g_{\rm s}}

\newcommand{\R}{\mathbbm{R}} 
\newcommand{\Z}{\mathbbm{Z}} 
\newcommand{\C}{\mathbbm{C}} 
 
\newcommand{\del}{\partial} 
\newcommand{\be}{\begin{equation}} 
\newcommand{\ee}{\end{equation}} 
\newcommand{\ba}{\begin{eqnarray}} 
\newcommand{\ea}{\end{eqnarray}} 
\newcommand{\bdm}{\begin{displaymath}} 
\newcommand{\edm}{\end{displaymath}} 
\newcommand{\ra}{\rangle} 
\newcommand{\la}{\langle} 
\newcommand{\pp}{\prime} 
\newcommand\fr[1]{\frac{1}{#1}} 
 
\newcommand{\wtil}{\widetilde} 
 
\newcommand{\one}{\mathbbm{1}}

\newcommand{\Dsm}{\,{\raisebox{0.5pt}{$/$} \hspace{-7pt} D}} 
 
\newcommand{\Asm}{\,{\raisebox{0.5pt}{$/$} \hspace{-7pt} A}}

\def\S{\Sigma} 
 
\def\b{\beta} 
\def\a{\alpha} 
\def\g{\gamma} 
\def\s{\sigma} 
 
\def\veps{\varepsilon} 
\def\vt{\vartheta} 
\def\bvt{{\bar\vartheta}} 
 
\def\adot{{\dot\alpha}} 
 
\def\d{\delta} 
\def\D{\Delta} 
\def\vp{\varphi} 
 
\newcommand{\im}{{\rm Im\,}} 
 
\newcommand{\nus}{\begin{displaystyle} 
    \bar\nu_{\rule{0pt}{2pt}}^{u[A}\nu^{B]}_u 
    \end{displaystyle}} 
\newcommand{\nut}{\begin{displaystyle} 
    \bar\nu_{\rule{0pt}{2pt}}^{u(A}\nu^{B)}_u 
    \end{displaystyle}}

\newcommand{\ie}{{i.e.\ }} 
\newcommand{\eg}{{e.g.\ }} 
\newcommand{\hepth}[1]{{\tt hep-th/#1}} 
 
\newcommand{\calA}{{\mathcal A}} 
\newcommand{\calB}{{\mathcal B}} 
\newcommand{\calC}{{\mathcal C}} 
\newcommand{\calD}{{\mathcal D}} 
\newcommand{\calE}{{\mathcal E}} 
\newcommand{\calF}{{\mathcal F}}

\newcommand{\calJ}{{\mathcal J}} 
\newcommand{\calK}{{\mathcal K}} 
\newcommand{\calL}{{\mathcal L}} 
\newcommand{\calM}{{\mathcal M}} 
\newcommand{\calN}{{\mathcal N}} 
\newcommand{\calO}{{\mathcal O}} 
\newcommand{\calP}{{\mathcal P}} 
\newcommand{\calQ}{{\mathcal Q}} 
\newcommand{\calR}{{\mathcal R}} 
\newcommand{\calS}{{\mathcal S}} 
\newcommand{\calT}{{\mathcal T}}

\newcommand{\calX}{{\mathcal X}} 
\newcommand{\calY}{{\mathcal Y}} 
\newcommand{\calZ}{{\mathcal Z}} 

\DeclareMathAlphabet{\mathpzc}{OT1}{pzc}{m}{it} 
\newcommand\hsp[1]{\hspace*{#1 cm}}

\newcommand{\mb}[1]{\mathbf{#1}} 
 
\def\bea{\begin{eqnarray}} 
\def\eea{\end{eqnarray}} 
\def\beas{\begin{eqnarray*}} 
\def\eeas{\end{eqnarray*}} 
\def\sla{\raise.15ex\hbox{$/$}\kern-.57em}

\def\bea{\begin{eqnarray}} 
\def\eea{\end{eqnarray}} 
\def\de{\partial}

\def\sla{\raise.15ex\hbox{$/$}\kern-.57em} 
\def\ap{{\alpha^\prime}}

\def\a{\alpha} 
\def\b{\beta} 
\def\g{\gamma} 
\def\d{\delta}

\def\s{\sigma}

\def\vt{\vartheta}

\def\cA{{\cal A}} 
 
\def\cC{{\cal C}} 
\def\cD{{\cal D}} 
\def\cE{{\cal E}} 
\def\cF{{\cal F}} 
\def\cG{{\cal G}} 
\def\cH{{\cal H}} 
 
\def\cJ{{\cal J}} 
 
\def\cL{{\cal L}} 
\def\cM{{\cal M}} 
\def\cN{{\cal N}} 
\def\cO{{\cal O}}

\def\cS{{\cal S}} 
 
\def\cU{{\cal U}}

\begin{document} 
 
\title*{Instantons and Supersymmetry} 
 
\author{Massimo Bianchi\inst{1,3} \and Stefano Kovacs\inst{2,4} 
\and Giancarlo Rossi\inst{1,5}} 
 
\institute{ 
University of Rome {\it Tor Vergata} and 
INFN Sez. di Roma {\it Tor Vergata} \\ 
Via della Ricerca Scientifica - 00133 Roma, Italy 
\and 
Trinity College Dublin \\ 
Dublin 2, Ireland 
\and 
\texttt{Massimo.Bianchi@roma2.infn.it} 
\and 
\texttt{kovacs@maths.tcd.ie} 
\and 
\texttt{giancarlo.rossi@roma2.infn.it} 
} 
 
\maketitle 
 
\begin{abstract} 
The role of instantons in describing non-perturbative aspects of 
globally supersymmetric gauge theories is reviewed. The cases of 
theories with ${\cal N}$=1, ${\cal N}$=2 and ${\cal N}$=4 
supersymmetry are discussed. Special attention is devoted to the 
intriguing relation between instanton solutions in field theory and 
branes in string theory. 
\end{abstract}

\vspace*{1cm}
\begin{center} To appear in the book {\it ``String theory and
fundamental interactions''} \\ 
published in celebration of the 65th birthday of Gabriele  Veneziano \\
Eds. M. Gasperini and J. Maharana \\
(Lecture Notes in Physics, Springer Berlin/Heidelberg, 2007) 
\end{center}

\vfill
\noindent
\rule{3cm}{0.4pt} \\
ROM2F/2007/08 \\
TCDMATH 07--08 \\
hep-th/0703142

\newpage


\section{Introduction} 
\label{sec:1} 
 
In this review we discuss the role of instantons in describing 
non-perturbative effects in globally supersymmetric gauge 
theories~\footnote{See the contribution by G. Shore to this book for 
applications to the non-supersymmetric case of QCD.}. 
 
Instantons (anti-instantons) are non-trivial self-dual 
(anti-self-dual) solutions of the equations of motion of the pure 
non-abelian Yang--Mills theory when the latter is formulated in a 
compactified S$_4$ Euclidean manifold. Instanton solutions are 
characterised by a topological charge (the Pontrjagin number, $K$) 
which takes integer values, $K \in \Z$. The integer $K$ represents the 
number of times the (sub)group $SU(2)$ of the gauge group is wrapped 
by the classical solution, while its space-time location spans the 
S$_3$-sphere at infinity. 
 
The presentation of the material of this review can be naturally split 
into three parts according to the number of supersymmetries endowed 
by the theory. In the first part (Sects.~\ref{sec:2}-\ref{sec:ELA}) we 
discuss (weak and strong coupling) computations of instanton dominated 
correlators in pure ${\cal N}=1$ Super Yang--Mills (SYM) and in some 
Super QCD-like (SQCD) and chiral extensions of it. A careful analysis 
of the latter case shows that with a suitable choice of the chiral 
matter flavour representations the interesting phenomenon of dynamical 
breaking of supersymmetry occurs in the theory, as a consequence of 
the constraints imposed by the Konishi anomaly equation. 
 
In the second part (Sects.~\ref{sec:N2-intro}-\ref{sec:instopenstr}) 
we move to the ${\cal N}=2$ Super Yang--Mills case. We will show how 
the highly sophisticated instanton calculus, developed in the years, 
is able to produce the correct coefficients that determine the exact 
expression of the ${\cal N}=2$ prepotential, derived in the famous 
Seiberg--Witten (SW) construction. We will also review the 
construction of the instanton solution in terms of branes with the 
purpose of illustrating the intriguing relation with string theory. 
 
In the third part (Sects.~\ref{sec:N4intro}-\ref{sec:inst-adscft}) we 
discuss the role of instantons in ${\cal N}=4$ Super Yang--Mills. 
Although there are no anomalous $U(1)$'s in this theory (with the 
consequence that there exist no chiral $U(1)$ selection rules that 
would limit the value of Pontrjagin number of the instanton solutions 
contributing to correlators, as instead happens in the ${\cal N}=1$ 
and ${\cal N}=2$ cases), instantons are crucial to check the validity 
of the Maldacena conjecture beyond the realm of perturbation theory. 
Furthermore, their correspondence with IIB string D-instantons gives 
us hope to understand the yet elusive Montonen--Olive duality between 
the weak and strong coupling regimes of ${\cal N}=4$ Super 
Yang--Mills. 
 
A detailed outline of this review is as follows. In Sect.~\ref{sec:2} 
we start with some introductory remarks about instantons and their 
interpretation as field configurations interpolating between classical 
Euclidean vacua (quantum tunneling) and we discuss in detail how 
semi-classical calculations are performed in the instanton background 
with special reference to the notion of collective coordinates. We 
also illustrate the simplifications that occur in supersymmetric 
theories. In Sect.~\ref{sec:SWTI} we derive from supersymmetry and 
holomorphicity the general structure of the Green functions with only 
insertions of lowest (highest) components of chiral (anti-chiral) 
superfields. We show that these Green functions do not depend on the 
operator insertion points and have a completely fixed dependence upon 
the parameters of the theory (like masses and coupling constant). We 
then move in Sect.~\ref{sec:IC} to the explicit semi-classical 
instanton computation of constant Green functions in pure SYM and in 
massive SQCD finding perfect agreement between the theoretical 
expectations spelled out in the previous section and the results of 
actual calculations. The main result of this analysis is that the 
perturbative non-renormalisation theorems of supersymmetry are 
violated in the semi-classical instanton approximation. The instanton 
calculus is then extended to encompass the more delicate cases of 
massless SQCD and to Georgi--Glashow type theories with matter in 
suitable non-anomalous chiral representations. In the first case 
certain inconsistencies are found between results obtained in the 
massless limit of the massive theories and what can be directly 
computed in the strictly massless situation. The problem is discussed 
in detail and the issue of ``strong coupling'' {\it vs} ``weak 
coupling'' instanton calculation strategy is addressed. In the second 
case conflicting results with the constraints imposed by the Konishi 
anomaly equation lead to the conclusion that in certain supersymmetric 
chiral theories supersymmetry is dynamically broken by 
non-perturbative instanton effects. In Sect.~\ref{sec:ELA} we give the 
expression of the effective action for all the cases for which we have 
obtained results in the semi-classical instanton approximation. A nice 
agreement between these two approaches is found, which gives support 
to instanton based computations. 
 
We then pass to discuss instanton effects in $\cN =2$ SYM theories.
After the introduction to the subject contained in
Sect.~\ref{sec:N2-intro}, we present some general discussion of their
properties in Sect.~\ref{sec:N2-gen}. We start by recalling their
supermultiplet content. We then describe the coupling of vector
multiplets to hypermultiplets and the structure of the classical and
effective actions. In Sect.~\ref{sec:SWanalysis} we review the
celebrated analysis of Seiberg and Witten and the derivation of the
expression of the  analytic prepotential in the case of pure $\cN =2$
SYM with $SU(2)$  gauge group. $\cN =2$ instanton calculus is argued
to provide a  powerful check of the SW prepotential in
Sect.~\ref{sec:SWcheck}. In  Sect.~\ref{sec:matrels} we describe
Matone's non-linear recursion  relations for the expansion
coefficients. The validity of the  recursion relations and thus of the
analytic prepotential itself is  checked against instanton
calculations for winding numbers $K=1$ and  $K=2$ in
Sect.~\ref{sec:constrcheck}. In order to go beyond these two  cases,
we follow the strategy advocated by Nekrasov and collaborators  which
is based on the possibility of topologically twisting $\cN =2$  SYM
theories and turning on a non-commutative deformation that  localises
the integration over instanton moduli spaces. After  reviewing the
strategy in Sect.~\ref{sec:topotwist}, we describe how  to couple
hypermultiplets in Sect.~\ref{sec:+hypers}. We then sketch  the
mathematical arguments that lead to the localisation of the  measure
in Sect.~\ref{sec:locK} and the computation of the residues  that
allows a non-perturbative check of the correctness of the SW
prepotential for arbitrary winding number in
Sect.~\ref{sec:residcheck}. In Sect.~\ref{sec:instopenstr} we change
gear and exploit the Veneziano model and D-branes in order to embed
(supersymmetric) YM theories in string theory. In particular we
outline the emergence of the ADHM data and the ADHM equations as a
result of the introduction of lower dimensional D-branes in a given
configuration with maximal $\cN =4$ supersymmetry. Finally we
describe in Sect.~\ref{sec:N2opstring} the truncation to $\cN =2$
supersymmetry and the derivation of the SW prepotential within this
framework in Sect.~\ref{SWprepfromstring}. 
 
In the final part of the review, starting in Sect.~\ref{sec:N4intro}, 
we discuss instanton effects in $\calN=4$ SYM, focussing in 
particular on the r\^ole of instantons in the context of the AdS/CFT 
correspondence. $\calN=4$ SYM is the maximally extended (rigid) 
supersymmetric theory in four dimensions and is believed to be 
exactly conformally invariant at the quantum level. The main 
properties of the model are reviewed in Sect.~\ref{sec:N4sym}. We give 
explicitly the form of the action and the supersymmetry 
transformations and we discuss the basic implications of conformal 
invariance on the physics of the theory, highlighting some of the 
features which make it special compared to the $\calN=1$ and 
$\calN=2$ theories considered in previous sections. General aspects 
of instanton calculus in $\calN=4$ SYM are presented in 
Sect.~\ref{sec:N4insteff}. We describe the general strategy for the 
calculation of instanton contributions to correlation functions of 
gauge-invariant composite operators in the semi-classical 
approximation, emphasising again the essential differences with 
respect to the $\calN=1$ and $\calN=2$ cases. In 
Sect.~\ref{sec:N4-1inst} we focus on the case of the $SU(N_c)$ gauge 
group, which is relevant for the AdS/CFT duality, and we discuss in 
detail the calculation of correlation functions in the one-instanton 
sector. We first construct a generating function which facilitates a 
systematic analysis of instanton contributions to gauge-invariant 
correlators and we then present some explicit examples. The 
generalisation of these results to multi-instanton sectors in the 
large-$N_c$ limit is briefly outlined in Sect.~\ref{sec:N4-multinst}. 
At this point we change somewhat perspective and we move to a 
discussion of the remarkable gauge/gravity duality conjectured by 
Maldacena, explaining how instanton calculus allows to test its 
validity beyond perturbation theory. In Sect.~\ref{sec:adscft-intro} 
we recall the basic aspects of the duality which relates $\calN=4$ 
SYM to type IIB superstring theory in an AdS$_5\times S^5$ 
background. Instanton effects in $\calN=4$ SYM are in correspondence 
with the effects of D-instantons in string theory. More precisely 
instanton contributions to correlators in $\calN=4$ SYM are related 
to D-instanton induced scattering amplitudes in the AdS$_5\times S^5$ 
string theory. In Sects.~\ref{sec:IIBeffact} 
and~\ref{sec:instvsDinst} we present the calculation of D-instanton 
contributions to the string amplitudes dual to the SYM correlation 
functions studied in Sect.~\ref{sec:N4-1inst}. The remarkable 
agreement between gauge and string theory calculations provides a 
rather stringent test of the conjectured duality. Finally in 
Sect.~\ref{sec:BMN-intro} we review the r\^ole of instantons in a 
particularly interesting limit of the AdS/CFT correspondence, the 
so-called BMN limit, in which the gravity side of the correspondence 
is under a better quantitative control beyond the low-energy 
supergravity approximation. We show how instanton effects provide 
again important insights into the non-perturbative features of the 
duality. 
 
Our notation and various technical details are discussed in a number 
of Appendices. 
 
Given the pedagogic nature of this review we refrain from drawing any 
conclusion or present any speculation. In Sect.~\ref{sec:CONCL} we 
try, instead, to summarise the crucial contributions given by Gabriele 
to the subject both as the father of open string theory and as one of 
the deepest and most original investigators of the non-perturbative 
aspects of gauge theories. We thus simply list a few lines of 
research activity where Gabriele's profound insight was precious to 
put existing problems in the correct perspective and help in solving 
them.

\section{Generalities about instantons} 
\label{sec:2} 
 
Instantons (anti-instantons) are self-dual 
(${F}_{\mu\nu}=\tilde{F}_{\mu\nu}$, 
$\tilde{F}_{\mu\nu}=\frac{1}{2}\epsilon_{\mu\nu\rho\sigma}F_{\rho\sigma}$) 
(anti-self-dual, ${F}_{\mu\nu}=-\tilde{F}_{\mu\nu}$) solutions of the 
classical non-abelian Yang--Mills (YM) Euclidean equations of motion 
(e.o.m.)~\cite{BEL,GTH}~\footnote{There are very many good reviews on 
the subject of instantons and their role in field theory. Some are 
listed in refs.~\cite{COL,AKMRV,REV,KHOREV}.}. They are classified by 
a topological number, the Pontrjagin (or winding) number, $K\in \Z$, 
which represents the number of times the (sub)group $SU(2)$ of the 
gauge group is wrapped by the classical solution, $A_\mu^{aI}(x)$, 
when $x$ spans the S$_3$-sphere at the infinity of the compactified 
S$_4$ Euclidean space-time~\footnote{In the following adjoint gauge 
(colour) indices will be indicated with early Latin letters, 
$a,b,c,\ldots$, and vector indices by middle Latin letters, 
$i,j,k,\ldots$. Thus for an $SU(N_c)$ gauge group we will have 
$a,b,c,\ldots=1,2,\ldots,N_c^2-1$ and 
$i,j,k,\ldots=1,2,\ldots,N_c$. Further notations are summarised in 
Appendix~\ref{sec:APPA}.}. Homotopy theory shows, in fact, that the 
homotopically inequivalent mappings S$_3\to SU(2)$ are classified by 
integers since $\Pi_3(SU(2))\sim \Pi_3({\rm S}_3)=\Z$~\cite{BOOK}. 
 
\subsection{The geometry of instantons} 
\label{sec:2.1} 
 
In the Feynman gauge ($\partial_\mu A_\mu^a=0$) the explicit 
expression of the gauge instanton with winding number $K=1$ for the 
$SU(2)$ gauge group (to which case we now restrict) is 
\beq 
A_\mu^{aI}=\frac{2}{g}\bar\eta^a_{\mu\nu}\frac{(x-x_0)_\nu}{(x-x_0)^2} 
\frac{\rho^2}{(x-x_0)^2+\rho^2}\, , 
\label{INST} 
\eeq 
where $\bar\eta^a_{\mu\nu}$ are the 't~Hooft 
symbols~\cite{GTH}~\footnote{They relate the generators of one of the 
two $SU(2)$ groups, in which the Euclidean Lorentz group, $SO(4)$, 
can be decomposed ($SO(4)\sim SU_L(2)\times SU_R(2)$), to the 
generators of the latter through the formula 
$\Sigma^L_{\mu\nu}=\half\bar\eta^a_{\mu\nu}\sigma_a$ with $\sigma_a$ 
the Pauli matrices. The similar coefficients for other $SU(2)$ group 
are the $\eta^a_{\mu\nu}$ symbols with 
$\Sigma^R_{\mu\nu}=\half\eta^a_{\mu\nu}\sigma_a$.}. In~(\ref{INST}) 
$x_0$ and $\rho$ are the so-called location and size of the 
instanton, respectively. They are not fixed by the YM classical 
e.o.m., neither is the orientation of the instanton gauge field in 
colour space. Consequently the derivatives of the instanton solution 
with respect to each one of these parameters (collective 
coordinates~\cite{GS}) will give rise to zero modes of the operator 
associated with the quadratic fluctuations of the gauge field in the 
instanton background~\cite{GTH,BER,YAF} (see Appendix~\ref{sec:APPB} 
for details). 
 
The winding number of a gauge configuration can be expressed in terms 
of the associated field strength through the (gauge invariant) formula 
\beq 
K=\frac{g^2}{32\pi^2} 
\int \dr^4x\, F^a_{\mu\nu}\tilde{F}^a_{\mu\nu}\, .\label{WNUM} 
\eeq 
{}For the action of a self-dual (or anti-self-dual) instanton 
configuration one then gets 
\beq 
S^I=\frac{8\pi^2}{g^2}|K|\, .\label{ACWIN} 
\eeq 
The topological nature of~(\ref{INST}) can be better enlightened 
by first recasting it in the form ($y\equiv x-x_0$) 
\beq 
A_\mu^I=\frac{i}{g}\frac{\rho^2}{y^2+\rho^2} 
\big{[}\Omega^{(1)\dagger}\partial_\mu\Omega^{(1)}\big{]}(y)\, , 
\label{WINST} 
\eeq 
where (see~(\ref{SIGMUNU})) 
\beqn 
\big{[}\Omega^{(1)\dagger}\partial_\mu\Omega^{(1)}\big{]}(x)= 
-\bar\sigma_{\mu\nu}\frac{x_\nu}{\sqrt{x^2}} \, .
\label{WINGAU} 
\eeqn 
In the previous equations 
\beq 
\Omega^{(1)}(x)=\sigma_\mu \frac{x_\mu}{\sqrt{x^2}} \label{WINTR} 
\eeq 
is a topologically non-trivial $SU(2)$ gauge transformation, since it 
does not tend to the group identity as $\sqrt{x^2}$ tends to infinity. 
To compute the winding number of the gauge configuration~(\ref{WINST}) 
it is convenient to gauge transform it by the transformation 
$\Omega^{(1)}$ itself. One gets in this way 
\beqn 
&&(A_\mu^I)^{N.S.}=\big{[}A_\mu^{I}\big{]}^{\Omega^{(1)}} 
=\frac{i}{g}\frac{y^2}{y^2+\rho^2} 
\big{[}\Omega^{(1)}\partial_\mu\Omega^{(1)\dagger}\big{]}(y) 
=\nn\\ 
&&=-\frac{i}{g}\sigma_{\mu\nu}\frac{y_\nu}{\sqrt{y^2}} 
\frac{y^2}{y^2+\rho^2}\, . 
\label{NSINST} 
\eeqn 
{}From the second equality we see that $(A_\mu^I)^{N.S.}$ tends at infinity 
to a non-trivial pure gauge. Inserting~(\ref{NSINST}) in~(\ref{WNUM}), one 
gets the expected result, $K=1$. 
 
The form~(\ref{INST}) (or~(\ref{WINST})) of the one-instanton field is 
called ``singular'' because the point where the non-vanishing 
contribution to the action integral comes from is at $x=x_0$, unlike 
the ``non-singular'' form~(\ref{NSINST}) in which this point has been 
brought to infinity. 
 
We recall in this context that the $F^a_{\mu\nu}\tilde{F}^a_{\mu\nu}$ 
density can be locally rewritten as the divergence of a gauge 
non-invariant vector through the formula 
\beq 
F^a_{\mu\nu}\tilde{F}^a_{\mu\nu}=2\partial_\mu K_\mu\, , \qquad 
K_\mu= \epsilon_{\mu\nu\rho\sigma}{\rm Tr}\big{[}A_\nu F_{\rho\sigma} 
+\frac{2ig}{3} A_\nu A_\rho A_\sigma\big{]}\, .\label{KCUR} 
\eeq 
Thus~(\ref{WNUM}) can be written 
\beq 
K=\frac{g^2}{16\pi^2} 
\int_{{\rm S}_3^\infty} \dr S_\mu K_\mu \, ,\label{XWIN} 
\eeq 
where S$_3^\infty$ is the three-sphere at infinity in S$_4$. Since, 
as we noticed above, $(A_\mu^I)^{N.S.}$ tends to a pure gauge at 
infinity and hence its field strength vanishes, (\ref{XWIN}) can be 
cast in the very expressive form 
\beq 
K=\frac{1}{24\pi^2} 
\int_{{\rm S}_3^\infty} \dr S_\mu \epsilon_{\mu\nu\rho\sigma}{\rm Tr}\big{[} 
\Omega^\dagger\partial_\nu\,\Omega \Omega^\dagger\partial_\rho\Omega\, 
\Omega^\dagger\partial_\sigma\Omega\big{]} \, ,\label{MAUCAR} 
\eeq 
in which we recognise the Cartan--Maurer formula. In general terms 
this quantity is an integer, which represents the winding number of 
the $SU(2)$ gauge transformation, $\Omega$~\footnote{It can be 
explicitly proved that, setting $\Omega_h=\exp[iT^ah^a]$, $K$ is 
invariant under the infinitesimal deformations $h^a\to h^a+\delta 
h^a$. Thus $K$ only depends on the homotopy class to which $\Omega_h$ 
belongs and can always be normalised so as to be an integer.}. 
 
\subsection{Quantum tunneling} 
\label{sec:2.2} 
 
The existence of instanton solutions in YM can be interpreted as an 
indication of quantum tunneling between different vacua, the latter 
being pure gauge configurations characterised by their winding 
numbers~\cite{GTH,TUN1,TUN2}. This fact can be illustrated in quite a 
number of ways. An easy, but heuristic argument is described below. A 
more sophisticated analysis is presented in Appendix~\ref{sec:APPC}. 
 
Consider again the asymptotic formula~(\ref{XWIN}), in which, however, 
the closed surface S$_3^\infty$ has been (smoothly) deformed to an 
other closed surface, which we take as a hyper-cylinder of length $T$, 
bounded at $t=-T/2$ and $t=T/2$ by three-dimensional compact spatial 
manifolds, S$_3$. In the limit $T\to\infty$~(\ref{XWIN}) can then 
be rewritten as the sum of three contributions 
\beqn 
&&K=\frac{ig^3}{24\pi^2}\lim_{T\to\infty}\Big{(}\int_{{\rm S}_3} 
\dr^3x\,\epsilon_{4ijk}{\rm Tr}\big{[}A_i A_j A_k\big{]}|_{t=T/2}+\nn\\ 
&&-\int_{{\rm S}_3}\dr^3x\,\epsilon_{4ijk}{\rm Tr} 
\big{[}A_i A_j A_k\big{]}|_{t=-T/2}+\nn\\ 
&&+\int_{-T/2}^{T/2}\int_{{\rm S}_L} \dr S_i\,\epsilon_{i\nu\rho\sigma} 
{\rm Tr}\big{[}A_\nu A_\rho A_\sigma\big{]}\Big{)}\, ,\label{KCYL} 
\eeqn 
where ${\rm S}_L$ is the three-dimensional lateral surface of the 
cylinder. It can be proved~\cite{SCIU} that one can always find a 
gauge where 1) $A_0=0$ on the lateral surface ${\rm S}_L$; 2) the 
(time-independent) gauge transformations $\Omega_\pm(\vec x)$ at 
$t=\pm T/2$ are such that at large $|\vec x|$ they are independent on 
the direction $\vec x/|\vec x|$. Under these conditions the third 
term in the r.h.s.\ of~(\ref{KCYL}) vanishes. The other two terms 
take integer values because they represent the winding number of the 
mapping $\vec x\to\Omega_\pm(\vec x)$ from the (manifold $\R^3$ 
compactified to the) S$_3$ sphere onto $SU(2)$~\footnote{As an example 
of such gauge transformations one can take $\Omega(\vec 
x)=\exp{[i\pi\vec\sigma\cdot\vec{x}/\sqrt{\vec x^2+1}]}$, in which all 
the point at large $|\vec x|$ are mapped into the group element 
$-\one$. In this way the three-dimensional space manifolds at $t=\pm 
T/2$ become topologically equivalent to S$_3$.}. The net result of 
these considerations is that~(\ref{KCYL}) can be cast in the form 
\beqn 
&&K=n_+-n_-\, ,\label{NPNM}\\ 
&&n_\pm=-\frac{1}{24\pi^2}\epsilon_{ijk}\int_{{\rm S}_3} 
\dr^3x\,{\rm Tr}\big{[}\Omega^\dagger_\pm\partial_i\Omega_\pm\, 
\Omega^\dagger_\pm\partial_j\Omega_\pm\,\Omega^\dagger_\pm 
\partial_k\Omega_\pm(\vec x)\big{]}\, ,\label{QDIF} 
\eeqn 
which shows that the instanton solution ($K\neq 0$) interpolates 
between vacuum states (pure gauge configurations) with different 
winding numbers. 
 
We refer the reader to Appendix~\ref{sec:APPC} for a more rigorous 
discussion of instanton induced tunneling effects in a YM theory. 
 
\subsection{Introducing fermions} 
\label{sec:2.3} 
 
In this subsection we want to briefly recall some elementary facts on 
how to deal with fermions in the functional language in general and in 
the semi-classical approximation in particular. 
 
\subsubsection{Fermionic functional integration} 
\label{sec:2.3.1} 
 
When fermions are introduced it is necessary to define integration 
rules for Grassmann variables. This is a beautiful piece of 
mathematics of which a simple account can be found in~\cite{BERE}. 
The well-known results of this analysis can be summarised as follows. 
 
1) The functional integration over the degrees of freedom of a Dirac 
fermion belonging to the representation ${\bf R}$ of the gauge group 
has the effect of adding to the gauge action a contribution which is 
formally given by 
\beqn 
&&\log\left\{\int {\cal D}\mu[{\bf R}](\psi,\bar\psi) 
\exp\big{[}\int \dr^4x\,(\bar\psi i\gamma_\mu D_\mu[{\bf R}]\psi)(x) 
\big{]}\right\}=\nn\\ 
&&=\log\left\{{\rm det}\big{[}i\gamma_\mu D_\mu[{\bf R}]\big{]}\right\}= 
{\rm Tr}\log\big{[}i\gamma_\mu D_\mu[{\bf R}]\big{]}\, ,\label{DETDIR} 
\eeqn 
where 
\beq 
iD_\mu[{\bf R}]\gamma_\mu=iD_\mu[{\bf R}]\left(\begin{array}{cc} 0 
& \sigma_\mu\\ 
\bar\sigma_\mu & 0\end{array}\right) 
\label{DIRAC} 
\eeq 
and the matrices $\sigma_\mu$ and $\bar\sigma_\mu$ are defined in 
Appendix~\ref{sec:APPA}. 
 
2) For a Weyl fermion, which has half the degrees of freedom of a
Dirac fermion (compare~(\ref{DIRAC}) and~(\ref{BSIGMU})), there  is
the subtlety that the Dirac--Weyl operator maps dotted indices  into
undotted ones, thus making problematic the definition of a
determinant for such an operator. In the literature many
prescriptions have been proposed to address this issue in a rigorous
way (see~\cite{MEU} and works quoted therein). Actually this
difficulty is not relevant in practice, because one can always imagine
to factor out the free operator and compute the determinant of the
resulting operator which is perfectly well defined~\cite{MRT}. The
contribution  from the free part is obviously irrelevant in the
computation of Green  functions as it will cancel with an identical
contribution from the  normalisation factor (see~(\ref{VEV})). Loosely
speaking, looking  at~(\ref{DIRAC}) and~(\ref{BSIGMU}), we may say
that {\it  ceteris paribus} the contribution of a Weyl fermion to the
functional  integral is the ``square root'' of that of a Dirac
fermion. 
 
\subsubsection{Fermionic zero modes} 
\label{sec:2.3.2} 
 
In computing the fermionic functional integral one is led to consider 
the decomposition of the associated spinor fields in eigenstates of 
the fermionic kinetic operator. As is well known, the existence of 
zero modes in certain background gauge fields, such as instantons, is 
of particular relevance for non-perturbative calculations both in 
ordinary and supersymmetric field theories~\cite{GTH,AKMRV,REV}. 
 
The number of zero modes of the Dirac operator in an external field is 
controlled by the famous Atiyah-Singer index theorem~\cite{AS}. The 
theorem states that ``the index of the Dirac operator~(\ref{DIRAC}), 
i.e.\ the difference between the number of left-handed ($n_L$) and 
right-handed ($n_R$) zero modes, is equal to twice the Dynkin index of 
the representation ${\bf R}$ times the Pontrjagin number of the 
background gauge field''. In formulae we write 
\beq 
{\rm ind}(D_\mu[{\bf R}]\gamma_\mu)\equiv n_L-n_R=2\ell[{\bf R}]K\, , 
\label{IND} 
\eeq 
where $\ell[{\bf R}]$ is the Dynkin index of the representation 
${\bf R}$. Let us now consider some interesting applications of this 
theorem. 
\begin{enumerate} 
\item Weyl fermion in the adjoint representation, ${\bf Adj}$, of 
the gauge group. We must distinguish between the left-handed 
($D_\mu[{\bf Adj}]\bar\sigma_\mu$) and the right-handed ($D_\mu[{\bf 
Adj}]\sigma_\mu$) Weyl operator. In the first case $n_R=0$ and the 
formula~(\ref{IND}) becomes 
\beq 
{\rm ind}(D_\mu[{\bf Adj}]\bar\sigma_\mu)= n_L=2 N_c K\, , 
\label{INDWL} 
\eeq 
because $\ell[{\bf Adj}]=N_c$. Since obviously $n_L$ is a non-negative 
number, there can exist zero modes of the left-handed Weyl operator 
only if the classical background instanton field has positive winding 
number, $K>0$. Similarly for the right-handed Weyl operator one gets 
\beq 
{\rm ind}(D_\mu[{\bf Adj}]\sigma_\mu)= - n_R=2 N_c K\, , 
\label{INDWR} 
\eeq 
implying that there can be zero modes only if $K<0$. 
\item Actually the number of zero-modes in the adjoint representation 
of any compact Lie group, $G$, is always given by twice the value of 
the quadratic Casimir operator, $2c_2({\bf Adj}(G))$. This result 
follows from the formula~\cite{BCGW} 
\beq 
{\bf Adj}(G)=4{\bf Adj}(SU(2))+n(G)\,{\bf 2}+s(G)\,{\bf 1}\, , 
\label{ADJDEC} 
\eeq 
which expresses how the adjoint representation of $G$ can be decomposed 
into irreducible representations of $SU(2)$. In~(\ref{ADJDEC}) we have 
introduced the definitions 
\beqn 
&&n(G)=2(c_2({\bf Adj}(G))-2)\, ,\label{NSG1}\\ 
&&s(G)=d({\bf Adj}(G))-4c_2({\bf Adj}(G))+5\, ,\label{NSG2} 
\eeqn 
where $d({\bf Adj}(G))$ is the dimension of the adjoint 
representation of $G$. The number of zero-modes is then 
$4+2(c_2({\bf Adj}(G))-2)=2c_2({\bf Adj}(G))$ 
\item Dirac fermion in the fundamental representation, ${\bf N_c}$, 
of the gauge group. Equation~(\ref{IND}) with $\ell[{\bf N_c}]=1/2$ gives 
\beq 
{\rm ind}(D_\mu[{\bf N_c}]\gamma_\mu)= n_L - n_R= K\, . 
\label{INDD} 
\eeq 
Again in the classical background instanton field there can be either 
left-handed fermionic zero modes, if $K>0$, or right-handed ones, if 
$K<0$. 
\item Fermion in the rank-two antisymmetric representation ${\bf 
N_c(N_c-1)/2}$. The number of zero modes (of definite chirality) is 
$(N_c-2)K$. 
\end{enumerate} 
Deriving the explicit expression of all these fermionic zero modes is 
beyond the scope of this review and we refer the interested reader to 
the general methods that, starting from the seminal paper 
of~\cite{ADHM}, have been developed in the 
literature~\cite{TEM,KHOREV}. However, for completeness we give their 
explicit expression for a few cases more relevant for this review in 
Appendix~\ref{sec:APPA}. 
 
\subsection{Putting together fermion and boson contributions} 
\label{sec:2.4} 
 
As we said, we are interested in computing expectation values of 
(multi-local) gauge invariant operators, by dominating the functional 
integral with the semi-classical contributions coming from the 
non-trivial minima (instantons) of the Euclidean action. The obvious 
question is whether this computational strategy leads to a reliable 
estimate of $\langle O\rangle$. 
 
\subsubsection{The general case} 
\label{sec:2.4.1} 
 
In order to prepare ourselves for this analysis, let us write down the 
formal result obtained by performing the integration over the 
quadratic fluctuations (semi-classical approximation, {\it s.c.}) 
around an instanton solution with winding number $K$. Including also 
the fermionic contribution in~(\ref{INTVEV}) and assuming for 
simplicity that there are no scalar fields in the theory~\footnote{The 
extension of the formulae of this section to the more general case 
where also elementary scalar fields are present is possible, but not 
completely trivial. See below and 
refs.~\cite{AKMRV,ADSALL1,RUS4,COR,KHOREV}}, one gets 
\beqn 
&&\hspace{-.4cm}\langle O\rangle\Big{|}_{s.c.} 
=\mu^{n_B-\kappa_F n_F}\frac{\er^{-\frac{8\pi^2}{g^2}|K|}}{Z|_{s.c.}} 
\label{INTBF}\\ 
&&\hspace{-.4cm}\int\!\prod_{j=1}^{n_F} \dr c_j 
\int\!\prod_{i=1}^{n_B} \dr\beta_i\frac{||a^{(i)}||}{\sqrt{2\pi}} 
\frac{({\rm det}'[{\cal M}_{\mu\nu}^{g.f.}])^{-\half} 
{\rm det}[-D^2(A^I)]}{({\rm det}[{\cal M}_{0;\mu\nu}^{g.f.}])^{-\half} 
{\rm det}[-\partial^2]}\frac{{\rm det}'[\Dslash(A^I)]}{{\rm det} 
[\,\slash\partial\,]}\,O(c;A^I)\, .\nn 
\eeqn 
where $\Dslash$ is the fermionic kinetic operator appropriate to the 
kind of fermion one is dealing with (Dirac or Weyl) and the prime on 
the determinants is there to mean that obviously only non-zero 
eigenvalues are to be included. Further observations about this 
formula are the following. 
 
$\bullet$ The factor $\kappa_F$ is 1 for a Dirac fermion and $\half$ 
for a Weyl fermion. 
 
$\bullet$ The residual fermionic integration $\prod_{j=1}^{n_F} dc_j$ 
is over the Grassmannian coefficients associated with the $n_F$ zero 
modes of the fermionic kinetic operator. We stress that in order not 
to get a trivially vanishing result, the Berezin~\cite{BERE} 
integration rules require a perfect matching in the number of 
fermionic zero modes between those of the fermion operators in the 
action and those contained in the operator $O$. 
 
$\bullet$ The extra $\mu$ dependence in front of the r.h.s.\ 
of~(\ref{INTBF}) (with respect to~(\ref{INTVEV})) is due (similarly 
to the case of the bosonic functional integration, see 
Appendix~\ref{sec:APPB}), to unmatched $\mu$ factors coming from the 
determinant of the fermion Pauli--Villars (PV) regulators~\footnote{In 
order to have a more readable formula we have not shown 
in~(\ref{INTBF}) the determinants of the various PV regulators.}. 
 
$\bullet$ The power $\kappa_F n_F$ is dictated by the way in which zero 
modes contribute to the fermionic mass term in the action and the nature of 
the Grassmannian integration rules. 
 
$\bullet$ No further factor comes from dealing with the fermionic zero 
modes, provided they are normalised to one, which we will always do 
(this is at variance with what happens for bosonic zero modes, each of 
which contributes a factor $||{\rm norm}||/\sqrt{2\pi}$ 
to~(\ref{INTBF})). 
 
$\bullet$ Generally speaking, the ratio of determinants 
in~(\ref{INTBF}) will be a function of the instanton collective 
coordinates as well as $\mu$. 
 
The computational strategy outlined above can be safely used if it can 
be convincingly argued that the classical minima (instantons) really 
dominate the integral. This is a delicate issue which can only be 
settled on a case by case basis. For instance, for the instanton 
contribution to dominate the functional integral one can imagine 
considering Green functions that are zero in perturbation theory. The 
argument here is that otherwise the non-perturbative instanton 
contribution, which is proportional to $\exp(-{8\pi^2}|K|/{g^2})$, 
would represent a completely negligible correction with respect to any 
perturbative term. This is the situation one is usually dealing with 
in ${\cal N}=1$ supersymmetric theories. 
 
In the ${\cal N}=2$ and ${\cal N}=4$ cases it is, instead, interesting 
to consider more general correlators which do not necessarily vanish 
in perturbation theory. In these cases instanton contributions, 
though comparatively exponentially small, can always be ``tracked'' if 
the theory $\vartheta$-dependence is followed (see 
Appendix~\ref{sec:APPC}). 
 
A second crucial question concerns the finiteness of the r.h.s.\ 
of~(\ref{INTBF}). In his beautiful paper 't~Hooft~\cite{GTH} has shown 
that in QCD the integration over the instanton collective coordinates 
around the classical instanton solution 
\beq 
A_\mu=A^I_\mu\, ,\qquad {\rm 
all\,\,other\,\,fields\,\,equal\,to\,\,zero} 
\label{ICL} 
\eeq 
does not lead to a finite result. The reason behind this fact is that the 
integration over the size of the instanton, $\rho$, which comes from 
the ratios of determinant in~(\ref{INTBF}) as well as from the norm of 
the bosonic zero modes, diverges in the infrared limit, i.e.\ for 
large values of $\rho$ (the integration near $\rho=0$ is, instead, 
convergent thanks to asymptotic freedom). This problem is not present 
in the supersymmetric case which we discuss next. 
 
\subsubsection{The supersymmetric case} 
\label{sec:2.4.2} 
 
Something really surprising indeed happens in the case of a 
supersymmetric theory. There, irrespective of the details of the 
theory (number of supersymmetries, gauge group, matter content, etc.), 
the whole ratio of (regularised) determinants is always exactly equal 
to 1~\cite{DADI}. This is because the eigenvalues of the various 
kinetic operators in the instanton background are, up to 
multiplicities, essentially all equal and, due to supersymmetry, there 
is a perfect matching between bosonic and fermionic degrees of 
freedom, leading to contributions that are one the inverse of the 
other. The formula~(\ref{INTBF}) thus becomes 
\beqn 
&&\langle O\rangle\Big{|}_{s.c.}^{\rm SUSY} 
=\mu^{n_B-\half n_F}\frac{\er^{-\frac{8\pi^2}{g^2}|K|}}{Z|_{s.c.}}\times\nn\\ 
&&\times\int\prod_{i=1}^{n_B} \dr\beta_i\frac{||a^{(i)}||}{\sqrt{2\pi}} 
\sum_{\{j_k\}}(-1)^{P_{\{j_k\}}}O(\prod_{k}f_{j_k};A^I)\, ,\label{INTSU} 
\eeqn 
where we have explicitly carried out the final integration over the 
Grassmannian variables $c_j$, ${j=1,2,\ldots,n_F}$. As a result the 
product of the $n_F$ fermionic fields contained in $O$ is simply 
replaced by the product of the wave functions, $f_j(x,\beta)$, of the 
$n_F$ zero modes. The sum over permutations is weighted by 
alternating signs because of Fermi statistics. Finally we have set 
$\kappa_F=\half$, because in supersymmetric theories fermions are 
always introduced as Weyl particles. 
 
Actually in the case $K=1$~(\ref{INTSU}) can be made even more 
explicit, because for a gauge invariant operator the only dependence 
on the collective coordinates is that on the size and position of the 
instanton. Using~(\ref{BOSZM}) of Appendix~\ref{sec:APPB} and 
the coset integration formula (derived in~\cite{BER}) necessary for 
the generalisation to the case of the $SU(N_c)$ group, one gets to 
leading order in $g$ (where $Z|_{s.c.}=1$) 
\beqn 
&&\hspace{-.8cm}\langle O\rangle\Big{|}_{s.c.}^{\rm SUSY} 
=V_{N_c}\mu^{4N_c-\half n_F} 
\frac{\er^{-\frac{8\pi^2}{g^2}}}{(g^2)^{2N_c}} 
\times\nn\\ 
&&\hspace{-.8cm}\times\int\!\frac{\dr\rho}{\rho^5}\,\dr^4x_0 (\rho^2)^{2N_c} 
\sum_{\{j_k\}}(-1)^{P_{\{j_k\}}}O(\prod_{k}f_{j_k};A^I)\, ,\label{INTSUN} 
\eeqn 
with 
\beq 
V_{N_c}=\frac{4}{\pi^2}\frac{(4\pi^2)^{2N_c}}{(N_c-1)!(N_c-2)!}\, . 
\label{KNC} 
\eeq 
Supersymmetry has in store another surprise for us. Recalling the 
multiplicity of the fermionic zero modes as given by the Atiyah--Singer 
theorem (see Appendix~\ref{sec:APPA}), one finds that for a 
supersymmetric theory 
\beqn 
&&4N_c -\half n_F=b_1\, ,\label{B1}\\ 
&&\beta=-\frac{g^3}{16\pi^2}b_1+{\rm O}(g^5)\, ,\label{BETA} 
\eeqn 
where $b_1>0$ is the first coefficient of the Callan--Symanzik 
$\beta$-function. To prove~(\ref{B1}) we recall the general formula 
\beq 
b_1=\frac{11}{3}\ell[{\bf Adj}]-\frac{2}{3}\sum_{{\bf R_F}} 
n_{{\bf R_F}}\ell[{\bf R_F}] 
-\frac{1}{3}\sum_{{\bf R_B}}n_{{\bf R_B}}\ell[{\bf R_B}]\, , 
\label{GENF} 
\eeq 
where $n_{{\bf R_F}}$ and $n_{{\bf R_B}}$ are the numbers of 
fermions and bosons in the representations ${\bf R_F}$ and ${\bf 
R_B}$, respectively. Since in a supersymmetric theory each fermion 
is accompanied by a bosonic partner belonging to the same 
representation ${\bf R}$, (\ref{GENF}) simplifies to 
\beq 
b_1=3\ell[{\bf Adj}]-\sum_{{\bf R}}n_{{\bf R}}\ell[{\bf R}]= 
3N_c-\sum_{{\bf R}}n_{{\bf R}}\ell[{\bf R}]\, ,\label{GENFSU} 
\eeq 
with $n_{{\bf R}}$ the number of chiral superfields in the 
representation ${\bf R}$. To be able to compare $b_1$ in the above 
equation with the combination that appears in the l.h.s.\ 
of~(\ref{B1}) we make use of the Atiyah--Singer theorem 
(see~(\ref{IND})). Separating out the contribution due to gluinos 
(the fermions in the gauge supermultiplet) which accounts for a 
$2N_c/2$ contribution, we can write the l.h.s.\ of~(\ref{B1}) in the 
form 
\beq 
4N_c -\half n_F=4N_c -N_c-\half 2\sum_{{\bf R}}n_{{\bf R}} 
\ell[{\bf R}]=3N_c-\sum_{{\bf R}}n_{{\bf R}}\ell[{\bf R}]\, , 
\label{B2} 
\eeq 
in agreement with~(\ref{GENFSU}). 
 
The interesting consequence of this equality is that we can combine 
the exponential of the instanton action with the explicit $\mu$ 
dependence to form the renormalisation group invariant 
$\Lambda$-parameter of the theory. Introducing the running coupling 
$g(\mu)$, we can thus write 
\beq 
\mu^{4N_c-\half n_F} \er^{-\frac{8\pi^2}{g(\mu)^2}}=\Lambda^{b_1}\, . 
\label{LAM} 
\eeq 
We will exploit this key observation in the following, making it more 
precise (Sect.~\ref{sec:ICSYM}). 
 
\section{Chiral and supersymmetric Ward--Takahashi identities} 
\label{sec:SWTI} 
 
Before embarking in explicit instantonic calculations of correlators, 
we want to spell out the constraints imposed on correlators by chiral 
and supersymmetric Ward--Takahashi identities (WTI's). We will show 
that in some interesting cases, when these ``geometric'' constraints 
are coupled to the requirement of renormalisability, the expression of 
certain Green functions is (up to multiplicative numerical constants) 
completely fixed. 
 
The special Green functions which enjoy this amazing property are the 
$n$-point correlation functions of lowest (highest) components of 
chiral (anti-chiral) gauge invariant composite superfields. Although 
this is a very limited set of correlators, we will see that their 
knowledge, when used in conjunction with clustering, is sufficient to 
draw interesting non-perturbative information about the structure of 
the vacuum and of its symmetry properties. For this reason in this 
section we will limit our consideration to such correlators. We will 
in particular concentrate on the case of ${\cal N}=1$ Super QCD 
(SQCD) (see Appendix~\ref{sec:APPA} for notations) with the purpose of 
exploring the properties of a sufficiently general theory in which also 
mass terms can be present. 
 
WTI's provide relations among different Green functions. They will be
worked out under the assumption that supersymmetry is not
spontaneously (or explicitly) broken, i.e.\ under the assumption that
the vacuum of the theory is annihilated by all the generators of
supersymmetry. Our philosophy will be that, if we find that some
dynamical calculation turns out to be in contradiction with
constraints imposed by supersymmetry, then this should be interpreted
as evidence for spontaneous supersymmetry breaking. 
 
As we explained above, we are now going to consider the $n$-point 
Green functions 
\beq 
G(x_1,\ldots,x_n)=\langle0|T\big{(}\chi_1(x_1)\ldots 
\chi_n(x_n)\big{)})|0\rangle \, ,\label{GRE} 
\eeq 
where each $\chi_k(x_k)$ is a local gauge invariant operator made out 
of a products of lowest components of the fundamental chiral 
superfields of the theory. Thus the operators $\chi_k$ are themselves 
lowest components of some composite chiral super field, $X_k$, for 
which we formally have the expansion 
\beqn 
&&X_k(x)=\chi_k(y)+\sqrt{2}\theta^\alpha\psi_{\alpha k}(y)+ 
\theta^2F_k(y)\, ,\label{SUPERF}\\ 
&&y_\mu=x_\mu+i\theta^\alpha\sigma_\mu^{\alpha\dot\alpha} 
\bar\theta_{\dot\alpha}\, .\label{YTHETA} 
\eeqn 
On these fields the $Q$ and $\bar Q$ generators of supersymmetry act 
as ``raising'' and ``lowering'' operators according to the 
(anti-)commutation rules 
\beqn 
&[\bar Q^{\dot\alpha},\chi_k(x)]=0\, ,\qquad & \{\bar Q^{\dot\alpha}, 
\psi^{\alpha}_k(x)\}= 
\sqrt{2}\,\bar\sigma_\mu^{\dot\alpha\alpha}\partial_\mu\chi_k(x)\, , 
\label{SCOM1}\\ 
&[Q_{\alpha},F_k(x)]=0\, ,\qquad & 
\{Q_{\alpha},\psi_{k\beta}(x)\}=\sqrt{2}\epsilon_{\alpha\beta}F_k(x)\,. 
\label{SCOM2} 
\eeqn 
 
\subsection{Space-time dependence} 
\label{sec:STD} 
 
The independence of the correlators of the form~(\ref{GRE}) from 
space-time arguments immediately follows from the (anti-)commutation 
relations~(\ref{SCOM1}). Taking, in fact, the derivative of $G$ with 
respect to $x_\ell$ and contracting with 
$\sqrt{2}\,\bar\sigma_\mu^{\dot\alpha\alpha}$, one gets 
\beqn 
&&\sqrt{2}\,\bar\sigma_\mu^{\dot\alpha\alpha} 
\frac{\partial}{\partial x_{\ell\mu}} 
G(x_1,\ldots,x_n)=\nn\\ 
&&=\langle 0|T\big{(}\chi_1(x_1)\ldots\{\bar Q^{\dot\alpha}, 
\psi^{\alpha}_\ell(x_\ell)\}\ldots\chi_n(x_n)\big{)}|0\rangle=0\, . 
\label{DER} 
\eeqn 
The last equality is a consequence of the fact that $\bar Q$ can be 
freely (first commutation rule in~(\ref{SCOM1})) brought to act on 
the vacuum state at the beginning and at the end of the string of 
$\chi_k$ operators and that, under the assumption that supersymmetry 
is unbroken, $\bar Q |0\rangle=0$. Contributions coming from the 
derivative acting on the $\theta$-functions that prescribe the 
time-ordering of operators in $G$ are zero because they give rise to 
the vanishing equal-time commutators, 
$[\chi_\ell(\vec{x}_\ell,t_\ell),\chi_k(\vec{x}_k,t_k)]\delta(t_k-t_\ell)=0$. 
Equation~(\ref{DER}) proves the constancy of $G$. A similar result 
clearly holds for $n$-point correlators, $G^*$, where only lowest 
components of antichiral superfields are inserted. 
 
We end this section with the important observation that all these 
correlators vanish identically in perturbation theory. Only 
non-perturbative instanton-like contributions can make them non-zero. 
 
\subsection{Mass and $g$ dependence} 
\label{sec:MGD} 
 
The following further properties hold for correlators of lowest 
components of chiral, $G$, (or antichiral, $G^*$) 
superfields~\cite{SWTI1,ARV,AKMRV} 
 
a) $G$ is an analytic function of the complex mass parameters $m_f$, 
i.e.\ it does not depend on $m_f^*$ (the opposite being true for 
$G^*$). 
 
b) The mass dependence of $G$ (and $G^*$) is completely fixed. 
 
c) When renormalisation group invariant operators are inserted, the 
dependence upon the coupling constant is, in a mass independent 
renormalisation scheme, fully accounted for by the renormalisation 
group invariant (RGI) quantities $\Lambda$ and $[m_f]_{\rm inv}\equiv 
\hat m_f$ (see below~(\ref{RGIQ})). 
 
It is important to remark that properties of this kind can be readily 
exported to the generating functional of Green functions, as they 
only follow from symmetry principles. They provide strong constraints 
on the form of the associated effective action. A celebrated example 
of application of this observation, though in a different context, 
can be found in the construction of the low energy effective action 
that describes the interaction of pions in QCD~\cite{WEIN,GL}. In 
supersymmetric theories the invariances are so tight that often the 
full expression of the effective superpotential is completely 
determined~\cite{VY,TVY,ADSALL1,ADSALL2} (see Sect.~\ref{sec:ELA}). 
 
\subsubsection{a) Mass analyticity} 
\label{sec:AMA} 
 
The statement a) follows from the supersymmetric relation (no sum over $f$) 
\beqn 
&&m^*_f\frac{\partial}{\partial m^*_f}G(x_1,\ldots,x_n)= 
m^*_f\langle 0|T\big{(}\chi_1(x_1),\ldots,\chi_n(x_n) 
\int \dr^4x\,F^{*f}_f(x)\big{)}|0\rangle=\nn\\ 
&&=m^*_f\int \dr^4x\,\langle 0|T\big{(}\chi_1(x_1)\ldots\chi_n(x_n) 
\{\bar Q_{\dot\alpha}\psi^{f\dot\alpha}_f(x)\}\big{)}|0\rangle=0\, , 
\label{MGRE} 
\eeqn 
with $F^{*f}_f$ the auxiliary field of the antichiral superfield 
$T^{*f}_{f}=(\chi^{*f}_f,\bar\psi^{*f}_f, F^{*f}_f)$. The first 
equality follows from the fact that, before the auxiliary field is 
eliminated by the e.o.m., $F^{*f}_f$ is the coefficient of 
$m^*_f$. The second is a consequence of the complex conjugate of the 
anticommutation relation in~(\ref{SCOM2}). Finally, since $\bar Q$ 
commutes with the $\chi_k$'s, it can be brought in contact with the 
vacuum state which is thus annihilated. 
 
\subsubsection{b) Mass dependence} 
\label{sec:BMD} 
 
In order to simplify this analysis we restrict to Green functions 
where only the gauge invariant composite operators 
\beqn 
&&\frac{g^2}{32\pi^2}\lambda^{\alpha a}(x)\lambda_\alpha^a(x)\equiv 
\frac{g^2}{32\pi^2}\lambda\lambda(x)\, ,\label{LL}\\ 
&&\tilde\phi^{fr}(x)\phi_{hr}(x)\equiv\tilde\phi^{f}\phi_{h}(x)\label{PSIER} 
\eeqn 
are inserted. They are the lowest components of chiral superfields
which will be called $S$ and $T^f_{h}$, respectively. Besides their
obvious complex conjugate fields, we will sometimes also consider the
composite operators
$\tilde\psi^{fr}_\alpha(x)\psi^\alpha_{hr}(x)=\tilde\psi^{f}\psi_{h}(x)$. In
general terms we will consider  correlators of the kind 
\beqn 
\hspace{-.5cm}&&G^{(p,q)f_1,\ldots,f_p}_{h_1,\ldots,h_p}(x_1,\ldots, 
x_p;x_{p+1},\ldots,x_{p+q})=\label{PQ}\\ 
\hspace{-.5cm}&&=\langle 0|T\big{(}\tilde\phi^{f_1}\phi_{h_1}(x_1) 
\ldots\tilde\phi^{f_p}\phi_{h_p}(x_p) 
\frac{g^2}{32\pi^2}\lambda\lambda(x_{p+1})\ldots\frac{g^2}{32\pi^2} 
\lambda\lambda(x_{p+q})\big{)})|0\rangle\, .\nn 
\eeqn 
The dependence of~(\ref{PQ}) upon the mass parameters can be 
established in the following way. First of all we notice that 
from~(\ref{MGRE}) we have 
\beqn 
m_f\frac{\partial}{\partial m_f}\,G^{(p,q)}= 
\Big{(}m_f\frac{\partial}{\partial m_f}-m^*_f 
\frac{\partial}{\partial m^*_f}\Big{)\,}G^{(p,q)}= 
\frac{1}{i}\frac{\partial}{\partial \alpha_f}\,G^{(p,q)}\, , 
\label{MSGRE} 
\eeqn 
where we have set 
\beq 
m_f=|m_f|\er^{i\alpha_f}\, . 
\label{MAL}\eeq 
In order to compute the derivative in the r.h.s.\ of~(\ref{MSGRE}) we 
perform the non-anomalous $U_A^f(1)$ transformation (see 
Appendix~\ref{sec:APPA}) 
\beqn 
&&(\tilde\psi^h,\psi_h)\to \er^{i\delta_{fh}\alpha_f/2}(\tilde\psi^h,\psi_h) 
\, ,\quad (\tilde\phi^h,\psi_h)\to 
\er^{i(\delta_{fh}-1/N_c)\alpha_f/2}(\tilde\phi^h,\psi_h)\, ,\nn\\ 
&&\lambda\to \er^{-i\alpha_f/2N_c}\lambda\, , 
\label{UUAF} 
\eeqn 
by means of which the $\alpha_f$ dependence of the action is 
eliminated, but it is brought in the fields appearing in 
$G^{(p,q)}$. This allows to carry out in an explicit way the 
$\alpha_f$ derivative, leading to the result 
\beqn 
&&m_f\frac{\partial}{\partial m_f}G^{(p,q)f_1,\ldots,f_p}_{h_1,\ldots,h_p}= 
q^{(f),f_1,\ldots,f_p}_{h_1,\ldots,h_p}G^{(p,q)f_1,\ldots,f_p}_{h_1,\ldots,h_p} 
\, ,\label{WTQ}\\ 
&&q^{(f),f_1,\ldots,f_p}_{h_1,\ldots,h_p}= 
\frac{p+q}{N_c}-\half\sum_{\ell=1}^{p}(\delta_{f_\ell,f}+ 
\delta_{h_\ell,f})\, ,\label{QCHAR} 
\eeqn 
where $q^{(f)}$ is the sum of all the $U_A^f(1)$ charges of the 
operators contained in $G^{(p,q)}$. The above differential equation is 
easily integrated and yields 
\beqn 
\prod_{\ell=1}^{p}(m_{f_\ell}m_{h_\ell})^\half 
G^{(p,q)f_1,\ldots,f_p}_{h_1,\ldots,h_p}= 
C^{(p,q)f_1,\ldots,f_p}_{h_1,\ldots,h_p}(\mu,g) 
\Big{(}\prod_{\ell=1}^{N_f}m_\ell\Big{)}^{\frac{(p+q)}{N_c}}\, . 
\label{GEXP} 
\eeqn 
The $\mu$ dependence of $C^{(p,q)f_1,\ldots,f_p}_{h_1,\ldots,h_p}(\mu,g)$ 
is trivially fixed by dimensional analysis and one finds 
\beq 
C^{(p,q)f_1,\ldots,f_p}_{h_1,\ldots,h_p}(\mu,g) \propto 
\mu^{(p+q)(3-N_f/N_c)}\, .\label{MUDEP} 
\eeq 
 
\subsubsection{c) $g$ dependence} 
\label{sec:CGD} 
 
The $g$ dependence of
$C^{(p,q)f_1,\ldots,f_p}_{h_1,\ldots,h_p}(\mu,g)$ is completely
determined by renormalisability. In fact, having factorised in the
l.h.s.\ of~(\ref{GEXP}) the mass factor
$\prod_{\ell=1}^{p}(m_{f_\ell}m_{h_\ell})^\half$, which precisely
serves the purpose of making the $\tilde\phi^{f}\phi_{h}$ operators in
$G^{(p,q)}$ behave like RGI insertions, the rest of the $g$ dependence
must all be expressed through the RGI quantities 
\beq 
\Lambda=\mu\exp\Big{(}-\int^g\frac{1}{\beta(g')}\dr g'\Big{)}\, ,\quad 
\hat m =m\exp\Big{(}-\int^g\frac{\gamma_m(g')}{\beta(g')}\dr g'\Big{)}\, , 
\label{RGIQ} 
\eeq 
where $\beta\neq 0$ and $\gamma_m(g)$ are the Callan--Symanzik 
function of the theory and the mass anomalous dimension of the matter 
superfield, respectively. This implies that the $g$ dependence must 
be of the form 
\beq 
C^{(p,q)f_1,\ldots,f_p}_{h_1,\ldots,h_p}(\mu,g) 
\propto \exp\Big{\{}-\int^g\frac{\dr g'}{\beta(g')}(p+q)\big{[}(3- 
\frac{N_f}{N_c})+\gamma_m(g')\frac{N_f}{N_c}\big{]}\Big{\}} 
\, ,\label{GDEP}\eeq 
in order to have 
\beqn 
&&\prod_{\ell=1}^{p}(m_{f_\ell}m_{h_\ell})^\half G^{(p,q)f_1, 
\ldots,f_p}_{h_1,\ldots,h_p}= \nn\\ 
&&=\Lambda^{(p+q)(3-N_f/N_c)}\Big{(}\prod_{\ell=1}^{N_f} 
\hat m_\ell\Big{)}^{\frac{(p+q)}{N_c}} 
t^{(p,q)f_1,\ldots,f_p}_{h_1,\ldots,h_p}\, ,\label{GFUL} 
\eeqn 
with $t^{(p,q)f_1,\ldots,f_p}_{h_1,\ldots,h_p}$ a dimensionless 
constant tensor in flavour space. 
 
The form of $t^{(p,q)f_1,\ldots,f_p}_{h_1,\ldots,h_p}$ is strongly 
constrained (and sometimes completely determined) by the pattern of 
unbroken flavour symmetries of the theory. Its explicit computation 
will be one of the main subjects of the next sections.

\subsection{The anomalous $U_\lambda(1)$ R-symmetry} 
\label{sec:ULUA} 
 
The integrated WTI associated with the anomalous $U_\lambda(1)$ 
R-symmetry (see Appendix~\ref{sec:APPA}, (\ref{ULAMM}), (\ref{CURC}) 
and~(\ref{ANOMALYC})) reads 
\beqn 
2iK N_c \langle O(x_1,\ldots,x_n)\rangle=\sum_{i=1}^{n}\langle 
\frac{\partial O^{(\alpha)}}{\partial \alpha(x_i)}(x_1,\ldots,x_n) 
\Big{|}_{\alpha=0}\rangle\, , 
\label{WTUL} 
\eeqn 
where $O^{(\alpha)}$ is the operator which is obtained by performing 
on $O$ a $U_\lambda(1)$ rotation of an angle $\alpha$. For the 
special Green function $G^{(p,q)f_1,\ldots,f_p}_{h_1,\ldots,h_p}$ 
(see~(\ref{PQ})) and~(\ref{WTUL}) simply becomes 
\beqn 
&&2 K N_c G^{(p,q)f_1,\ldots,f_p}_{h_1,\ldots,h_p}(x_1,\ldots,x_p;x_{p+1}, 
\ldots,x_{p+q})=\nn\\ 
&&=2(p+q)G^{(p,q)f_1,\ldots,f_p}_{h_1,\ldots,h_p}(x_1,\ldots,x_p;x_{p+1}, 
\ldots,x_{p+q})\, , 
\label{WTULG} 
\eeqn 
because the $U_\lambda(1)$ rotation of 
$G^{(p,q)f_1,\ldots,f_p}_{h_1,\ldots,h_p}$ is proportional to 
$G^{(p,q)f_1,\ldots,f_p}_{h_1,\ldots,h_p}$ itself through the factor 
$2(p+q)$. As a result only if 
\beq 
p+q=K N_c\, , 
\label{ISR} 
\eeq 
we can get a non-vanishing result. Notice that~(\ref{ISR}) implies 
$K > 0$ consistently with the fact that we are dealing with lowest 
components of chiral superfields. Negative values of $K$ will come 
into play in correlators with insertions of highest components of 
antichiral superfields. 
 
A particularly interesting situation arises if we insist that each 
flavour should appear exactly $K$ times. Then~(\ref{ISR}) requires 
$N_f\leq N_c$. At this point to simplify our treatment we restrict 
ourselves to the case $K=1$. Since in this situation $p=N_f$, the 
whole dependence on the bare mass parameters drops out from 
the Green function we are considering and we get 
\beq 
G^{(N_f,N_c-N_f)f_1,\ldots,f_{N_f}}_{h_1,\ldots,h_{N_f}} 
(x_1,\ldots,x_{N_f};x_{N_f+1},\ldots,x_{N_c})\propto \Lambda^{3N_c-N_f}\, . 
\label{GDIM} 
\eeq 
We expressly note that the exponent to which $\Lambda$ is raised 
in~(\ref{GDIM}) is not only the physical dimension of 
$G^{(N_f,N_c-N_f)}$, but it also coincides with the first coefficient of 
the $\beta$-function of SQCD (see the discussion and the formulae in 
Sect.~\ref{sec:2.4.2}). 
 
Among the Green functions of the type~(\ref{PQ}) which fulfill the 
further requirements spelled out in this subsection, we wish to 
specially mention here the one relevant in pure SYM where one gets the 
famous correlator~\cite{RUS1,RV} 
\beqn 
\hspace{-.5cm}G^{(0,N_c)}(x_1,\ldots,x_{N_c})= 
\langle \frac{g^2}{32\pi^2}\lambda\lambda(x_{1}) 
\ldots\frac{g^2}{32\pi^2}\lambda\lambda(x_{N_c})\rangle\, .\label{PQSYM} 
\eeqn 
 
\subsection{The Konishi anomaly} 
\label{sec:KA} 
 
The general need to regularise products of operator fields at the same 
point is at the origin of the axial anomaly~\cite{SABJ} (see 
Appendix~\ref{sec:APPA}) and of the anomalous contribution that 
appears in certain supersymmetric anticommutators. Starting from the 
supersymmetry graded algebra summarised in~(\ref{SCOM1}) 
and~(\ref{SCOM2}), it has been shown in~\cite{KON} that, after 
regularisation, in massive SQCD the following (anomalous) 
anticommutation relation holds 
\beq 
\frac{1}{2\sqrt{2}}\{\bar Q^{\dot\alpha},\bar\psi_{\dot\alpha}^f\phi_h(x)\}= 
-m_f\tilde\phi^f\phi_h(x)+\frac{g^2}{32\pi^2}\lambda\lambda(x)\delta^f_h\, , 
\label{KONANA} 
\eeq 
where besides the naive $m_f\tilde\phi^f\phi_h(x)$ term an extra 
contribution appears. This relation is what usually goes under the 
name of ``Konishi anomaly''. Clearly, if the vacuum of the theory is 
supersymmetric, by taking the v.e.v.\ of~(\ref{KONANA}) a 
proportionality relation between gluino and scalar condensates 
emerges, namely 
\beq 
m_f\langle \tilde\phi^f\phi_f\rangle = 
\frac{g^2}{32\pi^2}\langle\lambda\lambda\rangle\, ,\quad 
{\rm no\,\,sum\,\,over\,\,}f\, , 
\label{KONANA1} 
\eeq 
besides 
\beq 
\langle \tilde\phi^f\phi_h\rangle = 0\, ,\quad f\neq h \, .\label{KONANA2} 
\eeq

\section{Instanton calculus} 
\label{sec:IC} 
 
We want to show in this section that the Green functions considered 
in~(\ref{GDIM}) receive a non-vanishing computable one-instanton 
contribution. In other words, although zero in perturbation theory, 
they can be exactly evaluated in the semi-classical approximation by 
dominating the functional integral with the one-instanton 
saddle-point. A non-trivial result is obtained because the number of 
fermionic fields that are inserted in $G^{(N_f,N_c-N_f)}$ (either at 
face value or at the appropriate order in $g$) is precisely equal to 
the number of fermionic zero modes present in the $K=1$ instanton 
background. 
 
\subsection{Instanton calculus in SYM} 
\label{sec:ICSYM} 
 
The computation of the correlator~(\ref{PQSYM}) in the semi-classical 
one-instanton approximation is not too difficult by using the results 
we have recollected in Appendix~\ref{sec:APPB} (about bosonic zero 
modes and collective coordinate integration) and the explicit 
expression of the $2N_c$ gluino zero modes that can be found in 
Appendix~\ref{sec:APPA}~\cite{TEM,AKMRV}. 
 
We will consider the case of a pure SYM theory with gauge group 
$SU(N_c)$~\footnote{For SYM theories with other compact Lie group 
see~\cite{KP}}. The striking outcome of the calculation (which is 
based on equations from~(\ref{INTSU}) to~(\ref{KNC})) is that the 
apparently extremely complicated dependence of the correlator upon the 
space-time location of the inserted operators is completely washed out 
by the bosonic collective coordinate integration and, as expected, a 
space-time independent (constant) result is obtained in agreement with 
the supersymmetric WTI~(\ref{DER}). Explicitly one 
finds~\cite{RUS1,RV} 
\beqn 
\hspace{-.5cm}G^{(0,N_c)}(x_1,\ldots,x_{N_c})=C_{N_c} 
\big{(}\Lambda_{\rm SYM}^{\rm 2-loops}\big{)}^{3N_c} 
\, ,\label{NSYM} 
\eeqn 
where~\footnote{The constant $C_{N_c}$ differs from the similar 
constant appearing in~(4.9) of~\cite{AKMRV} by a factor 
$2^{N_c}$. This mistake was pointed out by various 
authors~\cite{FIPU,KHO1,COR} and was the consequence of an erroneous 
normalisation of the gluino zero modes.} 
\beqn 
&&C_{N_c}=\frac{2^{N_c}}{(3N_c-1)(N_c-1)!}\, ,\label{CNC}\\ 
&&\Lambda_{\rm SYM}^{\rm 2-loops}=\mu\, \er^{-8\pi^2/3N_c g^2} 
(g^2)^{-1/3}\, .\label{LAMB} 
\eeqn 
Equation~(\ref{LAMB}) follow from the known value of the 2-loop 
coefficient of the $\beta$-function of the theory and shows that 
dominating the functional integral by the semi-classical one-instanton 
saddle-point gives a (2-loop) RGI answer. 
 
{}From this result two important consequences can be derived, one 
concerning the form of the $\beta$-function of the theory and the 
second the structure of the vacuum. 
 
\subsubsection{The SYM $\beta$-function} 
\label{sec:SYMBF} 
 
One can argue that the result~(\ref{LAMB}) is valid to all loops in 
the sense that higher order power corrections in $g$ are indeed all 
vanishing. The argument goes as follows. As we remarked in 
Appendix~\ref{sec:APPB} just at the end of the first subsection, one 
can go on with perturbation theory around the instanton background by 
expanding in powers of $g$ terms cubic and quartic in the 
fluctuations, as well as terms coming from the Faddeev--Popov 
procedure. One should be finding in this way logarithmically divergent 
contributions which would be interpreted as higher order terms in the 
Callan--Symanzik $\beta$-function. In the present case, however, no 
such term can arise because there is no dimensionful quantity with 
which we might scale the (would-be) logarithmically divergent $\mu$ 
dependence. In fact, the only other dimensionful quantities are the 
relative distances $x_i-x_j$ of the operator insertion points. But the 
supersymmetric WTI~(\ref{DER}) prohibits any such dependence. 
 
We must conclude that in the regularisation and renormalisation scheme 
we work and in the background gauge, the $\Lambda$ parameter is 
``2-loop exact''. This observation is equivalent to the result of 
$\beta$-exactness first put forward in~\cite{RUS3}, which amounts to 
say that one has the exact formula 
\beqn 
\beta_{\rm SYM}(g)=-\frac{g^3}{16\pi^2} 
\frac{3N_c}{1-2{g^2N_c}/{16\pi^2}} \, .\label{EXB} 
\eeqn 
Introducing~(\ref{EXB}) in 
\beqn 
-\int^{g(\mu)}_{g(\mu_0)}\frac{\dr g'}{\beta_{\rm SYM}(g')}= 
\int^\mu_{\mu_0}\frac{\dr\mu'}{\mu'}\, .\label{GINT} 
\eeqn 
one gets, in fact, by a straightforward integration 
precisely~(\ref{LAMB}). We stress that no approximation (no expansion in 
powers of $g$) has been performed in the step from~(\ref{EXB}) to 
the formula~(\ref{LAMB}). 
 
Equation~(\ref{EXB}) can be generalised~\cite{RUS3} to encompass the case 
of extended supersymmetry with ${\cal N}=1,2,4$ supercharge multiplets 
through the simple formula 
\beqn 
\beta_{\cal N}(g)=-\frac{g^3}{16\pi^2}\frac{(4-{\cal N})N_c}{1-2(2-{\cal N}) 
{g^2N_c}/{16\pi^2}}\, ,\label{EXBN} 
\eeqn 
which incorporates the known facts that the ${\cal N}=2$ 
$\beta$-function is 1-loop exact and the ${\cal N}=4$ $\beta$-function 
just vanishes. 
 
\subsubsection{The structure of the SYM vacuum} 
\label{sec:SYMV} 
 
The space-time constancy of the result~(\ref{LAMB}) allows us to 
compute the expectation value of the composite operator 
${g^2}\lambda\lambda/{32\pi^2}$ by simply imagining that the 
separations $|x_i-x_j|$ are very large. Using clustering, it will be 
possible to write $G^{(0,N_c)}$ as the product of the v.e.v.'s of such 
operators (gluino condensate, in the following). 
 
The computation is straightforward if the vacuum of the theory is 
unique. Here the situation is more complicated because of the very 
fact that the gluino condensate is not vanishing. This means, in fact, 
that the residual $\Z_{2N_c}$ symmetry of the theory (see 
Appendix~\ref{sec:APPA}) is actually spontaneously broken down to 
$\Z_{2}$ with the consequence that there are $N_c$ degenerate 
vacua in which the theory can live, related by $\Z_{N_c}$ 
transformations. Incidentally we note that this result is perfectly 
consistent with the prediction based on the Witten index 
calculation~\cite{WITIND}. 
 
In the presence of many equivalent vacua the functional integral 
yields non-perturbative results where contributions coming from 
different vacua are averaged out. Thus in order to extract useful 
information from the clustering properties of the theory one has to 
take into account this phenomenon and go through a procedure called 
``vacuum disentangling''~\cite{RV,AKMRV}. All this simply means that 
we should write for $G^{(0,N_c)}$ the formula 
\beqn 
\hspace{-.5cm}G^{(0,N_c)}=\frac{1}{N_c} 
\sum_{k=1}^{N_c}\big{[}\langle\Omega_k| 
\frac{g^2}{32\pi^2}\lambda\lambda|\Omega_k\rangle\big{]}^{N_c}\, , 
\label{DIS} 
\eeqn 
with the gluino condensates transforming under $\Z_{2N_c}$ as 
\beqn 
\hspace{-.5cm}\langle\Omega_k|\frac{g^2}{32\pi^2}\lambda\lambda 
|\Omega_k\rangle=\er^{\frac{2\pi i k}{N_c}}\langle\Omega_0| 
\frac{g^2}{32\pi^2}\lambda\lambda|\Omega_0\rangle\, , 
\quad k=1,2,\ldots,N_c\, .\label{COND1} 
\eeqn 
This equation is telling us that the average in~(\ref{DIS}) is trivial 
and we get in the $k$-th vacuum 
\beqn 
\hspace{-.5cm}\langle\Omega_k|\frac{g^2}{32\pi^2}\lambda\lambda 
|\Omega_k\rangle=\er^{\frac{2\pi i k}{N_c}}\big{(}C_{N_c}\big{)}^{1/N_c} 
\big{(}\Lambda_{\rm SYM}^{\rm 2-loops}\big{)}^3\, , \quad 
k=1,\ldots,N_c \, .\label{COND2} 
\eeqn 
 
\subsubsection{Discussion of the results} 
\label{sec:DRSYM} 
 
The picture we got from the calculation presented in the previous
section looks rather convincing and physically sound. It perfectly
matches all our expectations and it has been carried out in a clean
and rigorous mathematical way. It is uniquely based on the assumption
that Green functions which can receive  contributions only from the
$K=1$ sector of the theory can be  reliably computed by dominating the
functional integral by the  one-instanton saddle-point. We have also
argued that the 2-loop RGI  result obtained in the semi-classical
approximation is exact in the  sense that it does not get further
perturbative corrections. 
 
Despite all these nice features, it has been argued in the literature 
that the method employed to get the result~(\ref{NSYM}) cannot be 
right because it seems to encounter a number of problems with other 
considerations. 
 
(1) The $N_c$ dependence of $C_{N_c}$ leads in the 't~Hooft limit 
($N_c\to\infty$ with $g^2N_c$ fixed) to an $N_c$ dependence of the 
gluino condensate~(\ref{COND2}) that it is not what one would 
expect from the fact that the gluinos (together with the gauge field, 
$A_\mu$) belong to the adjoint representation of the gauge 
group. Taking into account the $g^2$ factor that was introduced in 
front of the gluino bilinear, one would naively expect $\langle 
g^2\lambda\lambda\rangle\sim{\rm O}(g^2N_c^2)\sim{\rm O}(N_c)$ in the 
't~Hooft limit. From~(\ref{NSYM}) and~(\ref{CNC}), one finds 
instead $\langle g^2\lambda\lambda\rangle\sim{\rm O}(g^2N_c)\sim{\rm 
O}(1)$. 
 
(2) In~\cite{KHO1} the calculation of $G^{(0,N_c)}$ has been 
repeated in a fully supersymmetric formalism and the result summarised 
in~(\ref{NSYM}), (\ref{CNC}) and~(\ref{LAMB}) was confirmed (up 
to the correction for the factor $2^{N_c}$ that we already 
mentioned). Interestingly these authors have also been able to extend, 
in the large $N_c$ limit, the semi-classical instanton calculation to 
Green functions which receive contributions from topological sectors 
with winding numbers $K>1$, i.e.\ to Green functions with $K N_c$ 
insertions of the gluino bilinear. The result of this calculation, 
when clustering is used, is inconsistent with it, because it leads to 
a value of the condensate which is not independent of $K$. 
 
(3) The computation of $G^{(0,N_c)}$ can be indirectly done starting 
from the more complicated case where extra massive matter 
supermultiplets are added~\cite{ADSALL1,ADSALL2,SEI} and then exploit 
the notion of decoupling~\cite{APP,AMRV}. We recall that in a 
nut-shell decoupling is the property of a local field theory according 
to which when some mass becomes large, the corresponding matter field 
disappears from the low energy physics (see Appendix~\ref{sec:APPD}), 
modulo possible consistency conditions resulting from the requirement 
of anomaly cancellation~\cite{THOA}. 
 
{}From symmetry arguments it is often possible to determine, up to a 
multiplicative constant, the form of the effective superpotential of 
the enlarged theory in terms of the relevant composite operators (see 
Sect.~\ref{sec:ELA}). Then consistency arguments, following from 
sending to infinity each mass successively, supplemented by 
``constrained instanton'' calculations (see next paragraph), can be 
used to determine this constant (see Sect.~\ref{sec:NELSQCD}). Clearly 
checking its value is important for the self-consistency and the 
reliability of the various approaches (see point (1)) above). One finds 
that, if computed by looking at the effective superpotential 
calculations, the value of this constant does not agree with what one 
can deduce from the formulae~(\ref{NSYM}) to~(\ref{LAMB}). Of course 
the comparison was done after having properly matched the RGI 
parameters associated with the different regularisations employed in the 
various calculations~\cite{COR}. For instance, in the $SU(2)$ case one 
finds from the equations in Sect.~\ref{sec:ICSYM} $C_2=4/5$ instead 
of the result $C_2=1$ one would obtain from decoupling arguments. 
 
Despite a lot of work in the years that followed these findings, there 
is no clear understanding of why there are such discrepancies and 
where they come from. One line of 
arguments~\cite{RUS2,ADSALL1,ADSALL2,FS}, first prompted by the 
results described in (3), relies on the observation that when scalar 
fields are present other quasi-saddle-points exist, in which scalar 
fields get a non-vanishing v.e.v., which should be taken as background 
configurations in the semi-classical calculation of Green functions. In 
fact, the (partial or full) breaking of the gauge symmetry leads in 
the limit of very large v.e.v.'s to a weakly coupled theory, where 
semi-classical instanton calculations are expected to be reliable. In 
this context the non-trivial problem arising from the fact that in 
the presence of non-vanishing scalar v.e.v.'s the SQCD e.o.m.\ have 
no solution (owing precisely to the nature of the scalar boundary 
conditions) is circumvented by making recourse to the so-called 
``constrained instanton'' method~\cite{AFF}~\footnote{The theoretical 
foundation of the method is somewhat delicate (it relies on 
introducing in the functional integral a suitable ``constraint'' which 
breaks the integration measure into sectors of well defined instanton 
scale size) and its technical implementation requires a number of 
non-trivial mathematical steps. Its presentation is beyond the scope 
of this review, but can be found in the original literature. We 
recommend to the reader the nice work of~\cite{COR}.}. 
 
It is not completely clear to us whether the constrained instanton 
method (sometimes also called the ``weak coupling instanton'' (WCI) 
method, from which the nick-name ``strong coupling instanton'' (SCI) 
method was in opposition attributed to the approach described in 
Sect.~\ref{sec:ICSYM} and further employed in Sect.~\ref{sec:ICSQCD} 
below) can be considered as a completely satisfactory solution to the 
problems listed above. We now want to briefly discuss this question by 
illustrating some pro's and con's of the two approaches. 
 
$\bullet$ Certainly, if one accepts the WCI computational strategy, the 
problem mentioned in (1) disappears. As for the question of consistency 
with clustering (point (2) above), to date no check of the kind done 
in~\cite{KHO1} was carried out. Finally we do not see a really 
rigorous way to decide on the basis of the present knowledge whether 
$\sqrt{C_2}=2/\sqrt{5}$ or $\sqrt{C_2}=1$ is the correct answer for 
the constant in front of the gluino condensate. One possibility 
to settle this question could be to make recourse to a lattice 
formulation of SYM~\cite{CV} and directly measure in Monte Carlo 
simulations the gluino condensate. Up to now unfortunately severe 
technical difficulties have prevented such a measurement. For a recent 
review on the subject of supersymmetry on the lattice see~\cite{MONT}. 
 
$\bullet$ The whole idea of working in the Higgs phase of SQCD comes 
from the key observation that, in the massless limit, the 
superpotential possesses a complicated vacuum manifold (see 
Appendix~\ref{sec:APPE}). It is customary in the literature to speak 
about ``flat directions''~\cite{BD,ADSALL1,ADSALL2}, i.e.\ constant 
values of the scalar fields along which the $D$-term vanishes. It is 
in this situation that all the explicit WCI calculations have been 
carried out~\footnote{Besides the original papers in~\cite{RUS2}, the 
basic work from which all the old WCI calculations make reference to 
is the paper quoted in~\cite{COR}.}. Despite the fact that explicit 
instanton calculations have been carried out in the massless limit, 
their results and implications have been employed in the massive case. 
In this context it should be noted that the massless limit 
of the massive SQCD theory is a very 
delicate one. For instance, as we shall see, the SCI approach gives 
results in the massive case that, when extrapolated to vanishing 
mass, are not consistent with results directly obtained in the 
massless theory. This feature finds a natural explanation in the 
infrared structure of the theory which is such that the massless 
limit of the massive theory does not coincide with the strictly 
massless situation~\cite{AMRV}. 
 
$\bullet$ The WCI approach has found its most successful application 
in predicting the non-perturbative expansion coefficients of the 
SW~\cite{SW} expression for the effective prepotential of the ${\cal 
N}=2$ SYM theory (see Sect.~\ref{sec:SWcheck}). 
 
$\bullet$ On the other hand in ${\cal N}=4$ SYM, despite the fact that 
there are flat directions for the scalar potential, no scalar v.e.v.'s 
are assumed to be generated (as a non-vanishing v.e.v.\ would break 
the (super)conformal invariance of the theory) and all instanton 
calculations are performed in the SCI way we described in 
Sect.~\ref{sec:ICSYM}. Actually in ${\cal N}=4$ there is no running of 
the gauge coupling~(\ref{EXBN}) and one can always think that 
calculations are done at infinitesimally small values of $g$. Thus 
non-perturbative instanton calculations in ${\cal N}=4$ SYM do not 
seem to fall under the criticisms raised for the ${\cal N}=1$ and 
${\cal N}=2$ cases (see Sects.~\ref{sec:N4insteff}-\ref{sec:N4-1inst}). 
 
\subsection{Instanton calculus in SQCD} 
\label{sec:ICSQCD} 
 
In this section we move to SQCD. The action of SQCD is obtained by 
coupling (in a gauge invariant and supersymmetric way) to the SYM 
supermultiplet $N_f$ pairs of matter chiral superfields, $\Phi_{f}^r$ 
and $\tilde\Phi_{r}^f$ ($f=1,2,\ldots,N_f$, $r=1,2,\ldots,N_c$) 
belonging, respectively, to the ${\bf N_c}$ and ${\bf \bar N_c}$ 
representation of the gauge group (see Appendix~\ref{sec:APPA} for 
some detail and~\cite{KP} for an extension of these considerations to 
theories with different gauge groups and matter content). 
 
We want to identify and compute, according to the strategy developed 
in Sect.~\ref{sec:ICSYM} to deal with SYM, the Green functions that, 
besides being space-time constant, can be reliably evaluated by 
dominating the functional integral with the one-instanton 
saddle-point. We shall start by analysing the massive case, where the 
further information provided to us by the Konishi anomaly 
relation~\cite{KON} can be exploited and will allow to determine both 
the gluino and the scalar matter condensates and check the internal 
consistency of our calculations. In Sect.~\ref{sec:MLSQCD} we will 
discuss the puzzling features that arise when the limit $m\to 0$ is 
taken. 
 
\subsubsection{Massive SQCD} 
\label{sec:MVSQCD} 
 
Already looking at the general results derived in Sect.~\ref{sec:BMD} 
about the mass dependence of Green functions with only lowest 
components of chiral superfields, we see that their small mass limit 
is rather delicate, as infrared divergences seem to arise. To avoid 
hitting this difficulty we start by limiting the use of instanton 
calculus to the computation of the correlators that according 
to~(\ref{GFUL}) are mass independent. Among those we will 
concentrate here on the following three (see~(\ref{GDIM})) 
{\beqn 
\hspace{-1.cm}&& {\rm A)}\quad F^{(0,N_c)}(x_1,\ldots,x_{N_c})= 
\prod_{f=1}^{N_f}\frac{\partial}{\partial m_f}\,G^{(0,N_c)} 
(x_1,\ldots,x_{N_c})=\nn\\ 
\hspace{-1.cm}&&=\prod_{f=1}^{N_f}\frac{\partial}{\partial m_f}\, 
\langle \frac{g^2}{32\pi^2}\lambda\lambda(x_1)\ldots 
\frac{g^2}{32\pi^2}\lambda\lambda(x_{N_c})\rangle\, ,\label{FDER}\\ 
\hspace{-1.cm}&& {\rm B)}\quad G^{(N_f,N_c-N_f)}(x_1,\ldots,x_{N_c})=\nn\\ 
\hspace{-1.cm}&&=\langle \tilde\phi^{1}\phi_{1}(x_1)\ldots 
\tilde\phi^{{N_f}}\phi_{{N_f}}(x_{N_f}) 
\frac{g^2}{32\pi^2}\lambda\lambda(x_{N_f+1})\ldots 
\frac{g^2}{32\pi^2}\lambda\lambda(x_{N_c})\rangle\, ,\label{GNM} 
\eeqn} 
and, in the particular case $N_f=N_c$,\\ 
\beq 
\hspace{-6.cm}{\rm C)}\quad D(x,x')= 
\langle {\rm det}\phi(x){\rm det}\tilde\phi(x')\rangle\, ,\label{DET} 
\eeq 
\beqn 
{\rm det}\,\phi=\frac{1}{N_c!}\epsilon^{f_1,\ldots,f_{N_f}} 
\epsilon_{r_1,\ldots,r_{N_c}} 
\phi_{f_1}^{r_1},\ldots,\phi_{f_{N_f}}^{r_{N_c}}\, ,\label{DETP}\\ 
{\rm det}\,\tilde\phi=\frac{1}{N_c!}\epsilon_{f_1,\ldots,f_{N_f}} 
\epsilon^{r_1,\ldots,r_{N_c}} 
\tilde\phi^{f_1}_{r_1},\ldots,\tilde\phi^{f_{N_f}}_{r_{N_c}}\, , 
\label{DETPT} 
\eeqn 
 
(A) Let us start the discussion with $F^{(0,N_c)}$. We notice that it 
contains exactly the number of gluino fields necessary to match the 
number of zero modes that the theory possesses in the $K=1$ sector. 
We recall that, since at the moment we are considering the case in 
which the matter is massive, no zero modes associated with matter Weyl 
operators exist. In this situation we can safely compute the 
functional integral which defines the above correlator by dominating 
it with the one-instanton saddle-point. The calculation goes through 
the following steps. 
 
1) Every factor ${\partial}/{\partial m_f}$ can be replaced by the 
insertion of the action mass term 
\beq 
\frac{\partial}{\partial m_f}\to\int \dr^4x\,\big{[}\tilde\psi^f\psi_f(x)+ 
m^*_f(\phi^{*f}\phi_f(x)+\tilde\phi^f\tilde\phi^*_f(x))\big{]}\, , 
\label{INSM1} 
\eeq 
 
2) which, after integration over the matter supermultiplets, becomes 
\beq 
\frac{|m_f|^2}{\mu}{\rm Tr}\Big{[}\frac{2}{D^2-|m_f|^2}- 
{\rm tr}\big{(}\frac{1}{\Dslash\Dslash-|m_f|^2}\big{)}\Big{]}\, . 
\label{BECO} 
\eeq 
It is understood that the covariant operators $\Dslash$ and $D^2$ 
in~(\ref{BECO}) are computed in the one-instanton background field. 
The multiplicative mass factors in front of the trace have the 
following origin: i) the term $m^*_f$ comes from the expression of the 
matter propagators; ii) the ratio $m_f/\mu$ comes from what is left 
out from the ratio between the determinant of the matter Weyl 
operator and its regulator, after the supersymmetric cancellation of 
the non-zero mode contribution has been taken care of. 
 
3) A cancellation of modes also takes place between the two terms 
in~(\ref{BECO}). Only the fermionic mode with eigenvalue $m_f$ (i.e.\ 
the zero mode in the massless limit) contributes and one simply gets 
\beq 
\frac{\partial}{\partial m_f}\to\frac{|m_f|^2}{\mu}\frac{1}{|m_f|^2} 
=\frac{1}{\mu}\, . 
\label{INSM2} 
\eeq 
 
4) At this point the functional integration with respect to the gauge 
supermultiplet fields remains to be done. Since the sole effect of the 
matter integration is to yield the factor $\mu^{-N_f}$, we are left 
with exactly the same calculation we did in Sect.~\ref{sec:ICSYM}. We 
thus get 
\beq 
F^{(0,N_c)}(x_1,\ldots,x_{N_c})=C_{N_c} 
\frac{(\Lambda^{\rm 1-loop}_{\rm SQCD})^{3N_c-N_f}}{g^{2N_c}}\, , 
\label{FIST} 
\eeq 
from which, by integrating with respect to $m_f$, $f=1,\ldots,N_f$, we obtain 
\beqn 
G^{(0,N_c)}(x_1,\ldots,x_{N_c})=C_{N_c} 
\frac{(\Lambda^{\rm 1-loop}_{\rm SQCD})^{3N_c-N_f}}{g^{2N_c}} 
\prod_{f=1}^{N_f}m_f\, . 
\label{GINTM} 
\eeqn 
Two observations are in order here. First of all, by taking the 
$N_c$-th root of the above expression, one can determine the value of 
the gluino condensate in massive SQCD. One finds 
\beqn 
\hspace{-.5cm}\langle\Omega_k|\frac{g^2}{32\pi^2}\lambda\lambda| 
\Omega_k\rangle=\er^{\frac{2\pi i k}{N_c}}\Big{(}C_{N_c}\big{(} 
\Lambda_{\rm SQCD}^{\rm 1-loop}\big{)}^{3N_c-N_f} 
\frac{1}{g^{2N_c}}\prod_{f=1}^{N_f}m_f\Big{)}^{1/N_c} 
\, ,\label{LLSQCD} 
\eeqn 
which shows that, as in SYM, the discrete $\Z_{2N_c}$ symmetry 
is spontaneously broken down to $\Z_{2}$, living behind an 
$N_c$-fold vacuum degeneracy. This is the expected result, since the 
presence of massive fields cannot modify the value of the Witten 
index~\cite{WITIND}. 
 
Secondly, it can be checked that the formulae~(\ref{GINTM}) 
and~(\ref{LLSQCD}) define 2-loop RGI quantities, as it follows from 
the known expressions of the $\beta$ and $\gamma_m$ functions of the 
theory. Through O($g^5$) and O($g^2$) they read, respectively 
\beqn 
\hspace{-.8cm}&&\beta_{\rm SQCD}\!=\!-\frac{g^3}{16\pi^2}(3N_c-N_f)\! 
+\!\frac{g^5}{(16\pi^2)^2} 
(-6N_c^2+4N_cN_f-2\frac{N_f}{N_c})+{\rm O}(g^7)\, ,\label{BSQCD}\\ 
\hspace{-.8cm}&&\gamma_m=-\frac{g^2}{8\pi^2}\frac{N_c^2-1}{N_c}+ 
{\rm O}(g^4)\, .\label{GSQCD} 
\eeqn 
Actually it has been argued~\cite{RUS3} that the following ``exact'' 
formula holds 
\beqn 
\beta_{\rm SQCD}(g)=-\frac{g^3}{16\pi^2}-\frac{3N_c-N_f[1-\gamma_m(g)]} 
{{1-2{g^2N_c}/{16\pi^2}}}\, ,\label{BESQCD} 
\eeqn 
which generalises~(\ref{EXB}) to the SQCD 
case. Formula~(\ref{BESQCD}) perfectly fits with the previous ones to 
the order they are known and renders the expressions~(\ref{GINTM}) 
and~(\ref{LLSQCD}) RGI quantities to all orders. 
 
(B) The computation of the correlator~(\ref{GNM}) is much more subtle. 
First of all one notices that it vanishes to lowest order in $g$ 
because at the instanton saddle-point 
$\phi_f=\tilde\phi^f=0$. Secondly the number of inserted gluino 
fields does not appear to match the number of the existing zero 
modes. Finally the matter functional integration requires the 
knowledge of the massive fermion and scalar propagators, 
$(\Dslash\Dslash-|m|^2)^{-1}$ and $(D^2-|m|^2)^{-1}$, in the instanton 
background which is not available in closed form. 
 
The first and second problems are solved by observing that the 
integration over the scalar matter fields amounts to substituting 
$\phi_f$ and $\tilde\phi^f$ with the solutions of their classical 
e.o.m., which schematically read 
\beqn 
&&\phi_f=-i\sqrt{2}g\,(D^2-|m_f|^2)^{-1}\lambda\psi_f\, ,\label{EQMS}\\ 
&&\tilde\phi^f=i\sqrt{2}g\,(\tilde{D}^2-|m_f|^2)^{-1}\tilde\psi_f\lambda 
\, .\label{EQMST} 
\eeqn 
One easily checks that, at the expenses of going to higher order in 
$g$, in this fashion one ends up having the right number of inserted 
gluino fields. 
 
As for the last problem, we start by observing that the integration 
over the matter fermions has the effect of replacing for each flavour 
the $\tilde\psi^f(x)\psi_f(x')$ product with the corresponding 
fermionic propagator in the instanton background. After the matter 
integration one thus arrives at an extremely complicated integral over 
the collective instanton coordinates, where the unknown fermion and 
scalar background propagators appear. In order to proceed with the 
calculation we notice that the instanton semi-classical approximation 
respects supersymmetry and that consequently the correlators we are 
considering will come out to be constant in space-time and mass 
independent, as shown in Sect.~\ref{sec:SWTI}. The idea is then to 
perform the residual computation in the limit of very large masses 
(more precisely in the limit $m_f\gg|x_i-x_j|^{-1}\gg\Lambda_{\rm 
SQCD}$), where the fermion and scalar background propagators tend to 
their free-field expression. One ends up in this way with feasible 
integrals which yield the result (compare with~(\ref{GINTM})) 
\beqn 
G^{(N_f,N_c-N_f)}(x_1,\ldots,x_{N_c})=C_{N_c} 
\frac{(\Lambda^{\rm 1-loop}_{\rm SQCD})^{3N_c-N_f}}{g^{2N_c}}\, . 
\label{GINFNC} 
\eeqn 
We remark that this quantity is not RGI as it stands. To make it RGI we 
must renormalise the scalar fields. One way of doing this is to multiply 
both sides of~(\ref{GINFNC}) by the factor $\prod_f m_f$. 
 
(C) The computational strategy outlined above leads for the 
correlator~(\ref{DET}) to the simple result 
\beq 
D(x,x')=0\, .\label{DETV} 
\eeq 
 
{}From the results~(\ref{LLSQCD}) and~(\ref{GINFNC}) one can compute 
both the gluino and the scalar condensates. Recalling~(\ref{BSQCD}) 
and~(\ref{GSQCD}), one gets 
\beqn 
&&\langle\Omega_k|m_f\tilde\phi^f\phi_f|\Omega_k\rangle= 
\langle\Omega_k|\frac{g^2}{32\pi^2}\lambda\lambda|\Omega_k\rangle=\nn\\ 
&&=\er^{\frac{2\pi i k}{N_c}}\Big{(}C_{N_c} 
\big{(}\Lambda_{\rm SQCD}^{\rm 2-loop}\big{)}^{3N_c-N_f} 
\prod_{f=1}^{N_f} \hat m_f^{\rm 1-loop}\Big{)}^{1/N_c} 
\, .\label{LSSQCD} 
\eeqn 
Furthermore, one can derive the relations~\cite{ARV,AKMRV} 
\beqn 
&&\langle\Omega_k| \tilde\phi^f\phi_{h}|\Omega_k\rangle=0\, ,\quad 
f\neq h\, ,\label{FDH}\\ 
&& \langle\Omega_k|{\rm det}\,\tilde \phi|\Omega_k\rangle= 
\langle\Omega_k|{\rm det}\,\phi|\Omega_k\rangle=0\, .\label{DETZ} 
\eeqn 
All these results (see~(\ref{LSSQCD}) to~(\ref{DETZ})) are fully 
consistent with the WTI's of supersymmetry and with~(\ref{KONANA1}) 
and~(\ref{KONANA2}) implied by the Konishi 
anomaly relation~\cite{AKMRV}. 
 
The important conclusion of this thorough analysis is that the 
non-renormalisation theorems~\cite{NOREN} of supersymmetry are 
violated by instanton effects as it results from the fact that chiral 
(composite) operators acquire non-vanishing v.e.v.'s, while they are 
identically zero at the perturbative level. One way of 
understanding this surprising finding in the language of the 
effective theory approach of Sect.~\ref{sec:ELA} is to say that 
instantons generate a contribution to the effective superpotential 
which is non-perturbative in nature. 
 
\subsubsection{Massless SQCD} 
\label{sec:MLSQCD} 
 
We now consider the strictly massless ($m_f=0$, $f=1,\ldots,N_f$) SQCD 
theory. {}From the formulae we derived in the previous sections it 
should be already clear that the limit $m_f\to 0$ is not 
smooth. Indeed, we will see that a straightforward application of the 
instanton calculus rules, that we have developed in the massive case, 
to massless SQCD leads to results that do not agree with the massless 
limit of the massive formulae. 
 
The origin of this discrepancy is not completely clear. As we said, one 
possibility is that the $m_f\to 0$ limit of the massive theory does 
not coincide with the strictly massless theory, as a consequence of the 
fact that the small mass limit of massive SQCD is plagued by infrared 
divergences. Besides the divergences encountered if the massless 
limit of~(\ref{LSSQCD}) is taken, a simple analysis shows, in fact, 
that a (naive) small $|m_f|^2$ Taylor expansion gives raise to 
$|m_f|^2\times 1/|m_f|^2$ contributions that would be absent in the 
strictly massless SQCD theory. Another possibility, strongly advocated 
in refs.~\cite{RUS1} and~\cite{ADSALL1,ADSALL2}, is related to the 
observation that in the absence of mass terms the matter 
superpotential has a huge manifold of flat directions along which the 
exponential of the action does not provide any damping. In this 
situation it is not at all clear that the instanton 
solution~(\ref{ICL}) can be taken as the configuration which dominates 
the functional integral. Other types of quasi saddle-points, where 
scalar fields take a non-zero v.e.v., may be also relevant. The 
strategy suggested by these authors to deal with this situation will 
be discussed in Sect.~\ref{sec:ELA}. Here we want to first show what 
sort of results follow when the massive instanton calculus developed 
in Sect.~\ref{sec:ICSQCD} is blindly applied to massless SQCD. 
 
The Green functions that have the correct number of fermionic zero 
modes in the one-instanton background are restricted to 
\beqn 
\hspace{-.8cm}&&\bullet\, G^{(N_f,N_c-N_f)\{f\}}_{\{h\}} 
(x_1,\ldots,x_{N_c})=\quad\quad\quad\quad\quad {\rm for}\,\, 
N_c\geq N_f\label{GMLSQCD}\\ 
\hspace{-1.cm}&&=\langle \tilde\phi^{f_1}\phi_{h_1}(x_1)\ldots 
\tilde\phi^{{f_{N_f}}}\phi_{h_{N_f}}(x_{N_f}) 
\frac{g^2}{32\pi^2}\lambda\lambda(x_{N_f+1})\ldots 
\frac{g^2}{32\pi^2}\lambda\lambda(x_{N_c})\rangle\, ,\nn\\ 
\hspace{-.8cm}&&\bullet\, D(x,x')=\langle{\rm det}[\tilde\phi(x)] 
{\rm det}[\tilde\phi(x')]\rangle 
\, ,\quad\quad\quad\quad\quad {\rm for}\,\,N_c= N_f\, ,\label{DMLSQCD} 
\eeqn 
because now there exist zero modes also for the matter fermions, 
$\tilde\psi^f$ and $\psi_f$. A non-vanishing result is obtained if for 
each scalar field an appropriate Yukawa interaction term is brought 
down from the action. In this way $2N_c$ gluino zero modes, 
$\lambda_0$, together with the fermionic matter zero modes, 
$\tilde\psi_0^f$ and $\psi_{0f}$, $f=1,\ldots,N_f$, will appear 
simultaneously. At the same time, when scalars are contracted in 
pairs, the scalar propagator in the instanton background, $(D^2)^{-1}$ 
or $(\tilde{D}^2)^{-1}$, is generated which will act on the product 
$\lambda_0\psi_0$ or $\tilde\psi_0\lambda_0$, respectively. Unlike the 
massive case, closed expressions for $(D^2)^{-1}$ and 
$(\tilde{D}^2)^{-1}$ exist which allows to explicitly compute the form 
of the ``induced scalar modes'', by solving the field equations 
$D^2\phi+ig\sqrt{2}\lambda_0\psi_0=0$ and 
$\tilde{D}^2\tilde\phi-ig\sqrt{2}\tilde\psi_0\lambda_0=0$, 
respectively. 
 
The problem with the SCI computational strategy we have briefly 
described can already be seen by taking, for simplicity the case 
$N_c=2$ and $N_f=1$. In massless SQCD (after correcting for the usual 
factor $2^{N_c}$ with respect to result quoted in~\cite{AKMRV}), 
one gets 
\beq 
\langle\tilde\phi\phi(x_1)\frac{g^2}{32\pi^2}\lambda\lambda(x_2) 
\rangle\big{|}_{m=0}=\half\frac{(\Lambda_{2,1}^{{\rm 1-loop}})^5}{g^4}\, , 
\label{NCDNFUL} 
\eeq 
while for the same Green function in the massive case we got 
(see~(\ref{GINFNC})) 
\beq 
\langle\tilde\phi\phi(x_1)\frac{g^2}{32\pi^2}\lambda\lambda(x_2) 
\rangle\big{|}_{m\neq 0}= 
\frac{4}{5}\frac{(\Lambda_{2,1}^{{\rm 1-loop}})^5}{g^4}\, . 
\label{NCDNFUV} 
\eeq 
Apart from the numerical discrepancy visible between~(\ref{NCDNFUL}) 
and~(\ref{NCDNFUV}), what is more disturbing is 
that~(\ref{NCDNFUL}) is in conflict with the Konishi anomaly 
relation~(\ref{KONANA1}), which in the massless regime (and using 
clustering) implies the vanishing of the gluino condensate. An 
alternative to this conclusion would be to say that the scalar 
condensate can be infinite in massless SQCD (see the discussion in 
Sect.~\ref{sec:NCLNF}). 
 
Notice that for $N_f>1$ the massless SQCD action possesses a 
non-anomalous $SU_L(N_f)\times SU_R(N_f)\times U_V(1)\times U_{\hat 
A}(1)$ symmetry (see~(\ref{FTRA})). This means that in extracting 
the scalar condensates a vacuum disentangling step analogous to the 
one performed in Sect.~\ref{sec:SYMV} is necessary. Proceeding in this 
way, one again finds results for the condensates that do not agree 
with what was found in the massive case. 
 
Also the result for the correlator~(\ref{DMLSQCD}) is at variance 
with~(\ref{DETV}). We now find $D(x,x')\neq 0$, which implies (no 
disentangling is necessary here, as ${\rm det}\phi$ and ${\rm 
det}\tilde \phi$ are invariant under the chiral flavour group) 
\beqn 
\langle {\rm det}\,\tilde \phi\rangle\neq 0\, , \qquad 
\langle {\rm det}\,\phi \rangle\neq 0\, ,\label{DETNZ} 
\eeqn 
signaling the spontaneous breaking of the $U_V(1)$ symmetry. 
 
\subsection{The case of chiral theories} 
\label{sec:ICCT} 
 
In this section we wish to discuss the very interesting case of 
supersymmetric theories of the Georgi--Glashow type~\cite{GEGL}, 
where matter fermions are chiral. There is a quite remarkable 
literature on the subject. A selection of useful papers can be found 
in refs.~\cite{MV,ADSALL2,BGK1,BGK2,AKMRV}. 
 
In this review we will limit to consider $SU(N_c)$ gauge theories with 
matter in the fundamental, ${\bf N_c}$, and antisymmetric, ${\bf 
N_c(N_c-1)/2}$, representation. We recall that gauge anomaly 
cancellation requires the number of fundamentals, $n_{\rm fund}$, and 
antisymmetric, $n_{\rm anti}\equiv M$, representations to be related by 
\beq 
n_{\rm fund}=M(N_c-4)\, .\label{MN} 
\eeq 
The resulting $\beta$-function 
\beq 
\beta_{\rm GG}=-\frac{g^2}{8\pi^2}[3(N_c+M)-M N_c]+{\rm O}(g^5) 
\label{BGG}\eeq 
implies asymptotic freedom if $M<3N_c/(N_c-3)$. 
 
The composite operators that, besides ${g^2}\lambda\lambda/{32\pi^2}$, 
come into play are generically constructed in terms of the lowest 
components of the chiral matter superfields for which we introduce the 
notation 
\begin{equation} 
\begin{array}{ll} 
\Phi^I_r &\, , \quad I=1,2,\ldots,n_{\rm fund}\, ,\\\\ 
X^{rs}_i=-X^{sr}_i &\, ,\quad r,s=1,2,\ldots,N_c\, ,\,\,i=1,2,\ldots,M\, . 
\end{array} 
\label{CHIRMAT} 
\end{equation} 
 
Non-perturbative calculations are of special importance here, as 
Witten index arguments have so far been unable to make any definite 
statement about the nature of the vacua of the theory. Actually a 
variety of scenarios turn out to be realised according to the 
specific matter content of the action that can be summarised as 
follows. 
 
(I) {\it Unbroken supersymmetry with well defined 
vacua}~\cite{MV,ADSALL2}. One such example is the $SU(6)$ case with 
$M=1$ and correspondingly $n_{\rm fund}=2$. The allowed superpotential 
possesses no flat directions and the unique perturbative vacuum is at 
vanishing values of the scalar fields. There exist instanton dominated 
(constant) Green functions which upon using clustering give results in 
perfect agreement with the constraints coming from the Konishi anomaly 
relations. One finds that the discrete $\Z_{30}$ symmetry is 
spontaneously broken down to $\Z_{6}$, leaving behind 30/6=5 well 
defined supersymmetric vacua. We remark that here, unlike SYM and 
SQCD, the number of vacua is not equal to $N_c$. It should be noted 
that in this example vacuum disentangling can be trivially carried 
out. 
 
A more delicate situation occurs if we double the number of families, 
i.e.\ if we take $M=2$~\cite{BGK2}, because a non-trivial vacuum 
disentangling over the transformations of the complexification of the 
global symmetry group $SU(4)\times SU(2)$ is necessary here. When this 
is done, results from instanton calculations allow to determine all 
the condensates. In particular one finds that the discrete 
$\Z_{12}$ symmetry group is spontaneously broken down to $\Z_{3}$, 
leaving behind 12/3=4 well defined supersymmetric vacua. The 
interesting observation is that in this theory also the relations 
entailed by the Konishi anomaly equations allow to completely compute 
all the condensates. Reassuringly the two sets of results turn out to 
be in perfect agreement. For a discussion of these results from the 
complementary effective action point of view see Sect.~\ref{sec:SUS}. 
 
In both the above cases when the superpotential is switched off the 
vacuum becomes ill defined, because in this limit necessarily some of 
the condensates must ``run away'' to infinity. This is due to the fact 
that the relations among condensates involve (inverse) factors of the 
Yukawa couplings. 
 
(II) {\it Unbroken supersymmetry with ill defined vacua}. This 
situation occurs in theories based on a $SU(N_c)$ gauge group with 
$N_c$ even and larger than 8. Also in the presence of a non-vanishing 
superpotential, one finds that, in order to reconcile instanton 
results with the implication of the Konishi anomaly relations, one has 
to assume that some of the scalar condensates run away to 
infinity~\cite{BGK2}. Such a result is seen to be related to the 
existence of flat directions in the superpotential. In this respect 
the situation is similar to massless SQCD, where we had at the same 
time flat directions in the superpotential and infinite scalar 
condensate in order to avoid contradictory results between instanton 
calculations and the Konishi anomaly equation. 
 
(III) {\it Spontaneously broken supersymmetry}. This conclusion 
indirectly arises in Georgi--Glashow type models in which the gauge 
group is $SU(N_c)$ with $N_c$ odd, because of conflicting constraints 
between instanton calculations and relations implied by the Konishi 
anomaly equation. Several specific cases have been considered in the 
literature. We list here some interesting examples. 
\begin{enumerate} 
\item $N_c=5$, $M=1$ and consequently $n_{\rm fund}=1$. Although in 
this case no superpotential can be constructed, it can be shown that 
the theory does not admit any perturbative flat 
direction~\cite{ADSALL2}. Because of the absence of superpotential 
the Konishi anomaly relations imply that the gluino condensate must 
vanish. When this result is put together with the non-perturbative 
calculations of certain instanton dominated Green 
functions~\cite{MV,AKMRV}, one is led to the conclusion that the 
involved scalar condensate cannot take a finite value, if clustering 
is used and the vacuum is supersymmetric. The wandering to infinity of 
the scalar condensate in the absence of flat directions looks highly 
implausible and one should rather conclude that there is a dynamical 
breaking of supersymmetry owing to non-perturbative instanton 
effects. 
\item $N_c=5$, $M=2$ and consequently $n_{\rm fund}=2$. This time the 
theory admits a superpotential but still no flat directions 
exist. Unlike the previous case, from the Konishi anomaly equations 
one can prove that all condensates must vanish. This is in 
contradiction with the non-vanishing result given by the instanton 
calculation of the Green function where the product of these 
condensates appear, if clustering is invoked and the vacuum is 
supersymmetric. The most tempting conclusion is that supersymmetry is 
dynamically broken. One might object to this conclusion that actually 
the instanton calculation is performed in the absence of a 
superpotential, i.e.\ in a situation where disentangling is necessary 
and flat directions are present. This should not be a problem, 
however, because, unlike the case of the mass dependence, one expects 
the limit in which the superpotential coupling vanishes to be a smooth 
one. For a discussion of these results from the complementary 
effective action point of view see Sect.~\ref{sec:SUC}. 
\item $N_c\geq 7$, $M=1$. There exist many instanton dominated Green
functions which, after vacuum disentangling, yield an overdetermined
set of relations for the condensates~\cite{BGK2}. One can solve the
resulting equations finding full consistency with clustering. However,
the Konishi anomaly equations are such as to imply the vanishing of
several  condensates and thus through instanton results the run away
of others.  Because no (perturbative) flat directions exist in these
models, one  is led again to conclude that supersymmetry is
spontaneously broken. 
\end{enumerate} 
 
\section{The effective action approach} 
\label{sec:ELA} 
 
Symmetry properties of the action in the form of anomalous and 
non-anomalous WTI's together with explicit dynamical (instanton) 
calculations have taught us a lot about the nature of supersymmetric 
${\cal N}=1$ theories. A very useful and elegant way to recollect all 
these results is to make recourse to the notion of ``effective'', or 
``low energy'', action (sometimes also referred to as ``effective 
Lagrangian''). This notion, though with slightly different meanings 
and realm of application, has a long history. It was first introduced 
in the papers of~\cite{WEIN}, as a way of compactly deriving the 
soft pion theorems of Current Algebra, then fully developed for QCD 
with the inclusion of the $\eta'$ meson and the $U_A(1)$ anomaly in 
the works quoted in refs.~\cite{RDWN} and~\cite{GL}. A parallel road 
was opened by Symanzik~\cite{SYM} to deal with the lattice 
regularisation of QCD, which turned out to be crucial for 
understanding the approach to the continuum of the lattice theory. 
 
The extension of these ideas to supersymmetric theories was first 
proposed in refs.~\cite{VY,TVY}, where the cases of ${\cal N}=1$ pure 
SYM and SQCD were considered, and then expanded to a field of 
investigation of its own in~\cite{ADSALL1,FIPU,SEI}. Some review 
papers on the subject can be found in~\cite{REVET}. 
 
Effective actions for all the theories we have discussed in the 
previous sections have been constructed and many interesting results 
have been obtained. In the following we want to briefly review what 
was done with the main purpose of comparing with instanton results. 
 
\subsection{The effective action of SYM} 
\label{sec:ELSYM} 
 
The first step along the way of constructing the effective action, 
${\Gamma}_{\rm eff}$, describing the low energy dynamics of a theory 
is to identify the degrees of freedom relevant in the energy regime 
$E\ll \Lambda$, where $\Lambda$ is the theory RGI mass scale. In the 
pure SYM case, where confinement seems to hold, the obvious degrees of 
freedom can be collected in the (dimension three) superfield 
\beqn 
S=\frac{g^2}{16\pi^2}{\rm Tr}(W^\alpha W_\alpha)\, , \label{S} 
\label{WW} 
\eeqn 
whose lowest component is precisely the gluino composite 
operator~(\ref{LL}). 
 
The second step is the observation that the interesting piece of 
${\Gamma}_{\rm eff}^{\rm SYM}$ is not so much its kinetic contribution 
(a $D$-term which is non-holomorphic in $S$ and reduces to the standard 
kinetic terms as $g\to 0$), but rather the $F$-term which provides the 
correct anomalous transformation properties of the effective 
action. In the present case it is enough and convenient to make 
reference to the $U_R(1)$ symmetry (see~(\ref{UUR})) to fix the 
form of this term, which is often referred to with the name of 
``effective superpotential'' in the literature. 
 
Recalling the $U_R(1)$ transformation properties of the superfield $S$ 
(see Table in Appendix~\ref{sec:APPA}) 
\beq 
S(x,\theta)\to \er^{3i\alpha}S(x,\theta \er^{-3i\alpha/2})\, , 
\label{WWA}\eeq 
we are led to write for the full effective action the formula 
\beq 
{\Gamma}_{{\rm eff};N_c}^{\rm SYM}={\Gamma}_{\rm kin}^{\rm SYM}(S,S^*)+ 
\big{[}{W}_{{\rm eff};N_c}^{\rm SYM}(S)+{\rm h.c.}\big{]}\, . 
\label{LEFSYM}\eeq 
where 
\beqn 
&&{\Gamma}_{\rm kin}^{\rm SYM}(S,S^*)=k \big{[}(S^*S)^{1/3} 
\big{]}_D\, , \label{LKIN}\\ 
&&{W}_{{\rm eff};N_c}^{\rm SYM}(S)=-\Big{[}S\big{(}\log\frac{S^{N_c}} 
{(c\Lambda_{\rm SYM})^{3N_c}}-N_c\big{)}\Big{]}_F\, .\label{WESYM} 
\eeqn 
In the above equations we used the short-hand notation 
\beq 
\big{[}\big{(}\ldots\big{)}\big{]}_F\equiv 
\int \dr^4x\,\dr^2\theta\,\big{(}\ldots\big{)}\, ,\qquad 
\big{[}\big{(}\ldots\big{)}\big{]}_D\equiv 
\int \dr^4x\,\dr^2\theta\, \dr^2\bar\theta\,\big{(}\ldots\big{)}\, . 
\label{FNOT} 
\eeq 
The expression of ${\Gamma}_{\rm kin}^{\rm SYM}(S,S^*)$ 
in~(\ref{LKIN}) is in no way unique. It is only an example of a 
functional having the property that (with a suitable choice of the 
constant $k$) it reproduces the standard form of the kinetic term in 
the limit $g\to 0$. The other constant $c$ cannot be fixed by 
symmetry considerations only. A way of determining its value will be 
discussed in Sect.~\ref{sec:NELSQCD}. Equation~(\ref{WESYM}) is the 
famous Veneziano--Yankielowicz effective action~\cite{VY}. 
 
It is not difficult to prove that the second term in~(\ref{LEFSYM}) 
has the desired transformation properties under $U_R(1)$ 
(see~(\ref{RAN})). From~(\ref{WWA}) we get in fact (the 
$x$-dependence and the corresponding space-time integration is 
understood) 
\beqn 
\hspace{-.2cm}&&\bullet\!\int \!\dr^2\theta\,S(\theta)\to \er^{3i\alpha} 
\int \dr^2\theta\,S(x,\theta \er^{-3i\alpha/2})= \nn\\ 
\hspace{-.2cm}&&=\int \dr^2(\theta \er^{-3i\alpha/2}) 
S(\theta \er^{-3i\alpha/2})= 
\int \dr^2\theta\,S(\theta)\, ,\label{SINV}\\ 
\hspace{-.2cm}&&\bullet\!\int \dr^2\theta\,S(\theta)\log S(x,\theta)\to 
\er^{3i\alpha} \int \dr^2\theta\,S(\theta \er^{-3i\alpha/2})\big{(}3i\alpha+ 
\log S(\theta \er^{-3i\alpha/2})\big{)}=\nn\\ 
\hspace{-.2cm}&&=3i\alpha \int \dr^2(\theta \er^{-3i\alpha/2}) 
S(\theta \er^{-3i\alpha/2})+\!\!\int \dr^2(\theta \er^{-3i\alpha})S(\theta 
\er^{-3i\alpha}) \log S(\theta \er^{-3i\alpha/2})=\nn\\ 
\hspace{-.2cm}&&=3i\alpha\!\int \dr^2\theta\,S(\theta)+ 
\!\int \dr^2\theta\,S(\theta)\log S(\theta)\, . 
\label{PRO} 
\eeqn 
Useful information about the non-perturbative properties of the theory 
can be obtained from the formula~(\ref{LEFSYM}) by determining the 
values of $S$ which make ${\Gamma}_{\rm eff}^{\rm SYM}$ stationary. 
These are constant field configurations which minimise the effective 
action. Thus they yield the values of the v.e.v.\ of the gluino 
composite operator (gluino condensate). From~(\ref{LEFSYM}) one gets 
the result 
\beq 
\langle S\rangle =(c\Lambda_{\rm SYM})^{3}\er^{2i\pi k/N_c}\, , 
\quad k=1,\ldots,N_c\, . 
\label{GLUCON} 
\eeq 
If $c^3$ is identified with $(C_{N_c})^{1/N_c}$, then~(\ref{GLUCON}) 
becomes identical to~(\ref{COND2}). However, in connection with 
the comments we made in Sect.~\ref{sec:DRSYM}, we must remark here 
that there is a discrepancy between the number given by the above 
identification and the choice $c=1$ made in 
refs.~\cite{ADSALL2,COR,REVET}. The latter can be justified in the 
framework of the SQCD effective action approach, if in conjunction 
with WCI calculations, certain consistency relations following from 
decoupling (see Sect.~\ref{sec:NELSQCD}) are employed. 
 
A crystal clear way to resolve this puzzling discrepancy would be to 
arrive at an evaluation of the SYM effective action from first 
principles, i.e.\ {\it \'a la} Wilson--Polchinski~\cite{WP}. Many 
efforts have been made in this direction and a lot of interesting 
results have been obtained~\cite{KEN} insisting on the role of 
anomalies~\cite{ARKA,KON} in the construction of the Wilsonian 
action. A different and perhaps more promising road has been recently 
undertaken which uses the matrix model formulation of 
SYM~\cite{VW}. In this framework the form of the effective action 
could be derived~\cite{GEO,KY} and the result $c=1$ was obtained.

\subsection{The effective action of SQCD} 
\label{sec:ELSQCD} 
 
{}For the purpose of extending the previous considerations to SQCD it 
is convenient to distinguish among the three cases $N_c>N_f$, 
$N_c=N_f$ and $N_c<N_f$ and separately discuss the massive and the 
massless situation. 
 
\subsubsection{SQCD with $N_c>N_f$} 
\label{sec:NCLNF} 
 
\noindent $\bullet$ {\it The massive case} 
 
The generalisation of the previous formulae to massive SQCD is almost
immediate if one includes among the degrees of freedom that describe
the low energy dynamics of the theory also the composite
operators~(\ref{PSIER}) 
\beq 
T^f_{h}=\tilde\phi^{f}_r(x)\phi_{h}^r(x)\, .\label{TFF} 
\eeq 
Apart from the unessential (for this discussion) kinetic terms, one finds 
that the formula which extends~(\ref{LEFSYM}) is 
\beqn 
\hspace{-.8cm}&&{\Gamma}_{{\rm eff};N_c,N_f}^{\rm SQCD}(S,S^*;T,T^*)= 
\label{LESQCD}\\ 
\hspace{-.8cm}&&={\Gamma}_{\rm kin}^{\rm SQCD}(S,S^*;T,T^*)+ 
\big{[}{W}_{{\rm eff};N_c,N_f}^{\rm SQCD}(S;T)+ {\rm h.c.}\big{]}\, ,\nn\\ 
\hspace{-.8cm}&&{W}_{{\rm eff};N_c,N_f}^{\rm SQCD}(S;T)=\label{WESQCD}\\ 
\hspace{-.8cm}&&=\Big{[}-S\big{(}\log \frac{S^{N_c-N_f}{\rm det}T} 
{(c'\Lambda_{\rm SQCD})^{3N_c-N_f}}- 
(N_c-N_f)\big{)}+\sum_f m_f T^f_{f}\Big{]}_F\, .\nn 
\eeqn 
As before, the constant $c'$ cannot be fixed by symmetry 
considerations only. We will discuss the important issue of 
determining its precise value below. The minimisation 
conditions (also called $F$-flatness conditions) allow to determine 
all the condensates and one finds~\footnote{We recall the elementary 
formula $\dr\log [{\rm det}T]/\dr T^f_{h}=(T^{-1})_f^{h}$.} 
\beqn 
&&\langle S\rangle = \er^{\frac{2i\pi k}{N_c}}(c'\Lambda_{\rm SQCD})^{3} 
\prod_{f=1}^{N_f}\Big{(}\frac{m_f}{c'\Lambda_{\rm SQCD}}\Big{)}^{1/N_c}\, , 
\quad k=1,\ldots,N_c\, ,\label{GLU}\\ 
&&\langle T^f_{h} \rangle = \delta^f_{h}\frac{\langle S\rangle}{m_f}\, , 
\label{SCA} 
\eeqn 
viz.\ the same results that were obtained in~(\ref{LSSQCD}) up to 
the normalisation of the $\Lambda$ parameter. 
 
Equation~(\ref{WESQCD}) has a number of nice properties. 
 
1) It is mathematically meaningful not only for $N_c>N_f$, where 
instanton calculations are 
feasible, but for any value of $N_c$ and $N_f$, except $N_c= 
N_f$. Actually in the last case a further composite operator has to 
come into play, as already hinted at by the results of 
Sects.~\ref{sec:MVSQCD} and~\ref{sec:MLSQCD}. We will discuss in 
detail this case below. 
 
2) It obeys the expected decoupling theorem in the sense that when one 
of the flavours gets infinitely massive, (\ref{WESQCD}) precisely 
turns into the effective action for the theory with one less flavour, 
in which that particular flavour is absent, provided the $\Lambda$ 
parameters of the two theories, $\Lambda^{N_f}_{\rm SQCD}$ and 
$\Lambda^{N_f-1}_{\rm SQCD}$, are matched as described in 
Appendix~\ref{sec:APPD}. 
 
3) The massive $S$ field can be ``integrated out'', leaving a pure matter 
effective superpotential 
\beqn 
\hspace{-.8cm}&&{W}^{\rm SQCD}_{{\rm eff};N_c,N_f}(T)= 
\nn\\ 
\hspace{-.8cm}&&=\Big{[}(N_c-N_f)\Big{(} 
\frac{(c'\Lambda_{\rm SQCD})^{3N_c-N_f}}{{\rm det}T} 
\Big{)}^{\frac{1}{(N_c-N_f)}}+\sum_f m_f T^f_{f}\Big{]}_{F}\, , 
\label{LESSQCD}\eeqn 
which coincides with the Affleck--Dine--Seiberg~\cite{ADSALL1} 
effective superpotential. One has to remark, however, that the 
last formula is only meaningful for $N_c>N_f$.\\ 
 
\noindent $\bullet$ {\it The massless case} 
 
The discussion of the massless case is quite delicate (and 
controversial). This is to be expected looking at the results of the 
dynamical instanton calculations presented in Sect.~\ref{sec:MLSQCD}, 
which showed conflicting findings for the condensates when the 
massless limit of the massive results were compared to what one gets 
in the strictly massless case. 
 
Let us now see what are the implications of the massless SQCD 
effective action from using the formulae~(\ref{WESQCD}) 
or~(\ref{LESSQCD}). If we start from~(\ref{WESQCD}), we 
conclude that, when any of the masses is sent to zero, the gluino 
condensate must be taken to vanish for consistency with~(\ref{GLU}), 
while~(\ref{SCA}) does not contain sufficient information to 
determine any of the scalar condensates $\langle T^f_{h} \rangle$. If 
on the contrary all the masses are set to zero from the beginning, as 
was done in~\cite{ADSALL2}, and~(\ref{LESSQCD}) is used, one 
must conclude that the scalar condensates run away to infinity 
as the potential generated by the massless effective 
action~(\ref{LESSQCD}) monotonically decreases to zero in this limit.

\subsubsection{SQCD with $N_c=N_f$} 
\label{sec:NCENF} 
 
When $N_f=N_c$ new composite operators can be constructed which must 
appear among the fields of the effective action. They are the 
determinants, $X$ and $\tilde X$, over colour and flavour indices of 
the matter fields, whose lowest components are shown in~(\ref{DETP}) 
and~(\ref{DETPT}). The need for new field operators is confirmed by 
the observation that the action~(\ref{LESQCD}) does not contain a 
sufficiently large number of massless fermions to satisfy the 
't~Hooft anomaly matching conditions~\cite{THOA,SEI,REVET} associated 
with the non-anomalous $U_{\hat A}(1)$ symmetry defined 
by~(\ref{NONAM}) and~(\ref{FTRA}) (see also Table in 
Appendix~\ref{sec:APPA}).\\ 
 
\noindent $\bullet$ {\it The massive case} 
 
The most general form of effective action is 
\beqn 
&&{\Gamma}_{\rm eff}^{\rm SQCD}(S,S^*;T,T^*;X,\tilde X)= 
{\Gamma}_{\rm kin}^{\rm SQCD}(S,S^*;T,T^*;X,\tilde X)+\label{LESQCDNN}\\ 
&&+\Big{[}-S\big{(}\log \frac{{\rm det}T}{(c''\Lambda_{\rm SQCD})^{2N_c}}+ 
f(Z)\big{)}+\sum_f m_f T^f_{f}+{\rm h.c.}\Big{]}_{F}\, , 
\nn\eeqn 
where $f(Z)$ is a function of the ratio $Z=X\tilde X/{\rm det} T$. 
{}For $m_f\neq 0$ (all $f$) the stationarity conditions lead to the equations 
\beqn 
&&\log \frac{{\rm det}T}{(c''\Lambda_{\rm SQCD})^{2N_c}}+f(Z)=0\, , 
\label{M1}\\ 
&&m_f\delta^f_{h}= 
S\,\big{[}1-Z\frac{\partial f(Z)}{\partial Z}\big{]}(T^{-1})_f^{h}\, , 
\label{M2}\\ 
&&S\,\frac{\partial f(Z)}{\partial Z}\,\frac{\tilde X}{{\rm det} T}=0\, , 
\label{M3}\\ 
&&S\frac{\partial f(Z)}{\partial Z}\,\frac{X}{{\rm det} T}=0\, ,\label{M4} 
\eeqn 
which have the solution 
\beqn 
&&\langle X\rangle\!=\!\langle \tilde X\rangle \!=\!0\, ,\label{SQCDC1}\\ 
&&m_f\langle T^f_{h}\rangle\!=\!\delta^f_{h}{\langle S\rangle}\, , 
\label{SQCDC2}\\ 
&&\langle {\rm det} T\rangle\! = \!\er^{-f(0)}c'' 
\big{(}\Lambda_{\rm{SQCD}}\big{)}^{2N_c}\, ,\label{SQCDC3} 
\eeqn 
a result with exactly the same structure 
as~(\ref{LSSQCD})-(\ref{DETZ}).\\ 
 
\noindent $\bullet$ {\it The massless case} 
 
The massless case is, as usual, more subtle. Besides $\langle 
S\rangle=0$ (as implied by the massless limit of the Konishi anomaly 
equation~(\ref{SQCDC2})), by varying the effective action with respect 
to $S$ one still gets the constraint 
\beq 
\log \frac{{\rm det}T}{(c''\Lambda_{\rm SQCD})^{2N_c}}+f(Z)=0\, , 
\label{VARS} 
\eeq 
which (if $f(Z)\neq 0$) only fixes one combination of $\langle 
X\rangle$, $\langle \tilde X\rangle$ and $\langle {\rm det} T\rangle$, 
leaving the other two undetermined. This can be interpreted as the 
equivalent of the statement that for $N_f=N_c$ and $m_f=0$ the 
perturbative flat directions are not (all) removed, so vacua with 
arbitrarily large values of these condensates can occur. The 
effective action vanishes at the minimum and one is only left with 
the constraint~(\ref{VARS}). 
 
The explicit form of this constraint was worked out in~\cite{SEI} 
with the conclusion that the classical relation ${\rm det}T=X\tilde X$ 
is lifted by quantum correction to the formula 
\beq 
{\rm det}T-X\tilde X=(\Lambda_{\rm SQCD})^{2N_c}\, . 
\label{VARSQ} 
\eeq 
This was the first example of a by now well known phenomenon (see 
Sect.~\ref{sec:SWanalysis}) according to which the quantum theory can 
display a whole manifold of (degenerate) vacuum states where 
supersymmetry is unbroken. It is a complex K\"ahler manifold (often 
called the ``quantum moduli space'', ${\cal M}$) to which the point 
representing the classical vacuum not always belongs. We end this 
discussion by noticing that the constraint~(\ref{VARSQ}) can also be 
derived by a massless effective action of the type~(\ref{LESQCDNN}) 
if one simply takes 
\beq 
f(Z)=\log(1-Z)\, . 
\label{VARF} 
\eeq 
 
\subsubsection{SQCD with $N_f>N_c$} 
\label{sec:NCSNF} 
 
In this case neither dynamical instanton calculations are possible 
(see our discussion in Sect.~\ref{sec:IC}) nor the general 
considerations of~\cite{ADSALL2} apply. In principle one can 
imagine to go on with the effective action approach, guided by 
information on the relevant low energy degrees of freedom provided by 
the 't~Hooft anomaly matching conditions. 
 
$\bullet$ For instance, in the case $N_f=N_c+1$ the two baryon-like superfields 
\beqn 
&&B^f=\epsilon^{ff_1f_2\ldots f_{N_c}}\epsilon_{r_1r_2\ldots r_{N_c}} 
\Phi^{r_1}_{f_1}\Phi^{r_2}_{f_2}\ldots\Phi^{r_{N_c}}_{f_{N_c}}\, , 
\label{VARB}\\ 
&&\tilde B_f=\epsilon_{ff_1f_2\ldots f_{N_c}}\epsilon^{r_1r_2\ldots r_{N_c}} 
\bar\Phi_{r_1}^{f_1}\bar\Phi_{r_2}^{f_2}\ldots\bar\Phi_{r_{N_c}}^{f_{N_c}} 
\, .\label{VARBB} 
\eeqn 
must come into play in order to fulfill such conditions. They can 
combine with $T^f_{h}$ to give the term $B^fT_f^{h}\bar B_{h}$ in the 
effective action. The whole expression of the latter can then be 
argued to have the form 
\beq 
{\Gamma}_{\rm eff}^{\rm SQCD}= 
{\Gamma}_{\rm kin}^{\rm SQCD}+\Big{[}\frac{{\rm det}T-B^fT_f^{h}\tilde B_{h}} 
{(\Lambda_{\rm SQCD})^{b_1}}+{\rm h.c.}\Big{]}_F 
\label{EASQCD}\,, 
\eeq 
where $b_1=3N_c-N_f=2N_c-1$. As a consistency check, it can be shown 
that, if the $N_f+1$-th flavour is given a mass and decoupled, then 
the situation we described in the previous subsection where we had 
$N_f=N_c$ is recovered. 
 
It is interesting to remark that by solving the $F$-flatness equations 
implied by the effective action~(\ref{EASQCD}), one finds that, 
unlike the case $N_f=N_c$, the point corresponding to the classical 
vacuum $\langle T_f^{h}\rangle=\langle B^f\rangle=\langle \tilde 
B_{h}\rangle=0$ belongs to the moduli space of the theory. At this 
point the (non-anomalous) symmetry of the classical action, 
$SU_L(N_f)\times SU_R(N_f)\times U_V(1)\times U_{\hat A}(1)$, is fully 
unbroken. 
 
$\bullet$ For larger values of $N_f$ ($N_f>N_c+1$, but smaller than 
$3N_c$, where UV asymptotic freedom is lost) the above formulae have 
an obvious generalisation. One introduces the chiral superfields 
\beqn 
\hspace{-.8cm}&&B^{f_{1}\ldots f_{\tilde N_c}}= 
\epsilon^{f_1\ldots f_{\tilde N_c} 
f_{\tilde N_c +1}\ldots f_{N_f}}\epsilon_{r_1\ldots r_{N_c}} 
\Phi^{r_1}_{f_{\tilde N_c +1}}\ldots\Phi^{r_{N_c}}_{f_{N_f}}\, ,\label{BBB1}\\ 
\hspace{-.8cm}&&\tilde B_{f_{1}\ldots f_{\tilde N_c}}= 
\epsilon_{f_1\ldots f_{\tilde N_c}f_{\tilde N_c +1}\ldots f_{N_f}} 
\epsilon^{r_1\ldots r_{N_c}} 
\tilde\Phi_{r_1}^{f_{\tilde N_c+1}}\ldots\tilde\Phi_{r_{N_c}}^{f_{N_f}}\, , 
\label{BBB2}\eeqn 
where, following~\cite{SEI}, we have set 
\beq 
{\tilde N_c}=N_f-N_c\label{NFT}\, .\eeq 
In terms of the operators~(\ref{BBB1}), (\ref{BBB2}) and $T_f^{h}$ 
one may construct the effective action 
\beqn 
&&{\Gamma}_{\rm eff}^{\rm SQCD} = {\Gamma}_{\rm kin}^{\rm SQCD}+ 
\label{EASNT}\\ 
&&+\Big{[}\frac{{\rm det}T-B^{f_1f_2\ldots f_{\tilde N_c}} 
T_{f_1}^{h_1}T_{f_2}^{h_2}\ldots 
T_{f_{\tilde N_c}}^{h_{\tilde N_c}}\tilde B_{h_1h_2\ldots h_{\tilde N_c}}} 
{(\Lambda_{\rm SQCD})^{b_1}}+{\rm h.c.}\Big{]}_{F}\, ,\nn 
\eeqn 
where $b_1$ is the first coefficient of the $\beta$-function of the 
theory. 
 
The trouble with this analysis is that neither the 't~Hooft anomaly 
conditions are fulfilled, if only the above set of composite operators 
is considered, nor the superpotential has the correct quantum numbers 
to fit the anomalous symmetries of the theory. 
 
An inspiring and physically compelling interpretation of the situation 
was given in~\cite{SEI}, where it was argued that the theory admits 
also a ``dual'' description in terms of a SQCD-like action with the 
same global ``flavour'' symmetries, hence with quark fields $Q^f$ and 
$\tilde Q_{f}$ ($f=1,2,\ldots,N_f$), but with gauge group $SU(\tilde 
N_c)$ with $\tilde N_c=N_f-N_c$. This conclusion follows from 
the observation that the moduli space of the theory remains 
unmodified quantum mechanically for all values of $N_f>N_c+1$, at 
least up to $N_f=3N_c$. In turn this means that the classical vacuum 
at the origin (where all the expectation values of the composite 
fields which represent the degrees of freedom of the low energy 
theory vanish) preserves the original $SU_L(N_f)\times 
SU_R(N_f)\times U_V(1)\times U_{\hat A}(1)$ symmetry. Consequently in 
the dual theory there must necessarily be $N_f$ quark fields, though 
coupled to a different gauge group. In this theory the 
operators~(\ref{BBB1}) and~(\ref{BBB2}) are interpreted as composite 
operators of the form 
\beqn 
&&B^{f_{1}\ldots f_{\tilde N_c}}=\epsilon^{r_1\ldots r_{\tilde N_c}} 
\tilde Q_{r_1}^{f_1} 
\tilde Q_{r_2}^{f_2}\ldots \tilde Q_{r_{\tilde N_c}}^{f_{\tilde N_c}} 
\, ,\label{BBBT1}\\ 
&&\tilde B_{f_{1}\ldots f_{\tilde N_c}}=\epsilon_{r_1\ldots r_{\tilde N_c}} 
Q^{r_1}_{f_1}Q^{r_2}_{f_2}\ldots Q^{r_{\tilde N_c}}_{f_{\tilde N_c}} 
\, .\label{BBBT2} 
\eeqn 
An additional chiral, gauge invariant, supermultiplet, $M_f^{h}$, is 
assumed to exist, which is necessary for matching the 't~Hooft anomaly 
conditions. In terms of the above composite fields an effective 
superpotential can be written down. It reads $Q^f M_f^{h}\tilde 
Q_{h}$. The relation between this theory and the original theory is 
referred to as ``non-abelian electric-magnetic duality'' (or more 
simply as ``Seiberg duality'') and indeed it can be argued to be a 
duality relation in the sense that the dual of the dual is the 
original theory with the quarks and gluons of one description 
interpreted as solitons (magnetic monopoles) of the other. 
 
Summarising, according to~\cite{SEI,REVET}, we can briefly describe 
what happens to SQCD when $N_f$ increases at fixed $N_c$ beyond 
$N_c+1$ as follows. 
 
- {}For $N_c+2\leq N_f < 3N_c/2$ the asymptotic particles of the 
theory are the the dual quark fields $Q^f$ and $\tilde Q_{f}$ and the 
mesons $M_f^{h}$, which interact through an IR-free ($b_1=3\tilde 
N_c-N_f=2N_f-3N_c<0$) supersymmetric theory with gauge group 
$SU(\tilde N_c)$. This is how the SQCD theory we started with looks in 
terms of ``magnetic'' variables dual to the original ``electric'' 
variables (which are instead strongly coupled in this range of $N_f$ 
values). 
 
- As soon $N_f$ goes through the value $3N_c/2$, the first coefficient 
of the dual theory $\beta$-function changes sign and the theory is 
expected to flow to a non-trivial IR fixed point. This continues to be 
true for the whole range of values $3N_c/2<N_f<3N_c$. Both the 
original and the dual theory can be argued to be conformal theories 
of interacting quarks and gluons (we remark that 
$3N_c/2<N_f<3N_c\Longleftrightarrow 3\tilde N_c/2 < N_f < 3\tilde 
N_c$). However, as $N_f$ increases the electric variables tend to 
become more and more weakly coupled and the opposite happens for the 
dual magnetic variables. 
 
- At $N_f=3N_c$, where the original theory loses asymptotic freedom, 
the IR fixed fixed point comes to zero coupling. 
 
- {}For even larger values of $N_f>3N_c$ the original electric theory is 
an IR free theory of quarks and gluons. 
 
\subsubsection{Normalising the SQCD and SYM effective action} 
\label{sec:NELSQCD} 
 
As we have seen, the interesting piece of the SYM and SQCD effective 
actions can be fixed by symmetry arguments only up to a constant 
rescaling of their $\Lambda$ parameter. We want to show in this 
section how, exploiting the self-consistency requirement implicit in 
the decoupling theorem, one can fix these constants, if at the same 
time a dynamical (e.g.\ instanton based) information is available. We 
will develop the argument along the line of reasoning advocated in 
refs.~\cite{ADSALL2,COR,FIPU} and summarised in~\cite{REVET}.\\ 
 
\noindent$\bullet$ {\it The case of SQCD} 
 
Starting from~(\ref{LESSQCD}) with just one massive flavour, say the 
$N_f$-th one, we require that the effective superpotential of the theory 
with $N_f$ flavours, namely 
\beqn 
&&{W}_{{\rm eff};N_c,N_f}^{\rm SQCD}(T)= \nn\\ 
&&=\Big{[}(N_c-N_f)\eta(N_f)\Big{(} 
\frac{(\Lambda_{\rm SQCD}^{(N_f)})^{3N_c-N_f}} 
{{\rm det}T}\Big{)}^{\frac{1}{(N_c-N_f)}}+m_{N_f} 
T^{N_f}_{N_f}\Big{]}_F\, , 
\label{NFSQCD} 
\eeqn 
goes over to the effective superpotential of the theory with one 
flavour less when $m_{N_f}$ gets large after using~(\ref{EQLAM}). 
To simplify and clarify notations we have introduced the new constant 
$\eta(N_f)=(c')^{3N_c-N_f/N_c-N_f}$ with respect to what we had 
in~(\ref{LESSQCD}) and we have attached the extra superscript $N_f$ 
to the $\Lambda$ parameter of SQCD in order to trace the number of 
``active'' flavours in each theory. 
 
We now proceed to eliminate $T^{N_f}_{N_f}$ by using the $F$-flatness 
condition for $T^{N_f}_{N_f}$, which amounts to the stationarity 
equation $\partial {W}_{{\rm eff};N_c,N_f}^{\rm SQCD}(T)/\partial 
T^{N_f}_{N_f}=0$. We also notice that the analogous conditions for the 
$T^{f}_{N_f}$ and $T^{N_f}_{h}$ components imply the vanishing of 
their expectation value. After some algebra one finds that the r.h.s.\ 
of~(\ref{NFSQCD}) becomes 
\beqn 
\hspace{-.5cm}\Big{[}\big{(}\eta(N_f)\big{)}^{\frac{N_c-N_f}{N_c-N_f+1}} 
(N_c-N_f+1)\Big{(}\frac{m_{N_f}(\Lambda_{\rm SQCD}^{(N_f)})^{3N_c-N_f}} 
{{\rm det}\tilde T}\Big{)}^{\frac{1}{(N_c-N_f+1)}}\Big{]}_F\, , 
\label{NFIN}\eeqn 
where ${\rm det}\tilde T$ is the matter determinant with the $N_f$-th 
flavour missing. Since the decoupling condition~(\ref{EQLAM}) implies 
\beq 
m_{N_f}(\Lambda_{\rm SQCD}^{(N_f)})^{3N_c-N_f}= 
(\Lambda_{\rm SQCD}^{(N_f-1)})^{3N_c-N_f+1}\, , 
\label{LNF} 
\eeq 
we see that the expression~(\ref{NFIN}) becomes the formula for the 
effective superpotential of SQCD with $N_f-1$ flavours if $\eta(N_f)$ 
satisfies the equation 
\beq 
\big{(}\eta(N_f)\big{)}^{\frac{N_c-N_f}{N_c-N_f+1}}=\eta(N_f-1)\, . 
\label{ZETA}\eeq 
The most general solution of~(\ref{ZETA}) is 
\beq 
\eta(N_f)=\eta_0^{\frac{1}{N_c-N_f}}\, , 
\label{ZETAS} 
\eeq 
with $\eta_0$ a quantity which does not depend on $N_f$. The last 
observation is rather important as it can be exploited to simplify the 
calculation of $\eta_0$. In practice one can proceed in two ways. One 
is based on an explicit dynamical computation which was done in the 
WCI approach with the result 
$\eta_0=1$~\cite{RUS2,ADSALL1,ADSALL2,FS}. The calculation is 
performed in the especially simple case of SQCD with $N_f=N_c-1$ 
flavours, where the $SU(N_c)$ gauge symmetry is completely broken by 
non-vanishing scalar v.e.v.'s (see Appendix~\ref{sec:APPE}). In this 
situation the theory is weakly coupled for sufficiently large 
v.e.v.'s, thus constrained~\cite{AFF} instanton calculations are 
expected to be fully reliable. 
 
The second strategy~\cite{FIPU} consists in determining $\eta_0$ by 
means of a self-consistency constraint that fixes the value of the 
gluino condensate in an $SU(2)_1\times SU(2)_2$ gauge theory with 
matter in the $({\bf 2},{\bf 2})$ representation. The argument, which 
is quite elegant, exploits the knowledge of the effective 
superpotential of the theory, derived in~\cite{ILS}, and confirms the 
result $\eta_0=1$.\\ 
 
\noindent$\bullet$ {\it The case of SYM} 
 
Already the result mentioned at the end of the previous section is 
telling us that the normalising constant, $c$, in~(\ref{WESYM}) is 
to be taken equal to one, at variance with the direct SCI calculation 
which, if a factor $N_c=2$ is divided out, gave $c=1/\sqrt{5}$ 
(see~(\ref{COND2})-(\ref{CNC}) and the discussion in 
Sect.~\ref{sec:DRSYM}). 
 
There are other similar indirect ways to determine $c$. An elegant 
one is to start from a pure ${\cal N}=2$ SYM theory with the addition 
of a mass term for the chiral (matter) superfield which breaks 
supersymmetry down to ${\cal N}=1$. Decoupling the massive multiplet 
by sending the mass parameter to infinity leaves behind a pure ${\cal 
N}=1$ SYM theory. The reason to reach ${\cal N}=1$ in this somewhat 
complicated way is that for the effective action of pure ${\cal N}=2$ 
SYM we have the beautiful SW formula~\cite{SW} and a simple 
description of the theory in terms of low energy degrees of freedom 
(see Sect.~\ref{sec:SWanalysis}). 
 
To illustrate the method we wish to present here a simplified 
adaptation of the original argument given in~\cite{KHO1}. The 
starting point is the formula 
\beq 
\langle g^2\lambda^{\alpha a}\lambda_\alpha^a \rangle = -{16\pi}i \, 
\frac{\partial W_{\rm eff}^{{\cal N}=1}}{\partial \tau}= \frac{32\pi^2}{b_1} 
\,\Lambda\,\frac{\partial W_{\rm eff}^{{\cal N}=1}}{\partial \Lambda}\, , 
\label{LLEFF} 
\eeq 
where for the $SU(2)$ gauge group $b_1= 3 \times 2$ and we have set 
\beqn 
&&\tau=\frac{4\pi i}{g^2}+\frac{\vartheta}{2\pi}\, ,\label{TAU}\\ 
&&\Lambda=\mu \er^{2\pi i \tau(\mu)}\, .\label{LAMND} 
\eeqn 
We need to compute $W_{\rm eff}^{{\cal N}=1}$ with its correct 
normalisation. In principle $W_{\rm eff}^{{\cal N}=1}$ could be obtained 
from~(\ref{WESYM}), after integrating out the $S$ superfield. 
Precisely because at this stage the normalisation of the ${\cal N}=1$ 
effective action is unknown, we shall start from the well defined 
expression of the ${\cal N}=2$ effective superpotential which in the 
relevant strong-coupling regime takes the form 
\beq 
W_{\rm eff}^{{\cal N}=2}(A_D,M,\bar M)=\sqrt{2}\bar{M} A_D M +mU(A_D)\, , 
\label{NDSPW} 
\eeq 
where the chiral superfields $M, \bar M$ describe the monopole multiplet 
and $A_D$ is the dual Higgs superfield. In this regime the quantum modulus, 
$U$, is naturally expressed in terms of $A_D$ (and not of $A$). Solving the 
$F$-flatness conditions leads to the v.e.v.'s 
\beq 
a_D\equiv \langle A_D\rangle=0\, ,\qquad \langle M\rangle= 
\langle \bar M\rangle=\Big{(}-\frac{m}{\sqrt{2}}\, 
\frac{\dr U(x)}{\dr x}\Big{|}_{x=0}\Big{)}^\half\, . 
\label{VEVS} 
\eeq 
In this vacuum configuration one finds 
\beqn 
&&W_{\rm eff}^{{\cal N}=2}(0,\langle M\rangle,\langle \bar M\rangle)= 
m\, u(0)+\ldots =\nn\\ 
&&=m \big{(}\Lambda_{\rm SW}\big{)}^2+\ldots= 
2m\big{(}\Lambda^{{\cal N}=2}_{\rm PV}\big{)}^2+\ldots\, , 
\label{NDSPV} 
\eeqn 
where $u(0)=\langle U(0)\rangle =\Lambda^2_{\rm SW}$ (as it follows 
from the known relation between $a_D$ and $u=\langle 
U\rangle$~\footnote{We recall the SW formula 
$a_D=\frac{\sqrt{2}}{\pi}\int_{\Lambda^2_{\rm SW}}^{u}\dr x 
\sqrt{\frac{{x-u}}{{x^2-\Lambda^4_{\rm SW}}}}$, see 
Sect.~\ref{sec:SWanalysis}.}) and $\Lambda_{\rm SW}$ is the SW dynamical 
scale. In the last equality for the purpose of comparing with the 
rest of our formulae we have introduced the more standard 
Pauli-Villars scale (which we consistently used throughout this 
review) related to the former by the relation~\cite{FIPU} 
$\Lambda^{{\cal N}=2}_{\rm PV}=\Lambda_{\rm SW}/\sqrt{2}$. 
 
The last step of this quite elaborate argument consists in decoupling 
the matter superfield by sending $m$ to infinity while keeping fixed 
the combination (see~(\ref{EQLAM}) and~(\ref{EXBN})) 
\beq 
\Lambda^6=m^2\,\big{(}\Lambda^{{\cal N}=2}_{\rm PV}\big{)}^4\, . 
\label{LAMLAM} 
\eeq 
Inserting this relation in the last equality of~(\ref{NDSPV}) gives 
\beq 
W_{\rm eff}^{{\cal N}=1}=2\Lambda^3\, , 
\label{WNUEF} 
\eeq 
from which the equation 
\beq 
\langle\frac{g^2}{32\pi^2}\lambda^{\alpha a}\lambda_\alpha^a 
\rangle=\Lambda^3 
\label{LLFIN} 
\eeq 
follows. This calculation again yields the so-called WCI result $c=1$. 
 
Although, as we have developed the argument, this computation is 
enough to fix the normalisation of the effective potential for any 
number of colours, it would be nice to repeat a similar reasoning in 
the generic case of an $SU(N_c)$ gauge group in order to explicitly 
check the $N_c$ behaviour of the gluino condensate. This issue is of 
relevance for the interesting question of relating non-supersymmetric 
QCD-like gauge theories with supersymmetric ones in the large $N_c$ 
limit as proposed in the nice papers of~\cite{ASV}. 
 
\subsection{The effective action of Georgi--Glashow type models} 
\label{sec:ELCT} 
 
A number of interesting results have been obtained in the 
literature~\cite{BGK1,KV} for supersymmetric theories with chiral 
matter. Here, for brevity, we will only discuss two specific cases 1) 
$SU(6)$ with two matter superfields in the ${\bf 6}$ and one in the 
${\bf \bar{15}}$ representation and 2) $SU(5)$ with one matter 
superfields in the ${\bf 5}$ and another one in the ${\bf 
\bar{10}}$ representation, as prototypes of two different typical 
situations, namely unbroken supersymmetry with well defined vacua and 
dynamically broken supersymmetry, respectively (see the corresponding 
discussion in Sect.~\ref{sec:ICCT}). 
 
\subsubsection{$SU(6)$ with matter in $2\times {\bf 6} +{\bf \bar{15}}$} 
\label{sec:SUS} 
 
The construction of the effective action of this theory requires, 
besides the chiral composite superfields (see~(\ref{CHIRMAT})) 
\beqn 
&&S=\frac{g^2}{32\pi^2}W^\alpha W_\alpha\, ,\quad T= 
\epsilon_{IJ}\Phi^I_r\Phi^J_sX^{rs}\, ,\label{SUSCSF1}\\ 
&&U=\epsilon_{r_1s_1r_2s_2r_3s_3}X^{r_1s_1} X^{r_2s_2} X^{r_3s_3}\, , 
\label{SUSCSF2} 
\eeqn 
the real (vector) ones 
\beq 
R^J_I=\Phi_I^\dagger \er^{2gV({\bf 6})} \Phi^J\, , 
\quad Q=X^\dagger \er^{2gV({\bf \bar{15}})} X\, .\label{SUSR} 
\eeq 
The expression of the effective action which fulfills all the relevant 
anomalous and non-anomalous WTI's of the microscopic theory 
reads~\cite{BGK1} 
\beqn 
&&\Gamma_{\rm eff}^{{\rm GG}-SU(6)}=\Big{[}{\rm Tr}R+Q+S^\star S+\xi_R 
{\rm Tr}\log R+\xi_Q\log Q+\nn\\ 
&&+ {\rm Tr}(T^\dagger R^{-1}T) ({\rm det}R)^{-1}Q^{-1}+U^\star 
U Q^{-3}\Big{]}_D+\nn\\ 
&&+\Big{[}{\rm Tr}W^2_R+W^2_Q+S\log 
\frac{S^3XT}{\Lambda_{\rm GG}^{15}}+hT+h'U+{\rm h.c.}\Big{]}_F\, . 
\label{SUSEA} 
\eeqn 
where $\xi_R,\xi_Q$ are (in principle) calculable constants, 
$\Lambda_{\rm GG}$ is the RGI scale parameter of the theory and 
\beq 
W_{R\alpha}=-\frac{1}{4}\bar{D}^2R^{-1}D_\alpha R\, , 
\qquad W_{Q\alpha}=-\frac{1}{4}\bar{D}^2Q^{-1}D_\alpha Q \, .
\label{WRWQ} 
\eeq 
Despite its quite complicated form, the consequences of~(\ref{SUSEA}) 
are rather simple. One gets for the v.e.v.'s of the composite 
operators~(\ref{SUSCSF1}) and~(\ref{SUSCSF2}) 
\beqn 
&&\langle S\rangle_k=\big{(}hh'\big{)}^{1/5}\big{(} 
\Lambda_{\rm GG}\big{)}^{3}\er^{2\pi i k/5} 
\, ,\quad k=1,2,\ldots,5\, ,\label{VEVSUSS}\\ 
&&\langle T \rangle_k= \frac{\langle S\rangle_k}{h}\, , 
\qquad \langle U\rangle_k=\frac{\langle S\rangle_k}{h'} \, ,\label{VEVSUS} 
\eeqn 
in perfect agreement with instanton results and the constraints 
imposed by the Konishi anomaly equations. One finds that the discrete 
$\Z_{15}$ symmetry group is spontaneously broken down to $\Z_3$, 
leaving behind 15/3=5 well defined supersymmetric vacua. As 
we noticed in Sect.~\ref{sec:ICCT} point (I), for non-vanishing value 
of the Yukawa couplings $h,h'$ supersymmetry is unbroken and the 
vacuum states are well defined. Only when either $h$ or $h'$ go to 
zero and flat directions appear in the superpotential, some of the 
condensates run away to infinity. 
 
\subsubsection{$SU(5)$ with matter in ${\bf 5}+{\bf \bar{10}}$} 
\label{sec:SUC} 
 
This case is more interesting as the phenomenon of dynamical breaking 
of supersymmetry is seen to occur~\cite{KV}. The construction of the 
effective action which fulfills all the anomalous and non-anomalous 
WTI's of the microscopic theory again requires the introduction of the 
two real composite (vector) superfields 
\beq 
R=\Phi^\dagger \er^{2gV({\bf 5})} \Phi\, , 
\quad Q=X^\dagger \er^{2gV({\bf \bar{10}})} X\, .\label{SUCR} 
\eeq 
{}Furthermore, besides the chiral composite operators~(\ref{SUSCSF1}) 
and~(\ref{SUSCSF2}), the chiral superfields 
\beq 
Y\!=\!W_s^{r_1}W_r^s\Phi_t X^{tr}X^{r_2r_3}X^{r_4r_5} 
\epsilon_{r_1r_2r_3r_4r_5}\, ,\quad 
A\!=\!\frac{g^2}{16\pi^2}(W^2)^r_s\Phi_r\Phi_{s'}X^{s's}\label{SUCTH} 
\eeq 
must come into play in order to fulfill the 't~Hooft anomaly 
conditions. Finally the requirement of the absence of flat directions 
in the microscopic theory implies a judicious choice of the invariant 
kinetic terms. An expression of the effective action which satisfies 
all the above constraints is 
\beqn 
&&\Gamma_{\rm eff}^{{\rm GG}-SU(5)}=\Big{[}R+Q+S^\star S+\xi_1 
\log R+\xi_2\log Q+\nn\\ 
&&+(Y^\star R^{-1}T Q^{-3}Y)^{-1}+A^\star RQA\Big{]}_{D}+\nn\\ 
&&+\Big{[}\kappa_1 W^2_R+\kappa_2 W^2_Q+ 
S\log\frac{S^2 Y}{\Lambda_{\rm GG}^{13}}+{\rm h.c.}\Big{]}_{F}\, . 
\label{SUCEA} 
\eeqn 
The minimisation of $\Gamma_{\rm eff}^{{\rm GG}-SU(5)}$, displays the 
phenomenon of dynamical breaking of supersymmetry. One finds, in fact, 
that the minimum occurs at finite non-vanishing values of all the 
condensates (with the exception of $A$ for which a vanishing result 
is obtained) and that at this point the effective superpotential is 
positive. 
 
It is interesting to look at the spectrum of the low-lying states that
emerges from the analysis of the effective
potential~(\ref{SUCEA}). Together with supersymmetry, a non-anomalous
$U(1)$ is spontaneously broken by the v.e.v.\ of $Y$. Another
anomalous $U(1)$ remains instead unbroken and its triangle anomaly is
saturated by the composite fermion in $A$, which remains massless.
The only other massless fermion in the spectrum is the Goldstino
associated with the spontaneous breaking of supersymmetry. The latter
partially lies in the real vector fields $R$ and $Q$. In the spin zero
sector we find the massless Goldstone boson of the spontaneously
broken $U(1)$ mentioned above. Two more would-be Goldstone bosons are
eaten up {\it \'a la} Higgs to give mass to the vector bosons
belonging to $R$ and $Q$. It is the fact that the Goldstino partially
lies in the real superfields $R$ and $Q$ that prevents integrating out
their massive degrees of freedom, because if one does so the manifest
supersymmetry of the effective action is lost. 
 
The overall picture that is coming out is completely consistent with 
the symmetry breaking pattern that emerges from the dynamical 
instanton computation of the Green functions with only insertions of 
lowest components of chiral composite superfields (see 
Sect.~\ref{sec:ICCT} point (III) ${1.}$).


\section {$\cN =2$ SYM: Introduction} 
\label{sec:N2-intro} 
 
As shown in the previous discussion of $\cN =1$ SYM theories, the 
combination of instanton calculus with holomorphy of the $F$-terms in 
the (low-energy) effective action proves to be very powerful in that 
it allows to determine non-perturbative corrections to the 
superpotential and argue for dynamical supersymmetry breaking in a 
class of models. Unfortunately the spectrum of bound-states in 
supersymmetric vacuum configurations, if present, depends not only on 
the $F$-terms, encompassing the superpotential and gauge kinetic 
terms, but also on $D$-terms, encoding the kinetic terms for chiral 
multiplets and their couplings to vector multiplets. $D$-terms are 
determined by the K\"ahler potential $K(\Phi, \Phi^\dagger, V)$, a 
real non-holomorphic ``function'' of the (light) chiral multiplets and 
the vector multiplets, that in principle receives both perturbative 
and non-perturbative corrections~\footnote{For this reason the 
``exact'' $\beta$-function of~\cite{RUS1} should be properly seen as 
an elegant way to hide one's ignorance of the anomalous dimensions 
$\gamma$ of chiral multiplets.}. 
 
The situation significantly improves for $\cN =2$ SYM theories, since
the extra supersymmetry relates what in the $\cN =1$ description would
be unrelated, \ie the K\"ahler potential, the superpotential and the
gauge kinetic function~\cite{N2susy}. This is true not only when $\cN
=2$ vector multiplets are present but also when one couples the
resulting $\cN =2$ SYM to ``matter'' fields belonging to so-called
hypermultiplets, or hypers for short~\footnote{In fact it can even
improve if the hypermultiplets belong to some special representation
of the gauge group, whereby the theory becomes exactly superconformal
and thus UV finite so that the two derivative effective action does
not receive any correction either perturbatively or non
perturbatively. For instance, this is the case when one extra
hypermultiplet is added that belongs to the adjoint representation,
leading to the $\cN = 4$ SYM theory~\cite{N4susy}.}. $\cN =2$
supersymmetry allows only ($\cN =2$) minimal couplings of hyper to
vector multiplets, coded in ``tri-holomorphic moment maps'', and the
hypers are known to have vanishing anomalous
dimensions~\cite{Zeq0hypers}.  The (low-energy) effective theory is
thus  determined by an analytic prepotential $\cF$, which only depends
on  the $\cN =2$ vector multiplets, and a choice of gauging of
tri-holomorphic isometries of the hyperk\"ahler manifold described by
the hypers~\cite{N2frerev}. Vector multiplets are ``chiral'' in the
$\cN =2$ description. In turn the analytic prepotential is known to
receive only one-loop and non-perturbative corrections. In their
seminal paper~\cite{SW}, Seiberg and Witten were able to determine the
exact form of $\cF$ for pure $\cN =2$ SYM with gauge group $SU(2)$ by
a series of elegant arguments based on electric-magnetic
duality~\cite{emdual}.  In a subsequent paper~\cite{seibwitt2} they
extended  their arguments to the case of $\cN =2$ SQCD with gauge
group $SU(2)$  that arise after minimal coupling of $N_f$
hypermultiplets belonging  to the pseudo-real fundamental
representation of $SU(2)$. Later on,  these results have been
generalised to other gauge groups with hypers  in various
representations both in the Coulomb branch, corresponding  to turning
on v.e.v's of scalars in vector multiplets thus preserving  the rank
of the gauge group, and in the Higgs branch, corresponding to  turning
on v.e.v's of scalars in hypers thus generically reducing the  rank of
the gauge group. The possibility of having new and mixed  branches has
also been widely explored. 
 
Our aim is to first describe the structure of $\cN =2$ SYM theories
both at the microscopic level and at the macroscopic one, when they
are described in terms of Wilsonian low-energy effective actions. We
then review the arguments of Seiberg and Witten leading to the
identification of an auxiliary Riemann surface, \ie a ``complex
curve'', encoding the complexified gauge coupling $\tau$ in (the
ratio of) the derivatives of its two ``period'' integrals, eventually
arriving at the determination of $\cF$ in the simplest case of
$SU(2)$. We then discuss the non linear recursion relations satisfied
by the coefficients of the instanton expansion following the work of
Matone's~\cite{matone} and check the consistency of Matone's
relations and, thus, of the SW prepotential, with instanton calculus
in sectors with $K=1,2$~\cite{fuctrav}. In order to tackle the
general case, \ie arbitrary $K$ and generic gauge group (with
$SU(N_c)$ in mind), one may exploit the ``topological twist'' of $\cN
=2$ SYM theories~\cite{N2twist} that, combined with some
``non-commutativity'' parameter~\cite{seibwitt3}, in the form of the
so-called $\Omega$-background, allowed Nekrasov and
collaborators~\cite{nekra1,mns,nekra2,nekra3} to localise the
integrals over  instanton moduli spaces and compute recursively the
expansion  coefficients of the non-perturbative series. To this end we
sketch  these beautiful, but at the same time rather technical,
mathematical  arguments underlying the ADHM construction. We then turn
to the  string description of the ADHM construction and its
ramifications~\cite{douglas,witten}.  The astonishing feature of
string theory is  that the sophisticated algebro-geometric ADHM
construction becomes  rather transparent and intuitive once D-branes
and their open string  excitations are taken into
account~\cite{billo1,billo2}. In  particular, (supersymmetric) gauge
theories emerge as the low-energy  effective theories governing the
dynamics of stacks of D-branes~\cite{branetogaugerev}.  In this
setting instantons can be realised as  lower dimensional D-branes
within higher dimensional ones~\cite{douglas,witten}.  The structure
of the ADHM data emerge naturally  from the set of open strings
connecting the various stacks of  branes. Even Nekrasov's
$\Omega$-background admits a natural  description in terms of the
closed string graviphoton that couples to  D-branes and their open
string excitations~\cite{billo1,billo2}. Last  but not least, the long
sought for duality between gauge fields and  strings turns out to
emerge quite naturally, at least in the  maximally supersymmetric case
($\cN = 4$ SYM), in the form of  Maldacena's holographic
correspondence~\cite{malda,GKP,Witthol,adscftrevs,MBrev}.  We will
return to this  unprecedented achievement of string theory in
Sects.~\ref{sec:adscft-intro}-\ref{sec:inst-adscft}. Here we only
give a schematic description of how instanton effects can be computed
within string theory in a particular double scaling
limit~\cite{billo1,billo2}.

\section {$\cN =2$ SYM: generalities} 
\label{sec:N2-gen} 
 
$\cN =2$ SYM theories admit two kinds of massless multiplets, both 
containing four bosonic and as many fermionic degrees of freedom. 
Vector multiplets are described by chiral $\cN =2$ superfields that 
comprise a vector boson, two Weyl fermions (the gauginos) and a 
complex scalar all in the adjoint representation of the gauge 
group. $\cN=2$ vector superfields will be denoted by $A$ and their 
$\theta$ expansion schematically reads 
\be 
A(x,\theta) = a(x) + \theta^r_\alpha \lambda^\alpha_r(x) + 
{1\over 2} \theta^r_\alpha 
\sigma^{\mu\nu \alpha}{}_\beta \theta^\beta_r F_{\mu\nu}(x) + ... 
\,. \label{N2superf} 
\ee 
Higher order terms in $\theta^r$ with $r=1,2$ can be expressed as 
derivatives of the lower ones. In terms of $\cN =1$ supersymmetry, a 
$\cN =2$ vector multiplet can be decomposed into a vector superfield 
$V$ and a chiral superfield $\Phi$, both in the adjoint representation 
of the gauge group (see Appendix~\ref{sec:N2susygen} for notation). 
 
$\cN =2$ massless ``matter'' appears in hypermultiplets that consist 
of four real scalars and two Weyl fermions (the hyperinos), belonging 
to a real representation of the gauge group. In terms of $\cN =1$ 
supersymmetry, they can be decomposed into a chiral superfield $Q$ in 
an {\it a priori} complex representation ${\bf R }$ of the gauge group 
and a chiral superfield $\tilde{Q}$ in the conjugate representation 
${\bf {\bar R}}$. Among the massive representations, a special role is 
played by the 1/2 BPS representations that are shorter than generic 
massive representations in that they only involve eight bosonic and as 
many fermionic degrees of freedom (see Appendix~\ref{sec:BPSconf} for 
a brief explanation of this and related notions). Thanks to the 
relation between mass and ``central'' charge 
\be 
M = |Z| \, , 
\ee 
1/2 BPS states are indeed annihilated by half of the supersymmetry 
charges. 
 
The structure of classical $\cN =2$ SYM theories is tightly 
constrained by the large amount of (super)symmetry they are 
endowed with. The most general two-derivative classical action is 
completely determined in terms of an ``analytic'' prepotential 
$\cF$ that is {\it a priori} an analytic function of the $\cN =2$ 
vector multiplets $A$ and the complex coupling constant 
\be 
\tau = {\vartheta \over 2 \pi} + {4\pi i \over g^2} \, . 
\ee 
The $\cN =2$ 
hypermultiplet dynamics is described by a non linear $\sigma$ 
model on a hyperk\"ahler space (see Appendix~\ref{sec:diffgeo}). 
The coupling of $\cN =2$ hypermultiplets to vector multiplets is 
minimal in that the vector fields ``gauge'' (make local) the 
global hyperk\"ahler isometries of the hypermultiplet metric that 
preserve the three K\"ahler structures 
\be 
\omega^I = {1\over 2} \,\omega^I_{ij}\, \dr q^i \wedge \dr q^j \, , 
\ee 
where $\omega^I_{ij} = - \omega^I_{ji}$ with $I=1,2,3$ and $i, j 
=1,...,4n_{_H}$ are antisymmetric tensors such that $\dr\omega^I = 0$, 
where $\dr$ denotes the exterior differential in field space, \ie with 
respect to the scalar components $q^i$ of the $n_{_H}$ 
hypermultiplets. In the simple case of constant $\omega^I_{ij}$, 
writing $i = f + 4 r$ with $f=1,...,4$ and $r=0,...,n_{_H}-1$, one can 
choose 
\be 
\omega^I_{f + 4 r, f' + 4 r'} = \eta^I_{f f'} 
\delta_{r r'} \, , 
\ee 
where $\eta^I_{f f'}$ are the 't Hooft symbols~\cite{GTH}. 
 
The effect of ``gauging'' (hyperk\"ahler) isometries can be elegantly 
expressed through the minimal substitution 
\be 
\de_\mu q^i \rightarrow D_\mu q^i = \de_\mu q^i + 
g A_\mu^a \xi^i_a(q) \, , 
\label{mincoup} 
\ee 
where $a=1,...,n_{_V}$. A tri-holomorphic isometry generated by the 
vector field $\xi_a = \xi^i_a \partial / \partial q^i$ satisfies 
\be 
\cL_\xi \omega^I \equiv \iota_{\xi_a} 
\dr\omega^I + \dr (\iota_{\xi_a} \omega^I) = \dr (\iota_{\xi_a} 
\omega^I) = 0 \, , 
\ee 
where $\iota_{\xi_a}$ denotes contraction with the Killing vector 
field $\xi_a(q)$. As a consequence $\iota_{\xi_a} \omega^I = \dr 
\mu_a^I$ a tri-holomorphic Killing vector $\xi_a(q)$ admits 
hyperk\"ahler moment maps $\mu^I_a(q)$ since locally 
$\xi^i_a(q)\omega^I_{ij}(q) = \de_j \mu^I_a(q)$. The $\mu^I_a(q)$ may 
be thought of as some sort of $\cN =2$ auxiliary fields. In the $\cN 
=1$ notation, whereby a hypermultiplet with scalar components $q^f$ is 
described by two chiral multiplets with scalar components $\phi = q^1 
+ i q^2$ and $\tilde\phi = q^3 + i q^4$, one has 
\bea 
&& \mu^3_a(q) = D_a(\phi,\tilde\phi; 
\phi^\dagger,\tilde\phi^\dagger)\nn \\ 
&&\mu^+_a(q)= \mu^1_a(q) + i \mu^2_a(q) = F_a(\phi,\tilde\phi)\nn 
\\ &&\mu^-_a(q)= \mu^1_a(q) - i \mu^2_a(q) = 
\bar{F}_a(\phi^\dagger,\tilde\phi^\dagger)\, . 
\eea 
Indeed the contribution of the hypermultiplets to the potential is 
exactly given by 
\be 
V_H(q) = {1\over 2} \delta_{IJ}\im \tau_a(a) 
\mu^I_a(q)\mu^J_a(q) \, . \label{hyppot} 
\ee 
Notice that except for the minimal coupling~(\ref{mincoup}) and its 
$\cN =2$ completion, entailing~(\ref{hyppot}) and various Yukawa type 
interactions, there is no neutral coupling between the vector and the 
hyper multiplets. In particular, as indicated, the complexified gauge 
couplings $\tau_a(a)$ can only depend on the lowest scalar components 
$a$ of the $n_{_V}$ vector multiplets $A$. 
 
Quantum renormalisability drastically restricts the choice of $\cF(A)$ 
and $\xi(q)$, or equivalently of $\mu(q)$. In the microscopic 
fundamental theory, $\cF(A)$ is at most a quadratic function of $A$, 
while $\xi^i_a(q)$ are linear in the $q$'s, viz. 
\be 
\xi^i_a(q) = (T_a)^i{}_j q^j \, , 
\ee 
where $T_a$ are the generators of the gauge group in the ({\it a 
priori} reducible) representation spanned by the scalars in the 
hypermultiplets. Moreover the hyperk\"ahler metric is flat, up to 
global tri-holomorphic identifications $\R^{4n}/\Gamma$ (examples are 
the ALE spaces $\R^{4}/\Gamma_{ADE}$ where $n=1$ and $\Gamma_{ADE}$ is 
one of the Kleinian discrete subgroups of $SU(2)$, in the ADE 
classification, see \eg~\cite{BFMR}). As a result the tri-holomorphic 
moment maps $\mu^I_a(q)$ are completely determined in case of 
semi-simple gauge groups. When abelian factors are present in the 
gauge group, one can add constant tri-holomorphic Fayet-Iliopoulos 
terms $\zeta^I_a$, so that $\mu^I_a(q) = \hat\mu^I_a(q) + \zeta^I_a$, 
where $\hat\mu^I_a(q)$ is such that $\hat\mu^I_a(q=0)=0$. 
 
A different story applies to the Wilsonian effective 
action~\footnote{M.B. would like to thank M. Bochicchio for first 
pointing out the important difference between the ``non-local'' 1PI 
effective action, with an arbitrary number of derivatives, and the 
Wilsonian low-energy effective action, often considered only up to 
two derivatives.} for the light (massless) modes that survive, \ie do 
not acquire a mass after, partial or complete gauge symmetry breaking 
below the scale $\Lambda$~\cite{WP}. Here $\Lambda$ is the explicit 
cutoff in the Wilsonian effective action, such that all modes with 
mass or energy above this scale have been integrated out. It is not 
known how to explicitly perform this task, but the outcome of the 
``integrating out'' procedure is severely constrained by symmetries 
and one can often ``guess'' the correct result to lowest order 
approximation, which is nothing else but a book-keeping of the 
relevant degrees of freedom and the symmetries of the theory. 
 
In addition to the Coleman-Weinberg~\cite{CW} type logarithmic 
correction at one-loop 
\be 
\cF_{1-{\rm loop}}(A) = 
{i\over 8} b_1 A^2 \log{A^2 \over \Lambda^2} \, , 
\label{calFpert} 
\ee 
where $b_1 $ is the $ \beta$ function coefficient, \ie $b_1 = 2 N_c - 
N_f$ for $SU(N_c)$ with $N_f$ hypers in the fundamental 
representation, the prepotential can and in fact must acquire an 
infinite number of non-perturbative corrections. Indeed $\cN =2$ 
supersymmetry prevents further perturbative corrections, but the 
one-loop term violates positivity of the imaginary part of the 
effective gauge coupling 
\be 
\tau(a) = {\de^2 \cF (a) \over \de a^2} = {\vartheta (a) \over 
2 \pi} + {4\pi i \over g^2(a)} \, , 
\ee 
where $a$ denotes the lowest (scalar) component of the chiral 
superfield $A$ that describes the $\cN =2$ vector multiplet and has 
been defined in~(\ref{N2superf}).

\section {Seiberg--Witten analysis} 
\label{sec:SWanalysis} 
 
In their seminal paper~\cite{SW}, Seiberg and Witten have shown how 
one can exactly compute the analytic prepotential $\cF (A)$ in the 
case of an $SU(2)$ gauge theory without hypermultiplets. In another 
closely related paper~\cite{seibwitt2} they have shown how to 
incorporate $2 N_f$ half-hypermultiplets in the fundamental 
representation of $SU(2)$, leading to a theory that deserves to be 
called $\cN =2$ SQCD. The case $N_f = 2 N_c = 4$ is special since it 
corresponds to an exact quantum $\cN = 2$ superconformal theory. 
 
Clearly, in the Coulomb phase we are focussing on, higher derivative 
terms are generated by quantum effects with finite coefficients and 
are suppressed by inverse powers of the v.e.v.'s of the scalar 
fields. In the superconformal phase with vanishing v.e.v.'s the 
situation is much subtler. The relevant observables are correlation 
functions of gauge invariant operators. The modern tool to tackle 
this interesting issue is the AdS/CFT correspondence proposed by 
Maldacena that will be discussed in 
Sects.~\ref{sec:adscft-intro}-\ref{sec:inst-adscft}. 
 
After briefly reviewing the arguments of Seiberg and Witten's, based 
on symmetries, monodromies and duality, we will describe how to check 
the result by (constrained) instanton calculus. 
 
As we already mentioned, classical $\cN =2$ SYM admits an infinite 
tower of BPS saturated monopoles and dyons thanks to the existence 
of a complex scalar central charge $Z$ in the $\cN =2$ extended 
superalgebra. In the simple case of $SU(2)$, they are the 
supersymmetric analogue of the classical solutions found by 't 
Hooft~\cite{thooftmonop} and Polyakov~\cite{polyamonop} and by 
Julia and Zee~\cite{dyons}. Indeed, we recall that the potential 
of pure $\cN =2$ SYM, which reads 
\be 
V(a, a^{\dagger}) = {1 \over 2} \,\Tr([a,a^{\dagger}]^2) \, , 
\ee 
has flat directions identified by the condition 
$[a,a^{\dagger}]=0$. Up to gauge transformations, this means that both 
$a$ and $a^{\dagger}$ belong to the Cartan subalgebra generated by \eg 
$T_3 = \sigma_3/2$. In modern terms one says the theory admits a one 
complex dimensional moduli space of classical vacua parametrised by 
$a_3$ or rather by the gauge invariant composite 
\be 
u = \Tr(a^2) = {1 \over 2}a^2_3 \, . 
\ee 
Henceforth we denote $a_3$ by $a$ for simplicity. Along the flat 
direction the gauge group is broken to $U(1)$ and one automatically 
realises the Prasad--Sommerfield condition~\footnote{ The 
Prasad--Sommerfield condition of non vanishing scalar v.e.v.\ with 
zero potential is usually achieved by setting the scalar self-coupling 
$\lambda$ to zero in the potential $V(\varphi) = \lambda (\varphi^2 - 
\varphi_0^2)^2$ while keeping $|\varphi| = |\varphi_0|$ at infinity.} 
without the need of sending the scalar self-coupling to zero~\cite{PraSomm}. 
Monopole solutions that saturate the Bogomol'nyi bound 
\be 
M_M = |p \tau_0 a| \, , 
\ee 
where $\tau_0$ denotes the classical ``complexified'' coupling, 
that we have already encountered many times by now, and $p$ is the 
magnetic charge, can be explicitly constructed solving Nahm's 
equations~\cite{Nahm}. Notice the striking analogy with the Higgs 
formula for the mass of a $U(1)$ $W$-like boson with charge $q$ 
 
\be 
M_{_W} = |q_e a| \, . 
\ee 
In fact one can do better and show that 1/2 BPS saturated dyons have a 
mass spectrum given by the formula 
\be 
M_{_D} = |q_e a + q_m a_{_D}| = |Z| \, , 
\label{bpsdyons} 
\ee 
where we have introduced the notation $a_{_D} = \tau_0 a$ and $Z$ is 
the ``central'' extension of the $\cN = 2$ superalgebra, that being 
central, \ie commuting with all the remaining generators, has to 
be a c-number by Schur's Lemma~\cite{OliWitt,Osb}. In terms of the 
analytic prepotential $\cF(a) = \tau_0 a^2/2 + \cdots$ one is led to 
the identification 
\be 
a_{_D} = {\de \cF \over \de a} = \tau_0 a + \cdots \, , 
\ee 
where the dots take into account quantum corrections to $\cF(a)$ and 
thus to $\tau(a) = \tau_0 + \cdots$. The exact (quantum) 
identification $ a_{_D} = {\de \cF \over \de a}$ is tantamount to 
assuming that the classical formula~(\ref{bpsdyons}) for the central 
charge retains the same form in the quantum theory, as strongly 
suggested by consideration of $\cN = 2$ supersymmetry. Actually, the 
formula $|Z| = |q_e a + q_m a_{_D}|$ displays a remarkable symmetry 
under $SL(2,\Z)$ transformations acting on the electric and magnetic 
charges $q$ and $p$. In fact under 
\be 
q_e\rightarrow k q_e + l q_m \qquad q_m \rightarrow m q_e + n q_m 
\ee 
$Z$ is invariant if one simultaneously performs a ``monodromy'' 
transformation 
\be 
a \rightarrow n a - l a_{_D} \qquad a_{_D} \rightarrow 
- m a + k a_{_D} \, . 
\ee 
In this way $a, a_{_D}$ are seen as components of a section of an 
$SL(2,\Z)$ bundle over the moduli space of vacua parametrised by the 
gauge invariant composite $u$, for which we write $a=a(u), 
a_{_D}=a_{_D}(u)$. This geometrical description implies that the 
components $a=a(u)$ and $a_{_D}=a_{_D}(u)$ undergo non trivial 
transformations, \ie acquire non trivial monodromy, when one parallel 
transports them as functions of $u$ around some special points. As a 
result of their dependence of $a=a(u), a_{_D}=a_{_D}(u)$ on $u$, the 
complexified coupling that has been so far considered a function of 
$a$ can be considered a function of $u$ given by the ratio of the 
derivatives of $a_{_D}(u)$ and $a(u)$ through the chain rule 
\be 
\tau(a)= 
\tau(a(u))= \tau(u) = {\de^2 \cF (a) \over \de a^2} = {\de \over 
\de a} {\de \cF (a) \over \de a} = {\de a_{_D} \over \de a} = 
{{\dr a_{_D} \over \dr u} \over {\dr a \over \dr u}} \, , 
\ee 
with $\im\tau (u) > 0$ (for vacuum stability). Remarkably, at this 
point, the complexified effective coupling $\tau(u)$ can be considered 
as the modular parameter (the ``period'') of an auxiliary torus, a 
Riemann surface of genus one. The latter is also known as an 
``elliptic curve'', \ie a complex dimension one manifold whose periods 
are determined in terms of elliptic integrals. In fact determining 
this auxiliary elliptic curve, the so-called ``Seiberg-Witten 
curve'', allows one to compute its periods from the equations 
\be 
a_{_D}'(u)={\dr a_{_D} \over \dr u}\, , 
\quad a'(u)={\dr a_{_D} \over \dr u} 
\ee 
and, after integration w.r.t. $u$, $\cF (a)$ itself, since $a_{_D} = 
{\de \cF (a) \over \de a}$. 
 
In order to determine the SW curve one starts by computing the 
monodromy of the section ($a=a(u), a_{_D}=a_{_D}(u)$) at infinity 
where the theory, being asymptotically free ($b_1 = 2N_c = 4$ for 
$SU(2)$), is weakly coupled and (see~(\ref{calFpert})) 
\be 
\cF (a) \approx {i \over 2} a^2 \log {a^2 \over \Lambda^2} 
\ee 
with $\Lambda$ the RGI invariant scale 
\be 
\Lambda = M \exp(-8\pi^2 / b_1 g^2(M)) \, . \label{RGLam} 
\ee 
In this way one gets 
\be 
a_{_D} = \cF'(a) \approx {i \over 2} [ 2 a 
\log {a^2 \over \Lambda^2} + 2 a] \, . 
\ee 
Under $u\rightarrow \er^{2 \pi i } u$, one has $ a\rightarrow - a$ and 
$a_{_D} \rightarrow - a_{_D} + 2 a $. These considerations fix the 
monodromy of the section ($a=a(u), a_{_D}=a_{_D}(u)$) around the 
branch point at infinity. 
 
Perturbatively the other branch point of $\cF (a)$ is at $a=0$. If 
this were the full story, the theory would be inconsistent since $\im 
\tau$ could not possibly be positive throughout the moduli space, 
being $\tau$ holomorphic and thus $\im\tau$ harmonic. Seiberg and 
Witten argued that the non-abelian symmetry ($a=0$) is never restored 
at the quantum level and that this is consistent with assuming the 
existence of only two more singular points. They interpreted the 
singularities as due to the fact that some massive states become 
massless at each of the two additional singular points in the moduli 
space. In fact the two relevant states are a monopole $(q_e=0, q_m=1)$ 
and a dyon $(q_e=1, q_m=-1)$~\footnote{This particular choices of 
electric and magnetic charges simplify the notation in the 
following. Other choices are possible but require performing 
$SL(2,\Z)$ transformations}. In order to identify the location of 
these extra singularities, it is crucial to exploit a discrete $\Z_4$ 
symmetry of the quantum theory for $N_c = 2$ which is a remnant of the 
anomalous $U(1)_R$ subgroup of the $U(2)$ R-symmetry of the classical 
theory. Indeed classically $\cN = 2$ SYM is invariant under global 
$SU(2)\times U(1)_R$ transformations under which the gauge field is 
invariant, the gaugini rotate as a ${\bf 2}_{1/2}$ and the complex 
boson is a charge $+1$ $SU(2)$ singlet. The $U(1)_R$ symmetry is 
broken by the quantum anomaly that preserves a $\Z_{4 N_c} \approx 
\Z_{2 N_c} \times \Z_{2}$ where the latter factor is fermion parity 
and the former is the above mentioned $\Z_4$ under which $u 
\rightarrow - u$. This is enough to completely determine the SW curve 
\be 
\cE_{\rm SW}: \quad y^2 = (x^2 - \Lambda^2) (x^2 - u^2) \, , 
\label{SWcurve} 
\ee 
which is indeed singular when $u =\pm \Lambda$. A generic elliptic 
curve can be written as a double cover of the sphere $y^2 = 
(x-x_1)(x-x_2)(x-x_3)(x-x_4)$ with branch points at $x=x_i$. By the 
$SL(2,\C)$ symmetry of the sphere one can always put three of the 
branch points at, say, $0,1$ and $\infty$ so that the remaining 
complex parameter determines the shape of the torus (actually the 
ratio of the two periods). In order not to spoil the $\Z_2$ symmetry 
of quantum $\cN = 2$ SYM theory, it is however more convenient to fix 
only two branch points at, say, $\pm \Lambda$ (or $\pm 1$ after 
rescaling the variables). The remaining two branch points are set at 
$\pm u$. When $u$ reaches $\pm \Lambda$ the curve~(\ref{SWcurve}) 
representing the torus degenerates, \ie one of the two cycles and the 
corresponding period become zero signalling the presence of a 
singularity in the theory. 
 
The periods of $\cE_{\rm SW}$ can be expressed in terms of elliptic 
integrals and after identifying the cycles that correspond to 
$a_{_D}'(u)$ and $a'(u)$ one can eventually compute $\cF$. 
 
Making more precise the above geometrical considerations, we can say 
that the vector $(a_{_D}, a)$ is a section of a flat bundle over the 
moduli space parametrised by $u$ with monodromy group $\Gamma(2) 
\subset SL(2,\Z)$ generated by 
\be 
M_{-1} = \left(\begin{array}{cc} -1 & 2 \\ -2 & 3 
\end{array}\right) \qquad M_1 = \left(\begin{array}{cc} 
1 & 0 \\ -2 & 1 \end{array}\right) \qquad M_{\infty} = 
\left(\begin{array}{cc} -1 & 2 \\ 0 & -1 \end{array}\right) 
\label{monodmat} 
\ee 
such that $M_1 M_{-1} = M_{\infty}$. It can be checked that the 
monodromy transformations of $(a_{_D}, a)$ around $\pm \Lambda$ are 
indeed represented by the matrices $M_{\pm 1}$. 
 
If the modular parameter $\tau$ of $\cE_{\rm SW}$ were the ratio 
$a_{_D}/ a$ we would have completed our task, since $a_{_D}$ and 
$a$ would have coincided with the two ``canonical'' periods of the 
unique holomorphic~\footnote{Although it does not look holomorphic, 
$\omega$ is in fact holomorphic!} differential $\dr\omega = \dr 
x/y$. However $\tau = a_{_D}'(u)/ a'(u)$ and this means that 
$a_{_D}$ and $a$ are periods of a meromorphic differential whose 
derivative w.r.t. $u$ is the unique holomorphic differential 
$\dr\omega$. Seiberg and Witten identified the ``natural'' 
meromorphic differential $\dr\lambda(u)$ that prior to monodromy 
considerations is only determined up to a meromorphic differential 
independent of $u$. Setting $\Lambda = 1$ for simplicity, one 
finds that 
\be 
\dr\lambda(u) = {\sqrt{x-u}\over \sqrt{x^2 -1 }} \,\dr x 
\ee 
is the unique differential that satisfies all the requirements, \ie 
$\dr\lambda'(u) = \dr\omega$ and it is such that 
\be 
a_{_D}(u) = {\sqrt{2}\over \pi } \int^u_1 {\sqrt{x-u}\over 
\sqrt{x^2 -1 } } \,\dr x \, , \quad a(u) = {\sqrt{2}\over \pi } 
\int^1_{-1 } {\sqrt{x-u}\over \sqrt{x^2 -1 }} \,\dr x 
\ee 
have the correct monodromy when parallel transported around $\pm 1$ 
and $\infty$. Consistently with the surviving $\Z_4$ symmetry, one can 
write 
\be 
\cF (a) = \cF_{\rm pert} (a) + \cF_{\rm non-pert} (a) \,, 
\label{calFtot} 
\ee 
where 
\be 
\cF_{\rm pert} (a) = \cF_{\rm 
tree} (a) + \cF_{\rm 1-loop} (a) = {i \over 2} a^2 \log {a^2 \over 
\Lambda^2} \label{calFpert2} 
\ee 
and 
\be 
\cF_{\rm non-pert} (a) = 
a^2 \sum_{K=1}^\infty \cF_{_K} {\Lambda^{4K} \over a^{4K}} \, , 
\label{calFnonpert} 
\ee 
with the latter incorporating the contribution of instantons of 
increasing winding number $K$. 
 
A few comments are in order here. First $a=0$ is excised from the 
moduli space, \ie there is no value of $u$ such that $a(u)=0$. Second 
the singular points $u =\pm \Lambda$ correspond to $a_{_D}(u) = 0$ 
and $a_{_D}(u) = a$, respectively, and lie on the so called surface 
of marginal stability where $\im\tau=0$. This is the locus where the 
lattice of BPS states collapses and transitions of the form $|Z=Z_1 + 
Z_2\rangle \rightarrow |Z_1\rangle + |Z_2\rangle$ are allowed by both 
charge and mass conservation. 
 
The effective coupling $g^2(a) = 4\pi^2/\im\tau(u)$ is always 
semi-positive definite and never grows too much. At large $a$ this 
is due to asymptotic freedom. In the interior of the moduli space 
all charged vector bosons become extremely massive and the theory 
is essentially abelian. Near the singular points, one better 
switches to a dual magnetic or dyonic description whereby the 
abelian magnetic or dynonic photons are coupled to light monopoles 
and dyons. The effective coupling decreases with the 
renormalisation scale $\mu$ in the IR until it reaches the value 
\be \tilde{g}^2(\mu)\vert_{\mu = m} \approx {1 \over |\log 
(m/\Lambda)|} \, , \ee with $m$ the mass of the lightest charged 
state, be it a monopole or a dyon.  As stressed above, these 
states are arranged in hypermultiplets. Due to the presence of 
light charged particles in hypermultiplets coupled to abelian 
vector multiplets, the dual magnetic or dyonic theory is different 
from the original electric theory, that only involved non-abelian 
vector multiples. Yet electric-magnetic duality led us for quite a 
long way. Only for $\cN =4$ SYM and for other exactly 
(super)conformal invariant theories the dual magnetic or dyonic 
theory is expected to coincide with the electric 
theory~\footnote{More precisely in these cases the duality maps 
the electric theory into a magnetic theory with the same action 
but dual gauge group~\cite{gno}, $G^*$. The latter is obtained 
from the original gauge group of the electric theory, $G$, 
exchanging the role of the weight and root lattices. Therefore in 
the case of groups with simply-laced Lie algebras $G$ and $G^*$ 
are isomorphic. For groups with non simply-laced algebras this is 
not the case and one has the following pairs, $G\leftrightarrow 
G^*$: $SO(2n+1)\leftrightarrow Sp(n)$, $F_4\leftrightarrow 
F_4^\pp$, $G_2\leftrightarrow G_2^\pp$.}. 
 
Finally, at the singular points $u=\pm \Lambda^2$ new branches of 
the moduli space open up where monopoles or dyons can condense, 
\ie acquire a v.e.v., thus inducing (oblique) confinement of the 
chromo-electric charges and flux tubes due to the dual Meissner 
effect~\cite{dualMeissconf}. Adding $\cN =1$ supersymmetric mass 
terms to the adjoint chiral multiplet induces dynamical chiral 
symmetry breaking and confinement in a controllable way.

\section{Checking the SW formula by instanton calculations} 
\label{sec:SWcheck} 
 
Our next task is to check the SW prepotential against explicit
instanton computations. A one instanton check is not enough because
of ambiguities in the definition of $\Lambda$ that can rescale it by
a finite constant. The perspective for a two instanton check seems
{\it a priori } daunting but the calculation turns out to be
feasible~\cite{fuctrav,FIPU}. In fact one can do much better. Matone
has shown that the coefficients of the SW prepotential satisfy non
linear recurrence relations that can be checked to hold in instanton
calculus~\cite{matone}. Fucito and Travaglini have shown that
multi-instanton calculus precisely reproduces the desired
relations~\cite{fuctrav}. More recently the problem has been attacked
once again in a beautiful series of papers by Nekrasov and
collaborators~\cite{nekra1,mns,nekra2,nekra3}. Introducing suitable
deformations parameters ($\Omega$ background) one can localise the
measure over the multi-instanton (super)moduli space reducing the
calculation of the $\cF{_K}$ coefficients to a mere, though certainly
not trivial, combinatorial problem. We will have limited space to
discuss this fascinating issue and we refer the reader to the original
literature as well as to the accessible
reviews~\cite{nekra1,mns,nekra2,nekra3}.  We cannot resist saying
immediately  that the somewhat obscure deformation parameters
introduced by  Nekrasov admit a very natural explanation in a string
setting for the  problem whereby open string excitations of D-branes
account for the  gauge and matter light degrees of
freedom~\cite{douglas,witten,billo1}.  The $\Omega$-deformation is
equivalent  to turning on a background for a closed string state in
the so-called  ``Ramond-Ramond'' (R--R) sector, the graviphoton,
effectively  producing a non (anti-)commutative
superspace~\cite{billo2}, \ie a  superspace where the $\theta$
variables do not anti-commute very much  like the $x$ variables do not
commute. From this vantage point,  higher order terms in the
deformation, receiving instanton  corrections, are associated with
higher derivative gravitational  $F$-terms that appear in the type II
low energy effective actions  after compactification on Calabi-Yau
threefolds~\cite{gravFterms}.  A similar approach for the calculation
of the SW prepotential, based  on localisation on the instanton moduli
space, was proposed in~\cite{th1}.

\subsection{Matone relations} 
\label{sec:matrels}

Exploiting powerful results in the theory of uniformisation of 
Riemann surfaces, it was shown in~\cite{matone} that the non 
perturbative coefficients in the expansion of $\cF(a)$ satisfy 
certain recursion relations known as Matone's relations. In order 
to achieve this result it is convenient to consider the auxiliary 
function 
\be 
\cG(a) = \cF(a) - {a \over 2}{\de \cF(a) \over \de a} \, , 
\ee 
where $\cF(a)$ is defined in~(\ref{calFtot}) and consider the expansion 
\be 
\cG (a) = a^2 \sum_{K=0}^\infty \cG_{_K} 
{\Lambda^{4K} \over a^{4K}} \, . \label{calGexp} 
\ee 
We will momentarily see that $\cG(a)=u$. The expansion coefficients 
$\cG_K$ and $\cF_K$ are related by 
\be 
\cG_K = 2K \cF_K \, , 
\ee 
for $K\neq 0$ while $\cG_0 = 1/2$. 
 
As previously discussed, for $SU(2)$ the moduli space is parameterised 
by $u = \Tr(\phi^2)$ and turns out to be a Riemann sphere with three 
punctures at $u_1= - \Lambda$, $u_2= \Lambda$ and $u_3 =\infty$ with a 
symmetry $u \leftrightarrow -u$. We recall that $(a_{_D} (u), a(u))$ 
is a section of a flat bundle over the moduli space with monodromy 
group $\Gamma(2) \subset SL(2,\Z)$ generated by the three matrices 
$M_{-1}$, $M_{+1}$ and $M_{\infty}$ in~(\ref{monodmat}) with $M_{-1} 
M_{+1}=M_{\infty}$. 
 
Using the obvious integrability of the differential 
\be 
W(u)\, \dr u = a \,\dr a_{_D} - a_{_D} \dr a \, , 
\ee 
that being a complex function of the single variable $u$ is 
necessarily an exact differential, one can define the auxiliary 
function 
\be 
g(u) = \int_1^u \dr z \,W(z) \, . 
\ee 
This helps determining the behaviour of $\cF$ under monodromy (modular 
transformations). In fact by integrating 
\be 
\de_u \cF = a_{_D} \de_u a = {1\over 2}[\de_u 
(a_{_D} a) - W(u)] 
\ee 
one finds 
\be 
\cF(u) = {1\over 2}[ a_{_D} a - g(u)] + \cF_0 \, . 
\ee 
One can check that, under 
\ba 
&& a_{_D} \rightarrow \tilde{a}_{_D} = k {a}_{_D} - m a \nn \\ 
&& a \rightarrow \tilde{a} = -l {a}_{_D} + n a \, , \nn 
\ea 
with $kn - l m = 1$, one has 
\be 
\tilde{\cF}(\tilde{a}) = {\cF}({a}) + 
{1 \over 2} [ l m a_{_D} a - k l a_{_D}^2 - m n a^2] \, , 
\ee 
while $\cG(a)$, conveniently defined as above, turns out to be 
modular invariant, \ie 
\be 
\tilde{\cG}(\tilde{a}) = {\cG}({a}) \,, 
\ee 
since $u$ and hence $g(u)$ are invariant. By taking the ratio of 
$a_{_D}'(u)$ and $a'(u)$ and keeping in mind that $u$ is invariant, 
one also finds that 
\be 
\tau(a)= {\de^2 {\cF}({a}) 
\over \de{a}^2} = {a_{_D}'(u)\over a'(u)} \rightarrow 
\tilde{\tau}(\tilde{a}) = {\de^2 \tilde{\cF}(\tilde{a}) \over 
\de\tilde{a}^2} = { k \tau(a) -m \over - l \tau(a) + n } \, , 
\ee 
which is the expected projective transformation of the complexified 
coupling. 
 
By uniformisation arguments, \ie monodromy invariance and asymptotic 
behaviour at large $a$, Matone eventually showed that~\cite{matone} 
\be 
\cG(a) = -i\pi g(u)/2 = u \, . 
\ee 
The linear $u$ dependence of $g(u)$ is tantamount to saying that 
$W(u)$ is a constant, independent of $u$. In fact $W(u)= a(u) 
a_{_D}'(u) - a_{_D}(u) a'(u)$ is nothing but the Wronskian of the 
solutions of the second order differential equation satisfied by 
$a(u)$ and $ a_{_D}(u)$, which in canonical (Schr\"odinger-like) form 
reads 
\be 
(1-u^2)\,{\dr^2 \psi(u)\over \dr u^2} - {1\over 4} \psi(u) = 0 \, . 
\ee 
As a result of the uniformisation theorem of the moduli space of 
Riemann surfaces, $\cG(a)$ obeys a non-linear differential 
equation of the form 
\be 
(1-\cG^2) \,{\dr^2\cG\over \dr a^2} + {1\over4} a 
\left({\dr\cG\over \dr a}\right)^3 = 0 \, , 
\ee 
so that the coefficients of the expansion~(\ref{calGexp}) satisfy the 
sought for recursion relation 
\bea 
\cG_{K+1} &\!=\!& {1\over 8 \cG_{0}^2 ({K+1})^2} \times 
\left\{(2K-1)(4K-1)\cG_{K} + \rule{0pt}{14pt} \right. \nn \\ 
&\!+\!& 2\cG_{0}\sum_{N=0}^{K-1}c(N,K)\, 
\cG_{K-N}\,\cG_{N+1} - \nn \\ 
&\!-\!& \!\left. 2\sum_{L=0}^{K-1} \sum_{N=0}^{L+1} 
d(L,N,K)\,\cG_{K-L}\,\cG_{L+1-N}\,\cG_{N} 
\right\} \, , 
\eea 
where 
\bea 
&& c(N,K)= 2N(K-N-1) + K-1 \nn \\ 
&& d(L,N,K) = [2(K-L) -1][2K-3L-1 + 2N(L-N+1)] 
\eea 
and $\cG_0=1/2$. The first few coefficients read 
\be 
\cG_1 = {1 \over 2^2} \, , \quad \cG_2 = {5 \over 2^6} \, , 
\quad \cG_3 = {9 \over 2^7} \, , 
\ee 
in perfect agreement with the results of Seiberg and Witten. Moreover 
since $u = \cG(a)$ using the asymptotic behaviour of $\cG$ one can 
determine the constant value of $W$ that reads 
 
\be 
W = a_{_D}' a - a_{_D} a' = {2 i\over \pi} \, . 
\ee 
This relation is very useful in order to determine the ``critical'' 
curve where $\im\!(a_{_D}/ a)=0$. On this curve the lattice of BPS 
states collapses to a line, as already observed.

\subsection{(Constrained) instanton checks for $K=1,2$} 
\label{sec:constrcheck} 
 
Following Matone~\cite{matone}, Fucito and Travaglini~\cite{fuctrav} 
have been able to check the non perturbative relation 
\be 
\la \Tr(\phi^2) \ra(a) = u(a) = \cG(a) = 
\left(\cF(a) - {a \over 2}{\de \cF(a) \over \de a} \right) 
\ee 
for $K=1,2$ and show agreement with the SW prepotential. 
 
Using the relation between $\cF$ in~(\ref{calFtot}) and $\cG$ 
in~(\ref{calGexp}), one finds 
\be 
\la \Tr(\phi^2) \ra(a) = - {a^2 \over 2} 
- \sum_K \cG_K {\Lambda^{4K} \over a^{4K-2}} \, . 
\ee 
The calculation was carried out by making use of the ADHM 
construction, which we now briefly review in the $SU(2)$ case~\cite{ADHM}. 
In the ADHM approach~\cite{ADHM}, the gauge connection is 
written in the form 
\be 
A_\mu(x) = U^{\dagger}(x) \de_\mu U(x) \, . 
\ee 
The key observation is that $U(x)$ is not a unitary $SU(2)$ matrix but 
rather a $(1+K)\times 1$ ``array'' of quaternions, satisfying 
\be 
\Delta^{\dagger}(x) U(x) = 0 \, , 
\ee 
where 
\be \Delta(x) = a + b x \, , 
\ee 
with $x = x^\mu \sigma_\mu$ the position quaternion. Self-duality 
requires 
\be 
\Delta^{\dagger}(x) \Delta(x) = f^{-1}\otimes \one \, , 
\ee 
with $f$ an invertible $K\times K$ matrix and $\one$ the $2\times 2$ 
identity matrix. The projector on the kernel of $\Delta^{\dagger}(x)$, 
spanned by $U(x)$, reads 
\be 
P(x) = U(x) U^{\dagger}(x) = \one - \Delta f \Delta^{\dagger}(x) \, . 
\ee 
Gauge field zero-modes, that we here denote by $a_\mu$, are 
orthogonal to the gauge orbit and can be parametrised as~\cite{TEM,fuctrav} 
\be 
a_\mu(x) = U^{\dagger}(x)[C\bar\sigma_\mu f b^{\dagger} 
- b f \sigma_\mu C^{\dagger}] U(x) \, , 
\ee 
with $C$ a $(1+K)\times K$ ``matrix'' of quaternions satisfying 
\be 
\Delta^{\dagger}(x) C = (\Delta^{\dagger}(x) C)^T \, . 
\ee 
These conditions reduce the number of independent (quaternionic) 
components of $C$ from $(1+K)\times K$ to $(1+K)\times K - (K-1)\times 
K = 2K$, \ie $8K$ zero-modes as expected for $SU(2)$ 
instantons. Modulo symmetries, which are local $SU(2)$ and global 
$SO(K)$, the components of $C$ can be identified with the fluctuations 
of $\Delta$, $\delta \Delta$, \ie variations of the ADHM data, 
satisfying the self-duality condition 
\be 
(\Delta + \delta\Delta)^{\dagger}(x) (\Delta + 
\delta\Delta)(x) = f^{-1}\otimes \one 
\ee 
if non linear terms are neglected. Since $ \delta\Delta = C$ is linear 
in the gauge field zero-modes parametrised by $C$, one can identify 
zero-modes of the gauge fields with solutions of the linearised ADHM 
equations around a given self-dual solution. This is equivalent to 
identifying the bosonic zero modes as solutions of the equation $ 
S[A_\mu + a_\mu] = S[A_\mu] $ up to cubic terms. One can similarly 
determine the fermionic zero-modes that in the case of $\cN=2$ are as 
many as the bosonic zero-modes and are given by~\cite{TEM, 
fuctrav} 
\be 
\lambda^{(i)}_{\beta\dot{a}} =\sigma^\mu_{\beta\dot{a}} a^{(i)}_\mu \, , 
\ee 
with $i=1,..., 8$. In the presence of flat directions of the
classical scalar potential, the constrained instanton method entails
an expansion around a solution of the (approximate) coupled
equations~\footnote{The attentive reader may notice that these are not
the classical equations since the scalar induced source $J^\nu =
\phi^{\dagger} (D^\nu \phi) - (D^\nu \phi^{\dagger}) \phi$ is being
neglected. Exact topologically non-trivial solutions in the presence
of non-zero v.e.v.'s for the scalars are not
known~\cite{hyperinstmonoWitFre}.  The standard approach, which allows
to  control the fluctuations around the approximate solution, consists
in  adding to the action a ``constraint'' on the instanton size. The
resulting ``solution'' is thus known as a ``constrained
instanton''~\cite{AFF}.} 
\be 
D_\mu F^{\mu\nu} = 0 \, , \quad D^2 \phi = 0 
\ee 
with boundary condition at infinity, $\phi \rightarrow \phi_{\rm 
flat}$. For $SU(2)$ $\phi_{\rm flat} = a \sigma_3/2i$ modulo gauge 
transformations. 
 
For $K=1$, everything simplifies drastically. As discussed in 
detail in Sects.~\ref{sec:2.3},~\ref{sec:2.4} and Appendix~B, 
the bosonic measure (``integrated'' over $SU(2)/\Z_2$) reads 
\be 
\dr\mu_B = {4 \over \pi^2} \left( {2\pi\rho\mu \over g^2} 
\right)^8 {\dr^4x_0 \,\dr\rho \over \rho^5} \, . 
\ee 
Using the fermionic zero modes, that are not normalised, the fermionic 
measure is given by 
\be 
\dr\mu_F = \dr^4\eta \,\dr^4\bar\xi \left( 
{g^2 \over 32\pi^2\mu} \right)^4 \, . 
\ee 
Due to the presence of the scalar v.e.v., $a$, the classical action 
consists of various terms 
\be 
S_{\rm cl} = S_{\rm YM}+ S_{\rm scal} + S_{\rm ferm} + 
S_{\rm Yuk} + S_{\rm pot} \, . 
\ee 
After integration over the fluctuations of $\phi$ and $\phi^{\dagger}$ 
around their v.e.v., the Yukawa couplings produce an additional (to 
$\phi_{\rm harmonic}$) inhomogeneous term in $\phi$ of the form 
\be 
\phi^a_{\rm inhom} =\sqrt{2} [D^{-2}]^{a}{}_{b} \epsilon^{bcd} 
\lambda^{\alpha}_c \lambda_{d\alpha} = \sqrt{2} \zeta^{\alpha} 
\lambda^{a}_{\alpha} \, , 
\ee 
where $\zeta^{\alpha} = \eta^{\alpha} + x^\mu \sigma_\mu^{\a\dot\a} 
\bar\xi_{\dot\a} $. The absence of zero-modes with ``wrong'' chirality 
leads to 
\be 
S_{\rm Yuk} = \bar{a}^b\bar\xi_{\dot\alpha} \sigma_b^{\dot\alpha\dot\beta} 
\bar\xi_{\dot\beta} {g \over \sqrt{2}} \left( {g^2 \over 
32\pi^2\mu} \right)^{-1} \, . 
\ee 
Moreover 
\be 
S_{\rm scal} = 4\pi^2 |a|^2 \rho^2 
\ee 
and we set 
\be 
\Lambda^4 = \mu^4\er^{-8\pi^2/g^2} \, . 
\ee 
The explicit computation of $u$ (the v.e.v.\ of $\Tr(\phi^2)$) then yields 
\bea 
u &\!=\!& \la \phi^a 
\phi_a \ra_{K=1} = \Lambda^4 \int {4 \over \pi^2} \left( {2\pi 
\over g} \right)^8 {\dr^4x_0 \dr\rho} \, \rho^3 \,\er^{-4\pi^2 
|a|^2 \rho^2} F^a_{\mu\nu}F_a^{\mu\nu} 
\times \nn \\ 
&\!\times\!&\int \dr^4\eta \dr^4\bar\xi \left( {g^2 \over 32\pi^2} 
\right)^4 (\eta\eta)^2 \,\exp\left[{-\bar{a}^b \bar\xi \sigma_b 
\bar\xi {g \over \sqrt{2}} \left( {g^2 \over 32\pi^2\mu} 
\right)^{-1}}\right] \, . 
\eea 
Performing the integrations over the collective coordinates yields 
\be \la \phi^a \phi_a \ra_{K=1} 
= {2 \over g^4} {\Lambda^4 \over a^2} 
\ee 
in agreement with $\cG_1$. 
 
For $K=2$, the (constrained) instanton calculus is more laborious. 
The off-diagonal component $d$ of the lower sub-block of $\Delta$ is 
of the form 
\be d = {1 \over 2}{y \over y^2}(\bar{v}_2 v_1 
-\bar{v}_1 v_2 ) \, , 
\ee 
where $y=x_1 - x_2 \equiv 2 e$ and $x_0 = 
(x_1 + x_2)/2$, with $x_1$ and $x_2$ denoting the two instanton 
``centers'' and $v_1$ and $v_2$ the two extra quaternionic collective 
coordinates. Similar restrictions as before apply to $C = 
\delta\Delta$ so that the off-diagonal component $\gamma$ of the 
lower sub-block of $C$ is of the form 
\be 
\gamma = {y \over y^2}(2\bar{d}\eta + \bar{v}_2 \nu_1 
-\bar{v}_1 \nu_2 ) \, , 
\ee 
where $\eta$, $\nu_1$ and $\nu_2$ are quaternions that parametrise 
the independent fluctuations of the fermions. Separating the four 
collective coordinates associated with translations, $x_0$, and 
the four broken Poincar\'e supersymmetries, $\eta_0$, the relevant 
correlator reads 
\bea 
\la \phi^a \phi_a \ra_{K=2} &=& {\Lambda^8 
\over 16} \int \dr^4 v \,\dr^4 e \,\dr^4\bar{\xi}\,\dr^4\nu_1 \, 
\dr^4\nu_2 \left( {J_B \over J_F} \right)^{1/2} 
\er^{-S_{\rm Yuk}-S_{\rm scal}} \times \nn \\ 
&\!\times\!& \int \dr^4x_0 \dr^4\eta_0 \,(\eta_0 \eta_0)^2 
F^a_{\mu\nu}F_a^{\mu\nu} 
\eea 
where 
\be 
\left( {J_B \over J_F} 
\right)^{1/2} = {2^{10} \over \pi^{8}} {| \ |e|^{2} - |d|^{2} \ | 
\over |v_1|^{2} + |v_2|^{2} + 4 (|d|^{2}+ |c|^{2})} \, . 
\ee 
Performing all the many necessary integrations yields 
\be 
\la \phi^a \phi_a 
\ra_{K=2} = - {5 \over 4 g^8} {\Lambda^8\over a^6} 
\ee 
in agreement with $\cG_2$. 
 
Actually one can formally prove that Matone relations are satisfied by 
instanton calculus for any $K$~\cite{fuctrav}. 
 
Another elegant approach to derive the SW prepotential from first 
principles is based on the so called $\cN = 2^*$ theory. This is 
nothing else but $\cN = 4$ SYM theory deformed by the addition of 
a mass $M$ for the hypermultiplet in the adjoint representation, 
or equivalently the same mass $M_1 = M_2 = M$ for two of the three 
adjoint chiral multiplets in the $\cN = 1$ description of the $\cN 
= 4$ theory. Quite remarkably the hypermultiplet, $H = \{\Phi_1, 
\Phi_2\}$, appears quadratically in the microscopic action, 
\bea 
S[\Phi_{I=1,2}; \Phi_3,V] &=&\int \dr^2\theta \dr^2\bar\theta \, 
\Tr(\Phi^\dagger_I \er^{gV} \Phi^I) + \\ 
&+& \int \dr^2\theta \,g \Tr([\Phi_1,\Phi_2]\Phi_3) + {1\over 2} M 
\,\Tr(\Phi_I)^2 + {\rm h.c.} \, , \nn 
\eea
and can be integrated out in a Gaussian fashion. One ends up with an
effective action \`a la Wilson-Polchinski where $M$ plays the role of
an UV cutoff. The advantage of the approach is the UV finiteness of
$\cN = 4$ SYM theory which persists after the inclusion of the $\cN =
2$ supersymmetric mass terms. The resulting low energy effective
action is expected to coincide with the one resulting from the SW
prepotential. As we said in Sect.\ref{sec:ELA}, this has been
partially checked by means of the exact Renormalisation Group
in~\cite{KEN}.

\section{Topological twist and non-commutative deformation} 
\label{sec:topotwist} 
 
$\cN =2$ SYM theories admit an interesting reformulation which goes 
under the name of ``topological twist''~\cite{N2twist}. Although the 
topologically twisted version is not fully equivalent to the original 
(dynamical) theory, some of the observables coincide. In particular, 
one can suspect that the analytic prepotential $\cF$ could be one of 
these observables thanks to holomorphy. As we will see later on, this 
is not completely true. The topological theory cannot reproduce the 
logarithmic term generated by one-loop corrections. Yet, a properly 
defined partition function of the topological theory captures all the 
non-perturbative corrections to $\cF$ and more. Indeed, higher 
derivative ``gravitational'' $F$-terms can be reliably computed by 
means of its topologically twisted version if a suitable background 
inducing ``non-commutativity'' is turned on. After briefly reviewing 
the topological twist formalism, we will sketch the arguments leading 
to the derivation of $\cF_{\rm non-pert}$ from the topological 
partition function. 
 
The topological twist consists in bringing bosons and fermions to 
transform in the same way under the subgroup $SU(2)_L \times SU(2)_D 
\subset SU(2)_L \times SU(2)_R \times SU(2)_I $, where $D$ stands for 
the diagonal subgroup of $SU(2)_R \times SU(2)_I$, which is not to be 
confused with the (Euclidean) Lorentz group $SU(2)_L \times SU(2)_R$, 
since $SU(2)_I$ is part of the R-symmetry group $U(1) \times 
SU(2)_I$. Under $SU(2)_L \times SU(2)_D$ the two Weyl gaugini 
transform as a four-vector, $\psi_\mu \in (1/2, 1/2)$, a singlet, 
$\bar\eta\in (0,0)$, and a self-dual tensor, $\bar\chi^+_{\mu\nu}\in 
(0,1)$, where, adhering to standard notation, $(j_L, j_D)$ refers to 
the $SU(2)_L \times SU(2)_D$ spins rather than the dimension of the 
representation. Similarly the superspace variables dual to the eight 
supercharges are $\theta \rightarrow \theta_\mu, 
\bar\theta^+_{\mu\nu}, \bar\theta$, so that the chiral superfield 
$\Phi$ admits the newly looking decomposition 
\be 
\Phi = \phi + \theta^\mu \psi_\mu + 
{1\over 2} \theta^\mu\theta^\nu F_{\mu\nu} + \cdots \, , 
\ee 
where 
\be 
\theta^\mu = {1\over 2} \sigma^\mu_{\alpha, r} \theta^{\alpha, r} \, . 
\ee 
The supercharge $\bar{Q}= \varepsilon^{\dot\a r} \bar{Q}_{\dot\a r}$ 
is a scalar and plays the role of topological BRST charge. In the 
topologically twisted version, which we would like to stress is only a 
reformulation of $\cN =2$ SYM theories, the action reads 
\be 
S_{\rm top} = \int F\wedge F + \{\bar Q,\Psi\} \, , 
\ee 
where $\Psi = \phi \de_\mu \psi^\mu + F_{\mu\nu} \chi^{\mu\nu} + \eta 
[\bar\phi,\phi]$ is the ``topological gauge fermion''. 
 
For hyperk\"ahler manifolds, \ie manifolds with three closed K\"ahler
forms, the supercharges $\bar{Q}^+_{\mu\nu}$ can also be exploited in
order to perform the topological
twist~\cite{N2twist}. Nekrasov~\cite{nekra2,nekra3} proposed to also
use $Q_\mu$ or better deform  $\bar{Q}$ to 
\be 
\bar{Q}_E = \bar{Q} + E_a V^a_{\mu\nu} x^\mu Q^\nu \, , 
\ee 
where $V^a_{\mu\nu} = - V^a_{\nu\mu}$ are the six generators of the 
Euclidean rotation group $SO(4)$ and $E_a$ are constant 
parameters. This allows to define equivariant forms $\Omega(E) = 
\sum_p \Omega_p(E) = \sum_p \sum_{i_1,\ldots,i_p} {1\over p!} 
\Omega_{i_1,\ldots,i_p} \dr x^{i_1} \wedge \ldots \dr x^{i_p}$ such 
that 
\be 
R \Omega(E) = \Omega(R^{-1} E R) 
\ee 
for any $R\in SO(4)$. $\Omega(E)$ are naturally acted on by the 
equivariant exterior derivative 
\be 
\dr_E = \dr + \iota_{V(E)} \, , 
\ee 
where $\iota_{V(E)}$ denotes contraction with the vector field $V(E) = 
E_a V^a_{\mu\nu} x^\mu \de^\nu$, \ie 
\be 
\dr_E \Omega_p = \dr\Omega_p + \iota_{V(E)}\Omega_p \, . 
\ee 
As a result, acting with $\dr_E$ on a $p$-form generically yields both a 
$(p+1)$-form $\dr\Omega_p$ and a $(p-1)$-form $\iota_{V(E)}\Omega_p$.

One can check that the topological observable 
\be 
\cO_P^{\Omega(E)} = \int_{\R^4} \Omega(E) \wedge P(\Phi) 
\ee 
is $\bar{Q}_E$-closed iff $\Omega(E)$ is ``equivariantly'' closed, \ie 
iff $\dr_E \Omega(E)=0$. For generic choices of $E^a$ the set of 
equivariantly closed forms is empty. However one can consider 
$E^a\in U(2)_\omega \subset SO(4)$, where $U(2)_\omega$ is the 
stability group of a ``reference'' symplectic (K\"ahler and thus 
closed) form 
\beq
\omega = \dr x^1 \wedge \dr x^2 + \dr x^3 \wedge \dr x^4 \, , 
\label{OMEG}
\eeq
that by definition satisfies $\dr\omega = 0$. In this way from 
the condition of equivariance, that can be phrased in terms of the 
vanishing of the following Lie derivative 
\be 
\cL_{V(E)} \omega = 0 = \dr(\iota_{V(E)}\omega) + \iota_{V(E)}\dr \omega \, , 
\ee 
it follows that, at least locally, 
\be 
\iota_{V(E)}\omega = \dr\mu(E) \, , 
\ee 
or in other terms 
\be 
\dr_E (\omega - \mu(E)) = 0 \, . 
\ee 
Decomposing $\mu(E)$ along the four generators of the stability group 
$U(2)_\omega$, one finds 
\be 
h(x) \equiv \mu^0 =\delta_{\mu\nu} x^\mu x^\nu \quad \mu^a 
= \sum_{\mu <\nu} \eta^a_{\mu\nu} x^\mu x^\nu \, , 
\ee 
where $\eta^a_{\mu\nu}$ are 't Hooft symbols. Since $\omega$ defines a 
complex structure one can introduce complex coordinates $z_1, z_2$ 
such that $\omega = \dr z_1 \wedge \dr\bar{z}_1 + \dr z_2 \wedge 
\dr\bar{z}_2$. We also define 
\be 
H = \mu_{R}(E) = \epsilon_1 |z_1|^2 +\epsilon_2 |z_2|^2\, , 
\ee 
where $\mu_{R}(E) = {1\over 2} (\epsilon_1 + \epsilon_2) \mu^0(z,\bar 
z) + {1\over 2} (\epsilon_1 - \epsilon_2) \mu^3(z,\bar z)$ is an 
arbitrary linear combination of the ``real'' moment maps, the complex 
part being $\mu_C = \mu^1 + i \mu^2$. 
 
Relying on the equivariance properties of $\omega$ and $H$, one can 
define the generating function of $\bar{Q}_E$-closed observables by 
the formula 
\ba 
Z(a,\epsilon) = \left\la \exp\left\{{1\over (2\pi i)^2} \int_{\R^4 
\equiv\C^2}\left[\omega\wedge \Tr(\phi F + 
{1\over 2} \psi\wedge \psi) - \right. \right. \right. \nn \\ 
\left. \left. \left. - {1\over 2} H(x) \Tr(F \wedge 
F)\right]\right\} \right\ra_a \, , 
\ea 
where the suffix $a$  denotes the dependence on the scalar v.e.v.,
$a$. Supersymmetry,  which in this context is tantamount to
topological invariance  since $\bar Q_{E}$ is a linear combination of
the supercharges,  guarantees a perfect cancellation of all
perturbative  contributions between bosons and fermions. As a result
$Z(a,\epsilon)$ is saturated by instantons, viz. 
\be 
Z(a,\epsilon) 
= \sum_K q^K Z{_K}(a,\epsilon) \, , 
\ee 
where $q=\exp(2\pi i  \tau)$. Moreover the presence of $H$ suppresses
the contribution  of widely separated instantons and can be combined
with $\omega$  into 
\be 
\cH(x, \theta) = H(x) + {1\over 2} \omega_{\mu\nu} 
\theta^\mu \theta^\nu \, . 
\ee 
$\cH(x, \theta)$ represents a  manifestly supersymmetric regulator for
the holomorphic function  $\cF(a, \Lambda) $, where the explicit
presence of $\Lambda$ as an  argument is to denote the dependence of
$\cF$ on the  renormalisation group invariant scale. In turn the
latter gets  effectively replaced by $\cF(a, \Lambda \er^{-\cH})
$. Indeed  rescaling the metric of $\R^4 \equiv \C^2$ by a factor
$\lambda$  and taking the limit $\lambda \rightarrow \infty$, only the
last  term survives in the partition function, since all the other
terms  are suppressed by inverse powers of $\lambda$ that appear in
the  propagators needed for the contractions. Taking into account that
derivatives of $H$ with respect to $x^\mu$ or, equivalently, $z_1$
and $z_2$, are proportional to $\epsilon_{1,2}$ one finds 
\bea 
Z(a,\epsilon) &=& \exp\left\{{1\over 2(2\pi i)^2} \int_{\R^4} 
\omega\wedge\omega \, {\de^2 \cF(a, \Lambda \er^{-H}) \over 
\de\log\Lambda^2} 
\right\} + \cO(\epsilon) \nn \\ 
&\approx & \exp\left\{{\cF_{\rm inst} (a, \Lambda) + \cO(\epsilon) 
\over \epsilon_1 \epsilon_2 }\right\} 
\eea 
where 
\be 
\cF_{\rm inst} (a, \Lambda) = \int_0^\infty {\de^2 \cF(a, \Lambda \er^{-H}) 
\over \de H^2}\, H\,\dr H \label{HdH} \, . 
\ee 
Equation~(\ref{HdH}) makes the analytic properties of $Z$ and $\cF$ 
manifest. 

\subsection{Including Hypermultiplets} 
\label{sec:+hypers} 
 
In the presence of $N_f$ hypermultiplets in the fundamental 
representation with masses $m_f$, a possible 
parametrisation~\footnote{In order to make contact with the 
parametrisation used previously one has to perform the transformation 
$y = w - {1\over 2} P(z)$ and set $Q(z)$ to zero in the absence of 
hypers.} of the SW curve is~\cite{SWcurveNf} 
\be 
w + {\Lambda^{2N_c - N_f} Q(z) \over w } = P(z) \, , 
\ee 
where 
\be 
Q(z) = \prod_{f=1}^{N_f} (z + m_f) 
\ee 
and 
\be 
P(z) = \prod_{l=1}^{N_c} (z - \alpha_l) \, . 
\ee 
The $\alpha_l$'s are related to the v.e.v.'s of the adjoint scalars 
belonging to the Cartan subalgebra and are such that $\sum_l \alpha_l 
= 0$. The space of monic polynomials $P(z)$, \ie polynomials where the 
coefficient of the monomial of highest degree is 1, so that the 
coefficient of the monomial of next to highest degree is 0, is $\cU = 
\C^{N_c-1}$ and can thus be parametrised by the $N_c-1$ variables $u_n 
= \Tr(a^n)$, with $n=1,..., N_c-1$. The latter are symmetric 
polynomials in the $\alpha_l$ that can be identified with the $N_c-1$ 
Casimirs of $SU(N_c)$. The first two symmetric polynomials are 1 and 
$u_2 = \sum_l \alpha_l^2$ or, equivalently, 
$\sum_{l<l'}\alpha_l\alpha_{l'}$ since $\sum_l \alpha_l= 0$. The 
relation between $u_n$ and $\a_l$ can be similarly determined. We now 
discuss how to determine the relation between $u_n$ and the 
periods $a_l$ and $a_l^D$ of the SW curve. 
 
In the perturbative region, where $|\alpha_l|,|\alpha_l -\alpha_n| 
\gg |\Lambda|, |m_f|$, one can choose local coordinates 
\be 
a_l = {1\over 2\pi i} \oint_{A_l} {z \,\dr w \over w} 
\ee 
and 
\be 
a^D_l = {1\over 2\pi i} \oint_{B_l} {z \, \dr w \over w} \, , 
\ee 
where the $A_l$ cycles encircle the cuts in the $z$-plane from the 
point $\alpha_l^+$ to the point $\alpha_l^-$, where the points 
$\alpha_l^\pm$ are such that 
\be 
P(z=\alpha_l^\pm) = \pm 2 
\Lambda^{N_c - {N_f\over 2}} \sqrt{Q(z=\alpha_l^\pm)} \, . 
\ee 
Not all $A_l$ cycles are homologically independent since $\sum_l A_l 
\approx 0$ can be shrunk to zero. The $B_l$ cycles go through the cuts 
from $\alpha_l^+$ to $\alpha_{l+1({\rm mod}\:N)}^-$. Once again 
$\sum_l B_l \approx 0$ in homology. As a result, $\sum_l \dr a_l 
\wedge \dr a^D_l = 0$ and on local patches one can introduce the 
prepotential $\cF(a;m,\Lambda)$ such that 
\be 
\dr\cF(a;m,\Lambda) = \sum_l a^D_l \dr a_l \, . 
\ee 
$\cF(a; m,\Lambda)$ admits an expansion of the form 
\be 
\cF(a; m,\Lambda) = \cF_{\rm pert}(a; m,\Lambda) + 
\cF_{\rm inst}(a;m,\Lambda) \, , 
\ee 
where $\cF_{\rm inst}(a; m,\Lambda)$ encompasses the instanton 
contribution, that can be computed by the localisation techniques 
outlined below, and 
\ba 
\cF_{\rm pert}(a; m,\Lambda) &=& {1\over2} \sum_{l\neq l'} 
(a_l - a_{l'})^2 \log\left( a_l - a_{l'} \over\Lambda \right) + \nn \\ 
&-& \sum_{l, f} (a_l + m_f)^2 \log\left( a_l + m_f \over \Lambda 
\right) 
\ea 
encodes the logarithmic running of the gauge coupling with the mass 
scales at play. Indeed $a_l - a_{l'}$ are the masses of the $W$-bosons 
and $a_l + m_f$ are the masses of the charged hypermultiplets.

\subsection{Instanton measure and localisation for arbitrary $K$} 
\label{sec:locK} 
 
Following the ADHM construction~\cite{ADHM}, the moduli space 
$\cM_{K,N_c}$ of $K$ instantons in $SU(N_c)$ with fixed framing (\ie 
orientation in colour space) at infinity is a $4KN_c$ dimensional 
variety and can be viewed as the hyperk\"ahler quotient of the 
ADHM data $(B_1, B_2, I, J)$, where $B_{1,2}\in {\rm End}(V_K)$, 
$I \in {\rm Hom}(W_{N_c},V_K)$ and $J\in {\rm Hom}(V_K, W_{N_c})$, with 
respect to the action of $U(K)$. The corresponding formulae 
\be 
\mu_{\bf C} = [B_1, B_2] + IJ = 0 \label{adhmc} 
\ee and 
\be 
\mu_{\bf R} = [B_1, B_1^{\dagger}] + [B_2, B_2^{\dagger}] + 
II^{\dagger} - J^{\dagger} J = 0 \label{adhmr} 
\ee are the 
celebrated ADHM equations~\cite{ADHM} that indeed enjoy invariance 
under $U(K)$ transformations. 
 
As a result $\cM_{K,N_c}$ is neither compact in the UV (due to small 
size instantons) nor in the IR (due to the non compactness of $\R^4$). 
 
Various compactifications of $\cM_{K,N_c}$ have been
proposed~\cite{DonKron}.  The Uhlenbeck compactification
$\cM^U_{K,N_c}$  corresponds to the construction of a hyperk\"ahler
orbifold where the UV  problem is cured by including point-like
instantons, \eg gluing  subspaces of the form $\cM_{K-1,N_c} \times
\R^4$, $\cM_{K-2,N_c} \times  \R^8$, $\cM_{K-3,N_c} \times \R^{12}$
and so on. Alternatively,  according to Nekrasov and Schwarz the
singularities of $\cM^U_{K,N_c}$  can be blown up to a smooth space
$\cM^{NS}_{K,N_c}$ which includes  ``exceptional divisors'' in place
of the original singularities~\cite{nekra1,mns}.  This blowing up
relies on a non-commutative  extension of the gauge theory that
translates in the possibility of  deforming the ADHM
equations~(\ref{adhmc}) and~(\ref{adhmc})  to~\footnote{In principle
one can deform $\mu_{\bf C}$ as well. But  this deformation is
irrelevant as it can always be eliminated by a non  analytic change of
coordinates.} 
\be 
\mu_{\bf R} = \zeta_{\bf R} \one_K \, , \quad \mu_{\bf C} = 0 \, . 
\ee 
Deformed instanton calculus then boils down to computing equivariant 
volumes of $\cM^{NS}_{K,N_c}$, provided one uses in the definition of 
the integration measure the closed symplectic 2-form~\footnote{A 
symplectic 2-form is the generalisation of the familiar 2-form $\omega 
= \sum_i \dr p_i\wedge \dr q^i$ in phase space.} lifted from 
$\cM^U_{K,N_c}$, where the relevant symplectic form is the reference 
K\"ahler form. Since this symplectic form vanishes when restricted to 
the exceptional divisors it does not add contributions ``extraneous'' 
to the original ``commutative'' gauge theory. In order to localise the 
measure, \ie reduce the integrals to contour integrals that are 
calculable by the residue theorem, it is convenient to consider the 
combined action of $U(K)$, $G=SU(N_c)/\Z_{N_c}$ and ${\bf T}^2$, the 
latter representing the maximal torus, \ie the exponential of the 
Cartan subalgebra, of $SO(4)$. The use of this combined action is 
instrumental in deforming the symplectic K\"ahler form $\omega$ of 
$\R^4$ by the moment maps $\mu_G= \delta_G \cA^{^\Lambda} 
\omega^{\rm ADHM}_{_{\Lambda\Lambda'}}\cA^{^{\Lambda^\prime}}$, where 
$\cA^{^{\Lambda^\prime}}$ collectively denote the ADHM data, and 
$\mu_{T^2}=\epsilon_a x^i(V^a_{T^2})_i{}^j \omega_{jk} x^k$ and in 
constructing an equivariant form that localises the integrals on 
point-like abelian instantons. 
 
The partition function over the compactified instanton moduli 
space reads 
\be 
Z(a,\epsilon_1, \epsilon_2; q) = \sum_K q^K \oint_{\cM_K} 1 
\ee 
where $q = \er^{2\pi i \tau}$ and $\oint 1$ denotes the localisation 
of the integral to point-like instantons while $a=(a_1,...,a_{N_c})$ 
parametrise the Cartan subalgebra of $G= SU(N_c)$, \ie 
$\sum_{i=1}^{N_c} a_i =0$ and $\epsilon_1, \epsilon_2$ are deformation 
parameters corresponding to the $\Omega$ background, defined 
below. For the purpose of computing the integral it is convenient to 
rewrite the contour integral in the form 
\be 
Z_K = \oint_{\cM_K} 1 = \int_{\cM_K} \exp(\omega + 
\mu_G(a) + \mu_{T^2}(\epsilon)) \, , 
\ee 
due to topological BRST invariance. The non-perturbative contributions 
to the prepotential, but not the perturbative ones, are proportional 
to the logarithm of the topological partition function 
\be 
Z(a,\epsilon_1, \epsilon_2; q) = \exp\left( {1 \over \epsilon_1 
\epsilon_2}\cF_{\rm non-pert}(a,\epsilon_1, \epsilon_2; q)\right) \, , 
\ee 
as previously shown. Here we are only describing an efficient way to 
explicitly compute the contour integrals that yield $Z_K$, the 
coefficients of the expansion of the topological partition 
function. Using localisation, one can indeed derive an explicit 
expression for $Z(a,\epsilon_1, \epsilon_2; q)$. Taking for simplicity 
$\epsilon_1=- \epsilon_2= \hbar$ (the notation $\hbar$ suggests that 
some quantum non-commutativity is switched on as we will see!), one 
finds 
\be 
Z(a,\hbar, -\hbar; q) = 
\sum_{\vec{K}} q^{|\vec{K}|} \prod_{(m,n)\neq (i,j)} {a_{mi} + 
\hbar (K_{m,n} - K_{i,j} + j -n) \over a_{mi} + \hbar (j -n)} \, , 
\label{pippo} 
\ee 
where $a_{mi} = a_{m}-a_{i}$ and the sum is over the ``coloured'' 
partitions of the instanton numbers among the $N_c$ abelian factors 
$U(1)^{N_c}$ of the Cartan subalgebra of $U(N_c)$ 
\be 
\vec{K} = (\vec{K}_1, \ldots , \vec{K}_{N_c}) 
\ee 
with 
\be \vec{K}_n = \{ K_{n,1}\ge K_{n,2}\ge \cdots \ge 
K_{n,l_n}\ge K_{n,l_n + 1} = K_{n,l_n + 2}= \cdots = 0 \} \, , 
\ee 
while the product in~(\ref{pippo}) is over $1\le m, i \le N_c$ and 
$n,j\ge 1$. 
 
The theory can be enlarged by the addition of $N_f$ hypermultiplets in 
the fundamental ${\bf N_c} +{\bf N_c}^*$ with masses $m_1, ..., 
m_{N_f}$. The explicit expression of $Z(a_i, m_f, \hbar, -\hbar; q)$ 
in this case becomes 
\bea 
Z(a_i, m_f,\hbar, -\hbar; q) &=& \sum_{\vec{K}} (q \hbar^{N_f})^{|\vec{K}|} 
\! \prod_{(m,n)}\prod_{f=1}^{N_f}\! {\Gamma( {1\over \hbar} ( 
a_{m} +m_f) + 1 + K_{m,n} - n) \over a_{mi} + \hbar (j -n)} \times \nn \\ 
&&\times \!\!\prod_{(m,n)\neq (i,j)} {a_{mi} + \hbar (K_{m,n} - 
K_{i,j} + j -n) \over a_{mi} + \hbar (j -n)} \, . 
\eea

\subsection{Computing the residues and checking the instanton 
contributions} 
\label{sec:residcheck} 
 
In a remarkable paper, Moore, Nekrasov and Shatashvili~\cite{mns} have 
indeed been able to reduce the computation of $Z_K$ in the $K$-instanton 
sector to contour integrals of the form 
\ba 
Z_K(a;\epsilon_i) = {1 \over K!} {(\epsilon_1 + \epsilon_2)^K 
\over (2\pi i \epsilon_1 \epsilon_2)^K} \oint \prod_{I=1}^K 
{\dr\phi_I Q(\phi_I) \over 
P(\phi_I) P(\phi_I + \epsilon_1 + \epsilon_2)} && \!\!\!\times \nn \\ 
\times \prod_{1\le I < J \le K} {\phi_{IJ}^2(\phi_{IJ}^2 - (\epsilon_1 
+ \epsilon_2)^2) \over (\phi_{IJ}^2 - \epsilon_1^2) 
(\phi_{IJ}^2 - \epsilon_2^2)} && , 
\ea 
where the complex variables $\phi_{IJ} = \phi_{I} - \phi_{J}$ can be 
thought of as entries of a $K \times K$ matrix, $P(z)$ and $Q(z)$ were 
defined before and the integration contours run along the real axis. 
 
The variables $\phi_I$, $a_l$ and $\epsilon_{1,2}$ represent an 
infinitesimal deformation of the ADHM equations such that 
\ba 
&& [B_1, \phi] = \epsilon_{1} B_1 \quad [B_2, \phi] = \epsilon_{2} 
B_2 \nn \\ 
&& -\phi I + I a = 0 \quad -a J + J \phi = - (\epsilon_{1} + 
\epsilon_{2}) J \, . \ea In the bases of the $K$ dimensional 
vector space $V_{_K}$ and the $N_c$ dimensional vector space 
$W_{_{N_c}}$ of the ADHM construction, where the $K\times K$ matrix 
$\phi$ and the $N_c\times N_c$ matrix $a$, representing the scalar 
v.e.v.'s, are diagonal, one has \ba && (\phi_{IJ} + 
\epsilon_1)B_{1,IJ} = 0 \qquad 
(\phi_{IJ} + \epsilon_2)B_{2,IJ} = 0 \nn \\ 
&& (\phi_{I}-a_l) I_{I,l} = 0 \qquad (\phi_{I} + \epsilon_{1} 
+ \epsilon_{2}-a_l) J_{l,I} = 0 \, . 
\ea 
The poles at $\phi_{IJ}= \pm \epsilon_{1,2}$ should be avoided by 
deforming the contour or setting $\epsilon_{1,2} \rightarrow 
\epsilon_{1,2} + i \delta$. Similarly $a_{l} \rightarrow a_l + i 
\delta'$ in order to avoid the zeroes of $P$ in the denominator. 
The origin of the poles at $\phi_{IJ}= \pm \epsilon_{1,2}$ can be 
understood by means of the Duistermaat--Heckman (DH) formula 
\be 
{1 \over n!} \int_{X^{2n}} \omega^n \,\er^{-\mu[\xi]} = \!\! 
\sum_{P_f:V_\xi(P_f)=0} {\er^{-\mu[\xi](P_f)} \over \prod_{i=1}^n 
W_i[\xi](P_f)} \, , 
\ee 
where ${X^{2n}}$ is a symplectic manifold with symplectic form 
$\omega$ and $\mu$ is the moment map of a ``torus'' action generated 
by $\xi$ and represented by $V_\xi$, that has fixed points $P_f$ with 
``exponents'' $W_i[\xi](P_f)$. In the case of $X = \cM^{NS}_{K,N_c}$ 
the relevant torus action, that consists of the geometric 
transformations that form an abelian group, is $U(1)^{N_c-1}\subset 
SU(N_c)$ and $U(1)^{2}\subset U(2)_\omega$. Indeed $U(1)^{N_c-1}$ is 
the maximal torus of the gauge group, generated by the Cartan 
subalgebra and one cannot hope to get any larger torus action from the 
gauge group generators. Similarly $U(1)^{2}$ is the maximal abelian 
subgroup of the stability group of the symplectic K\"ahler form and 
one cannot get anything more from the Euclidean rotation group. For 
generic ADHM data, the deformed ADHM equations have solutions only in 
correspondence with the poles of the integrand, this means that 
$\phi_I$ and $\phi_{IJ} = \phi_{I} - \phi_{J}$ are uniquely specified 
in terms of $a_l$, $\epsilon_{1,2}$ and $P_f$. The last ingredient, 
$\prod_{i=1}^n W_i[\xi](f)$, in the DH formula can be related to the 
Chern character of the tangent bundle of $\cM^{NS}_{K,N_c}$ at the 
point $P_f$. 
 
Another important step in the computation of the contour integral is 
the classification of the residues in terms of Young 
tableaux~\footnote{Young tableaux are sets of boxes. The number of 
columns is $N_c$ for $U(N_c)$. Starting from the first column the 
number of boxes should not increase. Boxes in the same column 
correspond to antisymmetrised indices. Boxes in the same row 
correspond to symmetrised indices.} $\vec{Y} = (Y_1, ..., Y_{N_c})$, 
such that $\sum_l |Y_l| = K$. Indeed to each $Y_l$ with $0< K_l\le K$ 
boxes corresponds a partition 
\be 
K_{l,1}\ge \cdots\ge K_{l,n_l}\ge K_{l,n_l + 1} 
= K_{l,n_l + 2} = \cdots = 0 \, . 
\ee 
Then the pole corresponding to a given $\vec{Y}$ is located at 
\be 
\phi^{(r,s)}_I = a_l + \epsilon_{1} (r - 1) + \epsilon_{2} (s 
- 1) \, , 
\ee 
with the integers $r$ and $s$ such that $0\le r \le n_{l}$ and $0\le s 
\le K_{l,r}$. 
 
In more physical terms the fixed points of the action of $G\times T^2$ 
on the ``resolved'' $\cM^{NS}_{K,N_c}$ correspond to $U(N_c)$ 
non-commutative instantons that split into $U(1)^{N_c}$ 
non-commutative instantons such that the instanton charge $K$ is split 
into $K = \sum_l K_l$ with $K_l$ in the $l^{th}$ subgroup. The 
non-commutativity induced by the $\epsilon$-deformation prevents the 
instantons from coalescing one on top of the other. 
 
For a given $\vec{Y}$ the residue of the contour integral reads 
\bea 
&&R(\vec{Y}) = {1\over (\epsilon_1 \epsilon_2)^K} \times\\ 
&&\prod_{l=1}^{N_c} \prod_{r=1}^{n_l} \prod_{s=1}^{K_{l,r}} {T_l 
(\epsilon_1 (r - 1) + \epsilon_{2} (s - 1))\over 
(\epsilon(\ell_l(r,s) +1)- \epsilon_{2} h_l(r,s))(\epsilon_{2} 
h_l(r,s) -\epsilon\ell_l(r,s)) } \times \prod_{l<m}^{1,{N_c}} 
\prod_{r=1}^{n_l} \prod_{p=1}^{n_m} 
 \nonumber\\ 
&& \prod_{s=1}^{K_{l,r}} \prod_{t=1}^{K_{m,p}}\left[{(a_{lm}+ 
\epsilon_1 (t - K_{m,p}) - \epsilon_{2} (s - 1)) (a_{lm}+ 
\epsilon_1 t - \epsilon_{2} (s - 1- K_{l,r}))\over (a_{lm}+ 
\epsilon_1 (t - K_{m,p}) - \epsilon_{2} (s - 1- K_{l,r})) (a_{lm}+ 
\epsilon_1 t - \epsilon_{2} (s - 1) } \right]^2 , \nn 
\eea 
where $\epsilon = \epsilon_{1} + \epsilon_{2}$, $a_{lm} = a_{l} - 
a_{m}$, $\ell_l(r,s) = K_{l,r} - s$, $h_l(r,s) = K_{l,r} + K_{l,s} -r 
- s + 1$ and 
\be 
T_l (z) = {Q(z + a_l) \over \prod_{m\neq l}(z 
+ a_{lm})(z + \epsilon +a_{lm})} \, . 
\ee 
For future use it is convenient to define 
\be 
S_l (z) = {Q(z +a_l) \over \prod_{m\neq l}(z + a_{lm})^2} \, , 
\ee 
in terms of which the first two coefficients of the instanton 
expansion of the topological partition function are given by 
\ba 
Z_1 &=& {1\over \epsilon_{1} \epsilon_{2}} \sum_l S_l(0) \, , \\ 
Z_2 &=& {1\over (\epsilon_{1} \epsilon_{2})^2} \left[ {1\over 
4}\sum_l S_l(0)[S_l(\hbar) + S_l(-\hbar)] + {1\over 2} \sum_{l\neq 
m} {S_l(0)S_m(0) a_{lm}^4 \over (a_{lm}^2 - \hbar^2)^2}\right] \nn 
\ea 
and so on. Using the known relation between the topological partition 
function and the non-perturbative contribution to the holomorphic 
prepotential~(\ref{calFnonpert}) one gets 
\ba 
\cF_1 &=& \sum_l S_l(0) \, , \nn \\ 
\cF_2 &=& {1\over 4} \sum_l S_l(0)S_l''(0) + \sum_{l\neq m} 
{S_l(0)S_m(0)\over a_{lm}^2} + \cO(\hbar^2) 
\ea 
and so on. Formulae tend soon to become unwieldy but Nekrasov has been 
able to check agreement with previous results for the holomorphic 
prepotential up to five instantons~\cite{nekra2,nekra3}. The 
consistency among various independent approaches confirms the 
correctness of the result for the SW prepotential.

\section{(Constrained) instantons from open strings} 
\label{sec:instopenstr} 
 
One of the most astonishing features of critical strings is the 
presence of a massless vector boson in the open string spectrum and of 
a massless symmetric tensor in the closed string spectrum. The latter 
can be interpreted as the graviton. The former can be interpreted as 
the photon in the abelian case or as a gauge boson in the non-abelian 
one. Originally, a Yang-Mills group was introduced {\it ad hoc} 
through Chan--Paton (CP) factors. They respect the cyclicity of the 
Veneziano amplitude~\cite{veneziano}, that requires insertions of open 
string vertex operators on the boundary of a disk. In modern terms the 
group theory structure emerges from certain configuration of 
D$p$-branes (D standing for Dirichelet, $p$ for the number of spatial 
dimensions of the brane), \ie hypersurfaces where open strings can 
end~\cite{Dbranes}. 
 
In the supersymmetric case, \ie after GSO projection, the low-energy
world-volume dynamics of $N_c$ coincident D$p$-branes is governed by
the dimensional reduction from $d=10$ to $d=p+1$ of the $\cN = 1$ SYM
theory with gauge group $U(N_c)$~\cite{wittbranes}. In particular,
$p=3$ corresponds to the celebrated $\cN = 4$ SYM in $d=4$, some
(non-)perturbative properties of which will be discussed later
on. From a macroscopic view-point D$p$-branes are 1/2 BPS solitons of
type II or type I supergravities, in that they preserve one half of
the supersymmetries of the parent theory. Configurations with
different kinds of D$p$-branes are generically non supersymmetric
except for very special choices of embeddings, \ie dimensions and
orientation of the various branes w.r.t. one another. For our purposes
of relating strings to instanton calculus, it is crucial that a
configuration with $K$ D$(p-4)$-branes lying within a stack of $N_c$
D$p$-branes, \ie such that the branes have $p-4$ dimensions in common,
preserves 1/4 of the original supersymmetries. In fact this
configuration is a ``bound state'' at threshold~\cite{wittbranes}, \ie
the mass of the bound state is the sum of the masses of the
constituent branes. Moreover the D$(p-4)$-branes have all the right to
be considered as a ``gas'' of instantons within the
D$p$-branes~\cite{douglas}. 
 
We will exploit the fact that D$(p-4)$-branes behave as a gas of 
instantons within the D$p$-branes for the case $p=3$ that corresponds 
to $\cN = 4$ SYM and will indicate how to get instantons in gauge 
theories with less or no supersymmetries. We will also discuss how to 
tune the parameters, \ie the string tension $T= 1/2\pi \ap$ and the 
string coupling $g_s$ (related to the v.e.v.\ of the massless scalar 
dilaton) in order to decouple heavy string modes. We will not consider 
the cases $p\neq 3$. 
 
In the presence of $N_c$ D$3$ and $K$ D$(-1)$-branes there are three 
sectors of the open string spectrum. Strings that start and end on 
D$3$-branes provide the $U(N_c)$ gauge fields and their 
superpartners. Strings that start and end on D$(-1)$-branes yield 
$U(K)$ non-dynamical (background) gauge fields and their 
superpartners. Together they provide a subset of the (super) ADHM 
data, \eg the center of mass $x_{CM} = \sum_i M_i x_i / \sum_i M_i$, 
where $M_i$ are the masses of the brane constituents, and global SUSY 
parameters. Strings that start on D$3$-branes and end on 
D$(-1)$-branes or that start on D$(-1)$-branes and end on D$3$-branes 
provide the remaining (super) ADHM data. 
 
Suppressing CP factors for the moment, the vertex operators for gauge 
bosons, that belong to the Neveu--Schwarz (NS) sector, read 
\be 
V_A = A_M(p) \Psi^M \er^{-\varphi} \er^{i p\cdot X} \, , 
\ee 
where $X^M$ and $\Psi^M$, with $M=0,...,9$, denote the bosonic and 
fermionic string coordinates, respectively, and $\varphi$ the 
superghost boson. BRST invariance requires $p^2 = 0$ and $p\cdot A(p) 
= 0$, which is the form that the linearised Yang-Mills equations for 
$A^M(p)$ take in the transverse gauge. Vertex operators for gauginos, 
belonging to the Ramond (R) sector, read 
\be 
V_\Lambda =\Lambda^a(p) S_a \er^{-\varphi/2} \er^{i p\cdot X} \, , 
\ee 
where $S_a$, with $a=1,... 16$, is a chiral spin field that creates a 
cut for $\Psi^M$, \ie a line connecting two branch points of the 
polydromous fields $\Psi^M$. This means that the Operator Product 
Expansion (OPE) of $\Psi^M(z)$ with $S_a(w)$ contains half integer 
powers of $z-w$. BRST invariance requires $p^2 = 0$ and $p\cdot 
\Gamma_{ab} \Lambda^b(p) = 0$, which is the massless Dirac equation 
for $\Lambda^b(p)$.

After reduction to $d=4$, relevant for D$3$-branes, the gauge bosons 
in $d=10$ yield gauge bosons $A_\mu$ as well as six real scalars 
$A_i=\phi_i$. The $d=10$ gauginos yield four Weyl gauginos 
$\Lambda_\alpha^A$ and their antiparticles 
$\bar\Lambda^{\dot\alpha}_A$. The structure of the on-shell effective 
action can be extracted from the knowledge of the scattering 
amplitudes on the disk with D$3$-brane boundary conditions. In the 
low-energy limit, $\ap\rightarrow 0$ with the Yang--Mills coupling 
$g^2= 4\pi g_s$ fixed, the effective action coincides with $\cN =4$ 
SYM theory. 
 
After reduction to $d=0$, relevant for D$(-1)$-branes, also known as 
D-instantons, the gauge field vertex operator $V_A$ defined above 
yields 10 non-dynamical ``fields'', \ie matrices whose dynamics is 
governed by an action in 0 dimensions. Due to the breaking of the 
(Euclidean) Lorentz symmetry $SO(10)$ to $SO(4)\times SO(6)$ in the 
presence of D$3$-branes, it turns out to be convenient to split the 
ten ``gauge bosons'', $a_M$, into four gauge bosons, $a_\mu$, and six 
real ``scalars'', $\chi_i$. Similarly, the $d=10$ gauginos, 
$V_\Lambda$, produce four non-dynamical Weyl ``gauginos'', 
$\Theta_\alpha^A$, and their antiparticles, 
$\bar\Theta^{\dot\alpha}_A$. The structure of the on-shell effective 
action can be extracted from the scattering amplitudes on the disk 
with D$(-1)$-brane boundary conditions. In the low-energy limit, 
$\ap\rightarrow 0$ with the zero-dimensional Yang--Mills coupling 
$g^2_{0}= g_s/4\pi^3(\ap)^2$ fixed, the effective action for the low 
lying excitations of the D$(-1)$-brane reads 
\be 
\cS_{{\rm D}(-1)} = \cS_{\rm cub} + \cS_{\rm quart} \, , 
\ee where \ba \cS_{\rm cub} = {i\over 
g_0^2} \,\Tr_K \left( \bar\Theta_A \sigma^\mu [a_\mu,\Theta^A] - 
{1\over 2} \tau_i^{AB} \bar\Theta_A 
[\chi^i,\bar\Theta_B] - \right. && \nn \\ 
\left. - {1\over 2} \bar\tau^i_{AB} \Theta_A 
[\chi_i,\Theta_B] \right) && , 
\ea 
with $\Tr_K$ denoting the trace in the $K$-dimensional representation 
of $U(K)$ and 
\be 
\cS_{\rm quart} = {1\over 4 g_0^2}\,\Tr_K \! 
\left([a_\mu,a_\nu][a^\mu,a^\nu] + 2 
[a_\mu,\chi_i][a^\mu,\chi^i] + [\chi_i,\chi_j][\chi^i, 
\chi^j]\right) . 
\ee 
In what follows, it is crucial to replace $\cS_{\rm quart}$ with a 
cubic action $\cS'_{\rm cub}$, through the Hubbard--Stratonovich 
procedure, that entails the introduction of auxiliary fields 
$X_{\mu\nu} $, $Y_{\mu i}$ and $Z_{ij}$. Their vertex operators, 
bilinear in the fermions $\Psi$'s, are not BRST invariant and {\it a 
priori} one should not insert them as vertices in scattering 
amplitudes. Nevertheless, three-point amplitudes with one auxiliary 
field insertion are consistent and yield the correct interactions, 
because the BRST non invariant part decouples. In the end one replaces 
$\cS_{\rm quart}$ with 
\bea 
\cS'_{\rm cub} ={1\over 2 g_0^2} && \!\Tr_K\!\left({1\over 2} 
X_{\mu\nu} X^{\mu\nu} + Y_{\mu i}Y^{\mu i} + {1\over 2} Z_{ij}Z^{ij} 
\right. + \nn \\ 
&& \left. \rule{0pt}{14pt} + X_{\mu\nu}[a^\mu,a^\nu] 
+ 2 Y_{\mu i}[a^\mu,\chi^i] + Z_{ij}[\chi^i, \chi^j]\right) \ . 
\eea 
 
We now pass to consider open strings connecting D$3$-branes to 
D$(-1)$-branes. Vertex operators in this sector involve $\Z_2$ 
bosonic twist fields, $\sigma_{(\mu)}$, because one is changing the 
boundary conditions of the four (Euclidean) ``spacetime'' coordinates 
from Neumann (D$3$) to Dirichelet (D$(-1)$). Twist fields are local 
conformal primary operators that generate a cut in the bosonic 
coordinate field $X$ very much like spin fields, already encountered 
above, generate cuts in the fermionic coordinates $\Psi$. In the 
canonical superghost picture $q = -1$ for bosons, the vertex operators 
read 
\be 
V^{(-1)}_w =w_{\dot\alpha} \Sigma C^{\dot\alpha} \er^{-\varphi} 
T_{K,N_c} \, ,\qquad V^{(-1)}_{\bar{w}} = \bar{w}_{\dot\alpha} \Sigma 
C^{\dot\alpha} \er^{-\varphi} T_{N_c,K} \, , 
\ee 
where $\Sigma = \prod_\mu \sigma_{(\mu)}$ is a bosonic twist field of 
dimension $1/4=4\times 1/16$ and $C^{\dot\alpha}$ is an $SO(4)$ spin 
field of dimension 1/4. $T_{N_c,K}$ denote the $K\times N_c$ 
Chan--Paton ``matrices''. The supersymmetry partners, in the canonical 
$q=-1/2$ picture for fermions, have vertex operators of the form 
\be 
V^{(-1/2)}_\nu = \nu^A \Sigma C_A \er^{-\varphi/2} T_{K,N_c} \, , 
\qquad V^{(-1/2)}_{\bar{\nu}} = {\bar{\nu}}^A \Sigma C_A 
\er^{-\varphi/2} T_{N_c,K} \, , 
\ee 
where $C_A$ is an $SO(6)$ spin field. 
 
Computing amplitudes on disks with mixed boundary conditions allows 
one to extract the effective action for the ``twisted'' 
sector. Defining the $K\times K$ matrices 
\be 
W^a = (w \sigma^a\bar{w})_{_{K\times K}} \, , 
\ee 
the action that governs the dynamics of the light modes (or 
moduli) of the system of D$(-1)$-branes in the presence of 
D$3$-branes, takes the form 
\ba 
\cS_{\rm twist} = {2i\over g_0^2} \,\Tr_K \!\left( (w_{\dot\alpha} 
\bar{\nu}^A + \nu^A\bar{w}_{\dot\alpha}) \bar\Theta^{\dot\alpha}_A 
- X_a W^a + \rule{0pt}{14pt} 
\right. \nn \\ 
+ \left. {1\over 2} \chi_i \tau^i_{AB} \nu^A \bar{\nu}^B- i \chi_i 
w^{\dot\alpha} \bar{w}_{\dot\alpha}\chi^i\right), 
\ea 
where we have set $X_{\mu\nu} = X_a \bar\eta^a_{\mu\nu} + \bar{X}_a 
\eta^a_{\mu\nu}$, so that the three components $\bar{X}_a$ actually 
decouple because $\eta^a_{\mu\nu} \bar\eta_b^{\mu\nu}=0$. 
 
Combining with the previous terms and rescaling appropriately the 
fields, so as to get a non-trivial field theory limit, one finds for 
the {\it complete} action that governs the dynamics of the light modes 
(or moduli) of the system of D$(-1)$-branes in the presence of 
D$3$-branes 
\be 
\cS_{\rm moduli} = \cS_{\rm cub} +\cS'_{\rm cub} + \cS_{\rm twist} \ . 
\ee 
 
One can check that 
\be 
x^\mu_0 = \Tr_K (a^\mu) \quad {\rm and} 
\quad \theta_0^{\alpha A} = \Tr_K (\Theta^{\alpha A}) \, , 
\ee 
drop from the action, while varying w.r.t. $X_a$ and 
$\bar\Theta^{\dot\alpha}_A$ yields the super ADHM equations. The 
latter consist in $3 K\times K$ real bosonic equations 
\be 
W^a + i \bar\eta^a_{\mu\nu} [a^\mu, a^\nu] = 0 
\ee 
that, taking into account $U(K)$ invariance, impose $4 K\times K$ 
constraints on the ADHM data which implement the hyperk\"ahler 
quotient, and $8{K\times K}$ fermionic constraints (for $\cN =4$ 
supersymmetry) 
\be 
w_{\dot\alpha} \bar{\nu}^A + \nu^A \bar{w}_{\dot\alpha} + 
[\Theta^{\alpha A}, a_\mu] \sigma^\mu_{\alpha\dot\alpha} = 0 \, , 
\ee 
that reduce the number of independent fermionic zero modes. These 
ingredients, \ie the constrained ADHM superdata encoded in the various 
open string vertex operators and their interactions encoded in the 
scattering amplitudes, are sufficient to reconstruct the classical 
super instanton profile as well as to compute instanton contributions 
to correlation functions. In particular 
\be 
A^{\rm inst}_\mu(p; w, \bar{w}) = \langle\langle 
V_{\bar{w}}^{(-1)} U_\mu^{(0)}(-p) V_w^{(-1)}\rangle\rangle = 
(\bar{w} \sigma_a w)_{_{N_c\times N_c}} \bar\eta^a_{\mu\nu} p^\nu 
\er^{-i p\cdot x_0} \, , 
\ee 
where $U^{(0)}_\mu$ is the ``amputated'' vertex operator 
\be 
U_\mu^{(0)}(-p) = 2i (\de X_\mu -ip\cdot 
\Psi\Psi_\mu) \er^{-i p\cdot X} 
\ee 
in the $q = 0$ superghost picture. After Fourier transforming to $x$ 
space one obtains 
\ba 
A^{\rm inst}_\mu(x; w, \bar{w}) &=& \int {\dr^4p\over 4\pi^2 
p^2} A^{\rm inst}_\mu(p; w, \bar{w})\er^{ip\cdot x} = \nn \\ 
&=& (\bar{w} \sigma_a w)_{_{N_c\times N_c}} \bar\eta^a_{\mu\nu} 
{(x-x_0)^\nu \over (x-x_0)^4} \, , 
\ea 
which should coincide with the asymptotic behaviour of the  {\it
unconstrained} instanton at large distance in the {\it singular}
gauge.  Indeed, focussing on $K=1$ and $N_c=2$, if one sets  $2 \rho^2
= \bar{w} w$ by a global $SU(2)$ rotation, one finds 
\be 
A^{{\rm inst}, a}_\mu(x;\rho) \approx 2 \rho^2 \bar\eta^a_{\mu\nu} 
{(x-x_0)^\nu \over (x-x_0)^4} \label{AAA}\, , 
\ee 
which is the large distance term in the expansion of the celebrated 
BPST solution. To make contact with~(\ref{INST}) one clearly has to 
extract a factor $g$ from~(\ref{AAA}). Higher order terms in 
$\rho^2=\bar{w}w/2$ are sub-dominant at large distances and are 
anyway determined by solving the YM equations with the given 
asymptotic behaviour. By similar methods one can compute the 
classical asymptotic profiles of the other elementary fields 
(gauginos and scalars) that involve the 16 supersymmetry (8 
Poincar\'e and 8 superconformal) parameters broken by the D-instanton 
but preserved by the D$3$-branes (in the near horizon limit). These 
profiles enter the computation of instanton contributions to 
amplitudes. 
 
One can then embark in the computation of instanton dominated 
correlators. Denoting by $U_{\cO}(p)$ the unintegrated open string 
vertex operators corresponding to the SYM fields ${\cO}(-p)$, one 
schematically has to compute 
\bea 
&&\langle\cO_1(p_1) ... \cO_n(p_n) 
\rangle |^{\rm D-inst}_{\rm amp} = \\ 
&& = \int \dr\cM \langle\langle U_{\cO_1}(-p_1) \rangle_{\cD(\cM)} 
...\langle\langle U_{\cO_n}(-p_n) \rangle\rangle_{\cD(\cM)} 
\er^{-\cS(\cM)} \nn \ . 
\eea 
The simple ``product'' form of the integrand is due to the fact that 
the amplitude is dominated by disconnected disks with mixed boundary 
conditions $\cD(\cM)$ obtained by inserting the non-dynamical 
(super)moduli fields, which must include at least the 16 exact 
fermionic zero-modes. This is the most interesting part of the string 
construction of instantons. We have only devoted few lines to it 
because, once the ``super-instanton'' profile has been generated and 
the ``supermoduli'' have been correctly identified, one can repeat word 
by word what has been pedagogically said and carefully done in the 
discussion of $\cN =1$ SYM.

\subsection{$\cN = 2$ SYM from open strings} 
\label{sec:N2opstring} 
 
There are various ways to realise $d=4$ $\cN = 2$ SYM in string 
theory. The easiest way is to put a stack of D$3$-branes at an 
orbifold point~\footnote{Another possibility is to consider 
intersecting branes or brane with internal magnetic fluxes preserving 
$\cN = 2$ supersymmetry. Other configurations are possible in 
M-theory, \eg by wrapping M5-branes around Riemann surfaces producing 
SW curves, etc.}, let us say the origin of $\R^6/\Gamma$, such that 
the holonomy group~\footnote{If the holonomy group $\Gamma \subset 
SU(3)$ one has $\cN = 1$ SYM, when $\Gamma=1$ (trivial holonomy group) 
one has $\cN = 4$ SYM.} $\Gamma$ is a discrete subgroup of $ SU(2)$ 
of dimension $r$. As discussed in~\cite{BFMR, DM}, in the context of 
ALE instantons in string theory, there are essentially two kinds of 
branes one can consider. Regular branes are those that transform in 
the ``regular'' representation of $\Gamma$, \ie the (usually 
reducible) representation of dimension $r= \sum_i n_i^2$ equal to the 
dimension of $\Gamma$. For instance for the cyclic group $\Z_n$, the 
regular $n$-dimensional reducible representation is simply the direct 
sum of the $n$ one-dimensional irreducible representations. The 
D$3$-branes can be moved away from the orbifold point, where the 
curvature is concentrated, to the flat bulk in such a way that the $r$ 
images in the covering space actually correspond to one physical brane 
in $\R^6/\Gamma$. There can be other branes that transform under 
smaller (irreducible) representation of $\Gamma$, \eg any of the $n-1$ 
non trivial one-dimensional irreps of $\Z_n$, and are called 
``fractional'' branes in that they carry fractional R--R charge in 
$\R^6/\Gamma$, corresponding to integer charge in the covering 
space. Branes of this kind cannot be moved away from the orbifold 
point and give rise to gauge theories with lower supersymmetry than 
branes that can be moved into the flat bulk (here ``moving'' has 
exactly the same meaning as above). In orbifolds the curvature is 
concentrated at the singularity. If a (stack of) branes is displaced 
from the singular (orbifold) point and placed in the bulk, the 
effective field theory governing the dynamics of the light modes 
enjoys $\cN = 4$ SUSY. 
 
For definiteness, let us consider the case of $\Gamma = \Z_n\subset 
SU(2)$, corresponding to the A-series in the ADE classification of 
discrete subgroups of $SU(2)$ and thus the case of ALE instantons, see 
\eg~\cite{BFMR}. The regular representation is $n$-dimensional and 
reducible. One starts with $n$ stacks of $N_c$ branes each. The 
reduction of SUSY from $\cN = 4$ to $\cN =2$ is achieved by truncating 
the parent theory with gauge group $U(nN_c)$ to the sector which is 
invariant under the action of $\Z_n$. The natural action of 
$\Z_n\subset SU(2)$ on the gauge fields and complex scalars is given 
by 
\be 
A_\mu \rightarrow A_\mu \quad , \quad \phi_{3} \rightarrow 
\phi_{3}\quad , \quad\phi_{1} \rightarrow \omega \phi_{1}\quad , 
\quad \phi_{2} \rightarrow \bar\omega \phi_{2} \quad ,\label{Znfields} 
\ee 
where $\omega = \exp(2\pi i/n)$. Furthermore $\Z_n$ is taken to 
act on the gauge group $U(nN_c)$ via a discrete Wilson line 
\be 
W_{\rm reg} = (\one_{_{n\times n}}, \omega \,\one_{_{n\times n}}, 
\ldots, \omega^{n-1}\, \one_{_{n\times n}}) \, , 
\ee 
in such a way that 
\be 
T^a \rightarrow W T^a W^{-1} \label{Zngauge}\, . 
\ee 
Taking into account the combined action of $\Z_n$ 
in~(\ref{Znfields})-(\ref{Zngauge}), one concludes that the condition 
$W \Phi W^{-1} = \omega_\Phi \Phi$, where $\Phi$ collectively denotes 
the (bosonic) fields, truncates the theory to one with a vector boson 
and a complex scalar $\phi_{3}$ in the adjoint of $U(N_c)^n$ and two 
complex bosons $\phi_{1,2}$ in the bi-fundamental of adjacent 
$U(N_c)$'s. Since we have chosen precisely $\Z_n\subset SU(2)$, 
out of the 16 supersymmetry parameters associated with the $\cN = 4$ 
Poincar\'e supersymmetry, 8 are invariant and generate the $\cN = 2$ 
Poincar\'e supersymmetry. Indeed the $(\mathbf{2}_L,\mathbf{4})$ and 
$(\mathbf{2}_R, \mathbf{4}^*)$ spinors (that arise from dimensional 
reduction of the $\mathbf{16}$ of $\cN =1$ SYM in $d=10$) give rise to 
$(\mathbf{2}_L,\mathbf{2},\mathbf{1})$ and 
$(\mathbf{2}_R,\mathbf{2},\mathbf{1})$ spinors that are invariant under 
$SU(2)_{H}$ as well as to $(\mathbf{2}_L,\mathbf{1}, \mathbf{2})$ and 
$(\mathbf{2}_R,\mathbf{1},\mathbf{2})$ spinors that are not invariant 
under $SU(2)_{H}$. The resulting $\cN = 2$ Poincar\'e supersymmetry 
implies that each of the above bosons is accompanied by its fermion 
superpartner that promote the theory to $\cN = 2$ SYM coupled to 
hypermultiplets in the $({\bf N}_i, {\bf N}^*_{i+1})\oplus 
({\bf N}_i^*, {\bf N}_{i+1})$ representation. The one-loop beta function of 
$SU(N_c)^n$ turns out to be zero, because $2 N_c - 2 N_c = 0$, while 
the $U(1)\subset U(N_c)^n$ are IR free (as for any abelian gauge 
theory coupled to charged matter) and thus the $U(1)$ vector 
multiplets decouple at low energies. One is dealing with an exact $\cN 
= 2$ superconformal theory in the IR. In fact, one can turn on 
v.e.v.'s of the adjoint scalar (Coulomb branch) or of the 
bi-fundamentals (Higgs branch). The former generically breaks the 
group to $U(1)^{nN_c}$, the latter to $U(N_c)_{\rm diag}$ realising 
the expected simultaneous motion of the $n$ stacks of $N_c$ branes 
away from the fixed point into the bulk, where supersymmetry is 
enhanced to $\cN = 4$, since the hypermultiplets in the 
bi-fundamentals produce the extra adjoint of $U(N_c)_{\rm diag}$ 
needed to promote a $\cN = 2$ vector multiplet to a $\cN = 4$ vector 
multiplet. The diagonal $U(1)\subset U(N_c)_{\rm diag}$ is free and 
corresponds to the center of mass motion of the bound state of the 
various stacks of D-branes. 
 
If instead of choosing the ``regular'' embedding of $\Z_n$ in 
$U(nN_c)$ one takes another representation for $W$, one gets non 
superconformal theories that live on fractional branes. In the 
extreme case where $W = W_k$ with 
\be 
W_{k} = (\omega^k \one_{_{M\times M}}) \, , 
\ee 
and $\omega = \er^{2\pi i/n}$ for any $k=1,..., n-1$ one gets pure 
$\cN = 2$ SYM with gauge group $U(M)$ where $M$ is not necessarily a 
multiple of $n$, \ie $M\neq nN_c$ generically. Fractional branes are 
stuck at the fixed point, conventionally put at the origin of 
$\R^6/\Z_n$ and cannot move away from it. Referring to our 
previous notation, out of the six real $\phi_i$'s only two (one 
complex), $\phi$ and $\phi^{\dagger}$, survive the orbifold 
projection. The precise linear combination of the six original real 
scalar fields is determined by the choice of the embedding of $SU(2)$ 
into the rotation group of $\R^6$, $SO(6)\approx 
SU(4)$. Similarly, out of the four gaugini only two survive the 
projection, \ie the ones that are singlets of $SU(2)\supset \Gamma$ 
and transform as a charged doublet under the $SU(2)\times U(1)$ 
subgroup of $SO(6)\approx SU(4)$ commuting with $\Gamma$. The 
complexified gauge coupling of the surviving $\cN = 2$ SYM theory with 
gauge group $U(M)$ is determined by the closed string background, \ie 
the v.e.v.'s of the so-called blowing up modes of the orbifold fixed 
point. The blowing up modes are nothing but twist fields for the 
closed string coordinates, this means that the OPE of the bosonic 
coordinates $X(z, \bar{z})$ with the bosonic twist fields $\sigma(w, 
\bar{w})$ contains fractional powers. We have already encountered 
twist fields for the open string coordinates. Since closed string 
vertex operators are given by combinations of open string vertex 
operators for the left- and right-moving excitations of the closed 
string, blowing up modes are described by products of twist fields for 
the left and right movers, schematically $\sigma(z, \bar{z})= \sigma_L 
(z) \sigma_R (\bar{z})$. 
 
Indeed one may regard fractional D$3$-branes as D$5$-branes 
wrapped around homologically non-trivial cycles, sometimes called 
``exceptional divisors'', that are complex varieties of codimension 
one~\footnote{Recall that $\Gamma \subset SU(2)$ only acts on $\C^2 
\equiv \R^4 \subset \C^3 \equiv \R^6$.} in $\R^4/\Gamma\equiv 
\C^2/\Gamma$, that shrink to zero size, \ie to zero area in one's 
preferred units, at the fixed point in the orbifold limit, \ie prior 
to resolution of the singularity. For a $\Z_n$ singularity there 
are $n-1$ two-spheres that intersect according to the Cartan matrix of 
$A_{n-1}$. The complexified coupling is given by the ``period 
integrals'' of the 2-form $B_2 + i C_2$, with $B_2$ belonging to the 
Neveu--Schwarz -- Neveu--Schwarz (NS--NS) sector and $C_2$ belonging 
to the R--R sector. For regular branes, the gauge coupling of the 
diagonal subgroup, the one surviving when the branes move to the bulk, 
is given by $\Phi + i C_0$, where $\Phi$ is the NS--NS dilaton and 
$C_0$ is the R--R scalar ``axion''. Indeed one can show that the 
corresponding tadpoles precisely match the one-loop running of the 
couplings~\cite{angarm, mbjfm}! 
 
Essentially the same analysis applies to open strings with both 
ends on D$(-1)$-branes (D-instantons). Taking $K$ fractional 
D-instantons with 
\be 
W^{{\rm D-inst}} _{l} = (\omega^l\one_{_{K\times K}}) \, , 
\ee 
produces the truncation of the world-volume low-energy theory to pure 
(0-dimensional!) $\cN = 2$ SYM with gauge group $U(K)$. The surviving 
adjoint scalars will be denoted by $\chi$ and $\chi^{\dagger}$. The 
two associated non-dynamical fermions will be denoted by 
$\Theta^r_\alpha$ with $r=1,2$ and their conjugates by 
$\bar\Theta_r^{\dot\alpha}$. Setting 
\be 
a^\mu_{_{K\times K}} = x_0^\mu \one_{_{K\times K}} + 
y_g^\mu {\bf T}^g_{_{K\times K}} \, , 
\ee 
where ${\bf T}^g_{_{K\times K}}$ are the generators of $SU(K)$, and 
\be 
\Theta^{r\alpha}_{_{K\times K}} = \theta^{r\alpha}_0 
\one_{_{K\times K}} + \zeta_g^{r\alpha}{\bf T}^g_{_{K\times K}} 
\ee 
one can regard $x_0^\mu$ and $\theta^{r\alpha}_0$ as coordinates in 
$\cN = 2$ superspace. 
 
Open strings connecting $N_c$ fractional D$3$-branes to $K$ 
fractional D-instantons belong to the bi-fundamental $({\bf 
N_c},{\bf \bar K})$ representation of $U(N_c)\times U(K)$. The bosonic 
modes $w_{\dot\alpha}$ and $\bar {w}_{\dot\alpha}$ are as in the $\cN 
= 4$ case, while the fermionic modes are halved and will be 
consistently denoted by $\nu^r$ and $\bar\nu^r$. 
 
In the double scaling limit $\ap \rightarrow 0$, $g_0 \rightarrow 
\infty$ ($g_0$ has mass dimension +2) with $(4\pi^2\ap g_0)^2 = 
4\pi g_s = g^2$ fixed, the non dynamical moduli fields are 
governed by the action \be \cS^{\cN = 2}_{\rm moduli} = \cS_{\rm 
bose} + \cS_{\rm fermi} + \cS_{\rm ADHM} \, , \ee where \ba 
\cS^{\cN = 2}_{\rm bose} &=& \Tr_K\!\left(-2 [\chi^{\dagger}, 
a^\mu][\chi,a_\mu] + \chi w^{\dot\alpha} \bar {w}_{\dot\alpha} 
\chi^{\dagger} + \chi^{\dagger} 
w^{\dot\alpha} \bar {w}_{\dot\alpha} \chi \right) \nn \\ 
\cS^{\cN = 2}_{\rm fermi} &=& i {\sqrt{2} \over 2} 
\varepsilon_{rs} \Tr_K\!\left( \nu^r \bar{\nu}^s \chi^{\dagger} - 
\Theta^r [\chi,\Theta^s] \right) \\ 
\cS^{\cN = 2}_{\rm ADHM} &=& -i \Tr_K [ \bar\Theta^{\dot\alpha}_r 
( w_{\dot\alpha}\bar{\nu}^r + \nu^r \bar{w}_{\dot\alpha} 
\!+\![\Theta^{\alpha r}, a_\mu] \sigma^\mu_{\alpha\dot\alpha}) 
\!-\! X_a ( W^a + i \bar\eta^a_{\mu\nu} [a^\mu,a^\nu])] . \nn 
\ea 
Varying the action w.r.t. $X_a$ and $\bar\Theta^{\dot\alpha}_A$ 
yields the $\cN =2$ super ADHM constraints 
\be W^a + i 
\bar\eta^a_{\mu\nu} [a^\mu,a^\nu] = 0 
\ee 
and 
\be 
w_{\dot\alpha}\bar{\nu}^r + \nu^r \bar{w}_{\dot\alpha} + 
[\Theta^{\alpha r}, a_\mu] \sigma^\mu_{\alpha\dot\alpha} = 0 \, . 
\ee 
As before, one can perform a Hubbard--Stratonovich transformation and 
replace the quartic couplings in $\cS^{\cN = 2}_{\rm bose}$ with 
trilinear couplings to auxiliary fields $Y^\mu_{_{K\times K}}$ and 
$U^{\dot\alpha}_{_{N_c\times K}}$ and $\bar{U}^{\dot\alpha}_{_{K\times 
N_c}}$ and their conjugates. As a result one gets 
\bea 
\cS^{\cN = 2}_{\rm bose} &=& \Tr_K\!\left( 2 
Y^\mu Y_\mu^{\dagger}- 2 Y_\mu [\chi^{\dagger}, a^\mu] - 2 
Y_\mu^{\dagger} [\chi,a_\mu] \right. +\\ 
&+& \left. U^{\dagger}_{\dot\alpha} U^{\dot\alpha} + 
\bar{U}^{\dagger}_{\dot\alpha} \bar{U}^{\dot\alpha}+ 
\bar{U}^{\dagger}_{\dot\alpha} \bar{w}^{\dot\alpha} \chi + 
\bar{U}_{\dot\alpha} \bar{w}^{\dot\alpha} \chi^{\dagger} + \chi 
{w}^{\dot\alpha} U^{\dagger}_{\dot\alpha} + \chi^{\dagger} 
{w}^{\dot\alpha} U_{\dot\alpha} \right) . \nn 
\eea 
Computing amplitudes with insertions of the scalar field vertex 
operator 
\be V^{(-1)}_\phi(p) = \phi(p) \er^{-\varphi} \er^{i p\cdot X} 
\ee 
at $p=0$, that correspond to turning on a v.e.v.\ for $\phi$ in the 
Cartan subalgebra of $U(N_c)$, one can construct the relevant action 
for the moduli fields. By the invariance of the scattering amplitudes 
under the exchange of the dynamical field $\phi$ with the 
non-dynamical field $\chi$, the effect of the presence of a constant 
$\phi$ in the computation of instanton effects simply amounts to the 
replacements 
\be 
\chi_{{K\times K}}\otimes \one_{{N_c\times N_c}} \rightarrow \chi_{{K\times 
K}}\otimes \one_{{N_c\times N_c}} - \one_{{K\times K}} \otimes 
\phi_{{N_c\times N_c}} 
\ee 
and 
\be 
\chi^{\dagger}_{{K\times K}}\otimes \one_{{N_c\times N_c}} \rightarrow 
\chi^{\dagger}_{{K\times K}}\otimes \one_{{N_c\times N_c}} - 
\one_{{K\times K}}\otimes \phi^{\dagger}_{{N_c\times N_c}} \, . 
\ee 
It is crucial to observe at this point that $\phi$ and 
$\phi^{\dagger}$ do not enter the fermionic action in the same way, 
indeed the additional terms in the fermionic action of the $\cN = 2$ 
supermoduli read 
\be 
\Delta\cS^{\cN = 2}_{\rm fermi} = i {\sqrt{2} \over 2}\, 
\varepsilon_{rs}\, \Tr_K\!\left( \nu^r \phi^{\dagger}\bar{\nu}^s \right) . 
\ee 
As a consequence all $2K(N_c-2)$ zero-modes associated with 
$\bar{\nu}^r$ and $ \nu^s$ are lifted.

\subsection{SW prepotential from string instantons} 
\label{SWprepfromstring} 
 
Let us now specialise to the case of a $SU(2)$ gauge group. We are 
ready to accomplish the task of checking the SW prepotential, 
$\cF_{\rm SW}$, by means of Veneziano's open string theory! The 
Wilsonian effective action for the light neutral modes is 
\be 
S_{\rm eff}[\Phi] = \int \dr^4x \,\dr^4\theta \cF(\Phi) + {\rm 
h.c.} \, , 
\ee 
where $\Phi = \Phi_3 \sigma^3/2$ is the $\cN = 2$ vector superfield, 
\be 
\Phi(x,\theta) = \phi(x) + \theta^{\alpha}_r \lambda^r_{\alpha}(x) + 
{1 \over 2}\theta^{\alpha}_r\theta^{ 
\beta}_s(\varepsilon^{rs} 
\sigma^{\mu\nu}_{\alpha\beta} F_{\mu\nu}(x) + 
\sigma_a^{rs}\varepsilon_{\alpha\beta} X^a(x)) + \cdots \ . 
\label{eqz} 
\ee 
In~(\ref{eqz}) $\cdots$ stands for higher order terms in $\theta$'s 
that can be expressed in terms of the lowest components. We hope the 
reader does not get confused by the notation. In this section, $\Phi$ 
denotes an $\cN = 2$ chiral superfield (previously denoted by $A$), 
$\phi$ is its lowest component and $v$ denote the v.e.v.\ of $\phi$, 
while $a$ or more precisely $a_\mu$ are the non dynamical moduli 
fields. 
 
The contribution of the $K$-instanton sector to $S_{\rm eff}[\Phi]$ is 
given by 
\be 
S^{(K)}_{\rm eff}[\Phi] = \int_{\cM_K} \dr\mu_K \,\er^{-\cS_K(\Phi,\mu)} \, , 
\ee 
where $\mu$ collectively denotes the supermoduli parametrising 
$\cM_K$. Separating the collective coordinates $x^\mu_0$ and 
$\theta^{r\alpha}_0$ and, dropping the subscript $0$ for simplicity, 
one gets 
\be 
S^{(K)}_{\rm eff}[\Phi] = \int \dr^4x \,\dr^4\theta 
\int_{\hat\cM_K} \dr\hat\mu_K \,\er^{-\cS_K(\Phi,\hat\mu)} \, , 
\label{above} 
\ee 
so that comparison with the formula~(\ref{above}) yields 
\be 
\cF_K (\Phi) = \int_{\hat\cM_K} 
\dr\hat\mu_K \, \er^{-\cS_K(\Phi,\hat\mu)} \, , 
\ee 
where ${\hat\cM_K}$ denotes the supermoduli space of ``centered'' 
instantons. ${\hat\cM_K}$ describes configurations with fixed position 
of the center of mass of the various instantons, which in turn are 
parameterised by $\hat\mu$'s, \ie by the collective coordinates that 
do not move the position of the center of mass. Since 
$\Phi(x,\theta)$ may be taken to be a constant (slowly varying) 
superfield $\Phi(x,\theta) = \phi$ independent of the $\hat\mu$'s, one 
can compute $\cF_K (\phi)$ and then promote the argument $\phi$ to a 
chiral superfield by holomorphy in the low energy 
approximation. Indeed higher (super)derivatives would contribute to 
the 1PI effective action. Resumming the infinite number of such 
contributions should reveal the spectrum of stable particles (BPS 
monopoles and dyons), expected on the basis of the SW analysis. 
The study of this feature is beyond the scope of the present analysis. 
 
Following this strategy till the end, one finds 
\be 
\cF_K (\phi) =\cC_K \phi^2 \, {\Lambda^{4K}\over \phi^{4K}} \, , 
\ee 
where $\Lambda = v \exp(-8\pi^2/g^2(v)b_1)$ is the RG 
invariant scale, dynamically generated by dimensional transmutation, 
and $v$ is an arbitrary scale that can be taken to coincide with the 
v.e.v.\ of $\phi$. The coefficients of the $\phi$ expansion of $\cF$ 
can be computed by setting $\phi$ to any convenient value, including 
$\phi = 0$, and are given by 
\be 
\cC_K =\int_{\hat\cM_K} \dr\hat\mu_K \, \er^{-\cS_K(\phi = 0,\hat\mu)} \,. 
\ee 
The coefficients $\cC_K$ are also known as Gromov-Witten invariants. 
They have been explicitly computed for $K=1$ and $K=2$ by performing 
the integral over ${\hat\cM_K}$ and shown to match with the SW 
proposal and to reproduce Matone's relations, as previously reviewed. 
 
In fact, as previously shown in Sect.~\ref{sec:locK}, they can all 
be computed by exploiting powerful localisation properties of the 
integral over the (super)moduli space. Nekrasov and 
collaborators~\cite{nekra1,mns,nekra2,nekra3} 
have been able to localise the 
integrals over instanton moduli spaces by turning on the so-called 
$\Omega$-background, characterised by a constant self-dual 
antisymmetric tensor $\Omega_{\mu\nu}= \varepsilon_a 
\eta^a_{\mu\nu}$. From the string vantage point, the 
$\Omega$-background amounts to a constant R--R graviphoton 
field-strength in the (Euclidean) spacetime directions 
$f_{\mu\nu}= f_a \eta^a_{\mu\nu}$. The precise numerical factor is 
${1\over 2}$ so that $f_{\mu\nu}= {1\over 2} \Omega_{\mu\nu}$ or 
$f_a= {1\over 2}\varepsilon_a$. In the presence of such a 
background the D$3$-brane action gets modified to~\footnote{In 
order to expose the relative strength of the various terms in the 
action, we henceforth switch to the perturbative normalisation, 
whereby we drop the overall $1/g^2$.}  
\be  
S_{{\rm D}3} = S_{\rm SYM} - \int \dr^4x \,[2ig f_{\mu\nu}  
\Tr_{N_c}( \bar\phi F^{\mu\nu}) + g^2 f_{\mu\nu} f^{\mu\nu} \, 
\Tr_{N_c}(\bar\phi^2)]\, ,  
\ee  
where $\Tr_{N_c}$ denotes the trace over the ${N_c}$ dimensional 
representation of the $U({N_c})$ Chan--Paton group associated with the 
D$3$-branes. The modification of the effective action of the 
D$3$-branes after switching on the $\Omega$-background can be derived 
by the procedure of computing open string scattering amplitudes on the 
disk with an insertion of a closed string vertex operator for the R--R 
graviphoton. In the canonical $(-1/2,-1/2)$ superghost picture the 
vertex operator for the R--R graviphoton reads \be V_f = f_{\mu\nu} 
S^\alpha \sigma^{\mu\nu}_{\alpha\beta} \tilde{S}^\beta \Sigma 
\tilde{\Sigma}^{\dagger} \er^{-(\varphi + \tilde\varphi)/2} \ . \ee We 
observe that the only relevant amplitude is \be \langle\langle 
V^{(-1)}_A V^{(0)}_{\bar\phi} V^{(-1/2,-1/2)}_f \rangle\rangle \ . \ee 
In fact all other amplitudes, including the one with 
$V^{(0)}_{\bar\phi}$ replaced by $V^{(0)}_{\phi}$, either vanish or 
are irrelevant in the low energy limit, \ie produce higher derivative 
terms. The combined effect of the $\Omega$-background and the 
non-vanishing v.e.v.\ for $\phi$ is to replace the standard ADHM 
matrix $\Delta_{_{({N_c}+2K)\times 2K}}$ with  
\be 
\Delta_{_{({N_c}+2K)\times 2K}} \rightarrow 
\Delta_{_{({N_c}+2K)\times 2K}} + i \cA_{_{({N_c}+2K) \times 
2K}}(v,\varepsilon) + \cdots \ .  
\ee  
It is important to note hat the upper block of 
$\cA_{_{({N_c}+2K)\times 2K}}(v,\varepsilon)$ is given by  
\be  
\cA^{\rm up} _{_{{N_c}\times 2K}}(\phi,\varepsilon) =  
\phi^u{}_v w^v_{i \dot\alpha} - w^u_{j 
\dot\alpha} \chi^j{}_i \, ,  
\ee  
where $u,v = 1,... {N_c}$, $i,j=1,..., K$, and the lower block is  
\be  
\cA^{\rm low}_{_{2K \times2K}}(v,\varepsilon) = [\chi,a_\mu] 
\sigma^\mu_{\alpha\dot\alpha} + \varepsilon_a 
\sigma^a_\alpha{}^\beta a_\mu\sigma^\mu_{\beta\dot\alpha} \ .  
\ee 
As in the standard (commutative, in the absence of graviphoton 
background, \ie for $\Omega = 0$) case, the gauge field can be written 
in the convenient form $gA_\mu = U^{\dagger} \de_\mu U$, which, as 
before, is not a pure gauge because $U$ is not an ${N_c}\times {N_c}$ 
matrix. 
 
The fermionic zero modes can be parametrised as 
\be 
g^{1/2}\lambda^{\alpha r} = U^{\dagger} (\cL^r f(w,x) 
\bar{b}^\alpha + b^\alpha f(w,x) \bar\cL^r) U \, , 
\ee 
where $\cL^r, \bar\cL^r$ are $K\times K$ spinor matrices satisfying 
the super ADHM constraints and $b,\bar{b}$ are $({N_c}+2K)\times 2K$ 
constant spinor matrices with vanishing upper block and diagonal lower 
block. Moreover, one has 
\be 
f_{_{K \times K}}(w,x) = 
(\bar{w}_{\dot\alpha} w^{\dot\alpha} + (a - x \one )^2)^{-1} 
\ee 
as a consequence of the ADHM constraints. Finally, the scalar field 
profile in the presence of the non-commutative $\Omega$ background is 
given by 
\be 
\phi = i\sqrt{2}\epsilon_{rs} 
U^{\dagger} \cL^r f(w,x) \bar\cL^s U + U^{\dagger} \cJ U 
\ee 
and correctly satisfies 
\be 
D^2 \phi = - i\sqrt{2}\epsilon_{rs} 
\lambda^{\alpha r} \lambda^s_\alpha - ig \Omega_{\mu\nu} 
F^{\mu\nu} 
\ee 
to lowest order in $g$ with $F_{\mu\nu} = \tilde{F}_{\mu\nu}$. We note 
that $\cJ$ is an $({N_c}+2K) \times ({N_c}+2K)$ block diagonal matrix 
with the upper block $\cJ^{\rm up}_{_{{N_c} \times {N_c}}} = 
v_{_{{N_c} \times {N_c}}}$, where $v_{_{{N_c} \times {N_c}}}$ 
represents the v.e.v.\ of the dynamical scalar fields $\phi_{_{{N_c} 
\times {N_c}}}$, and the lower block $\cJ^{\rm low}_{_{2K \times 2K}} 
= \chi_{_{K \times K}}\otimes \one + \one \otimes \varepsilon_a 
\sigma^a$, where $\chi_{_{K \times K}}$ represents the non-dynamical 
scalar fields (moduli). 
 
In principle one can analyse by similar means gauge theories with 
lower ($\cN =1$) or no supersymmetry. This analysis is only in its 
infancy and goes beyond the scope of this review. It is the subject of 
very intense research activity at present, see \eg~\cite{N1stringinst}. 
We hope we have provided the interested and 
proficient reader with the necessary tools to enter the {\it arena} of 
this fascinating endeavor.


\section{Instanton effects in $\calN=4$ SYM} 
\label{sec:N4intro} 
 
In the following sections we shall review the calculation of instanton
effects in $\calN=4$ SYM~\cite{N4susy}. This is the maximally extended
(rigid) supersymmetric theory in four dimensions and possesses a
number of remarkable properties. It is ultra-violet
finite~\cite{finN4} and provides an example of four dimensional
quantum field theory with exact conformal invariance at the quantum
level. The theory is also believed to be invariant under a
strong$\leftrightarrow$weak coupling duality, known as S-duality,
which generalises the Montonen--Olive electric-magnetic
duality~\cite{gno,emdual}. Originally the interest in the theory was
driven by the discovery of its finiteness properties. In recent years
it has been extensively studied in the context of the AdS/CFT
duality~\cite{malda,GKP,Witthol}, which relates it to type IIB
superstring theory in an AdS$_5\times S^5$ background. 
 
As a conformal field theory $\calN=4$ SYM has rather different 
physical properties from those of the $\calN=1$ and $\calN=2$ theories 
previously discussed. However, instanton effects play a decisive 
r\^ole also here and the methods of supersymmetric instanton calculus 
described in the previous sections have recently been extensively 
applied to the study of non-perturbative aspects of the model. In 
particular, instantons are expected to be instrumental to the 
realisation of S-duality in $\calN=4$ SYM and the study of their 
contributions has led to some of the most striking tests of the 
validity of the AdS/CFT correspondence. 
 
After a brief overview of the structure of $\calN=4$ SYM and its 
notable properties, in 
Sects.~\ref{sec:N4insteff}-\ref{sec:N4-multinst} we shall describe the 
calculation of instanton contributions to correlation functions, 
essentially following the method that in Sect.~\ref{sec:2} was called 
the SCI method. In Sects.~\ref{sec:adscft-intro} 
and~\ref{sec:inst-adscft} we shall then discuss the r\^ole of 
instantons in the AdS/CFT duality.

\section{$\calN=4$ supersymmetric Yang--Mills theory} 
\label{sec:N4sym} 
 
The $\calN=4$ supersymmetric Yang--Mills theory was originally 
constructed in~\cite{N4susy} as the dimensional reduction of the 
ten-dimensional $\calN=1$ supersymmetric Yang--Mills theory on a 
six-torus. The field content of the theory consists of a gauge field, 
$A_\mu$, four Weyl fermions, $\lambda^A_\a$ ($A=1,\ldots,4$), and six 
real scalars, $\vp^i$ ($i=1,\ldots,6$). In terms of $\calN=1$ 
multiplets these fields combine into one vector and three chiral 
multiplets. All the fields are in the adjoint representation of the 
gauge group, which in most of the following will be taken to be 
$SU(N_c)$. The global supergroup of symmetries of the theory is 
$PSU(2,2|4)$, whose maximal bosonic subgroup is $SO(2,4)\times SU(4)$. 
The $SO(2,4)$ factor is the four-dimensional conformal group and 
$SU(4)$ is the R-symmetry group under which the fermions transform in 
the $\mb4$ (and their conjugates in the $\mb{\bar 4}$), the scalars in 
the $\mb6$ and the gauge field is a singlet. It is often convenient to 
label the scalars by an antisymmetric pair of indices in the $\mb4$, 
as $\vp^{AB}$, subject to the reality constraint 
\be 
\bar\vp_{AB} = (\vp^{AB})^\dagger = \half \veps_{ABCD}\vp^{CD} \, . 
\label{scalconstr} 
\ee 
The two parametrisations are related by 
\be 
\vp^i = \fr{\sqrt{2}}\,\S^i_{AB}\vp^{AB} \, , \qquad 
\vp^{AB} = \fr{\sqrt{8}}\,\bar\S^{AB}_i \vp^i \, , 
\label{scalrel} 
\ee 
where $\S_{AB}^i$ ($\bar\S_i^{AB}$) are Clebsch-Gordan coefficients 
projecting the product of two $\mb4$'s (two $\mb{\bar 4}$'s) onto the 
$\mb6$. They can be expressed in terms of the 't Hooft 
symbols~\cite{GTH} as 
\ba 
&& \S^i_{AB} = (\S^a_{AB},\S^{a+3}_{AB}) 
= (\eta^a_{AB},i\bar\eta^a_{AB}) \, , \label{defsig6} \\ 
&& \bar\S_i^{AB} = (\bar\S^a_{AB},\bar\S^{a+3}_{AB}) 
= (-\eta_a^{AB},i\bar\eta^{AB}_a) \, , \qquad 
i=1,\ldots,6 \, , \; a=1,2,3 \, . \rule{0pt}{14pt} \nn 
\ea 
The elementary fields are conveniently represented as colour matrices 
and the classical action of the theory, which is uniquely determined 
(up to the choice of gauge group) by $\calN=4$ supersymmetry, can be 
written as 
\ba 
S &\!\!=\!\!& \int \dr^4x \, \Tr\left\{ \half F_{\mu\nu}F^{\mu\nu} + 
2 D_\mu\vp^{AB}D^\mu\bar\vp_{AB} -2i \lambda^{\a A}\Dsm_{\a\adot} 
\bar\lambda_A^\adot - \right. \label{N4action} \\ 
&\!\!- \!\!& \left. 2g \lambda^{\a A}[\lambda^B_\a,\bar\vp_{AB}] 
-2g \bar\lambda_{\adot A}[\bar\lambda^\adot_B,\vp^{AB}] 
- 2g^2 [\vp^{AB},\vp^{CD}][\bar\vp_{AB},\bar\vp_{CD}] 
\rule{0pt}{14pt} \right\} . \nn 
\ea 
The action~(\ref{N4action}) is invariant under the supersymmetry 
transformations 
\ba 
&& \hsp{-1} \delta_\epsilon \vp^{AB} = \half \left(\lambda^{\a A} 
\epsilon_\a^B- \lambda^{\a B} \epsilon_\a^A\right) + 
\half\veps^{ABCD} \bar\epsilon_{\adot C} 
\bar\lambda^\adot_D \nn \\ 
&& \hsp{-1} \delta_\epsilon \lambda_\a^A = -\half F_{\mu\nu} 
\s^{\mu\nu\,\b}_\a \epsilon_\b^A + 4i \left(\Dsm_{\a\adot} 
\vp^{AB}\right) \bar\epsilon^\adot_B - 8g[\bar\vp_{BC}, \vp^{CA}] 
\epsilon_\a^B \label{susytransf} \\ 
&& \hsp{-1} \delta_\epsilon A^\mu = -i \lambda^{\a A} 
\s^\mu_{\a\adot} \bar\epsilon^\adot_A -i \epsilon^{\a A} 
\s^\mu_{\a\adot} \bar\lambda^\adot_A \, . \nn 
\ea 
Given the gauge group, the action~(\ref{N4action}) contains a single 
parameter, the coupling constant $g$~\footnote{In the following we 
will maintain the notation used in the previous sections, denoting the 
Yang--Mills coupling by $g$. The string coupling constant will be 
denoted by $\gs$.}. The absence of divergences in the theory implies 
that the corresponding $\b$-function vanishes. As discussed in 
Appendix~\ref{sec:ATT}, it is possible to add to the action a 
$\vartheta$-term. 
 
The $\calN=4$ SYM theory has a vacuum manifold parametrised by the 
v.e.v.'s of the six scalars which make the potential vanish. 
The resulting moduli space turns out to be 
\be 
\calM = \R^{6r}/\calS_r \, , 
\label{N4modsp} 
\ee 
where $r$ is the rank of the gauge group and $\calS_r$ is the group of 
permutations of $r$ elements. At a generic point of the moduli space 
the theory is in a Coulomb phase and the gauge group is broken down to 
$U(1)^r$. In this phase and in the presence of a $\vartheta$-term the 
theory contains BPS-saturated monopole and dyon states characterised 
by integer quantum numbers, $q_e$ and $q_m$, associated with their 
electric and magnetic charges~\cite{OliWitt,Osb}. The conjectured 
S-duality of $\calN=4$ SYM requires that the spectrum of such states 
be invariant under the action of $SL(2,\Z)$ transformations acting 
projectively on the complexified coupling, $\tau$ (defined 
in~(\ref{TAU})), 
\be 
\tau \to \frac{a\tau+b}{c\tau+d} \, , \qquad 
a,b,c,d \in \Z \, , \qquad ad-bc=1 \, , \label{Scoupling} 
\ee 
while simultaneously rotating the electric and magnetic quantum 
numbers according to 
\be 
\left( \begin{array}{c} q_e \\ q_m 
\end{array} \right) \to \left( \begin{array}{cc} -a & b \\ 
c & -d \end{array} \right) \left( \begin{array}{c} q_e \\ 
q_m \end{array} \right) \, . 
\label{Schargrot} 
\ee 
Significant evidence in support of this conjecture has been obtained 
using semi-classical methods~\cite{sen}. 
 
The conformal phase of the theory corresponds to the origin of the 
moduli space where all the scalar v.e.v.'s vanish. As already 
observed, at this point the classical (super)conformal symmetry is 
preserved at the quantum level, resulting in a non-trivial interacting 
conformal field theory. In this phase the fundamental observables are 
correlation functions of gauge-invariant composite operators 
constructed from the elementary fields in~(\ref{N4action}). Such 
operators are classified according to their transformation under the 
global symmetries and are organised in multiplets of the 
superconformal group, $PSU(2,2|4)$. Some properties of the $\calN=4$ 
superconformal group and its multiplets are reviewed in 
Appendix~\ref{sec:psu224}. Each operator is characterised by its 
quantum numbers with respect to the bosonic subgroup $SO(2,4)\times 
SU(4)$. These can be chosen to be two spins, $(j_1,j_2)$, and the 
scaling dimension, $\D$, identifying the transformation under the 
conformal group together with three Dynkin labels, $[k,l,m]$, 
identifying the $SU(4)$ representation under which the operator 
transforms. $\calN=4$ composite operators can be broadly divided into 
two classes, protected operators belonging to short or semi-short 
``BPS'' multiplets of the superconformal group and unprotected ones 
belonging to long multiplets~\cite{N4multi,MBrev}. Correlation 
functions of protected operators satisfy special non-renormalisation 
properties. A notable example of BPS multiplet is the one comprising 
the $PSU(2,2|4)$ conserved currents, \ie the energy-momentum tensor, 
$\calT_{\mu\nu}$, and the supersymmetry and R-symmetry currents, 
$\S^\mu_{\a A}$ and $\calJ_A^{\mu\,B}$, respectively. We give here the 
explicit form of the first few components of the $\calT_{\mu\nu}$ 
multiplet 
\ba 
&& \calQ^{[A_1B_1][A_2B_2]} = \Tr \left( 2 \vp^{A_1B_1}\vp^{A_2B_2} 
+ \vp^{A_1A_2}\vp^{B_1B_2} + \vp^{A_1B_2}\vp^{A_2B_1}\right) \nn \\ 
&& \calX^{A_1[A_2B_2]}_\a = \Tr\left( 2 \lambda^{A_1}_\a \vp^{A_2B_2} 
+ \lambda^{A_2}_\a \vp^{A_1B_2} - \lambda^{B_2}_\a \vp^{A_1A_2} 
\right) \nn \\ 
&& \calE^{(A_1A_2)} = \Tr\left( -\lambda^{\a A_1}\lambda^{A_2}_\a 
+ g\,t^{(A_1A_2)}_{CDEFGH} \, \vp^{CD}\vp^{EF}\vp^{GH} \right) 
\label{scmult} \\ 
&& \calB^{[A_1A_2]}_{\mu\nu} = \Tr \left( \lambda^{\a A_1} 
\s_{\mu\nu\,\a}{}^\b \lambda^{A_2}_\b + 2i F_{\mu\nu} 
\vp^{A_1A_2} \right) \nn \\ 
&& \Lambda^A_\a = \Tr \left\{\s^{\mu\nu\,\b}_{\,\a} 
F_{\mu\nu} \lambda^A_\b + g [{\bar\vp}_{BC},\vp^{CA}] 
\lambda^B_\a + \left(\Dsm_{\a\adot} {\bar\lambda}^\adot_B 
+ g[\lambda^C_\a, {\bar\vp}_{BC}] \right) \vp^{AB} 
\right\} . \nn 
\ea 
The operator $\calQ$ is the lowest component of the multiplet and 
transforms in the ${\bf 20^\prime}$ of the $SU(4)$ R-symmetry, $\calE 
$ and $\calB_{\mu\nu}$ are respectively in the ${\bf 10}$ and ${\bf 
6}$ and the fermionic operators $\calX_\a$ and $\Lambda_\a$ transform 
in the ${\bf 20}$ and ${\bf 4}$ respectively. The tensor 
$t^{(A_1A_2)}_{CDEFGH}$ in $\calE$ projects the product of three ${\bf 
6}$'s onto the ${\bf 10}$. In the next sections we shall study 
various examples of correlation functions involving the operators 
in~(\ref{scmult}). We shall also consider other BPS multiplets in the 
same class whose lowest component is a dimension $\ell$ scalar 
operator which, in terms of the $\vp^i$ scalars, takes the 
form~\footnote{We use curly brackets to denote symmetrisation with 
subtraction of traces. For simple symmetrisation and 
anti-symmetrisation we use parentheses and square brackets, 
respectively.} 
\be 
\calQ_\ell^{\{i_1i_2\cdots i_\ell\}} = \Tr\left[\vp^{\{i_1}\vp^{i_2} 
\cdots\vp^{i_\ell\}}\right] \, . 
\label{KKQop} 
\ee 
The first example of long (non-BPS) multiplet is the $\calN=4$ Konishi 
multiplet, whose lowest component is the dimension 2, $SU(4)$ singlet 
scalar 
\be 
\calK_{\mb1} = \veps_{ABCD}\,\Tr\left(\vp^{AB}\vp^{CD}\right) 
\, . \label{K1def} 
\ee 
Conformal invariance implies that, given a complete basis of operators 
in the theory, any correlation function is fully determined, via the 
Operator Product Expansion (OPE), by two sets of numbers, the scaling 
dimensions of the operators and the Wilson coefficients that couple 
triplets of operators. Both scaling dimensions and Wilson coefficients 
receive perturbative and non-perturbative quantum corrections, so they 
are non-trivial functions of the coupling, $g$, and the 
$\vartheta$-angle. 
 
The spectrum of scaling dimensions in $\calN=4$ SYM has been the 
subject of extensive study in the context of the AdS/CFT 
correspondence. The scaling dimensions of composite operators are 
determined by their two-point correlation functions. For a primary 
operator, $\calO(x)$, conformal invariance fixes the form of the 
two-point function $\la\calO(x)\calO^\dagger(y)\ra$ to be 
\be 
\la\calO(x)\calO^\dagger(y)\ra = \frac{c}{(x-y)^{2\D}} \, , 
\label{conf2ptf} 
\ee 
where $\D$ is the scaling dimension and $c$ is a constant which may, 
in general, depend on $g$ and the $\vt$-angle~\footnote{This is 
actually an over-simplification. In general, in a given sector 
characterised by certain quantum numbers, one needs to consider a 
complete set of operators and resolve their mixing. The resolution of 
the operator mixing diagonalises the matrix of two-point 
functions. Only after this step equations of the form~(\ref{conf2ptf}) 
determine the physical scaling dimensions.}. In general $\D$ receives 
quantum corrections, $\D=\D_0+\g$, where $ \D_0$ is the bare or 
engineering dimension and $\g$ the anomalous part. The latter has an 
expansion of the form 
\be 
\g(g,\vartheta) = \sum_{n=1}^\infty \g^{\rm pert}_n g^{2n} + 
\sum_{K>0}\sum_{m=0}^\infty \left[ \g^{(K)}_m g^{2m}\, 
\er^{(-8\pi^2/g^2+i\vartheta)K} + {\rm c.c.} \right] \, , 
\label{adimexp} 
\ee 
where the first series contains the perturbative contributions and the 
second double series the instanton and anti-instanton contributions. 
The two-point functions of protected operators are not renormalised 
implying that their bare dimensions are not corrected ($\D=\D_0$). 
 
The behaviour~(\ref{adimexp}) of the anomalous dimensions illustrates 
an important feature of $\calN=4$ observables. In general physical 
quantities receive contributions at all orders in perturbation theory 
and from all instanton sectors. Moreover in each instanton sector 
there exists an infinite series of perturbative corrections arising 
from fluctuations around the leading instanton semi-classical 
contribution. This is the consequence of the absence of chiral 
selection rules and marks an important difference with respect to the 
cases of $\calN=1$ and $\calN=2$ theories considered in the previous 
sections. Indeed, in $\calN=4$ SYM there are no anomalous $U(1)$'s. 
As a consequence, as will be discussed in the next sections, there 
exist no correlation functions which are dominated by the contribution 
of specific instanton sectors. 
 
In the conformal phase the field equations of $\calN=4$ SYM admit no 
(non-singular) monopole or dyon solutions and the conjectured 
S-duality has a different realisation. Specifically it requires that 
the spectrum of scaling dimensions of gauge-invariant operators be 
invariant under the $SL(2,\Z)$ transformations~(\ref{Scoupling}). 
This suggests that the scaling dimensions should naturally be written 
as functions of $\tau$ and $\bar\tau$ in the form 
\be 
\D = \D(\tau,\bar\tau) = \D_0 + \g(\tau,\bar\tau) \, . 
\label{tauadim} 
\ee 
We then conclude that instanton effects, which are the source of the 
$\vartheta$ dependence in~(\ref{adimexp}), must play a crucial r\^ole 
here. Similarly, it can be argued that instantons are important in 
determining the behaviour of correlation functions under S-duality. As 
will be discussed in Sects.~\ref{sec:adscft-intro} 
and~\ref{sec:inst-adscft}, the arguments outlined here also resonate 
with what is understood about the r\^ole of D-instantons in the dual 
type IIB string theory compactified on AdS$_5\times S^5$. 
Unfortunately, little is known beyond these qualitative considerations 
and the details of how the S-duality of $\calN=4$ SYM is implemented 
in the superconformal phase remain largely elusive, see, 
however,~\cite{VaWi} and~\cite{langlands} for recent progress.

\section{Instanton calculus in $\calN=4$ SYM} 
\label{sec:N4insteff} 
 
In this section we discuss some general features of instanton calculus
in the $\calN=4$ SYM theory, highlighting the differences with respect
to the $\calN=1$ and $\calN=2$ cases. In the next section we analyse
in more detail one-instanton contributions to some specific
correlation functions, focussing on the case of $SU(N_c)$ gauge group
which is particularly important for the applications to the AdS/CFT
correspondence. Multi-instanton configurations are described using
the ADHM formalism~\cite{ADHM} (see also
Sect.~\ref{sec:constrcheck}). A full account of the technical aspects
of the ADHM construction is beyond the scope of the present work. A
comprehensive review of multi-instanton calculus in supersymmetric
gauge theories can be found in~\cite{KHOREV}. The calculation of
multi-instanton contributions is extremely involved and their direct
evaluation is only possible for small instanton numbers (although
remarkable progress was made in~\cite{nekra2,nekra3}. See the
discussion in Sect.~\ref{sec:topotwist}). However, dramatic
simplifications occur in the large $N_c$ limit of relevance for the
AdS/CFT duality.  A brief review of the computation of multi-instanton
corrections to  $\calN=4$ correlators in this limit will be given in
Sect.~\ref{sec:N4-multinst} following the work of~\cite{KHETAL}. 
 
The calculation of correlation functions in $\calN=4$ SYM in the 
semi-classical approximation proceeds as described in a general 
setting in Sect.~\ref{sec:2}. The path integral is evaluated using a 
saddle point approximation around the instanton configuration and thus 
reduced to a finite dimensional integral over the collective 
coordinate manifold associated with bosonic and fermionic zero- modes. 
However, the form of the interaction terms in the $\calN=4$ 
action~(\ref{N4action}), and in particular the coupling to the scalar 
fields, requires a modification of the previous analysis. 
 
In principle, as done in Sect.~\ref{sec:2}, it is possible to use as 
saddle point configuration the solution in which the gauge field is 
given by the standard bosonic instanton with all the other fields 
vanishing, \ie 
\be 
A_\mu = A_\mu^I \, , \qquad \lambda^A_\a = \bar\lambda^\adot_A = 
\vp^{AB} = 0 \, . 
\label{purbosinst} 
\ee 
The fluctuations of the various fields in the background of this 
configuration, including fermions, should then be treated 
perturbatively. This results in an expansion which, in the case under 
consideration of $\calN=4$ SYM, is somewhat hard to handle beyond 
leading order. A more efficient approach consists in utilising as 
saddle point configuration a finite action solution of the complete 
set of coupled field equations of the theory including higher order 
corrections in $g$. This also provides a natural framework for 
implementing the supersymmetric generalisation of the ADHM 
construction. 
 
A generic $n$-point correlation function computed in the 
semi-classical approximation around such a saddle point has an 
expression analogous to~(\ref{INTBF})-(\ref{INTSU}) which we 
schematically rewrite in the form 
\be 
\la\calO_1(x_1)\cdots\calO_n(x_n)\ra = \int \dr\mu(\b,c) \, 
\er^{-S_{\rm inst}} \, \hat\calO_1(x_1;\b,c) \cdots 
\hat\calO_n(x_n;\b,c) \, , 
\label{npointsc} 
\ee 
where $\dr\mu(\b,c)$ is the integration measure over the bosonic 
($\b$) and fermionic ($c$) collective coordinates arising from the 
zero-mode fluctuations around the classical solution and $S_{\rm 
inst}$ is the action evaluated on the solution. With $\hat\calO_i$ we 
denote the classical expressions of the operators at the saddle 
point. The latter depend on the insertion points of the operators and 
the collective coordinates. 
 
For pure $\calN=1$ SYM there exists a whole manifold of saddle points 
(including the one in~(\ref{purbosinst})) which correspond to 
field configurations with $A_\mu=A_\mu^I$, a gaugino solution of the 
Weyl--Dirac equation 
\be 
\bar\Dsm^{\adot\a}\lambda_\a = 0 \, , 
\label{fzmeq} 
\ee 
and the anti-chiral fermion, $\bar\lambda_\adot$, identically zero. 
The resulting semi-classical expectation values involve integrals over 
the $n_B$ bosonic collective coordinates as well as the $n_F$ fermion 
zero-modes resulting from the index theorem and discussed in previous 
sections. 
 
The generalisation of this analysis to the $\calN=4$ case incurs in a 
serious obstacle: no exact topologically non-trivial solution of the 
coupled field equations is known except~(\ref{purbosinst}). The 
$\calN=4$ SYM field equations read 
\ba 
&& \hsp{-0.2} D_\mu F^{\mu\nu} +i\,g \{\lambda^{\a A} 
\s^\nu_{\a\adot},\bar\lambda_A^\adot \} + \half g 
[\bar\vp_{AB},D^\nu\vp^{AB}] = 0 \nn \\ 
&& \hsp{-0.2} D^2 \vp^{AB} + \sqrt{2}\,g [\{\lambda^{\a A}, 
\lambda^B_\a\} + \half\,\veps^{ABCD} 
\{\bar\lambda_{\adot C},\bar\lambda^\adot_D\}] 
-\frac{g^2}{2} [\bar\vp_{CD},[\vp^{AB},\vp^{CD}]] = 0 \nn \\ 
&& \hsp{-0.2} \bar{\Dsm}^{\adot\a}\lambda^A_\a + 
i\sqrt{2}\,g[\vp^{AB},\bar\lambda^\adot_B] = 0 
\label{fieldeqs} \\ 
&& \hsp{-0.2} \Dsm_{\a\adot}\bar\lambda_A^\adot - i\sqrt{2}\, 
g[\bar\vp_{AB},\lambda^B_\a] = 0 \rule{0pt}{16pt} \nn \,. 
\ea 
In the following discussion we denote by $\Phi^{(n)}$ a solution of
the classical equations of motion for the generic field $\Phi$ which
depends on $n$ zero modes of the Dirac operator in the instanton
background. As already observed, the equations~(\ref{fieldeqs}) are
solved by  the purely bosonic configuration~(\ref{purbosinst}), where
$A_\mu^I=A^{(0)}_\mu$ is a charge-$K$ instanton, with $n_B$ associated
collective coordinates. One can try to do better and solve iteratively
the full set of coupled equations~(\ref{fieldeqs}). Upon substituting
the  instanton solution~(\ref{purbosinst}), the equation for
$\lambda^A$ is  of the form~(\ref{fzmeq}) and admits $n_F$ independent
solutions for  each ``flavour'' $A=1,\ldots,4$. After this first step
of the  iteration, which generates a non-trivial solution,
$\lambda^{A(1)}_\a$, for the gauginos, one notices that the
configuration 
\be 
A_\mu = A^{(0)}_\mu \, , \quad \lambda^A_\a 
= \lambda^{A(1)}_\a \, , \quad \vp^{AB} = 
\bar\lambda^\adot_A = 0 \, , 
\label{solstep1} 
\ee 
unlike what happens in the cases of ${\cal N}=1$ and ${\cal N}=2$ SYM, 
is not a solution of~(\ref{fieldeqs}) and the process has to 
continue. The equation for the scalar fields, obtained inserting back 
$\lambda^{A(1)}_\a$, admits a solution which is bilinear in the 
fermion modes, $\vp^{AB(2)}$. Again the resulting configuration 
\be 
A_\mu = A^{(0)}_\mu \, , \quad \lambda^A_\a 
= \lambda^{A(1)}_\a \, , \quad \vp^{AB} = 
\vp^{AB(2)} \, , \quad \bar\lambda^\adot_A = 0 \, , 
\label{solstep2} 
\ee 
is not an exact solution of~(\ref{fieldeqs}). A further iteration 
generates a non-trivial field configuration, 
$\bar\lambda^{\adot(3)}_A$, for the anti-chiral fermions involving 
three zero-modes. At this point also the first equation 
in~(\ref{fieldeqs}) gets an extra term, so that at the next step a 
modification, $A^{(4)}_\mu$, of the standard bosonic instanton is 
necessary. One may notice that the field strength associated with 
$A^{(4)}_\mu$ is no longer (anti-)self-dual. 
 
In principle this recursive procedure can only stop when, after a
number of successive iterations, a field configuration involving a
number of fermion modes exceeding $n_F$ is generated. The first few
steps of the construction described above were explicitly carried out
in~\cite{bvnv}. However, a complete super-instanton multiplet, which
would exactly solve~(\ref{fieldeqs}) in closed form is not
known. Indeed it has been argued in~\cite{KHOREV} that for generic
gauge group such an exact solution may not exist. In spite of this
obstacle it is possible to consistently compute the semi-classical
contribution to correlation functions expanding the path integral
around an appropriately approximate solution. The crucial observation
is that successive steps in the iterative procedure outlined above
produce corrections to the solution which are suppressed by increasing
powers of $g$. Therefore in the weak coupling limit it is consistent
to employ as saddle point configuration an approximate truncated
solution of the equations of motion in which only terms up to a
certain power of $g$ are retained. Thus the idea is to solve the field
equations to leading order and include in the integration
in~(\ref{npointsc}) all the zero-modes of the truncated equations. For
the purpose of computing correlators in the semi-classical
approximation the relevant saddle point is determined by solving the
system 
\ba 
&& F_{\mu\nu} = \wtil F_{\mu\nu} \nn \\ 
&& \bar\Dsm^{\adot\a}\lambda^A_\a = 0 \label{ELeqs} \\ 
&& D^2\vp^{AB} = g\,\sqrt{2} \left(\lambda^{\a A}\lambda^B_\a 
- \lambda^{\a B}\lambda^A_\a \right) \nn \, , 
\ea 
with the integration measure in~(\ref{npointsc}) including all the 
associated fermion zero-modes. The action evaluated on the solution 
of~(\ref{ELeqs}) is not simply given by ${8\pi^2}/{g^2}$, but 
manifestly depends on a subset of the collective coordinates 
\be 
S_{\rm inst} = \frac{8\pi^2}{g^2}-i\vartheta + \tilde S_{\rm inst} 
(\tilde\b,\tilde c) \, , 
\label{corrinstaction} 
\ee 
where we have denoted with $\tilde\b$ and $\tilde c$ the collective 
coordinates associated with the ``non-exact'' zero-modes, \ie those 
which are zero-modes of the truncated equations~(\ref{ELeqs}), but not 
of the full coupled equations~(\ref{fieldeqs}). These non-exact 
zero-modes are said to be ``lifted'' by the interactions. As will be 
discussed more explicitly in the next section, in the case of gauge 
group $SU(N_c)$, the only fermion modes which remain unlifted are 
those associated with the Poincar\'e and special supersymmetries 
which are broken in the instanton background. All the remaining modes 
are lifted by the coupling to the scalars. 
 
The lifting of fermion zero-modes has important consequences for the 
properties of correlation functions which receive instanton 
contributions. Since some of the fermionic collective coordinates 
($\tilde c$ in the formula above) appear explicitly in the action, it 
is not necessary to saturate the corresponding fermionic integrations 
with the operator insertions in order to obtain a non-vanishing 
result. This implies that in the $\calN=4$ theory there are no 
(strict) selection rules determining which correlation functions 
receive contributions from which winding number sector, unlike what 
happens in the $\calN=1$ and $\calN=2$ cases. In particular 
non-vanishing correlators receive contributions from configurations 
with arbitrary instanton number. This result emerging from explicit 
calculations has its origin in the absence of an anomalous $U(1)$ 
R-symmetry in $\calN=4$ SYM. 
 
It must be stressed that the approach described here is consistent 
only if restricted to the calculation of the leading order 
contributions in $g$ (see~(\ref{ELeqs})). The reason is that in order 
to go to higher orders in a consistent way one should also take 
systematically into account all the quantum fluctuations that beyond 
the semi-classical approximation have been neglected.

\section{One-instanton in $\calN=4$ SYM with $SU(N_c)$ 
gauge group} 
\label{sec:N4-1inst} 
 
In the one-instanton sector the approach described in the previous 
section can be implemented in a straightforward way. We focus here on 
the case of $SU(N_c)$ gauge group, but the generalisation to 
orthogonal and symplectic groups is not too difficult. As explained 
above, we shall use as saddle point for the calculation of correlation 
functions in the semi-classical approximation the solution to the 
truncated equations~(\ref{ELeqs})~\footnote{In preparation for the 
forthcoming discussion on the AdS/CFT duality we shall from now on 
work with fields rescaled by a factor of $g$. Consequently the action 
of the $\calN=4$ theory will have an overall factor $1/g^2$ in 
front. This is the normalisation which arises naturally in the dual 
string theory and therefore this rescaling will simplify the 
comparison of string and gauge theory results.}. The resulting 
saddle-point field configuration is characterised by $4N_c$ bosonic 
collective coordinates. As we know they are the position, $x_0$, and 
size, $\rho$, of the instanton and its global gauge orientation 
parameters. The latter can be conveniently identified with the set of 
variables, $w_{u\adot}$ and $\bar w^{\adot u}$ (where $u=1,\ldots,N_c 
$ is a colour index and $\adot=1,2$ is a spinor index), parametrising 
the coset $SU(N_c)/SU(N_c-2)\times U(1)$ describing the $SU(2)$ colour 
orientation of the instanton and its embedding into 
$SU(N_c)$~\cite{BER}. Moreover, as we will be working with the 
approximate solution of~(\ref{ELeqs}), all the $8N_c$ fermionic 
collective coordinates associated with the zero modes of the Dirac 
operator will be included in the integration measure. These comprise 
the sixteen moduli associated with the Poincar\'e and special 
supersymmetries broken by the bosonic instanton and denoted 
respectively by $\eta^A_\a$ and $\bar\xi^A_\adot$ ($A=1,2,3,4$). For 
brevity we shall refer collectively to these modes as superconformal 
modes. The remaining fermion moduli, which can be thought of 
superpartners of the gauge orientation parameters, are described by 
$8N_c$ parameters, $ \nu^A_u$ and $\bar\nu^{Au}$, subject to the 
$2\times 8$ constraints 
\be 
\bar w^{\adot u}\nu^A_u = 0 \, , \quad \bar\nu^{Au}w_{u\adot} = 0 \, , 
\label{fermiadhm} 
\ee 
which effectively reduce their number to $8(N_c-2)$. 
 
As discussed in the previous section, only the sixteen superconformal 
modes remain exact zero modes to all orders in $g$. The $\nu^A_u$ and 
$\bar\nu^{Au}$ modes are lifted and appear explicitly in the instanton 
action. The solution of the coupled equations~(\ref{ELeqs}) generates 
a non-trivial configuration for the scalars, $\vp^{AB}$, which is 
bilinear in the $8N_c$ fermion zero-modes. Substituting this solution, 
together with $\bar\lambda^\adot_A=0$, in the action (truncated to the 
cubic couplings for consistency with our iterative procedure) gives 
\be 
S_{\rm inst} = -2\pi i \tau + S_{4F} 
= -2\pi i \tau + \frac{\pi^2}{2g^2\rho^2} 
\,\veps_{ABCD}\,\calF^{AB}\calF^{CD} \, , 
\label{instaction1} 
\ee 
where $\tau$ was defined in~(\ref{TAU}) and 
\be 
\calF^{AB}=\frac{1}{2\sqrt{2}}\left(\bar\nu^{Au}\nu_u^B - 
\bar\nu^{Bu}\nu_u^A \right) \, . 
\label{instaction2} 
\ee 
It is the four-fermion term, $S_{4F}$, arising from the Yukawa 
couplings $\lambda^A[\lambda^B,\bar\vp_{AB}]$ which is responsible 
for the lifting the of the $8(N_c-2)$ $\nu^A_u$ and $\bar\nu^{Au}$ 
modes. 
 
\subsection{A generating function} 
\label{sec:genf} 
 
In the following we only consider correlators of gauge-invariant 
operators for which the integration over the moduli describing the 
global orientation of the instanton simply produces a volume factor 
which can be absorbed in the measure~\cite{BER}. We denote by 
$\dr\mu_{\rm phys}$ the gauge-invariant, or physical, integration 
measure on the instanton moduli space obtained after integration over 
the gauge orientation parameters. The physical measure to be used in 
the calculation of expectation values in semi-classical approximation 
in the one-instanton sector is~\footnote{Here and in the following 
formulae we omit a ($N_c$-independent) numerical constant that will be 
reinstated in the final expressions.} 
\ba 
&& \int \dr\mu_{\rm phys} \, \er^{-S_{\rm inst}} = 
\frac{\pi^{-4N_c}g^{4N_c}\er^{2\pi i\tau}}{(N_c-1)!(N_c-2)!} \times 
\label{physmeasure} \\ 
&& \times \int \dr\rho\,\dr^4x_0 \prod_{A=1}^4 
\dr^2\eta^A\dr^2\bar\xi^A \, \dr^{N_c-2}\nu^A \dr^{N_c-2}\bar\nu^A 
\,\rho^{4N_c-13} \er^{-S_{4F}} \, , \nn 
\ea 
where the instanton action is given 
in~(\ref{instaction1})-(\ref{instaction2}) and the $\rho$, $g$ and 
$N_c$ dependence is the result of the normalisation of the collective 
coordinates as explained in Sect.~\ref{sec:2}. 
 
Following~\cite{KHETAL} the integration over the $8(N_c-2)$ non-exact 
modes, $\nu^A_u$ and $\bar\nu^{Au}$, can be reduced to a Gaussian 
form introducing auxiliary bosonic coordinates, $\chi^i$, 
$i=1,\ldots,6$, and rewriting the r.h.s. of~(\ref{physmeasure}) in 
the form 
\ba 
&& \hsp{-1} \frac{\pi^{-4N_c}g^{4N_c}\er^{2\pi i\tau}}{(N_c-1)! 
(N_c-2)!} 
\int \dr\rho\,\dr^4x_0 \,\dr^6\chi \,\prod_{A=1}^4 
\dr^2\eta^A\dr^2\bar\xi^A \, \dr^{N_c-2}\nu^A \dr^{N_c-2}\bar\nu^A 
\times\nn \\ 
&& \hsp{2.5} \times \rho^{4N_c-7} \exp\left[-2\rho^2\chi^i\chi^i 
+\frac{4\pi i}{g}\chi_{AB}\calF^{AB}\right] \, , 
\label{physmeaschi} 
\ea 
where $\chi_{AB}=\frac{1}{\sqrt{8}}\S^i_{AB}\chi^i$ and $\calF^{AB}$ 
is defined in~(\ref{instaction2}). 
 
The semi-classical contribution to a correlation function of local 
gauge-invariant operators is obtained by integrating the product of 
their profiles in the instanton background with the above measure. The 
integration over the Grassmann variables in~(\ref{physmeaschi}) 
requires that the classical expressions of the operators soak up the 
sixteen fermion modes $\eta^A_\a$ and $\bar\xi^A_\adot$ for the result 
to be non-zero. The Grassmann integrals over the $\nu^A$ and 
$\bar\nu^A$ modes are non-vanishing even if the operators do not 
contain any dependence on these variables. In the following we shall 
use the terminology introduced in~\cite{gk} and refer to correlation 
functions in which the operator insertions soak up only the sixteen 
exact modes as ``minimal''. Correlators in which the operators contain 
a dependence on more than sixteen modes will be referred to as 
``non-minimal''. In order to systematically study the instanton 
contributions to generic correlation functions it is convenient to 
construct a generating function. This allows to drastically simplify 
the combinatorics associated with the $\nu^A$ and $\bar\nu^A$ 
integrations in the non-minimal cases~\cite{gk}. For this purpose we 
introduce sources, $\bar\vt^u_A$ and $\vt_{Au}$, 
in~(\ref{physmeaschi}) coupled to those fermionic variables and define 
\ba 
&& \hsp{-0.2}Z[\vt,\bar\vt]\! = 
\!\frac{\pi^{-4N_c}g^{4N_c}\er^{2\pi i\tau}} {(N_c-1)!(N_c-2)!} 
\!\int \!\dr\rho \, \dr^4x_0 \, \dr^6\chi \prod_{A=1}^4 
\!\dr^2\eta^A \, \dr^2 {\bar\xi}^A\, 
\dr^{N_c-2}{\bar\nu}^A\, \dr^{N_c-2}\nu^A \times\nn \\ 
&&\hsp{-0.2} \times \rho^{4N_c-7} \exp\left[-2\rho^2 \chi^i\chi^i+ 
\frac{\sqrt{8}\pi i}{g}{\bar\nu}^{Au}\chi_{AB}\nu^B_u + 
{\bar\vt}^u_{A}\nu^A_u + \vt_{Au}{\bar\nu}^{Au} \right] . 
\label{genfunct} 
\ea 
Since the integrals over ${\bar\nu}$ and $\nu$ are Gaussian, they 
can be immediately computed. Introducing polar coordinates 
\be 
\chi^i \, \rightarrow \, (r,\Omega) \: , \quad 
\sum_{i=1}^6 (\chi^i)^2 = r^2 \, , 
\label{polaromega} 
\ee 
we find 
\ba 
Z[\vt,\bar\vt] &\!\!=\!\!& 
\frac{2^{-29}\pi^{-13}\,g^{8}\er^{2\pi i\tau}}{(N_c-1)!(N_c-2)!} 
\int \dr\rho \, \dr^4x_0 \, \dr^5\Omega \prod_{A=1}^4 \dr^2\eta^A \, 
\dr^2 {\bar\xi}^A \, \rho^{4N_c-7} \times\nn \\ 
&& \times \int_0^\infty \dr r \, r^{4N_c-3} \er^{-2\rho^2 r^2} 
\calZ(\vt,\bar\vt;\Omega,r) \, , 
\label{genfunctfin} 
\ea 
where all the numerical coefficients omitted in previous expressions 
have been reinstated and we have introduced the density 
\be 
\calZ(\vt,\bar\vt;\Omega,r) = \exp\left[-\frac{ig}{\pi r}\, 
{\bar\vt}_A^u \Omega^{AB} \vt_{Bu} \right] \, , 
\label{density} 
\ee 
where the symplectic form $\Omega^{AB}$ is given by 
(see~(\ref{polaromega})) 
\be 
\Omega^{AB}= \bar\Sigma^{AB}_i\Omega^i \, , 
\qquad \sum_{i=1}^6 \left(\Omega^i\right)^2 = 1 \, . 
\label{defOmega} 
\ee 
Notice that the angular variables, $\Omega^i$, introduced in the polar 
representation of the auxiliary coordinates, $\chi^i$, parametrise a 
five-sphere. This will play a very important r\^ole in comparing with 
string theory results in the context of the AdS/CFT correspondence. 
 
A $n$-point correlation function in the semi-classical approximation 
takes now the form (see~(\ref{npointsc})) 
\be 
\la \calO_1(x_1)\cdots\calO_n(x_n) \ra = 
\int \dr\mu_{\rm phys} \, \er^{-S_{\rm inst}}\, \hat\calO_1 
\cdots\hat\calO_n 
\, , \label{gensemicl} 
\ee 
where 
\be 
\hat\calO_i = \hat\calO_i(x_i;x_0,\rho,\eta^A,\bar\xi^A,\nu^A,\bar 
\nu^A) 
\label{genprof} 
\ee 
denotes the classical instantonic profile of the operator $\calO_i$ 
and generically depends on all the bosonic and fermionic moduli. In 
particular the $\nu^A$ and $\bar\nu^A$ modes appear in gauge invariant 
operators only in colour singlet bilinears, in either symmetric or 
anti-symmetric combinations belonging to the representation ${\bf 10}$ 
or ${\bf 6}$ of $SU(4)$, respectively, \ie in the combinations 
\ba 
(\bar\nu^A \nu^B)_{\bf {10}} &\equiv& 
\nut = ({\bar\nu}^{Au}\nu^B_u + {\bar\nu}^{Bu}\nu^A_u) \, , 
\label{10bilinear} \\ 
(\bar\nu^A \nu^B)_{\bf 6} &\equiv& \nus = 
({\bar\nu}^{Au}\nu^B_u -{\bar\nu}^{Bu}\nu^A_u) \, . 
\label{6bilinear} 
\ea 
The strategy is then to rewrite the dependence on these collective 
coordinates in each insertion in~(\ref{gensemicl}) in terms of 
derivatives with respect to the sources $(\vt_A,\bar\vt_A)$. In this 
way the dependence on the $\nu^A$'s and $\bar\nu^A$'s is traded for a 
dependence on the angular variables $\Omega^{AB}$. After this step the 
integration over the radial parameter, $r$, can be computed and one is 
left with an integration over the bosonic coordinates, $x_0$, $\rho$, 
$\Omega^{AB}$, and the sixteen coordinates associated with the exact 
modes, $\eta^A_\a$ and $\bar\xi^A_\adot$. In the next subsections we 
present some examples of such calculations.

\subsection{Minimal correlation functions} 
\label{sec:mincorr} 
 
A class of correlation functions which have been extensively studied 
in the context of the AdS/CFT correspondence are those involving the 
operators in the $\calN=4$ supercurrent multiplet~(\ref{scmult}). This 
is a 1/2-BPS supermultiplet and thus all its components are protected 
operators. Their two- and three-point functions are not renormalised 
and in particular do not receive instanton contributions. However, 
their four- and higher-point functions can be non-zero in an instanton 
background and turn out to contain interesting dynamical 
information. The first examples of correlators in this class were 
considered in~\cite{bgkr} in the case of $SU(2)$ gauge group. The 
calculations have then been generalised to $SU(N_c)$ in~\cite{dkmv} 
and to multi-instantons in the large $N_c$ limit in~\cite{KHETAL}. 
 
$\bullet$ The simplest minimal correlation function involves sixteen 
insertions of the fermionic operator~\footnote{Here and in the 
following we use for the composite operators the normalisation 
appropriate in the context of the AdS/CFT correspondence, which 
requires that their tree-level correlation functions be proportional 
to $N_c^2$ (see~(\ref{sugra4cont})-(\ref{sugra4exc})).} 
\begin{align} 
\Lambda^A_\a = \fr{g^2}\,\Tr&\left\{\s^{\mu\nu\,\b}_{\,\a} 
F_{\mu\nu} \lambda^A_\b + [{\bar\vp}_{BC},\vp^{CA}] 
\lambda^B_\a + \right. \nn \\ 
&\left.\;+ \left(\Dsm_{\a\adot} {\bar\lambda}^\adot_B 
+ [\lambda^C_\a, {\bar\vp}_{BC}] \right) \vp^{AB} 
\right\} \, , 
\label{Lam4def} 
\end{align} 
transforming in the ${\bf 4}$ of the $SU(4)$ R-symmetry group 
\be 
G_{16}(x_1,x_2,\ldots,x_{16}) = \la \Lambda^{A_1}_{\a_1}(x_1) \, 
\Lambda^{A_2}_{\a_2}(x_2)\cdots\Lambda^{A_{16}}_{\a_{16}}(x_{16}) 
\ra \, . 
\label{G16Lam} 
\ee 
For the calculation of~(\ref{G16Lam}) in the semi-classical 
approximation only the contribution of the first term 
in~(\ref{Lam4def}) to the classical profile of $\Lambda^A_\a$ is 
relevant. In fact by reinserting for a moment the powers of $g$ that 
were absorbed in the redefinition of the fields, it is immediately 
seen that all the other terms are of higher order in $g$. 
 
Substituting the solution for $A^{(0)}_\mu$ and $\lambda^{A(1)}_\a$ 
one obtains for the classical profile of $\Lambda^A_\a$ 
\be 
\hat\Lambda^A_\a(x) = \frac{96}{g^2}\,\rho^4\,\left[f(x)\right]^4 
\, \zeta^A_\a(x) \, , 
\label{Lamprof} 
\ee 
where the function $f(x)$ is defined in~(\ref{FX}) and $\zeta^A_\a(x)$ 
is the combination 
\be 
\zeta^A_\a(x) = \fr{\sqrt{\rho}}\left[\rho\,\eta^A_\a - (x-x_0)_\mu 
\s^\mu_{\a\adot}\bar\xi^{\adot A} \right] \, . 
\label{zetadef} 
\ee 
We explicitly note that $\hat \Lambda^A_\a$ is linear in the 
superconformal collective coordinates and does not depend on the 
$\nu^A$ and $\bar\nu^A$ modes. This means that in evaluating the 
correlator~(\ref{G16Lam}) the sources $\bar\vt^u_A$ and $\vt_{Au}$ 
can be set to zero. We thus get 
\ba 
&& G_{16}(x_1,\ldots,x_{16}) = \frac{2^{51}\,3^{16}\,\pi^{-13}\, 
\er^{2\pi i\tau}}{g^{24}\,(N_c-1)!(N_c-2)!} \int \dr\rho \, 
\dr^4x_0 \, \dr^5\Omega \prod_{A=1}^4 \dr^2\eta^A \, 
\dr^2 {\bar\xi}^A \times \nn \\ 
&& \times \int_0^\infty \dr r \, r^{4N_c-3} \er^{-2\rho^2 r^2} 
\rho^{4N_c-7} \prod_{i=1}^{16} \rho^4\left[f(x_i)\right]^4\, 
\zeta^{A_i}_{\a_i}(x_i) \, . 
\label{G16-1} 
\ea 
The $r$ integral is elementary and yields 
\be 
\int_0^\infty \dr r\,r^{4N_c-3} \er^{-2\rho^2r^2} 
= 2^{-2N_c} \,\rho^{2-4N_c} \Gamma(2N_c-1) \, , 
\label{rint1} 
\ee 
so that 
\ba 
\hspace{-.2cm}G_{16}(x_1,\ldots,x_{16}) &\!=\!& 
\frac{c_1(N_c)\,2^{51}\,3^{16}\,\pi^{-13}\, 
\er^{2\pi i\tau}}{g^{24}} \int \frac{\dr\rho\,\dr^4x_0}{\rho^5} 
\, \dr^5\Omega \prod_{A=1}^4 \dr^2\eta^A \, 
\dr^2 {\bar\xi}^A \times \nn \\ 
\hspace{-.2cm}&& \times \prod_{i=1}^{16} 
\frac{\rho^4}{[(x_i-x_0)^2+\rho^2]^4}\, 
\zeta^{A_i}_{\a_i}(x_i) \, , 
\label{G16-fin} 
\ea 
where 
\be 
c_1(N_c) = \frac{2^{-2N_c}\Gamma(2N_c-1)}{(N_c-1)!(N_c-2)!} ~ 
\raisebox{-6pt}{$\begin{array}{c} \displaystyle \sim \\ 
\scriptstyle N_c\to\infty \end{array}$} N_c^{1/2} \, . 
\label{G16Ndep} 
\ee 
The integration over the fermion modes selects terms with eight
$\eta^A$'s and eight $\bar\xi^A$'s in~(\ref{G16-fin}) and results in a
fully anti-symmetric tensor in the $SU(4)$ and spinor indices. As will
be discussed in Sect.~\ref{sec:inst-adscft}, the unintegrated
expression~(\ref{G16-fin}) suffices for the comparison with the
associated dual process in string theory. 
 
$\bullet$ As a second example of minimal correlation function we 
consider the four-point function 
\be 
G_4(x_1,\ldots,x_4) = \la \calQ^{A_1B_1C_1D_1}(x_1) \cdots 
\calQ^{A_4B_4C_4D_4}(x_4) \ra \, , 
\label{G4Q} 
\ee 
where the scalar operators $\calQ^{ABCD}$ belong to the ${\bf 
20^\prime}$ of $SU(4)$ and are given by 
\be 
\calQ^{ABCD} = \fr{g^2} \, \Tr \left( 2 \vp^{AB}\vp^{CD} 
+ \vp^{AC}\vp^{BD} - \vp^{AD}\vp^{BC}\right) \, . 
\label{Q20def} 
\ee 
When evaluated on the solution of the saddle point 
equations~(\ref{ELeqs}), the $\calQ^{ABCD}$'s contain four fermion 
modes. Unlike the fermions $\Lambda^A_\a$ they also involve the 
$\nu^A$ and $\bar\nu^A$ modes. However, in the minimal 
correlator~(\ref{G4Q}) this dependence can be neglected as all the 
fermion modes need to be of type $\eta^A$ and $\bar\xi^A$ to saturate 
the corresponding Grassmann integrals. The relevant terms giving the 
profile of $\calQ^{ABCD}$ are 
\be 
\hat\calQ^{ABCD} = \frac{96}{g^2}\, \rho^4\left[f(x)\right]^4 
\, \left[(\zeta^{\a A}\zeta^C_\a)(\zeta^{\b B}\zeta^D_\b) - 
(\zeta^{\a A}\zeta^D_\a)(\zeta^{\b B}\zeta^C_\b) \right] \, , 
\label{Q20prof} 
\ee 
with $\zeta^A_\a$ defined in~(\ref{zetadef}). 
 
Proceeding as for the sixteen-point function~(\ref{G16Lam}), one finds 
\ba 
&& \hsp{-0.5}G_4(x_1,\ldots,x_4) = c_1(N)\,2^{-9}\,3^4\,\pi^{-13} 
\,\er^{2\pi i\tau} \int \frac{\dr\rho \, \dr^4x_0}{\rho^5} \, 
\dr^5\Omega \prod_{A=1}^4 \dr^2\eta^A \, \dr^2 {\bar\xi}^A 
\times \nn \\ 
&& \hsp{-0.5} \times \prod_{i=1}^4 
\frac{\rho^4}{[(x-x_0)^2+\rho^2]^4}\, 
\left[(\zeta^{A_i}\zeta^{C_i})(\zeta^{B_i}\zeta^{D_i}) - 
(\zeta^{A_i}\zeta^{D_i})(\zeta^{B_i}\zeta^{C_i}) 
\right]\!\!(x_i) \, , 
\label{G4fin} 
\ea 
with $c_1(N)$ given in~(\ref{G16Ndep}). 
 
In the case of the four-point function~(\ref{G4Q}) the final integrals 
in~(\ref{G4fin}) have been explicitly computed in~\cite{bkrs1}. The 
result is a very complicated function of the distances 
$x_{ij}^2=(x_i-x_j)^2$, which, however, can be used to extract 
information about instanton contributions to the anomalous dimensions 
of certain operators via the OPE analysis. As discussed 
in~\cite{bkrs1}, the result shows, in particular, that the Konishi 
operator~(\ref{K1def}) does not acquire an instanton induced anomalous 
dimension. The study of the OPE also shows that there are $SU(4)$ 
singlet operators with $\Delta_0=4$ which do receive an instanton 
contribution (as well as possibly a perturbative one) to their 
anomalous dimension. We shall briefly return to the calculation of 
instanton corrections to the scaling dimensions of composite operators 
at the end of the next subsection and to the interpretation 
of~(\ref{G16-fin}) and~(\ref{G4fin}) in Sect.~\ref{sec:instvsDinst}.

\subsection{Non-minimal correlation functions} 
\label{sec:nonmincorr} 
 
The minimal correlators considered above are not dominated by the 
contribution of the one-instanton sector. Apart from ordinary 
perturbative corrections, they receive contributions from $K>1$ 
instanton configurations as well as from perturbative fluctuations in 
each instanton sector. The non-minimal correlation functions, in which 
the operator insertions soak up more than the minimal number of 
fermion zero-modes, have similar properties in this respect. However, 
their calculation presents new complications that will now be 
illustrated with explicit examples. In general one can distinguish two 
classes of such non-minimal correlation functions, based on the 
features that differentiate them from related minimal cases, \ie those 
involving additional insertions and those involving higher-dimensional 
operators. 
 
$\bullet$ An example of the first type is the twenty-point function 
\be 
G_{20}(x_1,\ldots,x_{20}) = \la \Lambda^{A_1}_{\a_1}(x_1)\cdots 
\Lambda^{A_{16}}_{\a_{16}}(x_{16})\,\calE^{B_1C_1}(y_1)\cdots 
\calE^{B_4C_4}(y_4) \ra \, , 
\label{G20LamE} 
\ee 
where $\Lambda^A_\a$ is defined in~(\ref{Lam4def}) and 
\be 
\calE^{BC} = \fr{g^2}\,\Tr\left( -\lambda^{\a B}\lambda^{C}_\a 
+ t^{(BC)}_{DEFGHL} \, \vp^{DE}\vp^{FG}\vp^{HL} \right)\, . 
\label{E10def} 
\ee 
For the calculation of~(\ref{G20LamE}) in the semi-classical 
approximation only the first term in~(\ref{E10def}) is relevant. Its 
contribution to the classical profile of $\calE^{BC}$ is 
\be 
\hat\calE^{BC} = -\frac{96}{g^2} \, \rho^4 \left[f(x)\right]^4 
\,\zeta^{\a B}\zeta^C_\a - \frac{2}{g^2} \,\rho^2\left[f(x)\right]^3 
(\bar\nu^{Bu}\nu^C_u+\bar\nu^{Cu}\nu^B_u) \, . 
\label{E10prof} 
\ee 
As explained in the previous subsection, the operator $\Lambda^A_\a$ 
does not depend on the fermion modes of type $\nu^A$ and $\bar \nu^A$ 
and thus in evaluating~(\ref{G20LamE}) we need to use for each 
$\calE^{BC}$ insertion the second term in~(\ref{E10prof}), as the 
$\Lambda^A_\a$ insertions already soak up all the superconformal 
modes. In this way we get 
\ba 
G_{20}(x_1,\ldots,x_{20}) &\!=\!& \fr{g^{40}} \int \dr\mu_{\rm phys}\, 
\er^{-S_{\rm inst}} \prod_{i=1}^{16} 96\rho^4\left[f(x_i)\right]^4 
\zeta^{\a_iA_i}(x_i) \times \nn \\ 
&& \hsp{2} \times \prod_{j=1}^4 2 \rho^2 \left[f(y_j)\right]^3 
(\bar\nu^{(B_j}\nu^{C_j)}) \, . 
\label{G20LamE-1} 
\ea 
Using the generating function~(\ref{genfunctfin}), this formula can be 
rewritten in the form 
\ba 
&& G_{20}(x_1,\ldots,x_{20}) = 
\frac{2^{55}3^{16}\pi^{-13}\,\er^{2\pi i\tau}} 
{g^{32}(N_c-1)!(N_c-2)!} \int\! \dr\rho \, \dr^4x_0 \, \dr^5\Omega 
\prod_{A=1}^4 \!\dr^2\eta^A \,\dr^2 {\bar\xi}^A \times \nn \\ 
&& \times \int_0^\infty \dr r \, r^{4N_c-3} \er^{-2\rho^2 r^2} 
\, \rho^{4N_c-7} \prod_{i=1}^{16} \rho^4\left[f(x_i)\right]^4 
\, \zeta^{\a_iA_i}(x_i) \prod_{j=1}^4 \rho^2 
\left[f(y_j)\right]^3 \times \nn \\ 
&& \times \left.\left[\frac{\d^8\calZ(\vt,\bvt,\Omega,r)} 
{\d\vt_{u_1(B_1}\d\bvt^{u_1}_{C_1)}\cdots\d\vt_{u_4(B_4} 
\d\bvt^{u_4}_{C_4)}} \right]\right|_{\vt=\bvt=0} . 
\label{G20LamE-2} 
\ea 
After evaluating the derivatives and eliminating the sources the 
integral over $r$ can be performed and one gets 
\ba 
&& G_{20}(x_1,\ldots,x_{20}) = \frac{2^{57}3^{16}\pi^{-17}\,c_2(N_c) 
\,\er^{2\pi i\tau}}{g^{28}} \int \frac{\dr\rho \, \dr^4x_0}{\rho^5} \, 
\prod_{A=1}^4 \dr^2\eta^A \,\dr^2 {\bar\xi}^A \times \nn \\ 
&& \times \prod_{i=1}^{16} \frac{\rho^4}{[(x_i-x_0)^2+\rho^2]^4} 
\,\zeta^{\a_iA_i}(x_i) \prod_{j=1}^4 \frac{\rho^3} 
{[(y_j-x_0)^2+\rho^2]^3} \nn \\ 
&& \times \int \dr^5\Omega \, \left[ \Omega^{B_1C_2} 
\Omega^{B_2C_1}\Omega^{B_3C_4}\Omega^{B_4C_3} 
+ \cdots \right] \, , \label{G20LamE-fin} 
\ea 
where the ellipsis in the last line stands for permutations of the 
$B_i,C_i$ indices and 
\ba 
c_2(N_c) &\hsp{-0.3}=\hsp{-0.3}& 
\frac{2^{-2N_c}(N_c-2)^2\Gamma(2N_c-3)}{(N_c-1)!(N_c-2)!} \nn \\ 
&\hsp{-0.3}\raisebox{-6pt}{$\begin{array}{c} \displaystyle \sim \\ 
\scriptstyle N_c\to\infty \end{array}$}\hsp{-0.3}& 
N_c^{1/2} \left[1-\frac{25}{8N_c} 
+{\rm O}(1/N_c^2)\right] \, . 
\label{G20LamENdep} 
\ea 
The factor of $(N_c-2)^2$ in the numerator of $c_2(N_c)$ comes from 
the contraction of colour indices in the $\vt_{Au}$'s and $\bvt_A^u$'s 
sources. The integration over the five sphere in~(\ref{G20LamE-fin}) 
gives the $SU(4)$ tensor 
\be 
t^{B_1C_1\cdots B_4C_4} = \veps^{B_1C_2B_2C_1}\veps^ 
{B_3C_4B_4C_3} 
+ {\rm permutations} \, . 
\label{5sphres} 
\ee 
The main difference to be noted with respect to the minimal cases is 
the non-trivial dependence on the angular variables parametrising the 
five-sphere. In general, as in the above expression, the five-sphere 
integral factorises and gives rise to $SU(4)$ selection 
rules. Specifically, a correlation function can receive a non-zero 
instanton contribution only if the $SU(4)$ flavour indices carried by 
the non-exact modes, $\nu^A$ and $\bar\nu^A$, appear in a combination 
containing the $SU(4)$ singlet representation. We shall re-examine the 
results~(\ref{G20LamE-fin})-(\ref{G20LamENdep}) in 
Sect.~\ref{sec:instvsDinst} in connection with the corresponding 
processes in the dual string theory. In particular we will see that 
the calculation of non-minimal correlators such as~(\ref{G20LamE}) 
leads to a puzzle: the $N$-dependence in the SYM result does not agree 
with that of the amplitudes which are naturally identified as their 
dual. The resolution of the puzzle will require taking into account 
further types of contributions which do not arise in the minimal case 
(see Sect.~\ref{sec:instvsDinst}). 
 
From the previous example and the form of the generating
function~(\ref{genfunctfin}) we can deduce some general features of
non-minimal correlation functions. The insertion of each
$(\bar\nu^A\nu^B)$ bilinear in a correlator corresponds to two
derivatives of~(\ref{density}) with respect to the sources. This,
besides producing a factor of $g$, also modifies the $r$ dependence
of the integrand, thus affecting the overall $N_c$-dependence,
see~(\ref{rint1}).  Moreover additional factors of $N_c$ are
associated  with the contraction of the colour indices carried by the
$\nu^A_u$'s  and $\bar\nu^{Au}$'s variables. From~(\ref{genfunctfin})
and~(\ref{density}) one checks that in general the insertion of any
$(\bar\nu^A\nu^B)_{\bf {10}}$ pair yields a factor $g$ and the
insertion of each $(\bar\nu^A\nu^B)_{\bf 6}$ pair a factor
$g\sqrt{N_c}$. 
 
Schematically for a generic non-minimal $n$-point correlation function 
containing $q$ $(\bar\nu^A\nu^B)_{\bf {10}}$ factors and $p$ 
$(\bar\nu^A\nu^B)_{\bf 6}$ bilinears one finds 
\ba 
\la \calO_1(x_1) \cdots \calO_n(x_n) \ra &\!\sim\!& g^{8+p+q}\, 
\er^{2\pi i \tau} \a(N_c) \int \frac{\dr\rho\,\dr^4x_0}{\rho^5} 
\,\dr^5\Omega \, \prod_{A=1}^4 \dr^2\eta^A \, \dr^2\bar\xi^A 
\times \nn \\ 
&& \hsp{1} \times \rho^{p+q} \prod_{i=1}^n\tilde\calO_i(x_i;x_0, 
\rho,\eta,\bar\xi,\Omega) \, , 
\label{genNdep} 
\ea 
where $\tilde\calO_i$ denote the profiles of the operators after the 
dependence on the non-exact modes has been re-expressed in terms 
of 
the $\Omega^{AB}$'s of~(\ref{defOmega}). For future use we give 
the 
expression of the coefficient $\a(N_c)$ at large $N_c$ 
\ba 
\a(N_c) &\!=\!& \frac{2^{-2N_c}\Gamma\left(2N_c-1-\frac{p+q}{2} 
\right)} {(N_c-1)!(N_c-2)!}N_c^{p+\frac{q}{2}} 
\left[1+{\rm O}(\frac{1}{N_c})\right] \nn \\ 
&\!\sim\!& N_c^{\frac{1}{2}+\frac{p}{2}}\left[1+{\rm O} 
(\frac{1}{N_c})\right] \, . \label{Ndepnu} 
\ea 
$\bullet$ All the operators in the supercurrent multiplet considered 
so far only involve the $(\bar\nu^A\nu^B)_{\mb{10}}$ 
bilinears. Anti-symmetric bilinears occur in higher dimension 
operators such as those in multiplets having as lowest component the 
scalars~(\ref{KKQop}) with $\ell\ge 3$. An example of correlation 
function containing such insertions is 
\be 
G_{16}(x_1,\ldots,x_{16}) = \la 
\Lambda^{A_1}_{\a_1}(x_1) \cdots 
\Lambda^{A_{14}}_{\a_{14}}(x_{14}) 
\tilde\Lambda^{B_1B_2B_3}_{\b_1}(y_1) 
\tilde\Lambda^{C_1C_2C_3}_{\b_2}(y_2) \ra \, , 
\label{G16LamKK} 
\ee 
where the operator $\tilde\Lambda^{B_1B_2B_3}_\a$, which 
belongs to the same multiplet as $\calQ_{\ell=3}$ and transforms 
in the ${\bf \overline{20}}$ of $SU(4)$, is 
\ba 
\tilde\Lambda^{B_1B_2B_3}_\a &\!=\!& \fr{g^3N_c^{1/2}}\, \Tr 
\left[ 2 \lambda^{B_1}_\a\left( \lambda^{\b\,B_2}\lambda_\b^{B_3} 
+ \lambda^{\b\,B_3}\lambda_\b^{B_2} \right) + \right. \nn \\ 
&& + \lambda^{B_2}_\a \left( 
\lambda^{\b\,B_1}\lambda_\b^{B_3} + \lambda^{\b\,B_3} 
\lambda_\b^{B_1} \right) + \lambda^{B_3}_\a 
\left(\lambda^{\b\,B_1}\lambda_\b^{B_2} 
+ \lambda^{\b\,B_2}\lambda_\b^{B_1} \right) + \nn \\ 
&& \!\! \left. + F_{mn}\s^{mn \, \b}_{\,\a} 
\left( \{ \lambda^{B_2}_\b, \vp^{B_1B_3}\} + 
\{ \lambda^{B_3}_\b, \vp^{B_1B_2} \} \right) + \cdots \right] , 
\label{Lam20def} 
\ea 
with the ellipsis referring to terms which are negligible in the 
semi-classical approximation. 
 
The profile of the operator~(\ref{Lam20def}) in the one-instanton 
background is 
\be 
\hat{\tilde\Lambda}^{B_1B_2B_3}_\a = 
\frac{24}{g^3N_c^{1/2}} \,\rho^4 
\left[f(x)\right]^5\left[\zeta^{B_2}_\a 
(\bar\nu^{[B_1}\nu^{B_3]})+\zeta^{B_3}_\a 
(\bar\nu^{u[B_1}\nu^{B_2]}_u) \right] \, . 
\label{Lam20prof} 
\ee 
The correlation function~(\ref{G16LamKK}) is computed by replacing 
each inserted operator with its classical profile and integrating over 
the moduli space. In particular the normalisation of the operators 
$\tilde\Lambda^{ABC}_\a$ is such as to compensate the additional 
factors $N_c$ associated with the $(\bar\nu^A\nu^B)_{\bf 6}$ bilinears 
and the final result behaves again like $N_c^{1/2}$ in the large-$N_c$ 
limit. Proceeding as in the previous cases, one finds 
\ba 
&& G_{16}(x_1,\ldots,x_{16}) = 
\frac{c_3(N_c)\,2^{48}\,3^{16}\, \er^{2\pi 
i\tau}}{\pi^{17}\,g^{24}} \int \frac{\dr\rho \, 
\dr^4x_0}{\rho^5} \, \dr^5\Omega \prod_{A=1}^4 \dr^2\eta^A \, 
\dr^2 {\bar\xi}^A \times \nn \\ 
&& \times \prod_{i=1}^{14} \frac{\rho^4}{[(x_i-x_0)^2+\rho^2]^4}\, 
\zeta^{A_i}_{\a_i}(x_i) \times \label{G16KK-fin} \\ 
&&\times \left[\frac{\rho^5}{[y_1-x_0)^2+\rho^2]^5} 
\,\zeta^{B_2}_{\b_1}(y_1)\,\Omega^{B_1B_3}\, 
\frac{\rho^5}{[y_2-x_0)^2+\rho^2]^5}\,\zeta^{C_2}_{\b_2}(y_2)\, 
\Omega^{C_1C_3}+\cdots\right] \, , \nn 
\ea 
where the $\cdots$ refers to symmetrisation in $(B_2,B_3)$ and 
$(C_2,C_3)$. The $N_c$ dependence is contained in the coefficient 
$c_3(N_c)$, where 
\be 
c_3(N_c) = \frac{2^{-2N_c}(N_c-2)^2\Gamma(2N_c-2)}{N_c(N_c-1)! 
(N_c-2)!} ~ 
\raisebox{-6pt}{$\begin{array}{c} \displaystyle \sim \\ 
\scriptstyle N_c\to\infty \end{array}$} N_c^{1/2} \, . 
\label{G16KKNdep} 
\ee 
The integration over the five-sphere in this case gives a single 
$\veps$-tensor 
\be 
\int\dr^5\Omega\,\Omega^{AB}\Omega^{CD} = \frac{\pi^3}{6} 
\,\veps^{ABCD} \, . 
\label{5sph1eps} 
\ee 
 
The example of~(\ref{G16LamKK}) allows to illustrate another feature 
of non-minimal correlators. Since not all the fields are employed to 
soak up the sixteen exact superconformal modes, there are 
contributions to the expectation value in which pairs of fields are 
contracted with an instantonic propagator. In the case 
of~(\ref{G16LamKK}) for instance it is possible to contract pairs of 
scalars in the two $\tilde\Lambda^{ABC}_a$ operators. Contributions 
of this type are of the same order in $g$ as those in which the extra 
insertions soak up $\nu^A$ and $\bar\nu^A$ modes, since with the 
normalisations we are using ($S\propto 1/g^2$) the scalar propagator 
is proportional to $g^2$. They are, however, sub-leading with respect 
to terms containing $(\bar\nu^A\nu^B)_{\bf 6}$ pairs at large 
$N_c$. The evaluation of the contributions with contractions is rather 
involved because they require the use of the propagator in the 
instanton background, which has a complicated 
expression~\cite{bccl,TEM}. We shall not discuss further these 
effects, but we stress that they are essential for the comparison with 
certain string theory amplitudes~\cite{gk}. 
 
There are many other interesting examples of non-minimal correlation 
functions in $\calN=4$ SYM which could be discussed. For lack of 
space we conclude this section with a brief list of some other notable 
cases, referring the reader to the original literature for further 
details. A comprehensive study of non-minimal correlators can be found 
in~\cite{gk}. 
\begin{itemize} 
\itemsep=0pt 
\item A special class of correlation functions in $\calN=4$ SYM are 
the so-called extremal correlators. These are $n$-point functions of 
operators of the type~(\ref{KKQop}) in which the dimension, $\ell_1$, 
of one of the operators equals the sum of the dimensions, $\ell_i$, 
$i=2,\ldots,n$, of the others ($\ell_1=\sum_{i=2}^n\ell_i$). The 
analysis of the associated dual amplitudes in supergravity led to the 
prediction that such correlation functions should not be 
renormalised~\cite{extr}. This was then confirmed by field theory 
calculations in~\cite{bk,ehssw}. In particular an argument for the 
absence of instanton corrections to extremal correlators, based on the 
analysis of fermion zero-modes, was given in~\cite{bk}. Similar 
results have been shown to hold for next-to-extremal correlation 
functions for which $\ell_1=\sum_{i=2}^n\ell_i-2$~\cite{nextextr}. A 
more complicated class are the near extremal correlators, 
characterised by the condition $\ell_1=\sum_{i=2}^n\ell_i-m$ with 
$m\le n-3$. These satisfy certain partial non-renormalisation 
properties~\cite{nearextr}, which have been argued in~\cite{gk} to 
survive the inclusion of instanton corrections. 
\item The Wilson loop is a particularly important operator in 
non-abelian gauge theories since it plays the r\^ole of order 
parameter characterising confinement. In pure Yang--Mills theory the 
Wilson loop is the expectation value 
\be 
W[\calC]= \fr{N_c}\la\Tr_{_{\bm{N_c}}} 
\calP\exp \left[i\oint_\calC \dr x^\mu \, A_\mu\right] \ra 
\label{pYMwilson} 
\ee 
of the holonomy associated with the closed contour $\calC$. A 
generalisation of this quantity in $\calN=4$ SYM has been constructed 
in~\cite{wilson} together with a proposal for the dual quantity in 
string theory to be associated with it. A special class of Wilson 
loops in $\calN=4$ SYM are circular BPS loops, which are annihilated 
by sixteen linear combinations of Poincar\'e and special 
supersymmetries. These Wilson loops are defined as 
\be 
W[\calC_R] = \fr{N_c} \la \Tr_{_{\bm{N_c}}}\calP\exp 
\left[i\oint_{\calC_R}\dr s\, \left(A_\mu\dot x^\mu + i 
\vp_in^i|\dot x| \right) \right] \ra \, , 
\label{bpswilson} 
\ee 
where $n^i$ is a constant unit vector on the five-sphere and $\calC_R$ 
is a circle of radius $R$. An elegant method for computing instanton 
corrections to~(\ref{bpswilson}) in the case of $SU(2)$ gauge group 
was devised in~\cite{bgk}. The BPS Wilson loop is a non-minimal 
correlator since it is non-polynomial in the fields. In the $SU(N_c)$ 
case its calculation in the instanton background is a formidable task 
and the $SU(2)$ analysis of~\cite{bgk} has not been extended to this 
more general case. 
\item In Sect.~\ref{sec:mincorr} we mentioned that certain results 
concerning instanton corrections to the anomalous dimensions of gauge 
invariant composite operators can be obtained from the OPE analysis of 
four-point functions such as~(\ref{G4Q}). On the other hand, as 
discussed in Sect.~\ref{sec:N4sym}, the anomalous dimensions can be 
computed directly from two-point functions after resolving the 
operator mixing. Depending on the bare dimension of the operators one 
is considering two-point functions can be minimal or non-minimal. A 
systematic study of instanton contributions to two-point functions of 
scalar operator was initiated in~\cite{instadim}. As discussed in 
Sect.~\ref{sec:N4sym}, general considerations, and in particular 
arguments based on S-duality, suggest that generically anomalous 
dimensions in $\calN=4$ SYM should receive both perturbative and 
non-perturbative contributions. A rather surprising result found 
in~\cite{instadim} is the absence of instanton corrections to the 
majority of scalar operators of bare dimensions $\D_0\le 5$. 
\item Finally an important class of non-minimal correlation functions 
are those relevant for the so-called BMN limit, which is the subject 
of Sect.~\ref{sec:BMN-intro}. 
\end{itemize}

\section{Generalisation to multi-instanton sectors} 
\label{sec:N4-multinst} 
 
The generalisation of the analysis presented in the previous section 
to multi-instanton sectors is technically very involved and requires 
the full machinery of the ADHM construction~\cite{ADHM}. A detailed 
description of this formalism and its generalisation to supersymmetric 
theories, as well as references to the original literature can be 
found in~\cite{KHOREV}. Due to space limits we shall only report an 
important result of~\cite{KHETAL}, where multi-instanton contributions 
to $\calN=4$ SYM correlation functions were explicitly evaluated in 
the large $N_c$ limit. 
 
In the generic $K$-instanton sector and with gauge group $SU(N_c)$ an 
instanton configuration in pure Yang--Mills theory is characterised by 
$4KN_c$ collective coordinates parametrising a hyper-K\"ahler 
manifold, $\calM_K$. A description of the moduli space associated with 
general (anti-)self-dual gauge configurations can be given using the 
ADHM construction~\cite{ADHM}. This is based on the introduction of an 
over-complete set of matrix-valued parameters on $\calM_K$, satisfying 
non-linear constraints which can be shown to be equivalent to the 
self-duality condition for the Yang--Mills field strength. The 
constraints can be implemented describing the moduli space, $\calM_K$, 
and the associated metric by means of what is referred to as a 
hyper-K\"ahler quotient construction~\cite{hklr}. In the case of the 
$SU(N_c)$ $\calN=4$ SYM theory there are also $8KN_c$ 
fermionic collective coordinates in the generic $K$ instanton 
sector. These can be included in the ADHM formalism as matrix-valued 
generalisations of the collective coordinates introduced in the 
one-instanton sector, subject to suitable constraints. 
 
As usual, the calculation of instanton contributions to correlation 
functions involves the integration over the instanton moduli 
space. This can be formally achieved integrating over the redundant 
set of bosonic and fermionic ADHM matrices and imposing the 
constraints via $\d$-functions. However, as already observed, the 
calculations are not feasible for generic $K$, since an explicit 
solution to the constraints is not known. 
 
In the case of $\calN=4$ SYM these calculations are also not 
particularly enlightening since, in general, correlators receive 
non-vanishing contributions from all instanton sectors. However, a 
dramatic simplification occurs in the large $N_c$ limit, making the 
calculation of $K$-instanton corrections to correlation functions 
feasible for arbitrary $K$~\cite{KHETAL}. The reason for this 
simplification is that in the large $N_c$ limit the integration over 
the (super) moduli space is dominated by a very special configuration 
and can be evaluated using a saddle-point approximation. After the 
introduction of a matrix generalisation of the auxiliary variables 
$\chi^i$, the saddle-point that dominates the $K$-instanton moduli 
space integration corresponds to a configuration in which the $K$ 
instantons have the same size and share the same location both in 
space-time and in the five-sphere directions parametrised by the 
$\chi^i$'s~\footnote{In the analysis of the fluctuations around the 
saddle-point it is also important that, as far as the global gauge 
orientation is concerned, the $K$ instantons lie in mutually 
orthogonal $SU(2)$ subgroups of $SU(N_c)$.}. As in the one-instanton 
sector only the sixteen exact fermion zero-modes associated with the 
broken superconformal symmetries remain exact. The physical moduli 
space integration measure obtained using the saddle-point 
approximation is 
\ba 
&& \int \dr\mu^{(K)}_{\rm phys} \,\er^{-S_{\rm inst}} 
\label{largeNsp} \\ 
&&\raisebox{-6pt}{$\begin{array}{c} \displaystyle \longrightarrow \\ 
\scriptstyle N_c\to\infty \end{array}$}~ 
\frac{N_c^{1/2}\,g^8\,\er^{2\pi i K 
\tau}}{K^3\,2^{17K^2/2-K/2+25}\, \pi^{9K^2/2+9}} 
\int\frac{\dr^4x_0\,\dr\rho}{\rho^5}\,\dr^5\Omega 
\prod_{A=1}^4\dr^2\eta^A\,\dr^2\bar\xi \, Z_K \, , \nn 
\ea 
where $Z_K$ contains the integration over the fluctuations around the 
saddle-point. These can be expressed in terms of $[K]\times[K]$ 
bosonic and fermionic matrices, $A_M$, $M=0,\ldots,9$ and 
$\Psi_r$, $r=1,\ldots,16$, by means of which the $Z_K$ factor 
in~(\ref{largeNsp}) takes the form of the partition function of a 
$SU(K)$ 
supersymmetric matrix model~\footnote{This is the dimensional 
reduction to zero dimensions of ten-dimensional $\calN=1$ SYM.}, 
\ie 
\be 
Z_K = \frac{1}{{\rm Vol}\;SU(K)}\int \dr^{10}A\,\dr^{16}\Psi \, 
\er^{-S(A,\Psi)} \, , 
\ee 
where 
\be 
S(A,\Psi) = -\half\,\Tr_K\left([A_M,A_N]^2 + \bar\Psi[\Asm,\Psi] 
\right) \, . \label{matrixaction} 
\ee 
The partition function $Z_K$ was computed in~\cite{mns,kns} with the 
result 
\be 
Z_K = 2^{17K^2/2-K/2-8}\,\pi^{9K^2-9/2}\, K^{-1/2} 
\sum_{m|K}\fr{m^2} \, , 
\label{mmpartf} 
\ee 
where the sum is over the positive integer divisors of $K$. 
 
Correlation functions of composite operators in the large $N_c$ limit 
are computed integrating their profiles in the $K$-instanton 
background with the measure~(\ref{largeNsp}). In particular, if the 
operator profiles do not depend on the collective coordinates 
parametrising the matrix model, the partition function $Z_K$ factors 
out. This is the case for minimal correlation functions of gauge 
invariant operators such as those considered in 
Sect.~\ref{sec:mincorr}. As an example of this type we consider the 
$K$-instanton contribution to~(\ref{G16Lam}). In the large $N_c$ limit 
the profile of the operator $\Lambda^A_\a$ in the $K$-instanton 
background does not depend on the matrix model coordinates. It is 
proportional to its one-instanton expression, namely 
\be 
\left.\hat\Lambda^A_\a\right|_{K{\rm-inst}} = \frac{96\,K}{g^2}\, 
\frac{\rho^4}{[(x-x_0)^2+\rho^2]^4}\,\zeta^A_a \equiv 
K \left.\hat\Lambda^A_\a\right|_{1{\rm-inst}}\, . 
\label{LamKprof} 
\ee 
Therefore one finds~\cite{KHETAL} 
\ba 
&& \la \Lambda^{A_1}_{\a_1}(x_1) \cdots \Lambda^{A_{16}}_{\a_{16}} 
(x_{16}) \ra_{K{\rm -inst}} = \frac{N_c^{1/2}\,K^{25/2}\, 
2^{47}\,3^{16}\,\pi^{-27/2}\,\er^{2\pi iK\tau}}{g^{24}} 
\sum_{m|K}\fr{m^2} \times \nn \\ 
&& \times\int \frac{\dr\rho \, \dr^4x_0}{\rho^5} \, \dr^5\Omega 
\prod_{A=1}^4 \dr^2\eta^A \, \dr^2 {\bar\xi}^A 
\prod_{i=1}^{16} \frac{\rho^4}{[(x_i-x_0)^2+\rho^2]^4}\, 
\zeta^{A_i}_{\a_i}(x_i) \, . 
\label{G16LamKinst} 
\ea 
The calculation of multi-instanton contributions to non-minimal 
correlation functions, even in the large $N_c$ limit, is much more 
complicated. In the non-minimal case the operator insertions depend 
on the one-instanton moduli and also on the matrix model variables and 
thus one cannot simply factor out $Z_K$. It is natural to expect that 
in these cases, instead of the partition function, the integration 
over the $A_M$ and $\Psi_r$ variables should be related to certain 
correlation functions in the matrix model giving rise to 
generalisations of~(\ref{mmpartf}).

\section{AdS/CFT correspondence: a brief overview} 
\label{sec:adscft-intro} 
 
As already mentioned, the recent renewed interest in $\calN=4$ SYM 
stems from the conjecture about the AdS/CFT 
correspondence~\cite{malda,GKP,Witthol}. In this section we provide a 
brief overview of the main concepts at the basis of this conjecture 
and in the following sections we review the r\^ole of instantons in 
this context. 
 
The idea of the AdS/CFT correspondence was presented in~\cite{malda} 
and a more concrete formulation was given 
in~\cite{GKP,Witthol}. Reviews can be found 
in~\cite{adscftrevs,MBrev}. In~\cite{malda} Maldacena proposed a 
remarkable duality relation connecting two completely different 
theories, $\calN $=4 SYM with $SU(N_c)$ gauge group and type IIB 
superstring theory in an AdS$_5\times S^5$ background. 
 
The type IIB superstring theory has $\calN=(2,0)$ supersymmetry in ten 
dimensions, \ie it is invariant under 32 supersymmetries. Its spectrum 
contains a finite number of massless states and an infinite tower of 
massive states. The massless spectrum is chiral. The bosonic degrees 
of freedom are divided into the so-called 
Neveu--Schwarz-Neveu--Schwarz (NS--NS) and Ramond--Ramond (R--R) 
sectors. The massless NS--NS sector contains a traceless rank-two 
symmetric tensor (the graviton, $g_{MN}$, $M,N=0,\ldots,9$), an 
anti-symmetric two-form ($B_{MN}$) and a scalar (the dilaton, 
$\phi)$. The massless R--R sector contains a scalar ($C_{(0)}$), an 
anti-symmetric two-form ($C_{MN}$) and an anti-symmetric four-form 
($C_{MNPQ})$, with self-dual field strength. The massless fermions 
are the spin 1/2 dilatino ($\lambda$) and the spin 3/2 gravitino 
($\psi_M$). These are complex Weyl spinors of opposite 
chiralities. The theory has two parameters, the coupling constant, 
related to the v.e.v.\ of the dilaton, $\gs=\er^{\la\phi\ra}$, and the 
inverse string tension, $\al$. The latter sets the scale for the 
massive states in the spectrum which have masses proportional to 
$1/\sqrt{\al}$. 
 
The background relevant for the correspondence with $\calN=4$ SYM, 
AdS$_5\times S^5$, is the product of a five-dimensional anti de Sitter 
space and a five-sphere. The non-compact factor, Lorentzian AdS$_5$, 
can be described as a hyperboloid embedded in six dimensions, \ie in 
terms of six Cartesian coordinates, $X^i$, $i=0,\ldots,5$, satisfying 
the constraint 
\be 
X_0^2-X_1^2-\cdots-X_4^2+X_5^2 = L^2 
\label{hyperbol} 
\ee 
where $L$ is the (constant) radius of curvature. This definition 
immediately shows that the AdS$_5$ space has isometry group 
$SO(2,4)$. So-called global coordinates for AdS$_5$ are introduced 
setting 
\ba 
&& X_0 = L \cosh\rho\,\cos t \, , \qquad X_5 = 
L \cosh\rho\,\sin t \, , \nn \\ 
&& X_r = L \sinh\rho\,\Omega_r \, , \qquad r=1,2,3,4 \, , 
\qquad \sum_r \Omega_r^2 = 1 \, . 
\label{adsglob} 
\ea 
In terms of these coordinates the metric reads 
\be 
\dr s^2 = L^2 (-\cosh^2\rho\,\dr t^2 + \dr\rho^2 + 
\sinh^2\rho\,\dr\Omega^2) \, . 
\label{globmetr} 
\ee 
Another convenient set of coordinates for AdS$_5$ are the so-called 
Poincar\'e coordinates, $(z_\mu,z_0)$. The $z_0$ coordinate 
parametrises the radial direction of AdS$_5$ and the four $z_\mu$ 
coordinates parametrise the directions parallel to the boundary 
located at $z_0=0$. In terms of these coordinates the metric is 
\be 
\dr s^2 = \frac{L^2}{z_0^2}\left(\dr z_\mu^2 + \dr z_0^2\right) \, . 
\label{adspoinc} 
\ee 
The AdS$_5\times S^5$ space is maximally supersymmetric if the two 
factors have the same radius of curvature, $L$. The non-vanishing 
components of the Ricci tensor are 
\be 
R_{mn} = -\frac{4}{L^2}\, g_{mn} \, , \qquad 
R_{ab} = \frac{4}{L^2}\, g_{ab} \, , 
\label{ricci} 
\ee 
where the indices $m,n$ span the AdS$_5$ directions and the indices 
$a,b$ the $S^5$ directions. Moreover the self-dual R--R five-form 
field strength has a non-vanishing background value 
\be 
F_{mnpqr} = \fr{L}\,\veps_{mnpqr} \, , \qquad 
F_{abcde} = \fr{L}\,\veps_{abcde} \, . 
\label{5form} 
\ee 
 
The conjectured duality has a holographic nature in that it relates 
the physics described by the string theory in the bulk of AdS$_5\times 
S^5$ to that of a gauge theory, $\calN=4$ SYM, living on the 
four-dimensional boundary of AdS$_5$. 
 
The first ingredient of the correspondence is a dictionary relating 
the parameters of the two theories. In $\calN=4$ SYM the parameters 
are the coupling, $g$, and the rank of the gauge group. In the string 
theory, besides the coupling constant, $\gs$, and the inverse string 
tension, $\al$, the radius of curvature, $L$, of the AdS$_5$ and $S^5$ 
spaces enters as an additional dimensionful parameter. The relations 
among the gauge and string theory parameters are 
\be 
g^2 = 4\pi\gs \, , \qquad L^4 = 4\pi\gs \al^2 N_c \, . 
\label{dict} 
\ee 
The second equation can be used to relate the dimensionless ratio 
$L^4/\al^2$ to the 't Hooft coupling, $\lambda=g^2 N_c$, 
\be 
\frac{L^4}{\al^2} = \lambda \, . 
\label{dictlam} 
\ee 
The $\vartheta$-angle, that can be turned on in the gauge theory, is 
related to the expectation value of the R--R scalar 
\be 
\frac{\vartheta}{2\pi} = \la C_{(0)} \ra \, . 
\label{thetaC0} 
\ee 
Given this dictionary for the parameters of the two theories, the 
correspondence is formulated in terms of two additional basic 
ingredients: 
\begin{itemize} 
\itemsep=4pt 
\item A map between the fundamental degrees of freedom of the two 
theories. 
\item A prescription for the computation the observables of one theory 
in terms of those of the other. 
\end{itemize} 
The map between degrees of freedom is dictated by the symmetries. The 
(super)isometries of the string background, under which the states in 
the string spectrum are classified, coincide with the (super)group of 
global symmetries of the gauge theory, which, as already discussed, is 
$PSU(2,2|4)$. The duality associates states in the string spectrum 
with gauge-invariant composite operators in $\calN=4$ SYM, which have 
the same quantum numbers under the $SO(2,4)\times SO(6)$ maximal 
bosonic subgroup of $PSU(2,2|4)$. Specifically, 1)~the Lorentz 
quantum numbers are identified, 2) the masses of the string states are 
related to the scaling dimensions of the dual operators and 3) the 
$SO(6)$ quantum numbers arising in the Kaluza--Klein (KK) reduction on 
$S^5$ of the string theory are related to the Dynkin labels 
characterising the transformation of the dual gauge theory operators 
under the $SU(4)$ R-symmetry. Supersymmetry then implies that entire 
multiplets are related. The simplest case of this relation is 
represented by the correspondence between the supergravity multiplet, 
which contains the graviton and its superpartners, and the $\calN=4$ 
supercurrent multiplet discussed in Sect.~\ref{sec:N4sym}. 
 
The prescription relating observables on the two sides of the 
correspondence is based on the identification of properly defined 
partition functions. The string partition function in AdS$_5\times 
S^5$ is a functional of the boundary values of the fields. The latter 
play the r\^ole of sources for the dual operators in the boundary 
gauge theory~\cite{GKP,Witthol} and one is led to propose the 
holographic formula 
\be 
Z_{\rm IIB}[\Phi|_{\del {\rm AdS}} = J] = \int [\dr A][\dr\lambda] 
[\dr\bar\lambda][\dr\vp]\,\exp\!\left(\!-S_{\calN=4}+\int 
\calO_\Phi J \!\right) . \label{Zid} 
\ee 
Here $\Phi$ denotes a generic field in the string theory and 
$\calO_\Phi$ is the dual composite operator in $\calN=4$ SYM 
according to the map previously described. 
 
The quantisation of string theory in an AdS$_5\times S^5$ background 
is not understood well enough to make really operative use 
of~(\ref{Zid}). However, interesting results can be obtained in 
certain limits. In particular in the weak coupling and small curvature 
limit on the gravity side, where 
\be 
\gs \ll 1 \, , \qquad \frac{L^2}{\al} \gg 1 \, , 
\label{sugralim} 
\ee 
classical supergravity becomes a good approximation. Based on the 
dictionary~(\ref{dict}) this limit corresponds to the limit of large 
$N_c$ and large 't Hooft coupling, $\lambda$, in the gauge theory. 
Since in the $N_c\to\infty$ limit $\lambda$ plays effectively the 
r\^ole of coupling constant, one obtains a duality between classical 
type IIB supergravity in AdS$_5\times S^5$ and the strong coupling 
limit of $\calN=4$ SYM in the planar approximation. This observation 
illustrates the strong/weak nature of the duality, which on the one 
hand makes it difficult to test, but on the other makes it a powerful 
tool for the study of strongly coupled gauge theories. 
 
In the limit~(\ref{sugralim}) the IIB partition function 
in~(\ref{Zid}) is well approximated by 
\be 
Z_{\rm IIB}[\Phi|_{\del {\rm AdS}} = J] \sim 
\er^{-S_{\rm IIB}[\Phi|_{\del {\rm AdS}} = J]} \, , 
\label{sgpf} 
\ee 
where $S_{\rm IIB}$ is the classical type IIB supergravity action in 
the AdS$_5\times S^5$ background. 
 
\begin{figure}[!htb] 
\begin{center} 
\includegraphics[width=0.33\textwidth]{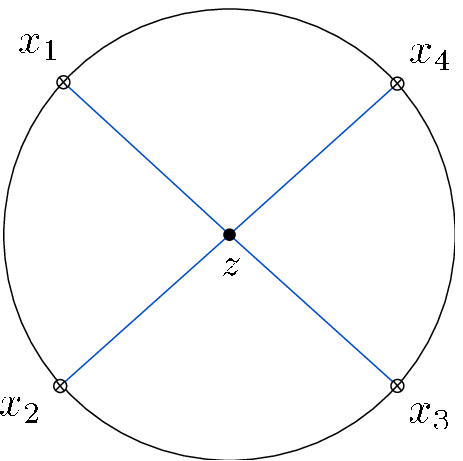} 
\hsp{1.5} 
\includegraphics[width=0.33\textwidth]{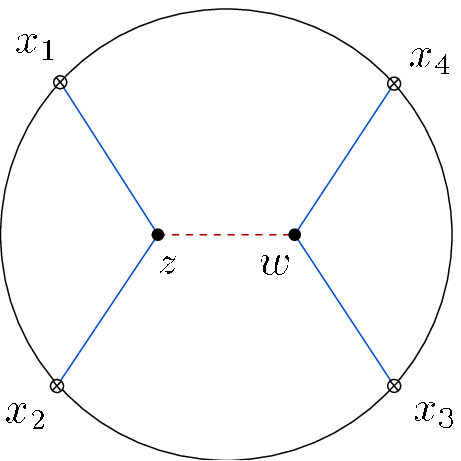} 
\end{center} 
\caption{Contact and exchange contributions to a four-point amplitude 
 in AdS$_5\times S^5$} 
\label{4pt-ads} 
\end{figure}

In this limit the relation~(\ref{Zid}) has a simple and intriguing 
interpretation. Correlation functions in the gauge theory are obtained 
taking functional derivatives with respect to the sources on the 
r.h.s.\ of~(\ref{Zid}). Using the approximation~(\ref{sgpf}), one 
finds that differentiating with respect to the sources is equivalent 
to solving the supergravity equations of motion with boundary 
conditions $\Phi|_{\del{\rm AdS}}=J$. Therefore the correspondence 
states that a $n$-point correlation function, 
$\la\calO_1(x_1)\cdots\calO_n(x_n)\ra$, in $\calN=4$ SYM is equal to 
an amplitude in which, for each $\calO_i$ insertion, the dual 
supergravity state, $\Phi_i$, is propagated from the bulk to the 
boundary point $x_i$. An intuitive graphical representation of this 
prescription was proposed in~\cite{Witthol}~\footnote{In the following 
we shall refer to processes of this type as scattering amplitudes in 
AdS.}. Figure~\ref{4pt-ads} represents the supergravity amplitudes 
contributing to the process dual to a SYM four-point function, 
$\la\calO_1(x_1)\cdots\calO_4(x_4)\ra$. The interior of the circle in 
Fig.~\ref{4pt-ads} represents the bulk of AdS$_5\times S^5$ and the 
circle itself is the four-dimensional boundary where the gauge theory 
lives. 
 
A normalisable solution of the free supergravity equations of motions 
satisfying the boundary condition $\Phi|_{\del{\rm AdS}}=J$ can be 
written as 
\be 
\Phi^{(i)}(z_\mu,z_0) = \int \dr^4x \, K_i(z_\mu,z_0;x^{(i)}_\mu)\, 
J_\Phi(x^{(i)}_\mu) \, , 
\label{b2bkernel} 
\ee 
where the function $K_i(z_\mu,z_0;x_\mu^{(i)})$ is a so-called 
bulk-to-boundary propagator, \ie the kernel that allows to express a 
supergravity field, $\Phi$, at the bulk point $(z_\mu,z_0)$ in terms 
of its boundary value, $J_\Phi$, at ($z_\mu=x_\mu^{(i)},z_0=0)$. 
Substituting the solution~(\ref{b2bkernel}) in the generating 
functional~(\ref{sgpf}) one obtains the following expressions for the 
two contributions in Fig.~\ref{4pt-ads} 
\ba 
\calA_{\rm cont}(x_1,\ldots,x_4) &\!=\!& N_c^2 \int 
\frac{\dr^4z\,\dr z_0}{z_0^5} \, \dr^5\omega \, 
\prod_{i=1}^4 K_i(z_\mu,z_0;x_\mu^{(i)}) \label{sugra4cont} \\ 
\calA_{\rm exc} (x_1,\ldots,x_4) &\!=\!& N_c^2 \int 
\frac{\dr^4z\,\dr z_0}{z_0^5} \, \dr^5\omega \, 
\sum_m \prod_{i=1}^2\prod_{j=3}^4 K_i(z_\mu,z_0;x_\mu^{(i)}) 
\times \nn \\ 
&& \hsp{2.25} \times G_m(z,w) \, K_j(z_\mu,z_0;x_\mu^{(j)}) \, , 
\label{sugra4exc} 
\ea 
where $G_m(z,w)$ in the second amplitude, corresponding to the 
exchange diagram, represents a bulk-to-bulk propagator in AdS$_5$ and 
the index $m$ runs over the set of all allowed intermediate states. 
In~(\ref{sugra4cont})-(\ref{sugra4exc}) we have used Poincar\'e 
coordinates, $(z_\mu,z_0)$, to parametrise AdS$_5$ and angles, 
$\omega_i$, for the five-sphere. In terms of these parameters the 
AdS$_5\times S^5$ metric becomes 
\be 
\dr s^2 = \frac{L^2}{z_0^2}\left(\dr z_\mu^2 + \dr z_0^2 + 
z_0^2 \,\dr \omega_5^2 \right) \, . 
\label{metric} 
\ee 
In~(\ref{sugra4cont})-(\ref{sugra4exc}) the overall factor of 
$N_c^2$ is obtained rewriting the coefficient in front of the IIB 
supergravity action in the string frame, namely $L^8/\al^4\gs^2$, in 
terms of Yang--Mills parameters using~(\ref{dict}). 
 
In the next sections we shall discuss the inclusion of instanton 
effects in this picture. In Sect.~\ref{sec:BMN-intro} we shall 
consider another notable limit, \ie the BMN limit, in which the string 
theory $\leftrightarrow$ field theory correspondence is under control 
beyond the supergravity approximation.

\section{Instanton effects in the AdS/CFT duality} 
\label{sec:inst-adscft} 
 
In the AdS/CFT correspondence the effects of Yang--Mills instantons in 
$\calN=4$ SYM are related to non-perturbative effects induced by 
D-instantons in the dual IIB string theory~\cite{bg}. In the low 
energy supergravity limit of string theory D-instantons arise as 
non-trivial solutions of the Euclidean field equations. In 
ten-dimensional Euclidean space they correspond to configurations in 
which the metric (in the Einstein frame) is flat and the dilaton and 
the R--R scalar have non-constant profiles while all the other fields 
vanish. As the ordinary Yang--Mills instantons, the supergravity 
D-instantons are characterised by their integer-valued charge. The 
supergravity action evaluated on a charge-$K$ D-instanton 
configuration is proportional to $K$, as in the Yang--Mills case, and 
inversely proportional to the string coupling, $\gs$. The 
D-instanton solution of the type IIB field equations in AdS$_5\times 
S^5$ has similar properties and can be obtained from the flat space 
solution~\cite{bgkr}. In string theory D-instantons are identified 
with D$(-1)$-branes, \ie point-like objects in Euclidean 
ten-dimensional space. Their world-volume is zero-dimensional and 
therefore open strings ending on D$(-1)$-branes carry no propagating 
degrees of freedom. They describe instead zero-modes associated with 
the D-instantons as discussed in 
Sect.~\ref{sec:instopenstr}. D-branes, and D-instantons in 
particular, can also be described in terms of closed string modes as 
collective excitations using the so-called boundary state 
formalism. This will be utilised in a special case in 
Sect.~\ref{sec:BMN-inst}. 
 
The discussion of the general principles of the AdS/CFT correspondence 
in Sect.~\ref{sec:adscft-intro} and specifically the fundamental 
relation~(\ref{Zid}) indicate that, in order to make contact with the 
calculation of instanton contributions to $\calN=4$ correlation 
functions, one should study D-instanton induced contributions to 
string scattering amplitudes in AdS$_5\times S^5$. In principle this 
involves including in the genus expansion of the closed string 
amplitudes the contribution of world-sheets with boundaries associated 
with the presence of D($-1$)-branes. However, as already explained, 
in the AdS$_5\times S^5$ background such calculations are not under 
control and one is restricted to a low energy supergravity analysis. 
In the supergravity approximation the inclusion of the effect of 
D-instantons requires a refinement of~(\ref{sgpf}) in which the 
classical supergravity action is replaced by the low energy effective 
action which incorporates the effect of the infinite tower of massive 
string excitations on the dynamics of the massless modes.

\subsection{The type IIB effective action} 
\label{sec:IIBeffact} 
 
The type IIB string theory effective action is expressed as a powers 
series in the inverse string tension, $\al$. It takes the form 
\be 
S^{\rm eff}_{\rm IIB} = \fr{\al^4} \, \left( S^{(0)} + \al^3 S^{(3)} 
+ \al^4 S^{(4)} + \cdots + \al^r S^{(r)} + \cdots \right) \, , 
\label{alprexp} 
\ee 
where $S^{(0)}$ denotes the classical action and the subsequent terms 
contain higher derivative couplings, which receive D-instanton 
contributions. The inclusion of such vertices in supergravity 
amplitudes in AdS$_5\times S^5$ gives rise to contributions which are 
in correspondence with the correlation functions discussed in 
Sects.~\ref{sec:N4-1inst} and~\ref{sec:N4-multinst}. 
 
The form of~(\ref{alprexp}) is in principle determined by 
supersymmetry. The terms appearing in the leading correction, 
$S^{(3)}$, have been extensively studied. The couplings arising at 
this level include the well known $\calR^4$ term and a large number of 
other terms related to it by supersymmetry. Schematically, in the 
string frame, the form of $S^{(3)}$ is 
\ba 
\al^3\,S^{(3)} &\!=\!& \fr{\al} \int\dr^{10}X\,\sqrt{-g}\, 
\er^{-\phi/2} \left[f_1^{(0,0)}(\tau,\bar\tau)\left( 
\calR^4+(G\bar G)^4+ \cdots\right) + \cdots + \nn \right. \\ 
&& \hsp{0.5}\left.+f_1^{(8,-8)}(\tau,\bar\tau) 
\left(G^8+\cdots\right) +\cdots + f_1^{(12,-12)}(\tau,\bar\tau) 
\lambda^{16}\right] \, . 
\label{S3terms} 
\ea 
The precise form of many of these couplings has been determined, see 
for instance~\cite{gg} where the $\calR^4$ coupling was studied. In 
the following we shall further discuss certain vertices which are 
relevant for the comparison with the Yang--Mills calculations of the 
previous sections. The coefficients in~(\ref{S3terms}) are functions 
of the complex scalar, $\tau = \tau_1+i\tau_2=C_{(0)}+i\er^{-\phi}$, 
where $ \phi$ is the dilaton and $C_{(0)}$ the R--R scalar. The 
effective action is invariant under $SL(2,\Z)$ transformations acting 
on $\tau$ as 
\be 
\tau \to \frac{a\tau+b}{c\tau+d} \, , 
\label{sl2tranf} 
\ee 
where the integers $a,b,c,d$ satisfy $ad-bc=1$. Under such 
transformations any supergravity field, $\Phi$, acquires a phase 
\be 
\Phi \to \left(\frac{c\tau+d}{c\bar\tau+d}\right)^{q_\Phi} \Phi \,, 
\label{sl2phase} 
\ee 
where $q_\Phi$ is the charge of $\Phi$ under the local $U(1)$ symmetry 
of~(\ref{alprexp}) which also rotates the two chiral 
supersymmetries~\cite{SIIB}. In particular the metric and the IIB self-dual 
five-form are not charged, the complex combination $G_{(3)} = (\tau\dr 
B_{(2)}+\dr C_{(2)})/\sqrt{\tau_2}$ (where $B_{(2)}$ and $C_{(2)}$ are 
the NS-- NS and R--R two-forms) has charge $1$, the fluctuation of the 
complex scalar, $\d\tau\equiv\hat\tau$, has charge $2$, the dilatino, 
$\lambda$, and the gravitino, $\psi_M$, have respectively charge $3/2$ 
and $1/2$. The coefficient functions in~(\ref{S3terms}) transform as 
modular forms with holomorphic and anti-holomorphic weights $(w,-w)$, 
so that 
\be 
f_1^{(w,-w)}(\tau,\bar\tau) \to 
\left(\frac{c\tau+d}{c\bar\tau+d}\right)^w 
f^{(w,-w)}_1(\tau,\bar\tau) \, . 
\label{modftranf} 
\ee 
Invariance under $SL(2,\Z)$ requires that the weight $w$ of the 
modular form in each term in the effective action be equal to half the 
sum of the $U(1)$ charges of the fields in the vertex. 
 
The modular forms $f^{(w,-w)}_1(\tau,\bar\tau)$ are obtained acting 
on $f^{(0,0)}_1(\tau,\bar\tau)$ with modular covariant derivatives 
\be 
f^{(w,-w)}_1(\tau,\bar\tau) = D_{w-1}D_{w-2}\cdots D_0 
f^{(0,0)}_1(\tau,\bar\tau) \, , 
\label{risew} 
\ee 
where $D_w = \tau_2\frac{\del}{\del\tau}-i\frac{w}{2}$. 
 
The modular form in front of the $\calR^4$ term, 
$f^{(0,0)}_1(\tau,\bar\tau)$, is given by a non-holomorphic Eisenstein 
series 
\be 
f^{(0,0)}_1(\tau,\bar\tau) = \sum_{(m,n)\neq(0,0)} 
\frac{\tau_2^{3/2}}{|m+n\tau|^3} \, . 
\label{nheseries} 
\ee 
It can be expanded in Fourier modes as 
\ba 
&& f_1^{(0,0)} (\tau, \bar \tau) = 
\sum_{K=-\infty }^\infty \calF^1_K(\tau_2) \, \er^{2\pi i K \tau_1} 
=2\zeta(3)\tau_2^{\frac{3}{2}} + \frac{2\pi^2}{3} 
\tau_2^{-\half} + \label{modfexp} \\ 
&& + 4 \pi \sum_{K\neq 0}|K|^{\half} \mu(K,1) \, 
\er^{-2\pi(|K|\tau_2-iK\tau_1)} 
\sum_{j=0}^\infty (4\pi K \tau_2)^{-j} 
\frac{\Gamma( j -1/2)}{\Gamma(- j -1/2) j!}\, , \nn 
\ea 
where the r.h.s. is the result of a further weak coupling (large 
$\tau_2$) expansion. 
 
The non-zero Fourier modes are interpreted as D-instanton 
contributions with instanton number $K$ ($K>0$ terms are D-instanton 
contributions while $K<0$ terms are anti D-instanton 
contributions). The measure factor, $\mu(K,1)$, is 
\be 
\mu(K,1) = \sum_{m|K} \fr{m^2} \, , 
\label{measK} 
\ee 
where the sum is over the positive integer divisors of $K$. The 
coefficients of the D-instanton terms in~(\ref{modfexp}), include an 
infinite series of perturbative fluctuations around any charge-$K$ 
D-instanton. The leading term in this series is the one of relevance 
for the comparison with the semi-classical Yang--Mills instanton 
calculations. In the case of $f^{(0,0)}_1(\tau,\bar\tau)$ this term is 
independent of $\tau_2$. From~(\ref{risew}) it follows that the 
leading D-instanton term in the modular form 
$f^{(w,-w)}_1(\tau,\bar\tau)$ behaves as $\tau_2^w=\gs^{-w}$. The zero 
D-instanton term, $\calF^1_0$, contains only two power-behaved 
contributions that arise in string perturbation theory as tree-level 
and one-loop contributions, with no higher-loop terms. 
 
Much less is known about higher order terms beyond $S^{(3)}$ in the 
string effective action, but various terms in $S^{(5)}$ are known and 
certain classes of terms at higher orders have been studied. Among 
the interactions at order $\al^5$ we have the following 
\ba 
\al^5 S^{(5)} &\!=\!& \al \int \dr^{10}X\,\sqrt{-g}\,\er^{\phi/2} 
\left[ f_2^{(0,0)}(\tau,\bar\tau)\,D^4\calR^4 + 
f_2^{(2,-2)}(\tau,\bar\tau)\,G^4\calR^4 + \right. \nn \\ 
&&\left. + f_2^{(12,-12)}(\tau,\bar\tau)\, \calR^2 \lambda^{16} 
+f_2^{(12,-12)}(\tau,\bar\tau)\, \calR^2 \lambda^{16} 
+ \cdots \right] \, . 
\label{al5action} 
\ea 
The modular forms, $f_2^{(w,-w)}(\tau,\bar\tau)$, appearing here are 
generalisations of those previously defined. More generally at higher 
orders in the $\al$ expansion one expects modular forms 
of the type 
\be 
f^{(0,0)}_l(\tau,\bar\tau) = \sum_{(m,n)\neq(0,0)} 
\frac{\tau_2^{l+\half}}{|m+n\tau|^{2l+1}} \, . 
\label{geneseries} 
\ee 
All these functions satisfy relations similar to~(\ref{risew}). 
The weak coupling expansion of $f_2^{(0,0}(\tau,\bar\tau)$ is 
\ba 
&& f_2^{(0,0)}(\tau,\bar\tau) = 2\zeta(5)\tau_2^{\frac{5}{2}} 
+ \frac{4\pi^4}{135}\tau_2^{-\frac{3}{2}} + \label{modfexpf2} \\ 
&& + \frac{8\pi^2}{3} \sum_{K\neq 0}|K|^{\frac{3}{2}} \mu(K,2)\, 
\er^{-2\pi(|K|\tau_2-iK\tau_1)} \left(1+\frac{3}{16\pi K}\tau_2^{-1} 
+\cdots \right) \, , \nn 
\ea 
where $\mu(K,2)=\sum_{m|K}1/m^4$. 
 
In the next subsection we shall discuss how the D-instanton induced 
terms appearing in the IIB effective action are related to instanton 
contributions to $\calN=4$ correlation functions. In the analysis of 
processes dual to non-minimal correlators it will also be important to 
include the effect of the fluctuations, $\hat\tau$, of the complex 
scalar in the modular forms $f_l^{(w,-w)}(\tau,\bar\tau)$. For 
instance re-writing the complex scalar as $\tau=\tau_0+\hat\tau$ 
(where $\tau_0$ is the constant background value of $\tau$), the 
expansion of the D-instanton exponential factor in 
$f_1^{(0,0)}(\tau,\bar\tau)$ gives rise to a series of the form 
\be 
\er^{2\pi iK\tau} = \er^{2\pi iK\tau_0} 
\sum_r \frac{(2\pi i K)^r}{r!}\,\hat\tau^r \, . 
\label{hattauexp} 
\ee 
Equations~(\ref{hattauexp}) and~(\ref{S3terms}) show that at order 
$\al$ in the string low energy action there are effective vertices of 
the form $\hat\tau^r\,\calR^4$, which can contribute to scattering 
amplitudes in the AdS$_5\times S^5$ background.

\subsection{D-instantons in AdS$_5\times S^5$ and comparison with 
Yang--Mills instantons} 
\label{sec:instvsDinst} 
 
The discussion in the previous subsection provides the background 
necessary to analyse the processes dual to the correlation functions 
computed in Sects.~\ref{sec:N4-1inst} and~\ref{sec:N4-multinst}. These 
are dual to supergravity amplitudes involving the D-instanton 
induced vertices in the IIB effective action. In order to make contact 
with $\calN=4$ SYM one needs to specialise the general expressions of 
the vertices in~(\ref{S3terms}) to the case of the AdS$_5\times S^5$ 
background. For this purpose we shall expand the ten-dimensional 
supergravity fields in harmonics on the five-sphere~\cite{krv} 
\be 
\Phi(X) = \sum_\ell \Phi^{I_\ell}(z) \, 
\calY^{(\ell)}_{I_\ell}(\omega) \, , 
\label{spherharm} 
\ee 
where the $\calY^{(\ell)}_{I_\ell}(\omega)$'s are spherical harmonics, 
with $\ell$ denoting the level and $I_\ell$ a set of $SO(6)$ 
indices. After expanding the supergravity fields in $S^{\rm eff}_{\rm 
IIB}$ in this way the amplitudes dual to SYM correlators are computed 
using the prescription described in Sect.~\ref{sec:adscft-intro}. 
 
In studying AdS amplitudes we distinguish again between minimal and 
non-minimal cases, characterising an amplitude as (non-)minimal if it 
is dual to a (non-)minimal Yang--Mills correlator.

\subsubsection{Minimal AdS amplitudes} 
\label{sec:minDinst} 
 
\begin{figure}[!htb] 
\begin{center} 
\includegraphics[width=0.35\textwidth]{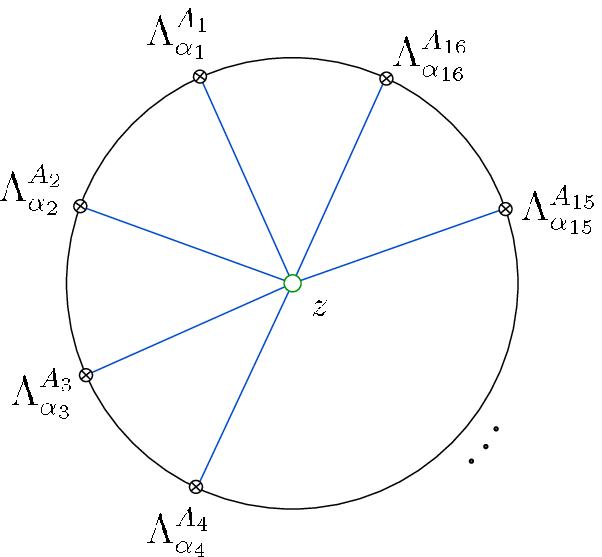} 
\hsp{1.5} 
\includegraphics[width=0.33\textwidth]{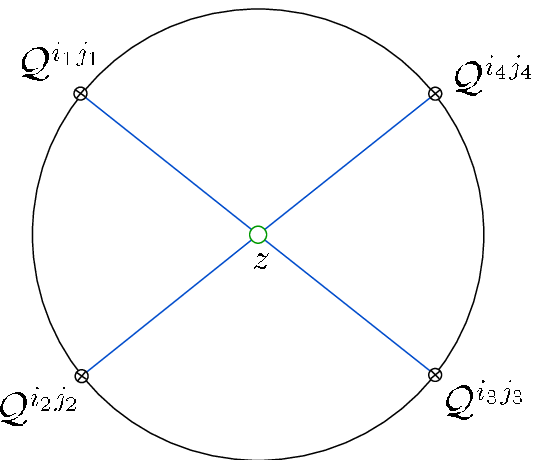} 
\end{center} 
\caption{D-instanton induced minimal amplitudes in AdS$_5\times S^5 
$.} 
\label{AdS-minimal} 
\end{figure}

The simplest minimal amplitude is the one dual to the sixteen-point
correlation function~(\ref{G16Lam}). The operator $\Lambda^A_\a$
in~(\ref{Lam4def})  is dual to the type IIB dilatino, $\lambda$, and
thus  according to the prescription explained in
section~\ref{sec:adscft-intro} we need to consider an amplitude with
sixteen dilatini propagating to the boundary. The vertex in the
effective action which contributes to such process is 
\be 
\fr{\al}\int \dr^{10}X\,\sqrt{-g}\,\er^{-\phi/2} 
f_1^{(12,-12)}(\tau,\bar\tau) \, t_{16}\,\lambda^{16} \, , 
\label{lam16vert} 
\ee 
where $t_{16}$ is a sixteen-index antisymmetric tensor contracting the 
spinor indices of the sixteen dilatini. The amplitude dual 
to~(\ref{G16Lam}) involves the leading D-instanton term 
in~(\ref{lam16vert}) (see~(\ref{risew})-(\ref{modfexp})), \ie 
\be 
\fr{\al}\int \dr^{10}X\,\sqrt{-g}\,2^{14}\pi^{13}\sum_{K>0}K^{25/2} 
\sum_{m|K}\fr{m^2}\, \er^{2\pi iK\tau}\er^{-25\phi/2} \,t_{16}\, 
\lambda^{16} \, . 
\label{Dlam16v} 
\ee 
The amplitude induced by this interaction is depicted on the l.h.s. of 
Fig.~\ref{AdS-minimal}: it is a contact amplitude in which the sixteen 
dilatini interact via the vertex~(\ref{Dlam16v}) and propagate to the 
boundary points $x_1,\ldots,x_{16}$. 
 
After introducing an explicit parametrisation for AdS$_5\times S^5$ 
and rewriting the string theory parameters, $\gs$ and $\al$, in terms 
of Yang--Mills parameters using the dictionary~(\ref{dict}), the 
amplitude in Fig.~\ref{AdS-minimal} becomes 
\ba 
\frac{N_c^{1/2}}{g^{24}} \sum_{K>0} K^{\frac{25}{2}}\sum_{m|K} 
\fr{m^2}\,\er^{2\pi iK\tau} \!\!\int \!\frac{\dr^4z\,\dr z_0}{z_0^5} 
\,\dr^5\omega\:t_{16} \times &&\nn \\ 
\times \prod_{i=1}^{16} \calY_F^{(0)}(\omega)\,K_{7/2}^F(z,z_0;x_i) 
&& , \label{lam16adsamp} 
\ea 
where overall numerical constants have been dropped and no indices 
have been indicated explicitly. In~(\ref{lam16adsamp}) we have used 
Poincar\'e coordinates, $(z_\mu,z_0)$, for AdS$_5$, with $z_\mu$ 
parametrising the directions parallel to the boundary and $z_0$ the 
radial direction. In terms of these coordinates and five angular 
variables for the $S^5$ factor, the AdS$_5\times S^5$ metric has the 
form~(\ref{metric}). The ten-dimensional dilatino has been expanded in 
spherical harmonics. In the expansion we have retained the ground 
state component, dual to the SYM operator $\Lambda^A_\a$. 
$\calY_F^{(0)}(\omega)$ denotes the corresponding harmonic 
function. In~(\ref{lam16adsamp}) $K_{7/2}^F$ denotes the 
bulk-to-boundary propagator for the dilatino, \ie a spin 1/2 fermion 
with AdS mass $-\frac{3}{2L}$ 
\be 
K_{7/2}^F (z,z_0;x) = K_4(z,z_0;x) \left[ \sqrt{z_0}\,\g_5- 
\fr{\sqrt{z_0}}\,(x-z)_\mu\g^\mu\right] \, , 
\label{K7/2prop} 
\ee 
where 
\be 
K_\D(z,z_0;x) = \frac{z_0^\D}{[(z-x)^2+z_0^2]^\D} \, . 
\label{KDprop} 
\ee 
Remarkably, the result~(\ref{lam16adsamp}), in its unintegrated form,
is in exact agreement with the multi-instanton contribution to the
correlation function~(\ref{G16Lam}) (cf. equation~(\ref{G16-fin}) and
its multi-instanton generalisation~(\ref{G16LamKinst})) after the
integration over the sixteen exact fermion zero-modes in the
latter. To compare the two results one identifies the AdS$_5$
coordinates, $z_\mu,z_0$, with the position and size of the instanton
and the $S^5$ angles with the auxiliary angular variables,
$\Omega^{AB}$, introduced in the gauge theory calculation. The
integration over the position of the interaction point in the
supergravity amplitude reproduces the integration over the $\calN=4$
moduli space, which, in the large $N_c$ limit and with the inclusion
of the auxiliary variables, is precisely one copy of AdS$_5\times
S^5$. The bulk-to-boundary propagators reconstruct the dependence on
the moduli contained in the profiles of the Yang--Mills
operators. Finally, although we have not kept track of all the
numerical factors, the dependence on the parameters, $g$ and $N_c $,
as well as on the instanton number, $K$, are in perfect agreement. The
factors of $\tau_2$ in the weak coupling expansion of the modular
form $f^{(12,-12)}_1(\tau,\bar\tau)$ give rise to the same $g$
dependence as in the Yang--Mills result. Similarly the matrix model
partition function is reproduced by the measure factor, $\mu(K,1)$, in
the modular form, see~(\ref{measK}). The power of $N_c$
in~(\ref{lam16adsamp})  follows from the application of the AdS/CFT
dictionary~(\ref{dict}), which gives 
\be 
\frac{\er^{-\phi/2}L^2}{\al} = 2 \pi^{1/2} \sqrt{N_c} \, . 
\label{N1/2} 
\ee 
The calculation of the amplitude dual to the four-point
function~(\ref{G4Q})  is completely analogous. The $\calN=4$ scalar
operator  $\calQ^{ABCD}$ in the $\mb{20^\prime}$ of SU(4) is dual to a
linear  combination of the trace part of the metric in the $S^5$
directions  and the $S^5$ components of the R--R four-form
potential. The scalar  in the supergravity multiplet, corresponding to
$\calQ^{ABCD}$, arises  at level $\ell=2$ in the expansion in
spherical harmonics. An  amplitude contributing to the process dual
to~(\ref{G4Q}) involves the  $\calR^4$ interaction in the bulk. This
is depicted on the r.h.s in  Fig.~\ref{AdS-minimal}. Proceeding as in
the case of the  sixteen-point amplitude one finds that this
four-point amplitude is 
\ba 
N_c^{1/2}\sum_{K>0} K^{1/2}\sum_{m|K}\fr{m^2}\, 
\er^{2\pi iK\tau} \int \frac{\dr^4z\,\dr z_0}{z_0^5}\,\dr^5\omega 
\times && \nn \\ 
\times \prod_{i=1}^4 \calY^{(2)}_B(\omega)\,K_4(z,z_0;x_i) && , 
\label{Q4adsampl} 
\ea 
where the bulk-to-boundary propagator, $K_4$, is now the one 
appropriate for a scalar of mass squared $-4/L^2$ in AdS and the 
$\calY^{(2)}_B$'s are $\ell=2$ scalar spherical harmonics. Again 
the result agrees perfectly with the Yang--Mills calculation. 
 
The examples described here illustrate the striking agreement between
instanton contributions to Yang--Mills correlation functions and
D-instanton induced supergravity amplitudes. The agreement found
represents one of the most convincing tests of the validity of the
AdS/CFT correspondence. The majority of the explicit tests of the
Maldacena conjecture compare protected quantities which do not depend
on the coupling constant and thus coincide with their free theory
expressions. In these cases the comparison is not affected by the
strong/weak coupling nature of the correspondence, but the
calculations simply test that the same non-renormalisation properties
are valid on both sides. The calculations reviewed above represent
instead one of the few instances in which a precise comparison is
possible for quantities which do receive non-trivial quantum
corrections~\footnote{The BPS Wilson loops mentioned in
Sect.~\ref{sec:nonmincorr} provide another notable example in
perturbation theory.}. Such a precise agreement is remarkable and
somewhat unexpected, since the calculations in this section and those
in Sect.~\ref{sec:N4-1inst} appear to have different regimes of
validity. It is natural to interpret the result of the comparison as
due to an underlying partial non-renormalisation property~\cite{gogr},
whose origin, however, remains unexplained.

\subsubsection{Non-minimal AdS amplitudes} 
\label{sec:nonminDinst} 
 
The discussion in the previous subsection has a natural generalisation 
to the case of amplitudes dual to the non-minimal correlation 
functions of Sect.~\ref{sec:nonmincorr}. In the non-minimal case the 
study of SYM correlators has not been generalised to multi-instanton 
sectors. In this section we show how the supergravity analysis gives 
results which are in qualitative agreement with those of the $\calN=4$ 
calculations in the one-instanton sector. We consider supergravity 
amplitudes related to the two main types of non-minimal correlators 
discussed in Sect.~\ref{sec:nonmincorr}. 
 
Correlators such as~(\ref{G16LamKK}) correspond to amplitudes
involving KK excited states in the spectrum of type  IIB supergravity
in AdS$_5\times S^5$. The $\calN=4$ operators in 1/2  BPS multiplets
with lowest component a scalar of the form~(\ref{KKQop})  with
$\ell>2$ are dual to KK excited states. In  particular, the fermionic
operator $\tilde\Lambda^{ABC}_\a$  in~(\ref{Lam20def}) is dual to the
first KK excitation of the  dilatino. Therefore the amplitude dual to
the correlation function~(\ref{G16LamKK})  is similar to the
sixteen-point amplitude considered  in the previous subsection, but
with two of the dilatini in the first  KK excited level. 
 
The other class of non-minimal correlators presented in 
Sect.~\ref{sec:nonmincorr} comprises higher-point functions which have 
a natural interpretation as dual to amplitudes induced by vertices in 
the string effective action at order $\al^5$ and higher. However, in 
these cases the comparison is less straightforward and as will be 
shown shortly there are subtleties that need to be taken into account. 
 
\begin{figure}[!htb] 
\begin{center} 
\includegraphics[width=0.35\textwidth]{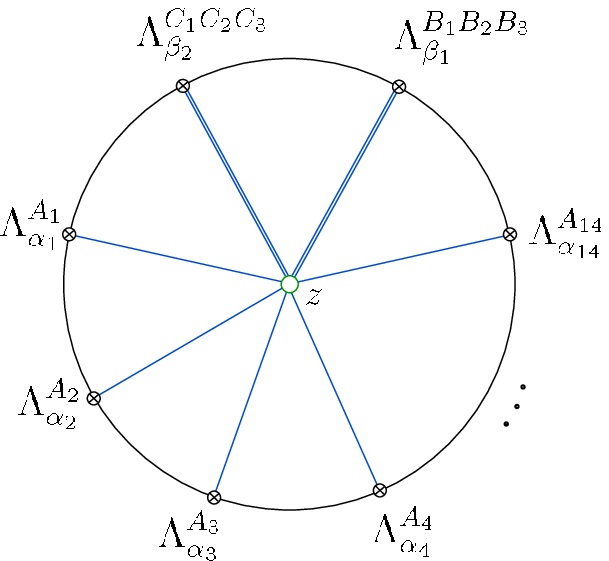} 
\end{center} 
\caption{Contact and exchange contributions to a four-point amplitude 
 in AdS$_5\times S^5$} 
\label{AdS-KK} 
\end{figure}

The amplitude dual to the sixteen-point correlator~(\ref{G16LamKK}) 
involves the same interaction as in~(\ref{lam16vert})-(\ref{Dlam16v}) 
with the only difference that upon reduction on the five-sphere one 
selects for two of the dilatini the first excited state instead of the 
KK ground state. The diagrammatic representation of the amplitude is 
given in Fig.~\ref{AdS-KK}, where the double lines indicate 
bulk-to-boundary propagators for the KK excited states. The dilatini 
in the first KK level are spin 1/2 fermions of mass $-\frac{5}{2L}$ 
for which the bulk-to-boundary propagator is 
\be 
K_{9/2}^F (z,z_0;x) = K_5(z,z_0;x) \left[ \sqrt{z_0}\,\g_5- 
\fr{\sqrt{z_0}}\,(x-z)_\mu\g^\mu\right] \, . 
\label{K9/2prop} 
\ee 
 
The resulting amplitude is 
\ba 
&& \frac{N_c^{1/2}}{g^{24}} \sum_{K>0} K^{\frac{25}{2}}\sum_{m|K}\fr 
{m^2}\, 
\er^{2\pi iK\tau} \!\!\int \!\frac{\dr^4z\,\dr z_0}{z_0^5}\,\dr^5 
\omega 
\,t_{16} \times \nn \\ 
&&\times \prod_{i=1}^{14} \calY_F^{(0)}(\omega)\,K_{7/2}^F(z,z_0;x_i) 
\prod_{j=1}^{2} \calY_F^{(1)}(\omega)\,K_{9/2}^F(z,z_0;y_i) \, . 
\label{lam16KKadsamp} 
\ea 
where $\calY_F^{(1)}(\omega)$ denotes the first excited fermionic
spherical harmonic. The result~(\ref{lam16KKadsamp}) is again in
agreement with the corresponding Yang--Mills
calculation~(\ref{G16KK-fin}). 
 
Non-minimal amplitudes of this type, involving KK excited states, are 
generalisations of the analogous minimal ones. Apart from the 
appearance of bulk-to-boundary propagators for fields of the 
appropriate mass, the only difference is in the five-sphere integrals, 
because of the presence of higher harmonics, which are necessary to 
reproduce the $\Omega$-dependence of the corresponding Yang--Mills 
expressions. 
 
\begin{figure}[!htb] 
\begin{center} 
\includegraphics[width=0.33\textwidth]{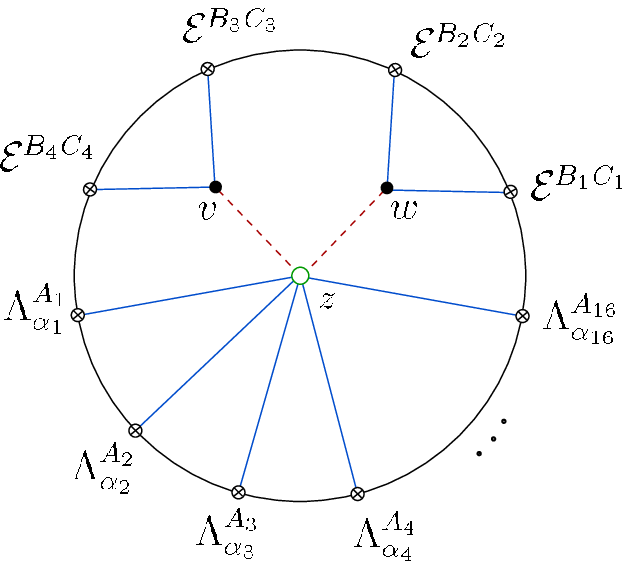} 
\hsp{1.5} 
\includegraphics[width=0.33\textwidth]{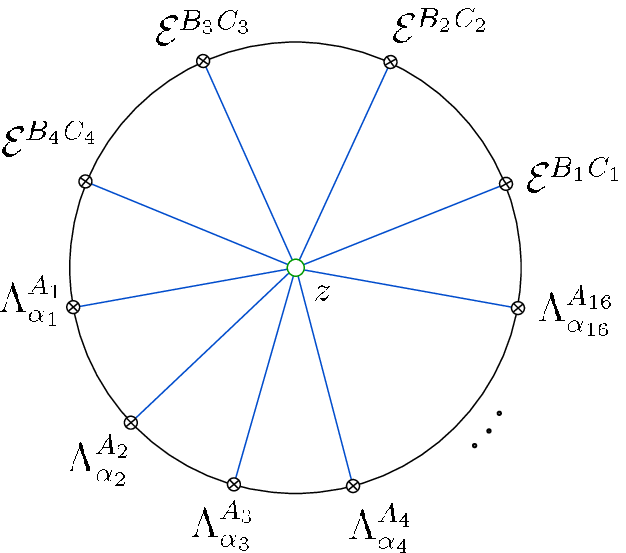} 
\end{center} 
\caption{Contributions to the twenty-point amplitude dual to the 
correlation function~(\ref{G20LamE}).} 
\label{AdS-Lam16E4} 
\end{figure}

The study of the other class of non-minimal amplitudes is more 
complicated and yields some surprises. In order to describe the main 
features of these amplitudes we focus on the example of the process 
dual to the correlation function~(\ref{G20LamE}) involving sixteen 
fermionic operators, $\Lambda^A_\a$, in the $\mb4$ of SU(4) and four 
scalar operators, $\calE^{AB}$, in the $\mb{10}$. As already observed, 
it is natural to expect that in these cases the amplitudes should 
involve couplings of order $\al^5$ and beyond. The operator 
$\calE^{AB}$ in~(\ref{E10def}) is dual to a linear combination of the 
NS--NS and R--R two-forms with indices in the internal directions. The 
corresponding field strength, $G_{(3)}$, was defined in 
Sect.~\ref{sec:IIBeffact}. Therefore a contact amplitude involving the 
vertex 
\be 
\al\int \dr^{10}X\,\sqrt{-g}\,\er^{\phi/2} 
f_2^{(14,-14)}(\tau,\bar\tau) \, G^4\,\lambda^{16} 
\label{lam16G4vert} 
\ee 
represents an obvious candidate for the dual of the
correlator~(\ref{G20LamE}).  This process is represented in the second
diagram of  Fig.~\ref{AdS-Lam16E4}. 
 
This interpretation, however, leads to a puzzle. Using the
dictionary~(\ref{dict})  the twelve-derivative couplings at order
$\al^5$ give rise  to contributions of order $N_c^{-1/2}$, in fact 
\be 
\frac{\al\,\er^{\phi/2}}{L} \sim N_c^{-1/2} \, , 
\label{N-1/2} 
\ee 
which is not the behaviour expected from the Yang--Mills analysis. The 
leading contribution to the correlation function~(\ref{G20LamE}) is in 
fact of order $N_c^{1/2}$, as follows 
from~(\ref{G20LamE-fin})-(\ref{G20LamENdep}). 
 
The resolution of this mismatch requires the inclusion in the 
supergravity analysis of contributions of a type not encountered in 
the calculation of minimal amplitudes. These are exchange diagrams 
involving a D-instanton induced vertex as well as additional 
perturbative couplings. The relevant D-instanton vertices are those 
of order $1/\al$ in the expansion of the effective action, so that the 
resulting amplitudes give rise to contributions of order $N_c^{1/2}$, 
see~(\ref{N1/2}). In order to generate contributions to non-minimal 
amplitudes one needs to include in the D-instanton vertices the 
fluctuations of the complex scalar, $\tau=\tau_0+\hat\tau$, as 
described at the end of Sect.~\ref{sec:IIBeffact}. 
 
In the case of the amplitude under consideration one needs to consider 
the vertex coupling sixteen dilatini with two additional insertions of 
$\hat\tau$ coming from the expansion of the exponential factor in the 
modular form $f^{(12,-12)}_1(\tau,\bar\tau)$. The non-perturbative 
part of the effective vertex is 
\be 
\fr{\al}\int \dr^{10}X\,\sqrt{-g}\,\er^{-25\phi/2}\, 
\er^{2\pi i\tau} \,\hat\tau^2 \,t_{16}\, \lambda^{16} \, , 
\label{Dlam16tau2v} 
\ee 
where only the $K=1$ contribution relevant for the comparison with the 
one-instanton sector in $\calN=4$ SYM has been included. The amplitude 
contributing to the dual of the twenty-point correlator~(\ref{G20LamE}) 
is depicted on the l.h.s.\ of Fig.~\ref{AdS-Lam16E4}. 
The two bulk-to-bulk lines joining the D-instanton vertex at point 
$z$ to the points $v$ and $w$ are $\la\hat\tau\hat{\bar\tau}\ra$ 
propagators and the two cubic vertices are $\hat{\bar\tau} GG$ 
couplings from the classical type IIB action. 
 
In evaluating this diagram, upon using dimensional reduction-like 
formulae as~(\ref{spherharm}), one has to sum over all the 
contributions associated with the exchange of the KK excitations of 
the complex scalar. The coupling to the external three-forms restricts 
this sum to the states allowed by the $SO(6)$ selection rules enforced 
by the integration over the five sphere. In the present case there is 
only one allowed contribution, corresponding to the exchange of a 
complex scalar in the second KK excited level, \ie a state in the 
representation $\mb{20^\prime}$ of $SO(6)\sim SU(4)$. 
 
At first sight the resulting amplitude does not resemble the $\calN=4$
SYM result: it is an exchange amplitude requiring integrations over
three bulk points. However, because of the specific coupling involved
the integrations over the positions of the two cubic couplings can be
performed. This is because, after using expressions such
as~(\ref{spherharm}),  one can integrate by parts the derivatives in
each of the three-form field  strengths onto the $\hat\tau$ scalar,
and the cubic couplings schematically  reduce to the form 
\be 
(\del^2+m_{\tau_{\mb{20^\prime}}})\hat\tau \, B_{ij}B^{ij} \, , 
\label{efftaubb} 
\ee 
where $B$ is the complex combination of the NS--NS and R--R two-forms 
and the mass term comes from the derivatives in the $S^5$ 
directions. After the integration by parts one thus reconstructs the 
AdS$_5$ wave operator acting on the internal bulk-to-bulk propagators 
which then yield five-dimensional $\d$-functions. The integrations 
at the points $v$ and $w$ in Fig.~\ref{AdS-Lam16E4} can thus be 
computed and the exchange diagram reduces to a contact 
contribution. Therefore the net effect of the exchange diagram is to 
give rise to a new coupling in the AdS$_5$ effective action of the 
form 
\be 
\fr{\gs^2\al}\int \frac{\dr^4z\,\dr z_0}{z_0^5} \,\er^{-25\phi/2}\, 
\er^{2\pi i\tau} \,t_{16}\, \lambda^{16} \, B^4 \, , 
\label{efflam16b4} 
\ee 
where the factor of $\gs^{-2}$ arises from the rescaling of the 
complex scalar, $\hat\tau$, needed to make its kinetic term 
canonically normalised. 
 
The amplitude induced by this vertex (expressed in terms of SYM 
parameters) takes schematically the form 
\be 
\frac{\sqrt{N_c}\,\er^{2\pi i\tau}}{g^{28}} \int 
\frac{\dr^4z\,\dr z_0}{z_0^5} \, \prod_{i=1}^{16} 
K_{7/2}^F(z,z_0;x_i) \prod_{j=1}^4 K_3(z,z_0;y_j) \, , 
\label{lam16e4ads} 
\ee 
which reproduces the leading large $N_c$ term in the $\calN=4$ 
result~(\ref{G20LamE-fin}) with the correct space-time dependence. 
 
The amplitude involving the order $\al^5$ vertex $\lambda^{16}G^4$ 
in~(\ref{lam16G4vert}) gives rise to a contribution with the same 
space-time dependence, but of order $N_c^{-1/2}$. This sub-leading 
contribution is interpreted as corresponding to the $1/N_c$ correction 
in the SYM result~(\ref{G20LamE-fin}). 
 
The example of the above twenty-point function illustrates some 
features common to many non-minimal amplitudes. In general, unlike in 
the minimal cases, the amplitudes dual to non-minimal $\calN=4$ 
correlation functions receive several contributions. Various effects 
such as those described in the previous example need to be taken into 
account to show agreement between the Yang--Mills and supergravity 
calculations. More details and other non-minimal examples are 
discussed in~\cite{gk}.

\subsection{Beyond supergravity: the BMN limit} 
\label{sec:BMN-intro} 
 
The AdS/CFT correspondence discussed in the previous sections is a 
very remarkable duality and the study of instanton effects has led to 
some of the most successful tests of its validity. In the formulation 
presented so far the duality has, however, some limitations. Because 
of our present limited understanding of the quantisation of string 
theory in non-trivial backgrounds such as AdS$_5\times S^5$ the study 
of the gravity side of the correspondence is restricted to the 
supergravity approximation. Moreover, even in this regime, the 
strong--weak coupling nature of the duality makes the direct 
comparison of the two sides problematic. In this section we briefly 
review a very interesting limit of the correspondence, the so-called 
BMN limit~\cite{bmn}, which allows to overcome both the above 
limitations~\footnote{Another limit that has attracted some attention 
is the highly ``stringy'' regime $\lambda \rightarrow 0$ where the 
theory exposes higher spin symmetry enhancement~\cite{HigherSpinLimit}. 
The bulk counterpart of the recombination of 
semi-short multiplets into long ones and the emergence of anomalous 
dimensions in the boundary theory is a pantagruelic Higgs mechanism 
termed {\it La Grande Bouffe}~\cite{GrandeBouffe}.}. The idea is to 
consider string theory in a background obtained from AdS$_5\times S^5$ 
via a special procedure known as Penrose limit~\cite{penlim}. The 
result of the limit is a background with the geometry of a maximally 
supersymmetric gravitational plane-wave~\cite{bfhp}. Remarkably, 
despite the non-flatness of the metric and the presence of a R--R 
background, it is possible to quantise string theory in this 
geometry~\cite{met,mt}. In~\cite{bmn} it has been proposed that 
strings propagating in this particular plane-wave background are dual 
to a certain sector of $\calN=4$ SYM. The latter, usually referred to 
as the BMN sector, comprises operators of large scaling dimension, 
$\D$, and large charge, $J$, with respect to a $U(1)$ subgroup of the 
$SU(4)$ R-symmetry group. The possibility of quantising string theory 
in the plane-wave background has made the comparison between string 
and gauge theory possible beyond the supergravity approximation, 
albeit only in a specific sector of $\calN=4$ SYM. Moreover, in this 
limit there exists a regime in which both sides of the correspondence 
are weakly coupled, so that the strong--weak coupling problem is also 
avoided. 
 
At the heart of the correspondence proposed in~\cite{bmn} is a 
relation between the energy, $E$, of plane-wave string states and a 
combination of the scaling dimension and R-charge of the dual 
operators, which reads 
\be 
\fr{\mu}\, E = \D - J \, , 
\label{bmnrel} 
\ee 
where the parameter $\mu$ is related to the value of the R--R 
self-dual five-form present in the background, see~(\ref{pp5form}). 
The validity of~(\ref{bmnrel}) has been 
successfully tested in perturbation theory in a number of 
cases. Reviews of these results can be found in~\cite{bmnrevs}. In 
this subsection we present a brief overview of the non-perturbative 
tests carried out in~\cite{gks1,gks2,gks3}. 
 
In order to take the Penrose limit that gives rise to the plane-wave 
background we start with the AdS$_5\times S^5$ metric written in 
global coordinates 
\begin{align} 
\dr s^2= L^2 &\left[-\cosh^2\rho\,\dr t^2 + \dr\rho^2 + 
\sinh^2\rho \,\dr \Omega_3^2 + \right.\nn \\ 
& \left. \; + \cos^2\theta \,\dr\psi^2+ 
\dr\theta^2 + \sin^2\theta \,\dr\tilde\Omega_3^2\right] , 
\label{globads} 
\end{align} 
where $\Omega_3$ and $\tilde\Omega_3$ refer to angles parametrising 
two three-spheres inside AdS$_5$ and $S^5$ respectively. In this 
coordinate system one can choose 
\be 
\tilde x^\pm = \pm\fr{\sqrt{2}} (t\pm\psi) 
\label{lcvar} 
\ee 
as light-cone variables and define the new coordinates 
\be 
x^+ = \fr{\mu}\tilde x^+ \, , \qquad 
x^- = \mu L^2\tilde x^- \, , \qquad \rho = 
\frac{r}{L} \, , \qquad \theta = \frac{y}{L} \, , 
\label{lccoord} 
\ee 
where $\mu$ is an arbitrary scale. The Penrose limit is 
obtained sending $L$ to infinity while keeping $x^{\pm}, \rho$ and 
$y$ ``fixed''. The resulting metric is that of a plane-wave 
\be 
\dr s^2 = 2\dr x^+\dr x^- - \mu^2 x^Ix^I(\dr x^+)^2 
+ \dr x^I\dr x^I \, , 
\label{ppwave} 
\ee 
where $x^I$, $I=1,\ldots,8$, are Cartesian coordinates such that 
$x^Ix^I = r^2 + y^2$. The original AdS$_5\times S^5$ background has 
also a non-zero self-dual R--R five-form~(\ref{5form}), which after 
the limit has non-vanishing components 
\be 
F_{+1234} = F_{+5678} = 2 \mu \, , 
\label{pp5form} 
\ee 
with indices $1,2,\ldots,8$ corresponding to the $x^I$ 
directions~\footnote{Notice that the metric~(\ref{ppwave}) is $SO(8)$ 
invariant, but the background value of the five-form breaks this 
symmetry down to $SO(4)\times SO(4)$.}. 
 
The plane-wave background preserves the same (maximal) amount of 
supersymmetry as the original AdS$_5\times S^5$. In fact at the level 
of the super-isometries the Penrose limit corresponds to an 
In\"on\"u--Wigner contraction~\footnote{More precisely this 
contraction should be called a Saletan 
contraction~\cite{saletan}.}. The supergroup of isometries resulting 
from the contraction of $PSU(2,2|4)$ is $PSU(2|2) \times PSU(2|2) 
\times U(1) \times U(1)$ with maximal bosonic subgroup $H(4)^2 \rtimes 
SO(4) \times SO(4) \times U(1) \times U(1)$, where $H(4)$ denotes the 
four-dimensional Heisenberg group~\cite{bfhp}. In the following we 
shall denote the two $SO(4)$ factors with $SO(4)_C$ and $SO(4)_R$, 
where the subscript refers to the fact that they are subgroups 
respectively of the conformal group and the R-symmetry group of the 
dual $\calN=4$ SYM theory. The states in the string spectrum can 
therefore be labeled by quantum numbers characterising their 
transformation under $SO(4)_C \times SO(4)_R \times U(1) \times 
U(1)$. These are identified with two pairs of spins, 
$(s_1,s_2;s_1^\pp,s_2^\pp)$, and the light-cone energy and momentum, 
$(p_+,p_-)$, associated with translations in the $x^+$ and $x^-$ 
directions. 
 
A very remarkable feature of the plane-wave background is that it 
allows the quantisation of string theory in the Green--Schwarz (GS) 
formalism~\footnote{By ``quantisation'' we here mean the determination 
of the spectrum of states with $p_- \neq 0$, with $p_-> 0$ for 
incoming and $p_-<0$ for outgoing states. Interactions and the 
spectrum of states with $p_- = 0$ are much subtler and not fully known 
even at tree level.} As shown in~\cite{met}, in the light-cone gauge 
the plane-wave GS string is described by a (massive) free world-sheet 
theory. This allows to carry out the quantisation essentially in same 
fashion as in flat space. The string action constructed in~\cite{met} 
is 
\ba 
S &\!=\!& \fr{2\pi\al}\int_{-\infty}^{+\infty}\dr \tau \int_0^{2\pi 
p_-} \dr\s \, \left[ \half \del_+X^I\del_-X^I - \half m^2 X^IX^I + 
\right. \nn \\ 
&& \left. \hsp{2} +i\left(S^a\del_+S^a+\tilde S^a\del_-\tilde S^a 
-2mS^a\Pi_{ab}\tilde S^b \right) \right] , 
\label{gsaction} 
\ea 
where the mass parameter $m=\mu p_-\al$ has been
introduced. In~(\ref{gsaction})  the $X^I$'s denote the transverse
coordinates of the  string and the index $I=1,\ldots,8$ is in the
${\mathbf 8_{\rm v}}$ of  $SO(8)$. The $S^a$'s and $\tilde S^a$'s,
$a=1,\ldots,8$, are GS  fermions. These are $SO(8)$ spinors of the
same chirality in the  ${\mathbf 8_{\rm s}}$. The matrix $\Pi$ is a
product of $SO(8)$  $\g$-matrices, $\Pi = \g^1\g^2\g^3\g^4$. As usual
in the light-cone  gauge the non-physical components have been
eliminated setting  $X^+(\s,\tau)=2\pi\al p_-\tau$, whereas
$X^-(\s,\tau)$ is expressed in  terms of the $X^I$'s using the
so-called Virasoro constraints which  follow from consistency with the
equation of motion for the  world-sheet metric. 
 
Since~(\ref{gsaction}) describes a free theory the equations of 
motions lead to a standard mode expansion. For instance for the 
transverse bosons one finds 
\ba 
X^I(\sigma,\tau) &\!=\!& \cos(m\tau)\,x_0^I + \fr{m}\sin(m\tau)\, 
p_0^I + \nn \\ 
&\!+\!& i\sum_{n\neq 0} \fr{\omega_n}\left( 
\er^{-i\omega_n\tau+2i\pi n\s}\a^I_n 
+ \er^{-i\omega_n\tau-2i\pi n\s} \tilde\a^I_n \right)\, , 
\label{modexp} 
\ea 
where 
\be 
\omega_n = {\rm sign}(n)\sqrt{m^2+n^2} \, . 
\label{disprel} 
\ee 
The GS fermions have a similar mode expansion with coefficients which 
will be denoted by $S^a_{\pm n}$ and $\tilde S^a_{\pm n}$. 
 
Upon quantisation the coefficients in the expansion of the world-sheet 
fields give rise to creation and annihilation operators for the states 
in the string spectrum. In order to construct the physical creation 
and annihilation operators the $SO(8)$ oscillators need to be 
decomposed under $SO(4)_C\times SO(4)_R$. Massive excitations of the 
string are associated with non-zero oscillators. The bosonic ones, 
$\a^I_{n}$ and $\tilde\a^I_{n}$, are in the $\mb{8_{\rm v}}$ 
which decomposes as $\mb{8_{\rm v}}\to (\mb4;\mb1)\oplus(\mb1; \mb4)$, 
and one obtains 
\be 
\a^I_{n} ~ \to ~ \a^i_{n} \,, ~~ \a^{\mu+5}_{n} \, , 
\qquad \tilde\a^I_{n} ~ \to ~ \tilde\a^i_{n} \,, ~~ 
\tilde\a^{\mu+5}_{n} \, , \qquad n\in\Z\, , \quad n\neq 0 \, , 
\label{boscdec} 
\ee 
where $i=1,2,3,4$ and $\mu=0,1,2,3$ are vector indices in $SO(4)_R$ 
and $SO(4)_C$ respectively. The fermions are in the $\mb{8_{\rm s}}$ 
which decomposes into $(\mb2_L;\mb2_L)\oplus(\mb2_R;\mb2_R)$. The 
fermionic oscillators are decomposed using the projectors $P_\pm = 
\half(1\pm\Pi)$ 
\be 
S^a_{n} ~ \to ~ S^\pm_{n} = P_\pm S^a_{n} \, , 
\qquad \tilde S^\pm_{n} = P_\pm \tilde S^a_{n} \, , 
\label{foscdec} 
\ee 
which yield spinors with $SO(4)_C\times SO(4)_R$ chiralities $(+,+)$ 
and $(-,-)$. 
 
The zero-modes in the expansion of the world-sheet fields are treated 
similarly. The bosonic ones, $x_0^I$ and $p_0^I$, associated with the 
transverse position and momentum of the string, are combined into 
\be 
a^I = \fr{\sqrt{2|m|}}(p_0^I-i|m|x_0^I) \quad 
{\rm and} \quad {a^I}^\dagger = 
\fr{\sqrt{2|m|}}(p_0^I+i|m|x_0^I) \, . 
\ee 
These are then decomposed as in~(\ref{boscdec}). The fermion 
zero-modes, $S_0$ and $\tilde S_0$, are combined into 
\be 
\theta = \fr{\sqrt{2}}(S_0+i\tilde S_0) \quad 
{\rm and}\quad \bar\theta = \fr{\sqrt{2}}(S_0-i\tilde S_0) 
\ee 
and then further decomposed as 
\be 
\theta_{L,R} = P_{+,-} \theta \, , \qquad 
\bar\theta_{L,R} = P_{+,-}\bar\theta \, . 
\label{fzmdec} 
\ee 
The Fock space of states is built on a vacuum, 
$|0\ra_{h}=|0,p_-\ra_h$, defined as the state annihilated by 
$\theta_{R}$, $\bar\theta_{L}$, $a^I$ and all the non-zero oscillators 
of positive frequency. This is a non-degenerate bosonic state of zero 
mass usually referred to as the BMN vacuum. The fermionic zero-modes 
$\theta_{L}$ and $\bar\theta_{R}$ are creation operators and generate 
the supergravity multiplet acting on $|0\ra_h$~\cite{mt}. The bosonic 
zero-modes ${a^{I}}^{\dagger}$ create Kaluza--Klein-like 
excitations. The massive string modes are created by combinations of 
non-zero oscillators with negative frequencies, $\a^{i}_{-n}$, 
$\tilde\a^{i}_{-n}$, $\a^{\mu+5}_{-n}$, $\tilde\a^{\mu+5}_{-n}$, 
$S^{\pm}_{-n}$ and $\tilde S^{\pm}_{-n}$, acting on the BMN 
vacuum. Physical states, $|s\ra_{\rm phys}$, are subject to the level 
matching condition 
\be 
\left( N - \tilde N \right) |s\ra_{\rm phys} = 0 \, , 
\label{levmatch} 
\ee 
where the left and right moving number operators are defined by 
\be 
\begin{array}{l} 
\displaystyle 
N = \sum_{n=1}^\infty \left(\frac{n}{\omega_n} \, 
\a^I_{-n}\a^I_n + n \,S^a_{-n}S^a_n \right) \\ 
\displaystyle 
\tilde N = \sum_{n=1}^\infty \left(\frac{n}{\omega_n} \, 
\tilde\a^I_{-n}\tilde\a^I_n + n \,\tilde S^a_{-n}\tilde 
S^a_n \right) \, . \rule{0pt}{20pt} 
\end{array} 
\label{occnumber} 
\ee 
The string theory Hamiltonian can be expressed in terms of the above 
oscillators as 
\ba 
2p_-\, H &\!=\!& m \left({a^I}^\dagger a^I+\theta_L^a\bar 
\theta_L^a+\bar\theta_R^a\theta_R^a\right) + \nn \\ 
&\!+\!& \sum_{k=1}^\infty \left[ 
\alpha^I_{-k}\alpha_k^I+\tilde\alpha^I_{-k}\tilde\alpha_k^I 
+\omega_k\left(S^a_{-k}S^a_{k}+ 
\tilde S^a_{-k}\tilde S^a_k\right)\right] \, . 
\label{hamilstring} 
\ea 
From the form of the Hamiltonian~(\ref{hamilstring}) it is 
straightforward to compute the free string spectrum. The mass of a 
generic massive string excitation is 
\be 
\fr{\mu}\,M = \fr{m} \sum_{n=1}^\infty \left(N+\tilde N\right) 
|\omega_n| \, , 
\label{massivespectr} 
\ee 
with $\omega_n$ defined in~(\ref{disprel}) and $m=\mu p_-\al$. 
 
In the following our discussion will focus on massive string states, 
which are more interesting in the context of the duality with 
$\calN=4$ SYM since their masses receive quantum corrections. For 
simplicity we shall restrict our attention to states created by the 
$\a^i_{-n}$ and $\tilde\a^i_{-n}$ oscillators 
\be 
|s\ra = \a^{i_1}_{-n_1}\cdots\a^{i_r}_{-n_r}\cdots 
\tilde\a^{j_1}_{-m_1}\cdots\tilde\a^{j_s}_{-m_s}\cdots |0\ra_h \, , 
\label{gensststate} 
\ee 
with $\sum_r n_r = \sum_s m_s$ to satisfy level-matching. 
 
As has been mentioned before, the states in the spectrum are 
characterised by the their $SO(4)_C\times SO(4)_R$ quantum numbers, 
besides their mass and light-cone momentum. As in the original 
formulation of the AdS/CFT correspondence the quantum numbers 
associated with the symmetries on the two sides also dictate the map 
relating string states to composite operators in $\calN=4$ 
SYM. Equation~(\ref{lccoord}) leads to the following identifications 
\be 
\begin{array}{l} 
\displaystyle 
\fr{\mu}\,H \equiv \fr{\mu}\,p_+ = -i\del_+ = 
-i(\del_t+\del_\psi) \to \calD - \calJ \\ 
\displaystyle 
p_- = -i\del_- = \frac{i}{2\mu L^2} (\del_t-\del_\psi) 
\to \fr{2\mu L^2}(\calD+\calJ) \, , 
\end {array} 
\label{p+p-DJrel} 
\ee 
where, as usual, $L^4=4\pi\gs N_c\al^2$. Equations~(\ref{p+p-DJrel}) 
relate the string light-cone energy and momentum to linear 
combinations of the dilation operator, $\calD$, and the generator, 
$\calJ$, of the $U(1)$ subgroup of the $SU(4)$ R-symmetry singled out 
in taking the Penrose limit. From the above relations it follows that 
in the $L\to\infty$ limit string states with finite energy and 
light-cone momentum correspond to SYM operators with values of $\calD$ 
and $\calJ$ satisfying 
\be 
\D\to\infty \, , \qquad J\to\infty \, , \qquad \D-J \quad {\rm finite} 
\, . \label{bmnsect} 
\ee 
Operators with these properties form the BMN sector of $\calN=4$ 
SYM. The explicit form of the operators dual to states in the 
plane-wave string spectrum was proposed in~\cite{bmn}. The starting 
point for the construction of such operators is the definition of the 
dual to the BMN vacuum which is identified with the operator 
\be 
\calO = \fr{\sqrt{JN_c^J}}\,\Tr\left(Z^J\right) \, , 
\label{bmnvacop} 
\ee 
where $Z$ is the complex combination of the $\calN=4$ scalar fields
with $J=1$ for which we choose $Z=2\vp^{14}$
(see~(\ref{scalconstr})-(\ref{scalrel})).  The
operator~(\ref{bmnvacop}) has  $\D-J=0$ as expected for the dual of a
zero energy state. Operators  corresponding to the other states in the
string spectrum are obtained  inserting in the trace
in~(\ref{bmnvacop}) ``impurities'', \ie other  elementary fields in
the $\calN=4$ fundamental multiplet. The action  of each creation
operator on the string side corresponds to the  insertion of an
impurity of a certain type~\footnote{For this reason  the string
excitations are also often referred to as impurities.}. In  particular
the operators dual to states of the  form~(\ref{gensststate}) are 
\ba 
\calO^{i_1\ldots i_k}_{J;n_1\ldots n_k} 
&\!=\!& \fr{\sqrt{J^{k-1}\left(\frac{g^2N_c}{8\pi^2}\right)^{J+k}}} 
\times \nn \\ 
&\!\times\!& \hspace*{-0.1cm} \begin{array}[t]{c} 
{\displaystyle \sum_{p_1,\ldots,p_{k-1}=0}^J} \\ 
{\scriptstyle p_1+\cdots+p_{k-1} \le J} 
\end{array} \hspace*{-0.2cm} \er^{2\pi 
i[(n_1+\cdots+n_{k-1})p_1+(n_2+\cdots+n_{k-1})p_2+ 
\cdots+n_{k-1}p_{k-1}]/J} \times \nn \\ 
&\!\times\!& \Tr\left(Z^{J-(p_1+\cdots+p_{k-1})}\vp^{i_1}Z^{p_1} 
\vp^{i_2}\cdots Z^{p_{k-1}}\vp^{i_k}\right) \, , 
\label{scalimpops} 
\ea 
where the integers $n_i$ correspond to the mode numbers of the string 
state~\footnote{Conventionally left-moving modes correspond to $n_i>0$ 
and right-moving ones to $n_i<0$.} and the action of the creation 
operators in~(\ref{gensststate}) is in correspondence with the 
insertion of impurities, $\vp^{i}$, for which we have the definitions 
\be 
\begin{array}{ll} 
\displaystyle 
\displaystyle \vp^1 = \frac{1}{\sqrt{2}} \left( -\vp^{13} 
+ \vp^{24} \right) \, , \quad & \displaystyle 
\vp^2 = \frac{1}{\sqrt{2}} \left( \vp^{12} + \vp^{34} \right) 
\, , \\ 
\displaystyle \vp^3 = \frac{i}{\sqrt{2}} \left( -\vp^{13} - 
\vp^{24} \right) \, , \quad & \displaystyle 
\vp^4 = \frac{i}{\sqrt{2}} \left( \vp^{12} - \vp^{34} \right) 
\, , \rule{0pt}{18pt} 
\end{array} 
\label{4scalars} 
\ee 
The four real scalars, $\vp^i$, transform in the $(\mb1;\mb4)$ of 
$SO(4)_C\times SO(4)_R$ and the map between the 
operators~(\ref{scalimpops}) and the string states~(\ref{gensststate}) 
is determined by the $SO(4)_R$ quantum numbers. 
 
To make the comparison between string theory in the plane-wave 
background and the BMN sector of $\calN=4$ SYM possible at a 
quantitative level one has to consider the large $N_c$ limit. The 
combination of the large $N_c$ limit with the limit of large $\D$ and 
$J$, implies that new effective parameters, $\lambda^\pp$ and $g_2$, 
arise~\cite{bmn,kpss,mit1}, which are related to the ordinary 't Hooft 
parameters, $\lambda$ and $1/N_c$, by a rescaling 
\be 
\lambda^\pp = \frac{g^2N_c}{J^2} \, , \qquad g_2 = 
\frac{J^2}{N_c} \, . 
\label{g2lambpdef} 
\ee 
These in turn are related to the parameters of the plane-wave string 
theory by 
\be 
m^2=(\mu p_-\al)^2 = \fr{\lambda^\pp} \: , \qquad 
4\pi\gs m^2 = g_2 \, . 
\label{paramid} 
\ee 
The double scaling limit defined by~(\ref{bmnsect}) and $N_c\to\infty$ 
with $J^2/N_c$ fixed connects the weak coupling regime of the gauge 
theory to string theory at small $g_s$ and large $m$. The property 
that in this limit physical quantities can be expanded in powers of 
the effective parameters, $\lambda^\pp$ and $g_2$, is referred to as 
BMN scaling.

\subsubsection{Instanton effects in the BMN limit} 
\label{sec:BMN-inst} 
 
Tests of the BMN limit of the AdS/CFT correspondence consist in 
verifying the validity of the relation 
\be 
\fr{\mu}\,H = \calD-\calJ \, . 
\label{bmnrelop} 
\ee 
This is an operator relation and it requires that the eigenvalues of 
the two sides be equal, \ie masses of states in the plane-wave string 
spectrum, rescaled by a factor of $\mu$, should equal the combination 
$\D-J$ for the dual operators. In general the comparison requires the 
resolution of a mixing problem, \ie the diagonalisation of the 
operators in~(\ref{bmnrelop}). 
 
The quantum corrections to the string mass spectrum are extracted from
two-point amplitudes. At the perturbative level calculations of such
amplitudes have been performed using string field theory  methods. The
non-perturbative corrections that we are  interested in are induced by
two-point amplitudes in which the  external states are coupled to
D-instantons. These quantum  corrections to the string masses should
be compared to instanton  corrections to the eigenvalues of the
operator $\calD-\calJ$, \ie to  the anomalous dimensions of BMN
operators since the charge $J$ is not  renormalised. 
 
\begin{figure}[!htb] 
\begin{center} 
\includegraphics[width=0.66\textwidth]{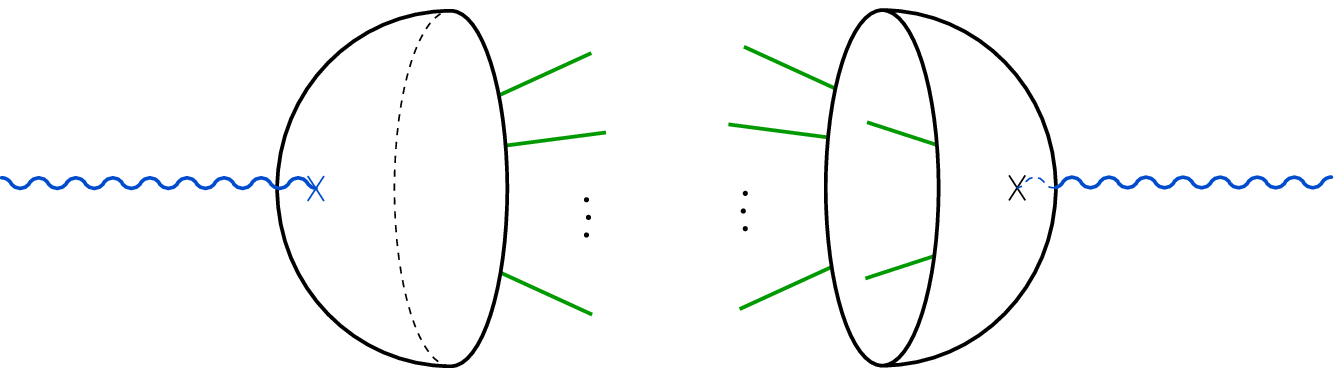} 
\end{center} 
\caption{Leading D-instanton contribution to a two-point scattering 
amplitude.} 
\label{D-2pampl} 
\end{figure}

The leading D-instanton contribution to a two-point amplitude is 
obtained coupling the external states to two disconnected disks, with 
Dirichlet boundary conditions, localised at the same space-time 
point. This is schematically depicted in Fig.~\ref{D-2pampl}. 
 
The D-instanton in the plane-wave string theory can be described as a 
collective excitation of elementary closed string oscillators using 
the boundary state formalism~\cite{gabgre}. The construction of the 
D-instanton boundary state in the plane-wave background follows 
closely the approach used in~\cite{gg2} for the light-cone GS string 
theory in flat space. The boundary state describing a D-instanton with 
transverse position $z^I$ will be denoted by $||z^I\ra\!\ra$. It is 
defined by the following gluing conditions 
\be 
\begin{array}{l} 
\displaystyle 
\left(\a^I_n-\tilde\a^I_{-n}\right)||z^I\ra\!\ra = 0 \, , \\ 
\displaystyle 
\left(S^a_n+iM^{ab}_n\tilde S^b_{-n}\right)||z^I\ra\!\ra = 0 
\rule{0pt}{17pt} \, , \qquad n\in\Z \, , 
\end{array} 
\label{Dgluing} 
\ee 
where the matrix $M_n$ is 
\be 
M_n = \fr{n}\left(\omega_n\one - m\Pi\right) \, , 
\label{Mdef} 
\ee 
with $\omega_n$ given in~(\ref{disprel}). The explicit expression 
for $||z^I\ra\!\ra$ is~\cite{gabgre} 
\be 
||z^I \ra\!\ra = (4\pi m)^2 
\exp \left( \sum_{k=1}^{\infty} 
\frac{1}{\omega_k} \alpha^I_{-k} \tilde\alpha^I_{-k} 
- i S^a_{-k}M_k\tilde{S}^a_{-k}\right) \, 
||z^I\ra\!\ra_0 \, , 
\label{boundone} 
\ee 
where $||z^I\ra\!\ra_0$ denotes the zero-mode part 
\be 
||z^I\ra\!\ra_0 = \er^{-|m| (z^I)^2 /2} \, \er^{i\sqrt{2|m|}z^I 
{a^I}^\dagger} \, \er^{\half {a^I}^\dagger{a^I}^\dagger} |0\ra_D 
\label{boundzero} 
\ee 
and $|0\ra_D=\theta^1_L\theta^2_L\theta^3_L\theta^4_L|0\ra_h$. 
 
The plane-wave background is maximally supersymmetric, \ie it is 
invariant under 32 supersymmetries. These are divided into sixteen 
kinematical supersymmetries, which do not commute with the string 
Hamiltonian, and sixteen dynamical ones, which commute with the 
Hamiltonian. The boundary state~(\ref{boundone})-(\ref{boundzero}) is 
annihilated by eight kinematical and eight dynamical supersymmetries 
as a consequence of the conditions~(\ref{Dgluing}). The other half of 
the supersymmetries acting on $||z^I\ra\!\ra$ generate the fermion 
zero-modes of the D-instanton. These are represented by the open 
strings attached to the boundary of the disks in 
Fig.~\ref{D-2pampl}. We shall denote the combinations of kinematical 
and dynamical supersymmetries which act non-trivially on the boundary 
state by $(\bar q_L,\bar q_R)$ and $Q^-$ respectively. The bosonic 
collective coordinates of the D-instanton correspond to its position 
in the ten-dimensional plane-wave geometry. 
 
In order to compute a two-point amplitude of the type represented in 
Fig.~\ref{D-2pampl} one needs to construct a state that includes the 
full dependence on the collective coordinates of the D-instanton and 
couples to two external states. We shall refer to such a state as a 
``dressed two-boundary state''. The latter describes the two disks in 
Fig.~\ref{D-2pampl} and is obtained considering the product of two 
boundary states associated with distinct Fock spaces but located at 
the same position, $z^I$. The ``dressing'' corresponding to the 
inclusion of the dependence on the bosonic and fermionic collective 
coordinates is achieved acting with the bosonic and fermionic 
generators of the broken symmetries. Denoting the two Fock spaces 
with indices 1 and 2, the dressed two-boundary state can be written as 
\ba 
&& \hsp{-0.5} ||V_2;\bm{z},\eta,\epsilon\ra\!\ra = 
\er^{iz^+(p_{1+}+p_{2+})}\er^{iz^-(p_{1-}+p_{2-})} 
\times \label{dressed2bs} \\ 
&& \times \left[\eta\left(Q^-_1+Q^-_2\right)\right]^8 
\left[\epsilon_L\left(q_{1L}+q_{2L}\right)\right]^4 
\left[\epsilon_L\left(q_{1L}+q_{2L}\right)\right]^4 
||z^I\ra\!\ra_1\otimes||z^I\ra\!\ra_2 \, , \nn 
\ea 
where $\bm{z}=(z^I,z^+,z^-)$ is the ten-dimensional location of the 
D-instanton and the $SO(8)$ spinors $\eta$ and 
$\epsilon\!=\!(\epsilon_L,\epsilon_R)$ denote the fermionic collective 
coordinates associated with the dynamical and kinematical 
supersymmetries broken by the D-instanton. 
 
The two-point amplitude that we are interested in is obtained coupling 
the dressed two-boundary state to a pair of external physical states 
and integrating over the bosonic and fermionic collective coordinates 
of the D-instanton. Denoting the incoming and outgoing states by 
$|s_1\ra$ and $|s_2\ra$, the one-particle irreducible part of the 
amplitude is 
\be 
\calA = c \,\gs^{7/2}\,\er^{2\pi i\tau} \int \dr^8 z \, \dr z^+ \, 
\dr z^- \, \dr^8\eta \, \dr^8\epsilon \, 
(\la s_1|\otimes\la s_2|)||V_2;\bm{z},\eta,\epsilon 
\ra\!\ra \, , 
\label{2ptDamplgen} 
\ee 
where $c$ is a numerical constant which has not been explicitly 
computed and the measure factor, $\gs^{7/2}\,\er^{2\pi i\tau}$, 
follows from the comparison with the D-instanton induced 
contributions to the low energy effective action. 
 
As an example of amplitude of the type~(\ref{2ptDamplgen}) we consider 
the leading D-instanton contribution to the case in which $|s_1\ra$ 
and $|s_2\ra$ are particular states in the 
class~(\ref{gensststate}). Specifically we consider $SO(4)_R$ singlet 
states with four impurities 
\ba 
\la s_1|\otimes \la s_2| &\!=\!& 
\fr{\omega_{n_1}\omega_{n_2}\omega_{m_1}\omega_{m_2}}\, 
\veps_{ijkl}\,\veps_{i^\pp j^\pp k^\pp l^\pp} \times \nn \\ 
&\!\times\!& {}_h\!\la 0|\a^{(1)i}_{n_1} \a^{(1)j}_{n_2} 
\tilde\a^{(1)k}_{n_1} \tilde\a^{(1)l}_{n_2} \otimes 
{}_h\!\la 0|\a^{(2)i^\pp}_{m_1} \a^{(2)j^\pp}_{m_2} 
\tilde\a^{(2)k^\pp}_{m_1} \tilde\a^{(2)l^\pp}_{m_2} \, , 
\label{4impses} 
\ea 
where the prefactors ensure that the states are normalised to one. A 
generic four impurity state has three independent mode numbers, after 
imposing the level matching condition. In~(\ref{4impses}) we have 
made a special choice: each of the states contains two left- and two 
right-moving excitations with pairwise equal mode numbers. This is 
because only states of this type couple to a D-instanton at leading 
order in $\gs$. This is a general property of D-instanton induced 
amplitudes: because of the way in which the creation operators enter 
in~(\ref{boundone})-(\ref{boundzero}), the boundary state couples only to 
states with the same number of left- and two right-moving oscillators 
with pairwise matched mode numbers. 
 
To proceed with the calculation of the leading D-instanton 
contribution we insert~(\ref{4impses}) into~(\ref{2ptDamplgen}). The 
general strategy for the calculation of such amplitudes consists in 
expanding the $||z^I\ra\!\ra$ factors in the dressed boundary state in 
a power series retaining only the terms which do not annihilate the 
product $\la s_1|\otimes\la s_2|$ on the left. The integration over 
the collective coordinates $z^+$ and $z^-$ imposes conservation of 
light-cone energy and momentum. Because of the non-linear dispersion 
relation~(\ref{disprel}) energy conservation requires that the mode 
numbers in the incoming and outgoing states be equal. Of the remaining 
integrations those over the eight transverse $z^I$'s and over the 
eight $\epsilon$ fermionic moduli are trivial in the case of external 
states of the type we are considering~\footnote{They give rise to 
$\d$-functions which in the present case simply integrate to 
one.}. The non-trivial part of the calculation is the integration over 
the eight fermion moduli $\eta$ 
\ba 
&& \hsp{-0.3} \la s_1|\otimes\la 
s_2| \int \dr^8 \eta \, \left[\eta(Q^-_1+Q^-_2) \right]^8 
\exp\left[\sum_n \fr{\omega_n} (\a^{(1)}_{-n} 
\tilde\a^{(1)}_{-n} + \right. 
\label{fourimpint} \\ 
&& \hsp{1} \left. + \a^{(2)}_{-n}\tilde\a^{(2)}_{-n}) 
-i(S^{(1)}_{-n}M_n\tilde S^{(1)}_{-n} + S^{(2)}_{-n} M_n 
\tilde S^{(2)}_{-n}) \right] |0\ra_1\otimes|0\ra_2 \, . \nn 
\ea 
These integrations induce a coupling between the two disks, since the 
dynamical supercharges, $Q^-$, which couple to the $\eta$'s depend 
non-trivially on the non-zero string oscillators. The calculation is 
greatly simplified when one considers the large $m$ limit relevant 
for the comparison with $\calN=4$ SYM at weak coupling. Since in this 
limit $M_n\sim m$, the dominant contribution to the amplitude with 
external states~(\ref{4impses}) is obtained retaining in the expansion 
of the boundary state two $SM\tilde S$ factors on each disk and 
distributing the eight $Q^-$'s evenly on the two disks. After some 
lengthy but straightforward algebra one obtains 
\be 
\calA(n_1,n_2) = \veps_{ijkl}\,\veps_{i^\pp j^\pp 
k^\pp l^\pp} \: \er^{2\pi i\tau} \, \gs^{7/2} \, m^8 \, 
\fr{n_1^2n_2^2} \, I^{ijkl,i^\pp j^\pp k^\pp l^\pp}_\eta 
\label{samplfin} 
\ee 
where 
\ba 
\hsp{-0.75} I^{ijkl,i^\pp j^\pp k^\pp l^\pp}_\eta 
&\!=\!& \int \dr^8\eta \, \eta^+\g^{ij}\eta^+ 
\eta^+\g^{kl}\eta^+ \eta^-\g^{i^\pp j^\pp}\eta^- 
\eta^-\g^{k^\pp l^\pp} \eta^- = \nn \\ 
&\!=\!& (\veps^{ijkl}+\d^{ik}\d^{jl}-\d^{il}\d^{jk}) 
(\veps^{i^\pp j^\pp k^\pp l^\pp}-\d^{i^\pp k^\pp} 
\d^{j^\pp l^\pp} +\d^{i^\pp l^\pp}\d^{j^\pp k^\pp}) 
\, . \label{etaint} 
\ea 
In the case of the amplitude~(\ref{samplfin}) the only contribution 
comes from the term containing the product of two $\veps$-tensors 
in~(\ref{etaint}) and one gets 
\be 
\calA(n_1,n_2) = 576 \: \er^{2\pi i\tau} \, \gs^{7/2} \, m^8 \, 
\fr{n_1^2n_2^2} \, . 
\label{samplfin2} 
\ee 
The same integral~(\ref{etaint}) arises in the calculation of 
two-point amplitudes between other $SO(4)_C\times SO(4)_R$ singlet 
four-impurity operators and in these cases the terms involving 
Kronecker $\d$'s in $I^{ijkl,i^\pp j^\pp k^\pp l^\pp}_\eta$ can 
contribute. 
 
The result~(\ref{samplfin}) is the leading non-perturbative correction 
to the one particle irreducible part of the two-point amplitude and 
thus it yields the D-instanton correction to the mass matrix for 
states of the form~(\ref{4impses}) 
\be 
\fr{\mu}\,\d M \sim \frac{\er^{2\pi i\tau}\,\gs^{7/2}\,m^7} 
{(n_1n_2)^2} = \frac{\er^{-\frac{8\pi^2}{g_2\lambda^\pp}+i\vartheta} 
g_2^{7/2}}{(n_1n_2)^2} \, . \label{smasscorr} 
\ee 
 
The SYM operators dual to the states in~(\ref{4impses}) are a special 
case of~(\ref{scalimpops}), \ie four impurity $SO(4)_C\times SO(4)_R$ 
singlets. They are given by 
\ba 
\calO_{n_1,n_2,n_3} &\!=\!& 
\frac{\veps_{ijkl}}{\sqrt{J^3(g^2N_c)^{J+4}}} 
\hspace*{-0.2cm} \begin{array}[t]{c} 
{\displaystyle \sum_{p,q,r=0}^J} \\ 
{\scriptstyle p+q+r\le J} 
\end{array} \hspace*{-0.2cm} \er^{2\pi i[(n_1+n_2+n_3)p 
+(n_2+n_3)q+n_3r]/J} \times \nn \\ 
&\!\times\!& \, \Tr\left[Z^{J-(p+q+r)} \vp^i Z^p \vp^j Z^q \vp^k 
Z^r \vp^l \right] \, . \label{4impsymop} 
\ea 
In order to compute the one instanton contribution to the matrix of 
anomalous dimensions for such operators one considers the two-point 
function 
\be 
G(x_1,x_2) = \la\calO_{n_1n_2n_3}(x_1)\,\bar\calO_{m_1m_2m_3} 
(x_2)\ra 
\, . \label{2ptfbmnop} 
\ee 
The calculation proceeds as in the case of the correlators discussed
in Sect.~\ref{sec:N4-1inst}. In the semi-classical approximation one
needs to compute the classical profiles of the operators
$\calO_{n_1n_2n_3}$ and $\bar\calO_{m_1m_2m_3}$ and integrate them
over the instanton moduli space. The profiles of the
operator~(\ref{4impsymop})  and of its conjugate contain $2J+8$
fermion  zero-modes each and thus~(\ref{2ptfbmnop}) is non-minimal
according  to the terminology introduced in Sect.~\ref{sec:N4-1inst}. 
 
Although the calculation of the two-point function~(\ref{2ptfbmnop})
presents no new conceptual difficulties, it involves rather
complicated combinatorics associated with the distribution of the
exact and non-exact fermion zero-modes in the two operators. Each of
the two operators should soak up eight of the sixteen superconformal
modes in the combination
$(\zeta^1)^2(\zeta^2)^2(\zeta^3)^2(\zeta^4)^2$, while the remaining
modes are of type $\nu^A$ and $\bar\nu^A$. Expanding the trace
in~(\ref{4impsymop})  and in the conjugate operator one obtains a
large  number of terms satisfying this requirement. The double limit
$N_c\to\infty$, $J\to\infty$, with $J^2/N_c$ fixed, simplifies
somewhat the analysis. The dominant contributions in this limit come
from certain specific distributions of the fermion modes. The large
$N_c$ limit requires that all the $\bar\nu^A\nu^B$ bilinears be in the
$\mb6$, see~(\ref{Ndepnu}). Moreover at large $J$ the leading
contributions to the operator profiles come from terms in which as
many of the superconformal modes as possible are provided by the $Z
$'s and $\bar Z$'s rather than by the impurities. This is because one
gets roughly a multiplicity factor of $J$ associated with every $Z$ or
$\bar Z$ providing one such mode. Taking into account these
simplifications the calculation of the profiles of $\calO_{n_1n_2n_3}$
and $\bar\calO_{m_1m_2m_3}$, albeit rather tedious, is
feasible. Eventually the dependence on the collective coordinates in
all the relevant terms in the profile of the operator
$\calO_{n_1n_2n_3}$ reduces to 
\be 
\frac{\rho^8}{[(x_1-x_0)^2+\rho^2]^{J+8}} 
\left(\bar\nu^{[1}\nu^{4]}\right)^J\left[\left(\zeta^1\right)^2 
\left(\zeta^2\right)^2\left(\zeta^3\right)^2\left(\zeta^4\right)^2 
\right]\!(x_1) \, . 
\label{O1-profile} 
\ee 
Similarly all the terms in the classical profile of 
$\bar\calO_{m_1m_2m_3}$ which contribute in the BMN limit contain the 
following factor 
\be 
\frac{\rho^8}{[(x_2-x_0)^2+\rho^2]^{J+8}} 
\left(\bar\nu^{[2}\nu^{3]}\right)^J\left[\left(\zeta^1\right)^2 
\left(\zeta^2\right)^2\left(\zeta^3\right)^2\left(\zeta^4\right)^2 
\right]\!(x_2) \, . 
\label{bO1-profile} 
\ee 
After factoring out the dependence on the collective coordinates the 
dependence on the mode numbers, $n_i$ and $m_i$, is determined by 
sums of the form 
\ba 
K(n_1,n_2,n_3;J) &=& \begin{array}[t]{c} 
{\displaystyle \sum_{p,q,r=0}^J} \\ 
{\scriptstyle p+q+r \le J} 
\end{array} \hsp{-0.1} \er^{2\pi i[(n_1+n_2+n_3)p 
+(n_2+n_3)q+n_3r]/J} \times \label{sum35} \\ 
&&\times \left[\frac{c_{1}}{4!}p(p-1)(p-2)(p-3) 
+\frac{c_{2}}{3!} q p(p-1)(p-2) +\cdots \right] \, , \nn 
\ea 
where each term contains combinatorial factors and 
$c_{1},c_{2},\ldots$ are numerical coefficients. 
 
The two-point function~(\ref{2ptfbmnop}) is thus 
\ba 
&& \la\calO(x_1)\,\bar\calO(x_2)\ra_{\rm inst} = \nn \\ 
&& = c(g,N_c,J) \int \frac{\dr^4 x_0\,\dr\rho}{\rho^5} 
\frac{\rho^{2J+16}}{[(x_1-x_0)^2+\rho^2]^{J+8} 
[(x_2-x_0)^2+\rho^2]^{J+8}} \times \nn \\ 
&& \times \int \dr^8\eta \, \dr^8\bar\xi \, \prod_{A=1}^4 
\left[\zeta^A(x_1)\right]^2 \left[\zeta^A(x_2)\right]^2 
\times \nn \\ 
&& \times \int \dr^5\Omega \, \left(\Omega^{14}\right)^J 
\left(\Omega^{23}\right)^J \, \left[K(n_1,n_2,n_3;J)\, 
K(m_1,m_2,m_3,J) \right] \, . \label{2ptfbmn1} 
\ea 
where $c(g,N_c,J)$ contains the dependence on the parameters arising
from the normalisation of the operators and the moduli space
integration measure, as well as the factors of $g\sqrt{N_c}$ obtained
rewriting the $(\bar\nu^A\nu^B)_\mb6$ bilinears in terms of the
angular variables $\Omega^{AB}$. In the large $J$ limit the sums
in~(\ref{sum35})  can be approximated with integrals. For instance the
first term becomes 
\ba 
&& \sum_{p,q,r=0}^J \er^{2\pi i[(n_1+n_2+n_3)p 
+(n_2+n_3)q+n_3r]/J} p(p-1)(p-2)(p-3) \label{J7} \\ 
&& \to J^7 \int_0^1\dr x\int_0^{1-x}\dr y 
\int_0^{1-x-y} \dr z \, \er^{2\pi i[(n_1+n_2+n_3)x+(n_2+n_3)y+n_3z]} 
\, x^4 \nn 
\ea 
From~(\ref{2ptfbmn1}) and~(\ref{J7}), recalling the analysis in 
Sect.~\ref{sec:N4-1inst}, one can deduce the dependence of the 
two-point function on the parameters. There are numerous sources of 
powers of $g$, $N_c$ and $J$ in the calculation, but remarkably 
the final result can be expressed only in terms of the parameters 
$g_2$ and $\lambda^\pp$, as required by BMN scaling. In detail one gets 
\ba 
&& \underbrace{\left(\fr{\sqrt{J^3(g^2N_c)^{J+4}}}\right)^2}_{ 
{\rm norm. ~ operators}} \times 
\underbrace{\er^{2\pi i\tau}g^8\sqrt{N_c}}_{{\rm measure}} 
\times 
\underbrace{\frac{\left(g\sqrt{N_c}\right)^{2J}}{J^2}}_{ 
\begin{array}{c} \scriptstyle \nu, \bar\nu \\ 
\scriptstyle {\rm integrals} \end{array}} 
\times \underbrace{\fr{J^2}}_{\begin{array}{c} 
\scriptstyle x_0, \rho \\ 
\scriptstyle {\rm integrals} \end{array}} \times 
\underbrace{\left(J^7\right)^2}_{{\rm sums}} \sim \nn \\ 
&& \sim \frac{J^7}{N_c^{7/2}}\,\er^{2\pi i\tau} = 
g_2^{7/2} \, \er^{-\frac{8\pi^2}{g_2\lambda^\pp}+i\vartheta} \, . 
\label{paramdep} 
\ea 
which is in agreement with the $\lambda^\pp$ and $g_2$ dependence of 
the string theory result~(\ref{samplfin}). 
 
The simple mode number dependence of the string two-point amplitude 
is more complicated to reproduce. In the SYM two-point function the 
dependence on the integers $n_i$ and $m_i$ is contained in the 
functions $K(n_1,n_2,n_3;J)$ and $K(m_1,m_2,m_3,J)$ defined 
in~(\ref{sum35}). Each term in these sums receives a large number of 
contributions resulting in very complicated expressions. However, 
combining all the contributions leads to impressive cancellations and 
a very simple result. In conclusion the one-instanton contribution to 
the two-point function~(\ref{2ptfbmnop}) can be written in the form 
\be 
G(x_1,x_2) = \frac{3^2\,(g_2)^{7/2} \, 
\er^{-\frac{8\pi^2}{g_2\lambda^\pp}+i\vartheta}}{2^{41}\,\pi^{13/2}}\, 
\fr{(n_1n_2m_1m_2)} \fr{(x_{12}^2)^{J+4}} 
\log\left[\Lambda^2 x_{12}^2\right] \, , 
\label{2ptfbmn2} 
\ee 
where $\Lambda$ is a scale that appears as a consequence of the 
logarithmic divergence in the $x_0$ and $\rho$ integrals, which 
signals a contribution to the matrix of anomalous dimensions. 
Notably the result is only non-zero if the mode numbers in the two 
operators are equal in pairs, again in agreement with string theory. 
 
From the coefficient of~(\ref{2ptfbmn2}) one can read off the 
contribution to the matrix of anomalous dimensions. The above 
calculation is not sufficient to determine the actual anomalous 
dimension of the operator~(\ref{4impsymop}) since this requires the 
diagonalisation of the matrix of two-point functions of all the 
operators with the same quantum numbers. However, all such two-point 
functions are expected to have the same dependence on $\lambda^\pp$ 
and $g_2$ found in~(\ref{2ptfbmn2}). Therefore one can conclude that 
the behaviour of the leading instanton contribution to the anomalous 
dimension of four impurity $SO(4)_C\times SO(4)_R$ singlet operators 
is 
\be 
\g_{\rm inst} \sim \frac{g_2^{7/2} \, 
\er^{-\frac{8\pi^2}{g_2\lambda^\pp}+i\vartheta}}{(n_1n_2)^2} \, , 
\label{bmnadim} 
\ee 
in agreement with~(\ref{smasscorr}). In view of the complexity of the 
calculation this result provides a striking test of the BMN proposal. 
 
A number of other two-point string amplitudes and their dual 
correlation functions have been studied in~\cite{gks1,gks2,gks3}. The 
many interesting results obtained in these papers can be summarised in 
the following statements. 
\begin{itemize} 
\item Four impurity operators in other representations of $SO(4)_R$ 
and the corresponding string states have two-point functions which 
behave as $(\lambda^\pp)^2(g_2)^{7/2} 
\exp(-8\pi^2/g_2\lambda^\pp+i\vartheta)$, \ie they are suppressed by two 
powers of $\lambda^\pp$ with respect to those in the singlet sector. 
\item Two impurity operators have the same suppression. The 
calculation of instanton contributions to two-point functions of two 
impurity operators in $\calN=4$ SYM is rather subtle because in order 
to saturate the integrations over the superconformal modes one needs 
to use the classical solution for the scalar fields involving six 
fermion modes, $\vp^{(6)\,AB}$. 
\item Supergravity states and their KK excitations do not couple to 
the D-instanton boundary state and thus, as expected, their masses 
do not receive non-perturbative corrections. This result is far from 
obvious in the gauge theory and requires non-trivial cancellations 
which have not been explicitly verified. 
\item (D-)Instantons contribute to the mixing of states in the 
NS--NS and R--R sectors of the plane string theory~\footnote{Unlike in 
flat space, in the plane-wave background this mixing occurs also in 
perturbation theory beyond tree-level~\cite{gks3}.}. 
\item Instanton contributions to two-point functions of certain 
operators dual to R--R string states, \ie operators with an even 
number of fermionic impurities, involve inverse powers of 
$\lambda^\pp$. Although this behaviour is rather surprising, it is not 
pathological in the $\lambda^\pp\to 0$ limit because the inverse 
powers of $\lambda^\pp$ are accompanied by the instanton weight 
$\exp(-8\pi^2/\lambda^\pp g_2)$. These two-point functions vanish in 
perturbation theory. 
\end{itemize} 
It is notable that many of these results can be straightforwardly 
obtained in string theory where they are easily deduced from 
properties of the D-instanton boundary state, whereas they are much 
more complicated to obtain from a field theoretical calculation in 
$\calN=4$ SYM.

\section{Conclusions} 
\label{sec:CONCL} 
 
We would like to conclude this long review by highlighting the many 
topics where Gabriele's contributions along the years have been at the 
heart of the theoretical developments that have made our understanding 
of non-perturbative effects of field theory so deep and powerful. 
 
Conceptually, perhaps the most important contributions in this 
direction have been his works on the foundation of the notion of 
effective action in a supersymmetric framework. The effective action 
for the ${\cal N}=1$ SYM theory~\cite{VY} and its extension to 
SQCD~\cite{TVY} are milestones along the way of dealing with the 
non-perturbative structure of field theory. These works appear as an 
immediate extension and generalisation of the approach established for 
the description of the low energy degrees of freedom of 
QCD~\cite{WEIN,GL}, as soon as the fundamental r\^ole of anomalies 
was recognized~\cite{WI,RDWN,DVNPV}. The validation of the famous 
Witten-Veneziano formula~\cite{WIVE} for the $\eta'$ mass, yielded by 
lattice simulations~\cite{GIU}, and the explicit instanton 
calculations, carried out in various instances in supersymmetric 
theories~\cite{AKMRV}, have beautifully confirmed the predictive 
power of the effective action approach both in a supersymmetric and 
in a non-supersymmetric context. 
 
Together with many other important, independently derived, 
results~\cite{ADSALL1}, these ideas have proved to be of enormous 
impact on the way we think today of possible extensions of the 
Standard Model. 
 
We cannot end this review without mentioning what we consider the most 
important step of modern physics beyond field theory, namely the 
construction of the dual Veneziano amplitude~\cite{veneziano}, which 
is expressed by the remarkably simple formula 
\beq 
A(s,t) = \int_0^1 \dr x\, x^{-\ap s -1} (1-x)^{-\ap t -1} 
\label{VEFO}\, . 
\eeq 
It is unanimously recognized that~(\ref{VEFO}) represents the founding 
paper of String Theory. It took some time to realise that the infinite 
tower of ``resonances'' exchanged in the $s$ and $t$ channel are the 
excitations of an open bosonic string living in 26 dimensions. Planar 
duality, $A(s,t) = A(t,s)$, and the UV softness of the amplitude are 
exposed quite neatly by its geometric interpretation in terms of 
vertex operators inserted on the boundary of a disk. The presence of a 
massless vector excitation has brought String Theory to be the most 
credited candidate for the unification of all interactions, including 
gravity. In this respect, the graviton comes in as the massless 
excitation of the closed string spectrum and its vertex operator is a 
sort of ``square'' of the vertex operator for the massless vector of 
the Veneziano amplitude. That open strings might be considered more 
fundamental than closed strings is something which seems to emerge in 
all modern approaches, where D-branes and their open string 
excitations are used to describe interactions mediated by gauge 
bosons. We want also to recall that in a somewhat more distant context 
string excitations have been shown to be able to account for the 
microscopic degrees of freedom of Black Holes, thus yielding what is 
considered today the only satisfactory solution to the holographic 
puzzle of Black Hole thermodynamics~\cite{VAST}. 
 
In the present review we have briefly sketched the enormous 
simplification that open strings bring into the ADHM construction of 
instantons. However, for lack of space we had no chance to stress the 
far-reaching consequences of ideas underlying the Veneziano 
amplitude in the quest for unification and in the process of 
clarification of the many puzzles of quantum gravity. We dare to 
conclude by saying that we expect the Veneziano amplitude to be among 
the basic blocks of any consistent formulation of the fundamental 
laws of Nature.

\subsection*{Acknowledgments} 
Discussions with Massimo Testa, Yassen Stanev and especially Michael 
Green are gratefully acknowledged. The work of SK was supported in 
part by a Marie Curie Intra-European Fellowship and by the EU-RTN 
network {\it Constituents, Fundamental Forces and Symmetries of the 
Universe} (MRTN-CT-2004-005104).

\appendix

\section{-- Notations} 
\label{sec:APPA} 
 
$\bullet$ {\bf Generalities} 
 
We work in Euclidean metric with $g^E_{\mu\nu}=\delta_{\mu\nu}$. 
Factors of the gauge coupling constant, $g$, will be explicit 
everywhere. We are interested in computing expectation values of 
gauge invariant (possibly multi-local) renormalisable, composite 
operators, i.e.\ functional integrals of the type 
\beq 
\hspace{-.05cm} 
\langle O\rangle\!=\!\frac{1}{Z}\! 
\int\!{\cal D}\mu(\psi,\bar\psi) {\cal D}A_\mu 
\exp\big{[}\!-S_{\rm YM}+\!\int \!\dr^4x \,\bar\psi(\Dslash+m)
\psi\big{]} O[\psi,\bar\psi,A_\mu]\, , 
\label{VEV} 
\eeq 
where $\Dslash$ can be either a Dirac or a Weyl--Dirac operator (see 
below) and $Z$ is a similar functional integral with $O$ replaced by 
the identity operator.\\ 
 
\noindent $\bullet$ {\bf Yang--Mills action} 
 
The pure Yang--Mills action has the form 
\beqn 
&&S_{\rm YM}=\frac{1}{2}\int \dr^4x\,{\rm Tr}[F_{\mu\nu}F_{\mu\nu}]= 
\frac{1}{4}\int \dr^4x \sum_a F^a_{\mu\nu}F^a_{\mu\nu}\, ,\label{PUGA}\\ 
&&F_{\mu\nu}=T^a F^a_{\mu\nu}\, ,\qquad F_{\mu\nu}= 
\partial_\mu A_\nu-\partial_\nu A_\mu +ig[A_\mu,A_\nu]\, . 
\label{DEFG} 
\eeqn 
 
\noindent $\bullet$ {\bf Some group theory formulae} 
 
In~(\ref{PUGA}) the matrices $T^a\equiv T^a_{{\bf N_c}}$, 
$a=1,2,\ldots,N^2_c-1$ are the $SU(N_c)$ generators in the fundamental 
representation, ${\bf N_c}$. In general the generators, $T_{{\bf 
R}}$, in the (irreducible) representation ${\bf R}$ are normalised 
according to the formula 
\beq 
{\rm Tr}\big{[}T_{{\bf R}}^a T^b_{{\bf R}}\big{]}=\ell[{\bf R}] 
\delta^{ab}\, , 
\label{NGEN} 
\eeq 
with $\ell[{\bf R}]$ the Dynkin index of the representation. It is 
customary to normalise generators in the ${\bf N_c}$, ${\bf \bar 
N_c}$ and ${\bf Adj}$ representations so that 
\beq 
\ell[{\bf N_c}]=\ell[{\bf \bar{N_c}}]=\frac{1}{2}\, ,\qquad 
\ell[{\bf Adj}]=N_c\, . 
\label{DYN} 
\eeq 
Taking the trace of the equation which defines the quadratic Casimir 
operator of the representation ${\bf R}$ 
\beq 
\sum_a T_{{\bf R}}^a T^a_{{\bf R}}=c_2[{\bf R}] 
\one_{{\rm dim}({\bf R})\times {\rm dim}({\bf R})}\, , 
\label{CASIM} 
\eeq 
and using~(\ref{NGEN}), one gets the useful relation 
\beq 
c_2[{\bf R}] {\rm dim}({\bf R})=\ell[{\bf R}] {\rm dim}(G)\, . 
\label{UREL} 
\eeq 
 
\noindent $\bullet$ {\bf Dirac fermions} 
 
The Euclidean action of a Dirac fermion, $\psi$, $\bar\psi$, in the 
representation ${\bf R}$ of the gauge group $SU(N_c)$ takes the form 
\beq 
S_{\rm DF}=\int \dr^4x \,\bar\psi_{r}D^{r}_{\mu s}[{\bf R}] 
\gamma_{\mu}\psi^{s}\, , 
\quad r,s=1,\ldots,{\rm dim}({\bf R}) \, ,
\label{DACT} 
\eeq 
where Dirac indices are understood and 
\beqn 
D^{r}_{\mu s}[{\bf R}]=\partial_\mu\delta^{r}_{s}- 
g(T^{a}_{{\bf R}})^{r}_{s} A_\mu^a\, .\label{DDCOV} 
\eeqn 
In~(\ref{DACT}) hermitean $\gamma$-matrices are used, satisfying 
the anti-commutation relations 
$\{\gamma_\mu,\gamma_\nu\}=2\delta_{\mu\nu}$. \\
 
\noindent $\bullet$ {\bf Weyl fermions} 
 
The Euclidean action of a Weyl fermion, 
$\lambda^{a}_{\alpha},\bar\lambda^{a}_{\dot\alpha}$ 
($\alpha,\dot\alpha=1,2$) belonging to the adjoint representation of 
the gauge group $SU(N_c)$ takes the form 
\beq 
S_{\rm WF}=\int \dr^4x \,\bar\lambda^{a}_{\dot\alpha} 
D^{ab}_\mu[{\bf Adj}]\bar\sigma_{\mu}^{\dot\alpha\alpha} 
\lambda^{b}_{\alpha}\, , 
\label{WACT} 
\eeq 
where 
\beqn 
&&D_\mu^{ab}[{\bf Adj}]=\partial_\mu\delta^{ab}-gf^{abc} 
A_\mu^c\, ,\label{DWCOV}\\ 
&&\bar\sigma_\mu =(\one, -i\sigma_k)\, ,\label{BSIGMU} 
\eeqn 
with $\sigma_k$ the Pauli matrices. It is also useful to introduce the 
matrices $(\sigma_{\mu})_{\alpha\dot\alpha}$ 
\beqn 
\sigma_\mu =(\one, i\sigma_k)\, .\label{SIGMU} 
\eeqn 
and the definitions 
\beqn 
&&(\sigma_{\mu\nu})_\alpha^\beta=\half(\sigma_{\mu\alpha\dot\alpha} 
\bar\sigma_{\nu}^{\dot\alpha\beta}- 
\sigma_{\nu\alpha\dot\alpha}\bar\sigma_\mu^{\dot\alpha\beta})\, , 
\label{SIGMUNU}\\ 
&&(\bar\sigma_{\mu\nu})_{\dot\alpha}^{\dot\beta}= 
\half(\bar\sigma_{\mu}^{\dot\beta\alpha}\sigma_{\nu\alpha\dot\alpha}- 
\bar\sigma_\nu^{\dot\beta\alpha}\sigma_{\mu\alpha\dot\alpha})\, . 
\label{SIGMUNUB} 
\eeqn 
 
\noindent $\bullet$ {\bf The SYM action} 
 
The Euclidean action of the minimal ${\cal N}=1$ supersymmetric gauge 
theory (Super Yang--Mills, SYM), $S_{\rm SYM}$, when written in 
components, is simply given by the sum of~(\ref{PUGA}) 
and~(\ref{WACT}). The classical action is invariant under the 
$U_\lambda(1)$ R-symmetry~\cite{RSYM} 
\beqn 
&&\lambda\to \er^{i\alpha}\lambda\, ,\qquad A_\mu\,\,{\rm untouched}\, . 
\label{ULAM} 
\eeqn 
Quantum mechanically this symmetry is anomalous with 
\beqn 
&&\partial_\mu J^{(\lambda)}_\mu=2i\ell[{\bf Adj}] 
\frac{g^2}{32\pi^2}F_{\mu\nu}^a\tilde F_{\mu\nu}^a= 
2iN_c\frac{g^2}{32\pi^2}F_{\mu\nu}^a\tilde F_{\mu\nu}^a\, , 
\label{ANOMALYC}\\ 
&&J^{(\lambda)}_\mu=\bar\lambda_{\dot\alpha}^a 
\bar\sigma_\mu^{\dot\alpha\alpha} 
\lambda_{\alpha}^a\, ,\label{CURL} 
\eeqn 
but has a non-anomalous discrete subgroup 
\beq 
\Z_{2N_c}=\{z_k=\er^{i\alpha_k}\, ,\alpha_k=2\pi k/2N_c\, , 
k=1,2,\ldots,2N_c\}\, .\label{ZDN} 
\eeq 
This important statement can be proved in different ways. An elegant 
proof makes use of the (natural) extension of SYM in which a 
$\vartheta$ term is added to the action (see last section of 
Appendix~C). In this situation under a $U_\lambda(1)$ rotation of the 
gluino fields in the functional integral, we get the (anomalous) WTI 
\beq 
\langle O_1(x_1)\ldots O_n(x_n)\rangle^{(\vartheta)}= 
\langle O_1(x_1)\ldots O_n(x_n)\rangle^{(\vartheta+2N_c\alpha)}\, . 
\label{UWTI} 
\eeq 
Since the theory is classically invariant under such a rotation (the 
transformation $u(\alpha)O_k 
u^\dagger(\alpha)=\exp{(i\eta_k\alpha)}O_k$, with $\eta_k$ the 
$U_\lambda(1)$ charge of $O_k$ leaves invariant the correlator if the 
vacuum is annihilated by the unitary operator $u(\alpha)$), the only 
effect of the transformation is to change the value of the $\vartheta$ 
angle. The change does not affect physics if 
\beq 
2N_c\alpha=2n\pi\, , \quad n\in \Z\, . 
\label{ANGL} 
\eeq 
Clearly this result holds for any value of $\vartheta$, thus also at 
$\vartheta=0$. 
 
When written in superfields, the SYM action takes the form~\cite{RSYM} 
\beqn 
&&S_{\rm SYM}=\int \dr^4x\,\dr^2\theta\, {\rm Tr}[W_\alpha W^\alpha]\, , 
\label{SYMWZ}\\ 
&&W_\alpha=-\frac{1}{4}\bar D^2 \big{(}\er^{-gV}D_\alpha \er^{gV}\big{)} 
\, , \label{WWZ}\\ 
&&V(x,\theta)= C(x)+\theta\mu(x)+\bar\theta\bar\mu(x)+ 
{1\over 2}\theta^2 S(x)+ {1\over 2}\bar\theta^2\bar{S}(x)+\nn\\ 
&&+\,\theta\sigma^\mu \bar\theta A_\mu(x) + {1\over 2} 
\bar\theta^2\theta\lambda(x)+ {1\over 2} \theta^2\bar\theta 
\bar\lambda(x)+ {1\over 4} \theta^2 \bar\theta^2 D(x)+\ldots\, , 
\label{VWZ} 
\eeqn 
where dots stand for terms that can be expressed as derivatives of the 
fields already present in~(\ref{VWZ}).\\ 
 
\noindent $\bullet$ {\bf The SQCD action} 
 
The Euclidean action of the ${\cal N}=1$ supersymmetric theory which 
more closely resembles QCD is obtained by coupling in a 
supersymmetric and gauge invariant way to the SYM supermultiplet 
$N_f$ pairs of matter chiral superfields fields ($f=1,\ldots,N_f$, 
$r=1,\ldots,N_c$) 
\beqn 
\hspace{-.5cm}&&\Phi_f^r(x)=\phi_f^r(y)+\sqrt{2}\theta^\alpha 
\psi^r_{\alpha f}(y) 
+\theta^2 F_f^r(y)\, ,\qquad y_\mu=x_\mu+i\theta\sigma_\mu\bar\theta 
\, ,\label{PHIN}\\ 
\hspace{-.5cm}&&\tilde\Phi_{r}^f(x)=\tilde\phi_{r}^f(y)+ 
\sqrt{2}\theta^\alpha\tilde\psi_{\alpha r}^{f}(y) 
+\theta^2 \tilde F_{r}^f(y)\, ,\label{PHINB} 
\eeqn 
belonging to the representations ${\bf N_c}$ and ${\bf \bar N_c}$, 
respectively, of the gauge group. In this way the gauge invariant 
mass term 
\beqn 
&&S_{\rm SQCD}^{\rm mass}=\sum_{f}\Big{[}m_f\int \dr^4x\sum_{\alpha r} 
\tilde\psi^{\alpha f}_r\psi_{\alpha f}^r+ 
m^{*}_f\int \dr^4x\sum_{\dot\alpha r}\bar\psi_{\dot\alpha r}^f 
{\bar{\tilde{\psi}}}^{\dot\alpha r}_f+\nn\\ 
&&+|m_f|^2\int \dr^4 x\sum_r(\phi^{*f}_r\phi^r_f+ 
\tilde\phi^{f}_r\tilde\phi^{*r}_f)\Big{]}\label{MTERM} 
\eeqn 
can be constructed. The rest of the action is completely standard 
and can be found in any textbook or, for instance, in~\cite{AKMRV}.\\ 
 
\noindent $\bullet$ {\bf The ``flavour'' symmetries of the SQCD action} 
 
I) The classical SQCD action is invariant under the $U_\lambda(1)$ 
R-symmetry~\cite{RSYM}~\footnote{$U_\lambda(1)$ is sometimes also 
called  $U_A^{PQ}(1)$~\cite{SWTI1}, where PQ stands for 
Peccei--Quinn~\cite{PEQU},  because it is anomalous and classically 
unbroken even at non-vanishing masses.} which now transforms 
in a non-trivial way gluinos and scalars according to 
\beqn 
&&\lambda\to \er^{i\alpha}\lambda\, ,\quad \phi\to \er^{i\alpha}\phi\, , 
\quad \tilde\phi\to \er^{i\alpha}\tilde\phi\, ,\quad +\quad 
{\rm complex\,\, conjugate\, ,}\nn\\ 
&&A_\mu\, ,\quad \psi,\bar\psi\, ,\quad \tilde\psi,\bar{\tilde\psi}\, , 
\quad {\rm untouched}\, .\label{ULAMM} 
\eeqn 
The $U_\lambda(1)$ R-symmetry of SQCD is anomalous with the same 
anomaly as in SYM (see~(\ref{ANOMALYC})). Again only the 
$\Z_{2N_c}$ subgroup is unbroken. With respect to the SYM case 
$J^{(\lambda)}_\mu$ must now be augmented with the inclusion of the 
matter contribution and reads 
\beqn 
J^{(\lambda)}_\mu=\bar\lambda_{\dot\alpha}^a 
\bar\sigma_\mu^{\dot\alpha\alpha}\lambda_{\alpha}^a+ 
\sum_f\Big{\{}\big{[}i\phi^{f*}\!\buildrel\leftrightarrow 
\over\partial_\mu\!\phi_f\big{]} 
+\big{[}\phi_f\to \tilde\phi^f\big{]}\Big{\}}\, . 
\label{CURC} 
\eeqn 
 
II) The massless theory with $N_f$ flavours possesses a global 
$SU(N_f)\times SU(N_f)\times U_V(1)\times U_A(1)$ symmetry. The chiral 
$SU(N_f)\times SU(N_f)$ symmetry is broken by matter mass terms. For 
instance, if all masses are equal ($m_f=m$), the unbroken subgroup is 
the diagonal vector group $U_V(N_f)$, while, if all masses are 
different ($m_1\neq m_2\neq\ldots\neq m_f$), the leftover unbroken 
subgroup is $U(1)^{N_f}$. 
 
III) The $U_A(1)$ transformation 
\beqn 
&&(\phi,\psi)\to \er^{i\alpha}(\phi,\psi)\, ,\quad 
(\tilde\phi,\tilde\psi) \to 
\er^{i\alpha}(\tilde\phi,\tilde\psi)\, ,\quad + 
\quad{\rm complex\,\, conjugate\, ,}\nn\\ 
&&\lambda\, , \quad A_\mu\, ,\quad{\rm untouched}\, .\label{UAUN} 
\eeqn 
is classically a symmetry at vanishing masses, but it is 
quantum-mechanically anomalous with 
\beqn 
\hspace{-.5cm}&&\partial_\mu J^{(A)}_\mu= 
2iN_f\frac{g^2}{32\pi^2}F_{\mu\nu}^a\tilde F_{\mu\nu}^a\, , 
\label{ANOMALYF}\\ 
\hspace{-.5cm}&&J^{(A)}_\mu=\sum_f\Big{\{}\big{[}\bar\psi_{\dot\alpha}^f 
\bar\sigma_\mu^{\dot\alpha\alpha}\psi_{\alpha f}+ 
i\phi^{*f}\!\buildrel\leftrightarrow\over\partial_\mu\!\phi_f\big{]} 
+\big{[}(\phi_f,\psi_f)\to (\tilde\phi^f,\tilde\psi^f)\big{]}\Big{\}} 
\, .\label{CURF} 
\eeqn 
 
IV) Often, instead of $U_\lambda(1)$, the linear combination 
\beq 
U_R(1)=\frac{3}{2}U_\lambda(1)-\half U_A(1) 
\label{UUR}\eeq 
is introduced because the associated current belongs to the current 
supermultiplet~\cite{FZPS}. This classical symmetry is anomalous and 
from~(\ref{ANOMALYC}) and~(\ref{ANOMALYF}) we see that the 
associated current obeys the anomaly equation 
\beq 
\partial_\mu{J}_\mu^{(R)}=i(3N_c-N_f)\frac{g^2}{32\pi^2}F_{\mu\nu}^a 
\tilde F_{\mu\nu}^a\, .\label{RAN} 
\eeq 
 
V) It is possible to construct a non-anomalous, exactly conserved 
current, ${J}_\mu^{({\hat A})}$, out of the two anomalous currents 
$J^{(A)}_\mu$ and $J^{(\lambda)}_\mu$ (see~(\ref{CURF}) 
and~(\ref{CURC})). One finds 
\beq 
{J}_\mu^{({\hat A})}=-N_f J^{(\lambda)}_\mu+N_c J^{(A)}_\mu\, .\label{NONAM} 
\eeq 
The transformations induced on the fields by the associated $U_{\hat A}(1)$ 
symmetry~(\ref{NONAM}) are 
\beqn 
\hspace{-.5cm}&&(\phi,\tilde\phi)\to 
\er^{i(N_c-N_f)\alpha}(\phi,\tilde\phi) 
\, , \,\,\,(\psi,\tilde\psi)\to \er^{iN_c\alpha}(\psi,\tilde\psi)\, , 
\,\,\,+\,\,\,\,{\rm complex\,\, conjugate}\, ,\nn\\ 
\hspace{-.5cm}&&\lambda\to \er^{-iN_c\alpha}\lambda\, ,\quad 
+\quad{\rm complex\,\, conjugate}\, .\label{FTRA} 
\eeqn 
 
VI) In the Table below we recollect for convenience the charges of 
elementary and composite gauge and matter fields under the various 
$U(1)$'s we have introduced. In the last two rows we report the 
coefficient of the anomaly of the associated current in units of 
$Q=\frac{g^2}{32\pi^2}F_{\mu\nu}^a\tilde F_{\mu\nu}^a$ and we indicate 
whether the conservation of the current is broken by mass terms or 
not. \\ 
\begin{table} 
\begin{center} 
\begin{tabular}{|c|c|c||c|c|c|c|c|} 
\hline 
Field & $SU(N_f)$ & $SU(N_f)$ & $U_A(1)$ & $U_R(1)$ & $U_{\hat A}(1)$ & 
$U_{\lambda}(1)\!=\!U_A^{PQ}(1)$ & $U_V(1)$ \\ 
\hline 
\hline 
$\lambda$                 & ${\bf 1}$&${\bf 1}$         & 0 & 3/2 
 & -$N_f$       & 1  & 0  \\ 
$\psi$                     &${\bf N_f}$&${\bf 1}$      & 1 &  1 
& $N_c$        & 0  & 1  \\ 
$\phi$                     &${\bf N_f}$&${\bf 1}$      & 1 &  -1/2 
& $N_c-N_f$    & 1  & 0  \\ 
$\tilde\psi$               &${\bf 1}$&${\bf N_f}$      & 1 &  1 
& $N_c$        & 0  & -1 \\ 
$\tilde\phi$               &${\bf 1}$&${\bf N_f}$      & 1 &  -1/2 
& $N_c-N_f$    & 1  & 0  \\ 
\hline 
$S|_{\theta=0}$            &${\bf 1}$&${\bf 1}$         & 0 &   3 
& $-2N_f$      & 2  & 0  \\ 
$T^f_{h}|_{\theta=0}$      &${\bf N_f}$&${\bf N_f}$   & 2 &   2 
& $2(N_c-N_f)$ & 2  & 0  \\ 
\hline 
\hline 
Anomaly              &/ & /        & $2N_f$ & $3N_c-N_f$ & 0   & 
$2N_c$ 
& 0 \\ 
\hline 
Mass term        &Yes & Yes      &  Yes   &  Yes       & Yes & 
No   & No \\ 
\hline 
\end{tabular} \vspace{.2cm} 
\caption{The $SU(N_f)\times SU(N_f)$ quantum numbers and 
$U(1)$-charges of elementary and composite fields of SQCD. The anomaly 
associated with each $U(1)$ current is given in units of 
${g^2}/{32\pi^2}F^a_{\mu\nu}\tilde{F}^a_{\mu\nu}$. In the last line 
by ``Yes'' (``No'') we mean that the corresponding symmetry is (is 
not) broken by the presence of mass terms.} 
\label{tab:cpu} 
\end{center} 
\end{table}

\noindent $\bullet$ {\bf Gluino zero modes} 
 
The explicit expression of the $2N_c$ gluino zero modes endowed with 
the correct normalisation $\int \dr^4x\,\sum_a\lambda^{a\alpha}(x) 
\lambda^{a*}_\alpha(x)=1$ is (the index counting the $2N_c$ zero modes 
is indicated in parenthesis) 
\beqn 
\hspace{-.9cm}&& (\lambda^{\alpha}_{(k)})^s_r(x)=\frac{1}{\pi} 
(\delta^\alpha_r\delta^s_k-\epsilon^{\alpha s}\epsilon_{rk}) 
\rho^2(f(x))^2\, ,\quad k=1,2\, ,\label{ZMG1}\\ 
\hspace{-.9cm}&& (\lambda^{\alpha (\dot\alpha)})^s_r(x)= 
-\frac{i}{\sqrt{2}\pi}\bar\sigma^{\dot\alpha\beta}_\mu(x-x_0)_\mu 
(\delta^\alpha_r\delta^s_\beta-\epsilon^{\alpha s} 
\epsilon_{r\beta})\rho(f(x))^2\, ,\quad \dot\alpha=1,2\, ,\label{ZMG2}\\ 
\hspace{-.9cm}&& (\lambda^{\alpha}_{(\pm i)})^s_r(x)=\frac{1}{\sqrt{2}\pi} 
(\delta^\alpha_r\delta^s_i\pm\epsilon^{\alpha s}\delta_{ri})\rho 
(f(x))^{3/2} \, ,\quad i=1,\ldots,N_c-2\, ,\label{ZMG3} 
\eeqn 
with 
\beq 
f(x)=\frac{1}{(x-x_0)^2+\rho^2}\, .\label{FX} 
\eeq 
The first four modes are $SU(2)$ triplets, while the last $2(N_c-2)$
are doublets. The four triplets can be directly generated from the
expression~(\ref{ICL}) by acting on it with anyone of two
supersymmetric and two superconformal transformations that are
unbroken in the background instanton field. They are often called
``exact zero modes'' in the literature, see for
instance~\cite{KHETAL} and references therein. This name  originates
from the following observation. Effectively the overall  field
configuration which is relevant for the kind of computations we  have
presented in Sect.~\ref{sec:IC} (see Sect.~\ref{sec:N4insteff} for
further applications) is given by the gauge instanton solution, the
associated set of fermionic zero modes and the expression of the
scalar fields that are obtained by solving their linearised classical
e.o.m., i.e.\ the e.o.m.\ that result upon neglecting the quartic
scalar self-interaction terms. The reason for neglecting such terms
is that the latter would give rise to contributions of higher order
in $g$ compared to the leading ones we have been keeping. When the
action of the theory is computed in this approximation and on the
above field configuration, it just happens that the result does not
depend on the fermionic collective coordinates associated with the
four $SU(2)$ triplet zero modes of~(\ref{ZMG1})  and~(\ref{ZMG2}). The
other fermionic zero modes will give origin to  quartic terms in the
remaining $2(N_c-2)$ fermionic collective  coordinates (see
Sect.~\ref{sec:N4insteff}).

\section{-- Bosonic collective coordinates and functional 
integration} 
\label{sec:APPB} 
 
In this Appendix we want to explain how one can compute the pure gauge 
part of the functional integration in the semi-classical approximation 
around a non-trivial instantonic background. We will follow the method 
of~\cite{YAF}, which neatly explains how to deal with the problem of 
bosonic zero modes and the consequent need of introducing collective 
coordinates. 
 
In the semi-classical approximation one starts by expanding the (gauge) 
action around the (instanton) classical solution, keeping only terms 
up to quadratic fluctuations. Setting 
\beq 
A_\mu=A^I_\mu+Q_\mu\, ,\label{FLUCT} 
\eeq 
one gets in this way 
\beq 
S_{\rm YM}=S^I-\half\int \dr^4x\,{\rm Tr}\big{[}Q_\mu 
{\cal M}_{\mu\nu} (A^I)Q_\nu\big{]}+{\mbox O}(Q^3)\, , 
\label{QUADR} 
\eeq 
where 
\beqn 
&&S^I=\frac{8\pi^2}{g^2}|K|\, ,\label{INTAC}\\ 
&&{\cal M}_{\mu\nu}(A^I)=-D^2(A^I)\delta_{\mu\nu}+ 
D_\mu(A^I)D_\nu(A^I)-2[F_{\mu\nu}^I,\cdot\,]\, ,\label{GOPE}\\ 
&&D_\mu(A^I)=\partial_\mu+g[A^I_\mu,\cdot\,]\, .\label{DGCOV} 
\eeqn 
The operator ${\cal M}_{\mu\nu}(A^I)$ has quite a large manifold of 
(normalisable and non-normalisable) zero modes. Not only it is 
annihilated by all the functions of the form $D_\nu(A^I)F(x)$, as a 
consequence of gauge invariance, but also by the $4|K| N_c$ 
normalisable vectors that are obtained by differentiating the 
instanton field configuration with respect to the $4|K| N_c$ 
parameters, $\beta_i$, $i=1,\ldots,4|K| N_c$ (bosonic collective 
coordinates in the following), upon which the most general classical 
solution depends~\footnote{We are referring here to the $SU(N_c)$ 
gauge group case. In the one-instanton sector, $|K|=1$, the collective 
coordinates are the size and the location of the instanton and its 
$4N_c-5$ ``orientation angles'' in colour 
space~\cite{BER,TEM,COR}.}. The existence of such zero modes is 
immediately proved by noticing that by differentiating the classical 
instanton e.o.m., $(\delta S_{\rm YM}/\delta A_\mu)_{A^I_\mu}=0$, with 
respect to $\beta_i$, one gets 
\beqn 
&&\int \dr^4x'\,\Big{(}\frac{\delta^2 S_{\rm YM}}{\delta A_\mu(x) 
\delta A_\nu(x')}\Big{)}_{A^I_\mu} 
\frac{\partial A^I_\nu(x',\beta)}{\partial \beta_i}=\nn\\ 
&&={\cal M}_{\mu\nu}(A^I)\frac{\partial A^I_\nu(x,\beta)} 
{\partial \beta_i}=0 
\, , \quad i=1,\ldots,4|K|N_c\, . 
\label{ZM} 
\eeqn 
The most elegant way to deal with an operator with such a kernel was 
worked out some time ago in~\cite{YAF}. The idea is to functionally 
integrate over all fluctuations, 
$Q_\mu(x)=A_\mu(x)-A^{I}_\mu(x,\beta)^U$, that are orthogonal to the 
manifold described by $A^{I}_\mu(x,\beta)^U$ when $U$ spans the space 
of topologically trivial gauge transformations, ${\cal G}_0$, and the 
parameters $\beta_i$ are let to move in their allowed range of 
variation. In more mathematical terms the latter manifold is called the 
``instanton moduli space''. 
 
The orthogonality conditions~(\ref{ZM}) are imposed by a 
straightforward generalisation of the usual Faddeev--Popov (FP) 
procedure~\cite{FP} which consists in introducing in the functional 
integral the identity 
\beqn 
\hspace{-.2cm}&&1=\Delta_{\rm FP}\int_{{\cal G}_0}\prod_{a,x} 
\delta h^a(x)\int_{\cal M} 
\prod_i \dr\beta_i\, \delta\Big{(} <(A_\mu-A^{I}_\mu(\beta)^{U_h}), 
\frac{\delta A^{I}_\mu(\beta)^{U_h}}{\delta h^a(x)}>\Big{)}\times\nn\\ 
\hspace{-.2cm}&&\times \delta\Big{(}<(A_\nu-A^{I}_\nu(\beta)^{U_h}), 
\frac{\partial A^{I}_\nu(\beta)^{U_h}}{\delta \beta_i}>\Big{)}\, , 
\label{FPDD} \eeqn 
where $\Delta_{\rm FP}$ is the FP determinant. In~(\ref{FPDD}) we have used 
the short-hand notation 
\beq <f_\mu,g_\mu>\,=\frac{1}{2}\int \dr^4x\,{\rm Tr}_{{\bf Adj}} 
[f_\mu(x) g_\mu(x)] \label{PRSC} 
\eeq 
for the scalar product $<\cdot,\cdot>$ induced in the space of 
functions by the form of the gauge action. After some algebra 
(see~\cite{TRAV} for details)~(\ref{FPDD}) can be cast in the more 
expressive form 
\beqn 
\hspace{-.2cm}&&1=\Delta_{\rm FP}\int_{{\cal G}_0}\! 
\prod_{a,x}{\cal D} \mu[h^a(x)]\int_{\cal M} 
\!\prod_i \dr\beta_i\, \delta\Big{(}{\rm Tr}[T^a D_\mu(A^{I}) 
(A_\mu(x)^{U_h^\dagger}\!-\!A^{I}_\mu(x,\beta))]\Big{)}\times\nn\\ 
\hspace{-.2cm}&&\times \delta\Big{(}<(A_\nu^{U_h^\dagger}-A^{I}_\nu(\beta)), 
\frac{\partial A^{I}_\nu(\beta)}{\delta \beta_i}>\Big{)}\, , 
\label{FPDD1} 
\eeqn 
which shows that we are naturally brought to work in the instanton 
background gauge. $\Delta_{\rm FP}$ can be shown to have in the 
semi-classical approximation the expression 
\beq 
\Delta_{\rm FP}={\rm det}_{a,b}^{x,y}\big{[}-D^{2}(A^I)^{ab} 
\delta(x-y)\big{]}{\rm det}_{i,j}\big{[}<a^{(i)}(\beta),a^{(j)} 
(\beta)>\big{]}\, , 
\label{DFP} 
\eeq 
where the $a^{(i)}$'s ($i=1,2,\ldots,2|K|N_c$) are the mutually 
orthogonal (see next subsection) vectors 
\beq 
a^{(i)}_\mu(x,\beta)=\Big{[}\delta_{\mu\nu}- 
D_\mu(A)[D^2(A)]^{-1}D_\nu(A)\big{|}_{A_\mu=A_\mu^I} 
\Big {]}\frac{\partial A^{I}_\nu(x,\beta)}{\partial \beta_i}\, . 
\label{CZM} 
\eeq 
We will indicate by $||a^{(i)}||$ their norm in the metric induced by 
the scalar product~(\ref{PRSC}). The vectors $a^{(i)}_\mu(x,\beta)$ 
are not exactly the functions ${\partial 
A^{I}_\mu(x,\beta)}/{\partial \beta_i}$. They differ from the latter 
by a term which makes them to fulfill the equation 
\beq 
D_\mu(A^I)a^{(i)}_\mu(x,\beta)=0\, , 
\label{TZM} 
\eeq 
i.e.\ which makes them transverse with respect to the covariant 
derivative in the instanton background. 
 
Putting everything together and noticing that the orthogonality 
condition among the vectors~(\ref{CZM}) makes immediate the 
computation of the factor ${\rm 
det}_{i,j}[<a^{(i)}(\beta),a^{(j)}(\beta)>]$ in $\Delta_{\rm FP}$, one 
finally gets for the v.e.v.\ of a gauge invariant operator, $O(A)$, 
in the semi-classical approximation around an instanton configuration 
with winding number $|K|$ the expression 
\beqn 
\langle O\rangle\Big{|}_{s.c.}&=&\frac{\er^{-\frac{8\pi^2} 
{g^2}|K|}}{Z|_{s.c.}}\int{\cal D}Q_\mu\prod_i \dr\beta_i 
\frac{||a^{(i)}||}{\sqrt{2\pi}} \times \label{VEVSM}\\ 
&\times& \er^{-\frac{1}{2}\int \dr^4x\dr^4y\, Q_\mu 
{\cal M}_{\mu\nu}^{g.f.}Q_\nu} 
{\rm det}[-D^2(A^I)]\delta(D_\mu^{ab}(A^I)Q_\mu^b) O(A^I)\nn\, , 
\eeqn 
where 
\beqn 
&&Z|_{s.c.}=\int{\cal D}Q_\mu 
\er^{-\frac{1}{2}\int \dr^4x\dr^4y\, Q_\mu {\cal M}_{0;\mu\nu}^{g.f.}Q_\nu} 
{\rm det}[-\partial^2]\delta(\partial_\mu Q_\mu^a)\, ,\label{PTZ}\\ 
&&{\cal M}_{\mu\nu}^{g.f.}= -D^2(A^I)\delta_{\mu\nu}-2\, 
[F_{\mu\nu}^I,\cdot\,]\, ,\label{MMN}\\ 
&&{\cal M}_{0;\mu\nu}^{g.f.}=-\partial^2\delta_{\mu\nu}\, .\label{MOMN} 
\eeqn 
$Z|_{s.c.}$ is the necessary normalisation factor which, in order to 
be consistent with the approximation we are working in, must be 
evaluated by expanding the action around the trivial solution of the 
field e.o.m.\ keeping only terms quadratic in the fluctuations. Note 
that to make more transparent analogies and differences 
between~(\ref{VEVSM}) and~(\ref{PTZ}) we have named $Q_\mu$ the 
integration variable also in~(\ref{PTZ}). ${\cal 
M}_{\mu\nu}^{g.f.}$ (${\cal M}_{0\mu\nu}^{g.f.}$) is the gauge fixed 
operator that governs the quadratic fluctuations of the gauge field in 
the instanton (trivial) background and ${\rm det}[-D^2(A^I)]$ (${\rm 
det}[-\partial^2]$) is the associated FP determinant. 
 
One can formally perform the gauge functional integrations in the 
r.h.s.\ of~(\ref{VEVSM}), getting 
\beqn 
\langle O\rangle\Big{|}_{s.c.}&=&\mu^{n_B} 
\frac{\er^{-\frac{8\pi^2}{g^2}|K|}}{Z|_{s.c.}} 
\int\prod_{i=1}^{n_B} \dr\beta_i\frac{||a^{(i)}||}{\sqrt{2\pi}} 
\times \nn \\ 
&\times&\frac{({\rm det}'[{\cal M}_{\mu\nu}^{g.f.}])^{-\half} 
{\rm det}[-D^2(A^I)]}{({\rm det}[{\cal M}_{0;\mu\nu}^{g.f.}])^{-\half} 
{\rm det} [-\partial^2]}\,O(A^I)\, ,\label{INTVEV} 
\eeqn 
where $n_B=4|K|N_c$ is the number of bosonic zero modes and $\mu$ is 
the subtraction point (see below). The prime on ${\rm det}'[{\cal 
M}_{\mu\nu}^{g.f.}]$ is to mean that the determinant should be taken 
in the space orthogonal to the manifold spanned by the zero 
modes~(\ref{CZM}). 
 
A number of observations are in order here. 
 
1) As is seen from the above equations, by the method of~\cite{YAF} 
one is naturally led to the background gauge fixing condition 
$D_\mu^{ab}(A^I)Q_\mu^b=0$. 
 
2) One must imagine that the above functional integral has been 
computed in some regularisation. In these instantonic computations it 
is customary to work in the Pauli--Villars (PV) 
regularisation~\cite{GTH}, where a ghost-like field with mass $\mu$ 
(but opposite statistics) is introduced for each fundamental field in 
the action (gluons, FP-ghosts and, if present, fermions). The net 
effect of the presence of PV regulators is that the result of the 
functional integration over quadratic fluctuations will have the form 
of a product of factors, with each term being the ratio of the 
determinant of each particle quadratic operator divided by the 
associated PV-ghost determinant (raised to the appropriate power 
according to multiplicity and statistics). 
 
3) When the limit $\mu\to \infty$ is taken, the only left-over 
$\mu$ dependence is the multiplicative factor $\mu^{n_B}$, $n_B=4|K|N_c$. 
This factor comes about because of the following reason. There is a 
one-to-one correspondence between the eigenvectors (and the 
eigenvalues) of analogous operators in each ratio of determinants, 
except for the zero modes. There is, in fact, a mismatch between 
numerator and denominator in the sense that there are some (actually 
$n_B=4|K|N_c$) eigenvalues in the PV-denominator that do not have 
their counterpart in the ``primed'' determinant in the 
numerator. This leaves out precisely a factor $(\mu^2)^{\half}$ for 
each bosonic collective coordinate (and actually a factor 
$\mu^{-\half}$ for each Weyl fermion zero mode, see~(\ref{INTBF}) 
in Sect.~\ref{sec:2.3}). 
 
4) The factor $1/\sqrt{2\pi}$ for each bosonic zero mode appears for 
a similar reason. In fact, the integrations that give rise to the 
product of eigenvalues finally leading to the various bosonic 
determinants are all Gaussian in the (quadratic, i.e.\ semi-classical) 
approximation in which we are working. Since, as we noticed above, 
there is a one-to-one correspondence between physical modes and PV 
modes, all the factors $\sqrt{2\pi}$ will compensate between the 
numerator and the denominator, except for the factors coming from the 
integration over the PV-modes that are in correspondence with the 
bosonic zero modes. The reason is that no Gaussian integration is 
associated with the bosonic zero modes, as the latter were replaced by 
integrations over the related collective coordinates. In this way a 
factor $1/\sqrt{2\pi}$ for each bosonic zero mode will be left in the 
denominator. 
 
5) In principle one can go beyond the semi-classical 
formulae~(\ref{VEVSM}) and~(\ref{INTVEV}), including perturbatively 
${\mbox O}(Q_\mu^3)\sim{\mbox O}(g)$ and ${\mbox 
O}(Q_\mu^4)\sim{\mbox O}(g^2)$ corrections that were neglected 
before. As is well known, perturbation theory in an external field is 
perfectly well defined and fully renormalisable.

\subsection*{Bosonic zero modes} 
\label{sec:APPBS} 
 
We close this Appendix by reporting in the case $N_c=2$ and $K=1$ the 
explicit expression of the $4 |K| N_c=8$ ``transverse'' bosonic zero 
modes and of their norms. One finds ($y=x-x_0$) 
\beqn 
\begin{array}{lll} 
&a_\mu^{(\nu)}=F_{\mu\nu}^I(y)\, ,\qquad &||a_\mu^{(\nu)}||= 
\frac{2\sqrt{2}\pi}{g} 
\, , 
\\\\&a_\mu^{({\rm dil.})}=A^I_{\mu}(y){\frac{2y^2}{\rho(y^2+\rho^2)}}\, , 
\qquad &||a_\mu^{({\rm dil.})}||={\frac{4\pi}{g}}\, , 
\\\\&a_\mu^{(a)}=D_{\mu}(A^I)\big{[}\frac{T^a}{g} 
\frac{y^2}{y^2+\rho^2}\big{]}\, , 
\qquad &||a_\mu^{(a)}||=\frac{2\pi\rho}{g}\, . 
\label{BOSZM} 
\end{array} 
\eeqn 
One can check that these vectors are mutually orthogonal. 
 
\section{-- Quantum tunneling} 
\label{sec:APPC} 
 
The emergence of the quantum tunneling phenomenon in the presence of 
instantons is most easily and rigorously explained in the 
Schr\"odinger functional formalism~\cite{FHSCHR}, where the theory is 
formulated in terms of a ``propagation kernel'' which expresses the 
probability amplitude to find the gauge field configuration $\vec 
A^{(2)}(\vec x)=(A_i^{(2)}(\vec x)\, ,i=1,2,3)$ at the final time 
$t=T/2$, if the gauge field configuration at the initial time $t=-T/2$ 
was $\vec A^{(1)}(\vec x)=(A_i^{(1)}(\vec x)\, ,i=1,2,3)$. 
 
\subsection*{The Schr\"odinger functional in the temporal gauge} 
\label{sec:SFTG} 
 
The Schr\"odinger kernel is most expressively written in the temporal 
gauge~\cite{PREV}. As a result of making use of the Faddeev--Popov 
procedure, one can show that in the formal continuum language it takes 
the form~\footnote{For the lattice regularised formulation of the 
Schr\"odinger kernel - more commonly called Schr\"odinger functional 
in that context - see~\cite{LSCH}.} 
\beqn 
&&{\cal K}[\vec A^{(2)},\vec A^{(1)};T]=\int_{\hat{\cal G}_0} 
\prod_{\vec x}{\cal D}\mu[h(\vec x)] 
\int_{\vec A^{(1)}(\vec x)}^{[\vec A^{(2)}(\vec x)]^{U_0[h(\vec x)]}} 
 \!\!\!\prod_{a,\vec x}\prod_{-\frac{T}{2}<t<\frac{T}{2}}\,{\cal D} 
\vec A^a(\vec x,t)\times\nn\\ 
&&\times \exp \big{[}-S_{YM}[\vec A, A_0=0]\big{]}\, , 
\label{SKER} 
\eeqn 
where $U_0[h]=\exp(iT^ah^a)\in{\hat{\cal G}_0}$ with ${\hat{\cal 
G}_0}$ the group of the time-independent, topologically trivial gauge 
transformations (i.e. those that tend to the group identity at 
spatial infinity) and ${\cal D}\mu[h(\vec x)]$ is the invariant Haar 
measure over $SU(N_c)$ at each spatial point $\vec x$. The integration 
over the spatial components of the gauge field is extended to all 
configurations that satisfy the boundary conditions $\vec A(\vec 
x,T/2)=[\vec A^{(2)}(\vec x)]^{U_0[h(\vec x)]}$ and $\vec A(\vec 
x,-T/2)=\vec A^{(1)}(\vec x)$. 
 
The gauge integration over ${\hat{\cal G}_0}$ plays a crucial role in 
the formalism as it has the effect of projecting out from the kernel 
all the states that do not satisfy the Gauss' law constraint. In 
fact, since the Gauss' law operator is the generator of the 
time-independent topologically trivial gauge transformations, only the 
states annihilated by it will appear in the spectral decomposition of 
${\cal K}[\vec A^{(2)},\vec A^{(1)};T]$~\cite{PREV}, for which we can 
then formally write 
\beq 
{\cal K}[\vec A^{(2)},\vec A^{(1)};T]=\sum_n \er^{-E_n T} 
\Psi_n[\vec A^{(2)}](\Psi_n[\vec A^{(1)}])^*\, ,\label{SPEC} 
\eeq 
where 
\beqn 
&&{\cal H}\,\Psi_n[\vec A]=E_n \Psi_n[\vec A]\, ,\label{HPSI}\\ 
&&D_i(\vec A)^{ab}\,\frac{\delta}{\delta A_i^b(\vec x)}\, 
\Psi_n[\vec A]=0\, .\label{GAUS} 
\eeqn 
The last equation is indeed the statement that the eigenstates of the 
Hamiltonian appearing in the spectral decomposition~(\ref{SPEC}) are 
left untouched by time-independent gauge transformations that tend to 
the identity at infinity. In fact, from the invariance property 
\beqn 
&&{\cal U}_{0}[h] \Psi_n[\vec A]=\Psi_n[\vec A^{U_0[h]}]=\Psi_n[\vec A] 
\, ,\label{GAUINV}\\ 
&&{\cal U}_0[h]=\exp\Big{(}-\int \dr^3x \,(D_i^{ab}h^b(\vec x)) 
\frac{\delta}{\delta A_i^a(\vec x)}\Big{)}\, , 
\label{OPINV} 
\eeqn 
the Gauss' law~(\ref{GAUS}) follows by expanding~(\ref{GAUINV}) in 
powers of $h(\vec x)$, if the latter function vanishes as $|\vec 
x|\to\infty$, i.e.\ precisely if $U_0[h]\in {\hat{\cal G}_0}$. An 
equivalent way to prove this statement is to observe that the 
Schr\"odinger kernel enjoys the invariance properties 
\beq 
{\cal K}[(\vec A^{(2)})^{U_0},(\vec A^{(1)});T]={\cal K}[\vec A^{(2)}, 
\vec A^{(1)};T] 
={\cal K}[\vec A^{(2)},(\vec A^{(1)})^{U_0};T]\, .\label{INVU1} 
\eeq 
The first equality follows from the invariance of the Haar measure, as 
the $U_0$ gauge transformation can be reabsorbed in the integration 
measure over $\hat{\cal G}_0$. The second equality is an immediate 
consequence of the previous equation and the invariance property 
\beq 
{\cal K}[(\vec A^{(2)})^{U},(\vec A^{(1)})^{U};T]= 
{\cal K}[\vec A^{(2)},\vec A^{(1)};T] 
\, ,\quad U\in\hat{\cal G}_0\label{INVU} 
\eeq 
which in turn follows from the observation that any time-independent 
gauge transformation acting on the boundary fields can be reabsorbed 
by the change of variables $\vec A \to \vec A'=\vec A^U$ 
in~(\ref{SKER}). The invariance property~(\ref{INVU}) can be used to 
show that the $\hat{\cal G}_0$ gauge integration in~(\ref{SKER}) 
can be equally well performed over the time-independent gauge 
transformations acting on the boundary field $\vec A^{(1)}$ at 
$t=-T/2$. 
 
\subsection*{Emergence of the $\vartheta$ angle} 
\label{sec:ETA} 
 
We finally notice that the states $\Psi_n$ also support a unitary 
representation, ${\cal U}_{_K}$, of the abelian homotopy group 
$\Pi_3(SU(2))\sim \Pi_3({\rm S}_3)=\Z$. Since the Hamiltonian, 
${\cal H}$, and ${\cal U}_{_K}$ commute, they can be simultaneously 
diagonalised. Thus on their common eigenvectors (for a while we will 
keep calling them $\Psi_n$) we have 
\beq 
{\cal U}_{_K}\Psi_n[\vec A]=\Psi_n[\vec A^{U_{_K}}]= 
\er^{-i\vartheta_{_K}}\Psi_n[\vec A]\, .\label{UNIT} 
\eeq 
Consistency with the group property 
\beq 
{\cal U}_{_K} {\cal U}_{_{K'}}={\cal U}_{_{K+K'}}\label{GPROP} 
\eeq 
implies 
\beq 
\vartheta_{_K}=K \vartheta\, ,\label{TETA} 
\eeq 
naturally leading to the emergence of a $\vartheta$-angle. States 
should (and will) then be indicated by $\Psi_n^{(\vartheta)}[\vec A]$ 
in the following. 
 
\subsection*{Classical vacua and quantum tunneling} 
\label{sec:CVQT} 
 
The classical vacua of the theory are immediately identified as the 
gauge configurations for which the classical Hamiltonian 
\beq 
H=\int \dr^3x\,\Big{(}\frac{1}{2}\dot{A}^a_i\dot{A}^a_i+ 
\frac{1}{4}{F}^a_{ij}{F}^a_{ij}\Big{)}\label{HCLAS} 
\eeq 
vanishes, thus as time-independent ($\dot{A}^a_i=0$) pure gauges 
(${F}^a_{ij}=0$). This simple argument shows that there are infinitely 
many ``vacua'' labeled by an integer, $K\in\Z$, which is telling 
us which homotopy class the $K$-th vacuum belongs to. 
 
In Euclidean time the one-instanton ($K=1$) solution interpolates 
between adjacent minima, i.e.\ between pure gauge configurations with 
winding number differing by one unit~\footnote{Multi-instanton 
solutions with $|K|>1$ connect vacua with winding numbers differing by 
exactly $|K|$ units. They have an action (see~(\ref{INTAC})) which 
is exponentially small with respect to the one-instanton action. In 
the approximation we are working, their contribution is automatically 
taken care of by the exponentiation of the one-instanton contribution 
implicit in the spectral formula~(\ref{SPEC}). Incidentally this is 
the way in which within the Schr\"odinger functional formalism the 
``dilute gas'' approximation~\cite{COL,TUN2} is recovered.}. The 
formulae~(\ref{NPNM}) and~(\ref{QDIF}) can then be immediately 
proved. Since $A_0=0$, one successively gets, in fact 
\beqn 
\hspace{-.2cm}&&K\!=\!\frac{g^2}{32\pi^2} 
\!\int \! \dr^4x\, F^a_{\mu\nu}\tilde{F}^a_{\mu\nu}(x)\!=\! 
\frac{g^2}{16\pi^2}\!\int\! \dr^4x\,\partial_\mu K_\mu(x)\!=\! 
\frac{g^2}{16\pi^2}\!\int \! \dr^4x\,\partial_0 K_0(\vec x,t)=\nn\\ 
\hspace{-.2cm}&&=\frac{g^2}{16\pi^2}\Big{[} 
\int \dr^3x\,K_0(\vec x,+\infty)-\int \dr^3x\,K_0(\vec x,-\infty) 
\Big{]}=n_+-n_-\, . 
\label{QDIFZ} 
\eeqn 
The last equality follows remembering that at very large (positive and 
negative) times $K_0\propto\epsilon_{ijk}{\rm Tr}[A_i A_j A_k]$ with 
$\vec A$ a pure gauge. 
 
The classical vacuum degeneracy is removed by the quantum mechanical 
tunneling between adjacent minima occurring with an amplitude 
$\Gamma^I\propto \exp(-S^I)=\exp(-8\pi^2/g^2)$. A band spectrum is 
generated with the lowest energy eigenstates and eigenvalues given by 
\beqn 
&&\Psi_0^{(\vartheta)}[\vec A]=\sum_{K\in\Z} 
\er^{-iK\vartheta}\Psi_0^{(K)}[\vec A]\, ,\label{QTU}\\ 
&&E_0(\vartheta)=\alpha_0+\beta_0\cos\vartheta\, ,\label{BSPE} 
\eeqn 
where, at the leading order in $\Gamma^I$, $\Psi_0^{(K)}[\vec A]$ 
is the perturbative vacuum state functional ``centered'' around the 
$K$-th minimum of the energy, i.e.\ around a pure gauge field with 
winding number $K$ and $\alpha_0$, $\beta_0$ are computable constants 
proportional to the spatial volume of the system. 
 
It is not too difficult to prove the result~(\ref{BSPE}). We start by 
observing that, once quantum tunneling has been recognised to take 
place, the spectral decomposition of the Schr\"odinger kernel can be 
written in more informative form ($\sum_n\to\int \dr\vartheta\sum_\ell$) 
\beq 
{\cal K}[\vec A^{(2)},\vec A^{(1)};T]=\int_0^{2\pi}\dr\vartheta 
\sum_\ell \er^{-E_\ell(\theta) T}\Psi_\ell^{(\theta)}[\vec A^{(2)}] 
(\Psi_\ell^{(\theta)}[\vec A^{(1)}])^*\, .\label{SPECTETA} 
\eeq 
For the purpose of our calculation it is enough to take $\vec A^{(2)}$ 
and $\vec A^{(1)}$ as pure gauges. At this point only their winding 
number matters and we can simplify our notation by writing the 
Schr\"odinger kernel and the associated state functionals in the form 
${\cal K}[K^{(2)},K^{(1)};T]$ and $\Psi_\ell^{(\vartheta)}[K]$, 
respectively. In this notation~(\ref{UNIT}) becomes 
\beq 
{\cal U}_{_K}\Psi_\ell^{(\vartheta)}[0]=\Psi_\ell^{(\vartheta)}[K]= 
\er^{-i\vartheta K}\Psi_\ell^{(\vartheta)}[0] 
\, ,\label{NUNIT}\eeq 
where, we stress, ``$0$'' means a pure gauge configuration with $K=0$. 
 
To leading order in the instanton tunneling amplitude, we only need to 
evaluate the kernels ${\cal K}[K,K;T]$, ${\cal K}[K,K+1;T]$ 
and ${\cal K}[K+1,K;T]$, as all the others should be considered 
exponentially small to this order 
\beq 
{\cal K}[K,K';T]=0\, ,\qquad|K-K'|>1\, .\label{KMN}\eeq 
In order to proceed further we first note the relation 
\beqn 
\hspace{-.2cm}&&{\cal K}[K,K+1;T]=({\cal K}[K+1,K;T])^*=\nn\\ 
\hspace{-.2cm}&&=\int_0^{2\pi} \dr\vartheta \sum_\ell 
\Psi_\ell^{(\vartheta)}[K](\Psi_\ell^{(\vartheta)}[K+1])^* 
\er^{-E_\ell(\vartheta)T}=\nn\\ 
\hspace{-.2cm}&&=\int_0^{2\pi} \dr\vartheta\, \er^{i\vartheta}\sum_\ell 
\Psi_\ell^{(\vartheta)}[0](\Psi_\ell^{(\vartheta)}[0])^* 
\er^{-E_\ell(\vartheta)T} 
\, ,\label{KNNP}\eeqn 
that follows from~(\ref{NUNIT}). Since we are interested in 
computing the energy of the lowest lying state, we shall take $T$ very 
large, keeping however $T\exp(-8\pi^2/g^2)< 1$. Expanding the exponential 
of the energy up to terms linear in $T$, one finds 
\beqn 
\hspace{-.2cm}&&{\cal K}[K,K+1;T]= 
\!\int_0^{2\pi}\!\!\dr\vartheta \er^{i\vartheta}| 
\Psi^{(\vartheta)}_0[0]|^2 (1-E_0(\vartheta)T+{\mbox O}(T^2))=\nn\\ 
\hspace{-.2cm}&&=|\Psi^{\rm P.T.}_0[0]|^2\!\!\int_0^{2\pi} 
\!\!\frac{\dr\vartheta}{2\pi} 
\er^{i\vartheta}(1-E_0(\vartheta)T+{\mbox O}(T^2))=\nn\\ 
\hspace{-.2cm}&&=-T|\Psi^{\rm P.T.}_0[0]|^2\!\!\int_0^{2\pi} 
\!\!\frac{\dr\vartheta}{2\pi} 
\er^{i\vartheta} E_0(\vartheta)+{\mbox O}(T^2) 
\, ,\label{KNNPT}\\ 
\hspace{-.2cm}&&{\cal K}[K+1,K;T]= 
-T|\Psi^{\rm P.T.}_0[0]|^2\!\int_0^{2\pi}\!\!\frac{\dr\vartheta}{2\pi} 
\er^{-i\vartheta} E_0(\vartheta)+{\mbox O}(T^2) \, ,\label{KNPNT}\\ 
\hspace{-.2cm}&&{\cal K}[K,K;T]= 
|\Psi^{\rm P.T.}_0[0]|^2-T|\Psi^{\rm P.T.}_0[0]|^2\!\int_0^{2\pi} 
\!\!\frac{\dr\vartheta}{2\pi} 
E_0(\vartheta)+{\mbox O}(T^2)\, , 
\label{KNNT}\eeqn 
where the first equality in~(\ref{KNNPT}) follows from the fact 
that in the semi-classical approximation one has 
\beq 
|\Psi^{(\vartheta)}_0[0]|^2=\frac{1}{2\pi}| 
\Psi^{\rm P.T.}_0[0]|^2\, .\label{PPT} 
\eeq 
We conclude from~(\ref{KMN}), (\ref{KNNPT}), (\ref{KNPNT}) 
and~(\ref{KNNT}) that the coefficient of the terms linear in $T$ has 
only the three non-vanishing Fourier components of order $\pm 1, 
0$. Thus $E_0(\vartheta)$ is precisely of the form~(\ref{BSPE}). 
 
\subsection*{Adding a $\vartheta$-term} 
\label{sec:ATT} 
 
It is instructive to see what happens if a $\vartheta$-term is added 
to the gauge action. In this case the contribution 
\beq 
i\vartheta\frac{g^2}{32\pi^2} \int \dr^4x\, F^a_{\mu\nu} 
\tilde{F}^a_{\mu\nu}(x) 
\label{TTT}\eeq 
should be included in~(\ref{PUGA})~\footnote{Notice the presence 
of the imaginary unit in front of this term even in Euclidean metric.}. 
It is easy to prove that such an action describes a world with a well 
defined $\vartheta$-angle (obviously equal to the value appearing 
in~(\ref{TTT})). From the formula (see~(\ref{SKER})) 
\beqn 
&&{\cal K}^{(\vartheta)}[\vec A^{(2)},\vec A^{(1)};T]= 
\int_{\hat{\cal G}_0}\prod_{\vec x}{\cal D}\mu[h(\vec x)] 
\tilde{{\cal K}}^{(\vartheta)}[(\vec A^{(2)})^{U_0[h]}, 
\vec A^{(1)};T]\, ,\label{KTIL}\\ 
&&\tilde{{\cal K}}^{(\vartheta)}[\vec A^{(2)},\vec A^{(1)};T] 
=\int_{\vec A^{(1)}(\vec x)}^{\vec A^{(2)}(\vec x)}\! 
\prod_{a,\vec x}\prod_{-\frac{T}{2}<t<\frac{T}{2}}\,{\cal D} 
\vec A^a(\vec x,t)\nn\\ 
&&\exp{\big{[}-S_{YM}[\vec A, A_0=0]-i\vartheta\frac{g^2}{32\pi^2} 
\int \dr^4x\, F^a_{\mu\nu}\tilde{F}^a_{\mu\nu}(x)\big{]}}\, , 
\label{SKERT}\eeqn 
one checks, in fact, that under a homotopically non-trivial 
(time-independent) gauge transformation with winding number $K$, 
acting, say, on the boundary gauge field at $T/2$, the Schr\"odinger 
kernel is not invariant (recall the situation in the absence of a 
$\vartheta$-term,~(\ref{INVU})), rather one has 
\beqn 
{{\cal K}}^{(\vartheta)}[(\vec A^{(2)})^{U_{_K}},\vec A^{(1)};T] 
=\er^{-iK\vartheta}{\cal K}^{(\vartheta)}[\vec A^{(2)},\vec A^{(1)};T]\, . 
\label{KERT} 
\eeqn 
This result (which incidentally implies that physics is invariant if 
we replace $\vartheta$ with $\vartheta+2\pi$) follows from the fact 
that the exponent in~(\ref{SKERT}) is not invariant under such 
gauge transformation. Obviously the YM action is invariant, but the 
second term is not. The reason can be traced back to the fact that the 
vector $K_\mu$ in~(\ref{KCUR}) is not gauge invariant. Under the 
time independent gauge transformation $U_{_K}$ in~(\ref{KERT}) one 
finds, in fact (recall that we are in the temporal gauge) 
\beqn 
&&\frac{g^2}{32\pi^2}\int \dr^4x\, \big{[}F^a_{\mu\nu} 
\tilde{F}^a_{\mu\nu}\big{]}^{U_{_K}}= 
\frac{g^2}{16\pi^2}\int \dr^3x\,\big{[}K_0[(\vec A^{(2)})^{U_{_K}}]- 
K_0[\vec A^{(1)}]\big{]}=\nn\\ 
&&=\frac{g^2}{16\pi^2}\int \dr^3x\,\big{[}K_0[\vec A^{(2)}]- 
K_0[\vec A^{(1)}]\big{]}+\label{KNG}\\ 
&&+\frac{\epsilon_{ijk}}{24\pi^2}\int_{{\rm S}_3}\! 
\dr^3x\,{\rm Tr}\big{[}U_{_K}^\dagger\partial_i U_{_K}\, 
U_{_K}^\dagger\partial_j U_{_K}\,U_{_K}^\dagger\partial_k U_{_K} 
\big{]}\!=\!\frac{g^2}{32\pi^2} 
\int\! \dr^4x\, \big{[}F^a_{\mu\nu}\tilde{F}^a_{\mu\nu}\big{]}\!+\!K 
\, .\nn 
\eeqn 
{}From the spectral decomposition of ${\cal K}^{(\vartheta)}$, one 
concludes that~(\ref{NUNIT}) holds for each state appearing in it, 
thus proving the announced statement. 
 
\section{-- Decoupling} 
\label{sec:APPD} 
 
The physical content of the Applequist--Carazzone theorem~\cite{APP} 
is that in an asymptotically free theory a heavy particle (i.e.\ a 
particle with $m_f\gg\Lambda$) should ``decouple'', that is to say, it 
should not influence physics at energies $E\ll m_f$. 
 
The most important (for us) consequence of this statement is that one 
can relate the $\Lambda$ parameter of an $SU(N_c)$ gauge theory with 
$N_f$ flavours to that of the theory with $N_f-1$ dynamically active 
flavours, which is obtained after the mass of one of the flavours (say 
the $N_f$-th one) has been sent to infinity. 
 
The running of the coupling constant of the two theories is guided at 
1-loop by the evolution equations (the dependence of $b_1$ on $N_c$ is 
understood) 
\beqn 
&&\frac{g^2_{N_f}(\mu)}{8\pi^2}=\frac{1}{b_{1,N_f} 
\log\mu/\Lambda^{(N_f)}}\, , 
\label{EVOL1}\\ 
&&\frac{g^2_{N_f-1}(\mu)}{8\pi^2}=\frac{1}{b_{1,N_f-1} 
\log\mu/\Lambda^{(N_f-1)}} 
\label{EVOL2}\, .\eeqn 
A necessary implication of decoupling is that for 
$m_f\gg\Lambda^{(N_f-1)},\Lambda^{(N_f)}$ the running of $g^2$ in the 
theory with $N_f$ flavour must change from the behaviour 
in~(\ref{EVOL1}) - when $\mu$ is sufficiently larger than $m_f$ - to 
that in~(\ref{EVOL2}) - when $\mu$ is well below it. The equality 
of the coupling constants at $\mu\sim m_f$ (required by smoothness) 
leads to the sought relation 
\beq 
\Big{(}\frac{m_f}{\Lambda^{(N_f)}}\Big{)}^{b_{1,N_f}}= 
\Big{(}\frac{m_f}{\Lambda^{(N_f-1)}}\Big{)}^{b_{1,N_f-1}}\, . 
\label{EQLAM} 
\eeq 
Notice that, since we are assuming that $m_f$ is larger than both 
$\Lambda^{(N_f-1)}$ and $\Lambda^{(N_f)}$, from~(\ref{EQLAM}) it 
follows $\Lambda^{(N_f-1)}>\Lambda^{(N_f)}$. This relation is 
phenomenologically quite important. It is telling us that, when the 
energy scale, $E$, of a process goes through the production threshold 
of a particle of mass $m_f$, since the running of the coupling 
constant switches from that of~(\ref{EVOL1}) to that of~(\ref{EVOL2}), 
it just happens that the value taken by the effective coupling constant, 
$g^2_{\rm eff}(E)$, that controls the process is always the largest 
between $g^2_{N_f-1}(E)$ and $g^2_{N_f}(E)$ for all values of $E$. 
 
\section{-- Flat directions of massless SQCD} 
\label{sec:APPE} 
 
In this Appendix we want to elucidate the structure of the vacuum 
manifold of massless SQCD. The theory possesses the (non-anomalous) 
symmetry group (see Table~\ref{tab:cpu})
\beq 
G=SU_L(N_f)\times SU_R(N_f)\times U_V(1)\times U_{\hat A}(1)\, . 
\label{SYMG} 
\eeq 
Any field configuration of the type 
\beqn 
&&A_\mu=\lambda=\psi=\bar\psi=\tilde\psi=\bar{\tilde{\psi}}=0\, , 
\label{VACCON1}\\ 
&&D^a=\phi_{f}^{r\,\dagger} (T^a)_r^{r'}\phi_{r'}^{f}- 
\tilde\phi_{f}^{r} (T^a)_r^{r'}\tilde\phi_{r'}^{f\,\dagger}=0 
\label{VACCON2} 
\eeqn 
has vanishing energy, thus it is to be interpreted as a classical 
vacuum state. Non-renormalisation theorems ensure that this 
configuration is stable against perturbative corrections (but, as we 
have seen, not against non-perturbative instantonic corrections). 
 
For the applications it is important to determine the solutions 
of~(\ref{VACCON2}). In order to simplify the discussion, it is 
convenient to separately examine the case $N_f<N_c$ and $N_f\geq N_c$. 
 
$\bullet$ For $N_f<N_c$ it can be easily seen that (up to symmetry 
operations) the most general solution of~(\ref{VACCON2}) is given by 
\beq 
\langle \phi_{r}^{f}\rangle =\langle \tilde\phi_{r}^{f\,\dagger}\rangle = 
\left\{\begin{array}{cc} v_r\delta^{rf} &\quad 1\leq r\leq N_f\\ 
0& \quad{\rm otherwise}\end{array}\right. 
\label{SCNF1} 
\eeq 
If the $v$'s are all non-vanishing the gauge symmetry is broken from 
the original $SU(N_c)$ group down to $SU(N_c-N_f)$. In the special 
case $N_f=N_c-1$ the gauge symmetry is completely broken. Among the 
quark superfields, $(2N_c-N_f)N_f$ of them become heavy owing to the 
super-Higgs mechanism, while the remaining $N_f^2$ will contain the 
Goldstone bosons of the various broken global symmetries, as well as 
their superpartners. The pattern of surviving symmetries will depend 
upon the detailed values assumed by the $v_r$'s in~(\ref{SCNF1}). 
$\bullet$ For $N_f\geq N_c$ the analysis of~(\ref{VACCON2}) is a 
bit more involved. The result is that (up to symmetry operations) the 
most general pattern of scalar v.e.v.'s that makes $D^a$ vanish is
\beqn 
&&\langle\phi_{r}^{f}\rangle\,=\,\left\{\begin{array}{cc} 
v_r\delta^{rf} &\quad 1\leq f\leq N_c\\ 
0& \quad{\rm otherwise}\end{array}\right.\label{SCNF2}\\ 
&&\langle\tilde\phi_{r}^{f\,\dagger} \rangle = 
\left\{\begin{array}{cc}(|v_r|^2-b^2)^\half\delta^{rf} & 
\quad 1\leq f\leq N_c\\ 
0& \quad{\rm otherwise}\end{array}\right.\label{SCNF3} 
\eeqn 
where $b$ is an arbitrary constant. For non-zero $v_r$ the gauge 
symmetry is completely broken and $N_c^2-1$ quarks become massive by 
the super-Higgs mechanism. Again the detailed pattern of surviving 
symmetries depend on the particular values taken by the scalar 
v.e.v.'s~(\ref{SCNF2}) and~(\ref{SCNF3}). 
 
We wish to conclude with a comment. As we have seen, the vacuum 
manifold is not compact. This is due to the fact that the symmetry of 
the set of supersymmetric vacua is a certain complexification of the 
symmetry group of the Lagrangian~\cite{OWLK}. In fact, any rescaling 
of the massless scalar fields, although not a symmetry of the theory, 
when applied to a vacuum configuration leads to another acceptable, 
physically inequivalent, vacuum.

\section{-- $\cN = 2$ Lagrangian and supersymmetry transformations} 
\label{sec:N2susygen} 
 
Rigid $\cN = 2$ supersymmetric theories consist of two kinds of 
massless multiplets. Vector multiplets and hypermultiplets. 
 
Vector multiplets contain a vector $A_\mu$, two spinor gaugini 
$\lambda^r_\alpha$ and a complex scalar $\phi$ all transforming in the 
adjoint representation of the gauge group. Vector multiplets are 
described by chiral superfields usually denoted by $\cA$ whose 
$\theta$ expansion schematically reads 
\be 
\cA(x,\theta) = a(x) + \theta^r_\alpha \lambda^\alpha_r(x) 
+ {1\over 2} \theta^r_\alpha \sigma^{\mu\nu \alpha}{}_\beta 
\theta^\beta_r F_{\mu\nu}(x) + \cdots \, . 
\ee 
Higher order terms in $\theta^r$ with $r=1,2$ can be expressed as 
derivatives of the lower ones. 
 
The Lagrangian of pure $\cN = 2$ SYM theory is given by 
\be 
L = \int \dr^4\theta \,\cF(\cA) \, , 
\ee 
where $\cF(\cA)$ is a group invariant function of the chiral 
superfields. Renormalisability restricts $\cF(\cA)$ to be quadratic 
\be 
\cF(\cA) = {1\over 2} \tau_0 \cA^2 \, , 
\ee 
so that 
\be 
\tau_0 \delta_{ab} = {\partial^2 \cF(\cA)\over \partial \cA_a 
\partial \cA_b} 
\ee 
and 
\bea 
L_{\cN = 2} &=& \im\tau_0 \,\Tr \left({1\over 4} 
F_{\mu\nu} F^{\mu\nu} + i \lambda^r \sigma^\mu D_\mu \bar 
\lambda_r + D_\mu \phi D^\mu\phi^\dagger + \right.\nn \\ 
&+&\left. \rule{0pt}{13pt}\![\phi, \phi^\dagger]^2 + 
\lambda^r [\phi^\dagger,\lambda_r] + \bar\lambda^r 
[\phi,\bar\lambda_r]\right) \ . 
\eea 
The Lagrangian $L_{\cN = 2}$ is invariant under Poincar\'e 
transformations (up to total derivatives), under $U(2)_R$ R-symmetry 
transformations and under the global ${\cN = 2}$ supersymmetry 
transformations 
\bea 
&& \delta A_\mu = \eta^r \sigma_\mu 
\bar\lambda_r + \bar\eta^r \bar\sigma_\mu \lambda_r \nn \\ 
&& \delta \lambda^r = ({1\over 2} F_{\mu\nu} \sigma^{\mu\nu} + 
[\phi, \phi^\dagger] ) \eta^r + i \sigma^\mu D_\mu \phi 
\bar\eta^r \nn \\ 
&& \delta\phi = \eta^r \lambda_r \, , 
\eea 
where $A_\mu$ is the gauge potential associated with the field 
strength $F_{\mu\nu}$. R-symmetry indices are raised and lowered with 
the symplectic matrix $\varepsilon^{rs}$.

\section{-- BPS configurations} 
\label{sec:BPSconf} 
 
The acronym BPS, for Bogomol'nyi--Prasad--Sommefield, was initially 
introduced to designate certain solitonic solutions in 
non-supersymmetric quantum field theories. The simplest configuration 
of this type is a symmetric monopole arising in the Georgi--Glashow 
model~\cite{GeoGla} describing a $SU(2)$ gauge field coupled to a 
scalar field in the adjoint representation. The Lagrangian of the 
model is 
\be 
\calL = -\fr{4}\,F^a_{\mu\nu}F^{a\mu\nu} + 
\half\,D_\mu\Phi^a D^\mu\Phi^a - \frac{\lambda}{4}(\Phi^a\Phi^a 
- v^2)^2 \, , \qquad a=1,2,3 \, ,\label{gglagr} 
\ee 
with gauge coupling constant $e$. As shown by 't 
Hooft~\cite{thooftmonop} and Polyakov~\cite{polyamonop}, in the 
Coulomb phase, \ie in the presence of a v.e.v.\ for the adjoint scalar, 
the theory admits monopole solutions characterised by an 
integer-valued topological charge. Static finite energy configurations 
have vanishing scalar potential at spatial infinity. The condition for 
the vanishing of the potential defines a two-sphere, 
$\sum_a\Phi^a\Phi^a = v^2$. Therefore for such configurations the 
scalar field provides a map from the two-sphere at spatial infinity 
into the two-sphere of the Higgs vacuum. This map defines the second 
homotopy group of $S^2$, $\Pi_2(S^2)\equiv\Z$. As a result, the 
magnetic charge, $g$, associated with a solution of the field 
equations satisfies a Dirac quantisation condition. Denoting by 
$\vec{B}$ the non-abelian magnetic field with components $B^{ai} = 
-\half\veps^{ijk}F^a_{jk}$, one finds 
\be 
g = \int_{S^2_\infty} \vec{B}\cdot\dr\vec{\S} = \fr{2ev^3}\int \dr\S^i\, 
\veps^{ijk}\veps^{abc} \Phi^a\del^j \Phi^b\del^k\Phi^c = 
\frac{4\pi n}{e} \, , \label{diracquant} 
\ee 
where the integer $n$ is the winding number determined by the 
behaviour of the Higgs field at spatial infinity. 
 
No exact solution to the complete non-linear field equations is know 
explicitly, even in the simplest case of gauge group $SU(2)$. However, 
the analysis can be drastically simplified taking advantage of the 
implications of a general bound on the mass of field configurations 
with non-vanishing winding number known as the {\em Bogomol'nyi 
bound}. For a static field configuration with vanishing electric 
field, $E^{ai} = -F^{a0i} = 0$, the energy (mass) satisfies 
\ba 
M &\!=\!& \int \dr^3r \, \half\left[\vec{B^a}\cdot\vec{B^a} + 
\vec{D}\Phi^a\cdot\vec{D}\Phi^a + V(\Phi) \right] \ge \nn \\ 
&\!\ge\!& \half \int \dr^3r \left(\vec{B^a} - \vec{D}\Phi^a\right)^2 
+ vg \, , 
\label{massbound} 
\ea 
implying the bound 
\be 
M\ge v g \, . 
\label{bogbound} 
\ee 
Minimal energy configurations saturate the bound and thus should have 
vanishing potential and should satisfy the first order Bogomol'nyi 
equation 
\be 
\vec{B^a} = \vec{D}\Phi^a \, . 
\label{bogeq} 
\ee 
The first explicit example of solution to~(\ref{bogeq}) with $M=vg$ is 
the spherically symmetric one constructed by Prasad and 
Sommerfield~\cite{PraSomm}. Following their analysis the expression 
{\em BPS saturated} has been used to designate solutions to the field 
equations saturating the Bogomol'nyi bound. For dyons with electric 
and magnetic charges $e$ and $g$ respectively the bound generalises to 
\be 
M \ge v(e^2+g^2)^{1/2} \, . 
\label{dyonbound} 
\ee 
A comprehensive review of the physics of solitons and monopoles in 
gauge theory can be found in~\cite{tong}. 
 
\section{-- Extended superalgebras, central charges and 
multiplet shortening} 
\label{sec:susyZ} 
 
In the context of supersymmetric theories certain short multiplets 
which correspond to special representations of the supersymmetry 
algebra (see also Appendix~\ref{sec:APPA}) are referred to as {\em 
BPS multiplets}. States in such multiplets saturate a generalisation 
of the Bogomol'nyi bound~(\ref{bogbound}), which relates their mass, 
$M$, to their ``central charge''. 
 
The $\cN =1$ supersymmetry algebra in $D=4$, 
\be 
\{ Q_\alpha, \bar{Q}_{\dot\alpha}\} = i \sigma^\mu_{\alpha\dot\alpha} 
P_\mu \, ,\qquad \{ Q_\alpha, {Q}_{\beta}\} = 0 \, , 
\label{N1salg} 
\ee 
admits no central extension. Generalised (non scalar) central charges 
associated with the existence of domain wall configurations may 
appear, but they carry Lorentz indices. 
 
Extended supersymmetry algebras, on the other hand, admit non trivial 
{\it bona fide} central charges, usually denoted by $Z$. The $\cN =1$ 
superalgebra~(\ref{N1salg}) can be generalised to 
\be 
\{Q^A_\alpha, \bar{Q}_{\dot\alpha B}\} = i \delta^A{}_B 
\sigma^\mu_{\alpha\dot\alpha} P_\mu \, , \qquad \{ Q^A_\alpha, 
{Q}^B_{\beta}\} = Z^{AB} \epsilon_{\alpha\beta} \, , 
\label{Nextsa} 
\ee 
where $A, B = 1,...,\cN$ are supersymmetry indices and the central 
charges, $Z^{AB}$, satisfy $Z^{AB} =- Z^{BA}$. In particular for $\cN 
= 2$ there is only one complex central charge, $Z \equiv Z^{12}$, 
while for $\cN = 4$ there are six complex central charges satisfying a 
(self) duality condition $Z^{AB} = \half \epsilon^{ABCD} 
\bar{Z}_{CD}$, very much as the elementary scalar fields in the 
theory. By means of a R-symmetry transformation, the central charges 
can be skew diagonalised and shown to satisfy 
\be 
M \ge |Z_1| \ge |Z_2| \ge \cdots \ge 
|Z_{i}|\ge \cdots \ge 0 \, , \label{susybb} 
\ee 
where $M^2 = P_\mu P^\mu$ is one of the Casimirs of the representation 
one is considering, for a proper ordering of the skew eigenvalues 
$Z_i$, $i=1,...,[\cN/2]$. For $\cN$ odd, one eigenvalue is necessarily 
zero by Binet's theorem. The relation~(\ref{susybb}) represents a 
generalisation of the Bogomol'nyi bound. 
 
Irreducible ``massive'' representations of the supersymmetry 
algebra are indeed constructed by going to the rest frame. This 
reduces the form of the algebra to that of a Clifford algebra and 
one can split the $4\cN$ supercharges into $2\calN$ creation 
operators and $2\calN$ annihilation operators by considering 
suitable linear combinations~\cite{RSYM}. This means that, in 
general, a massive multiplet consists of $2^{2\cN}$ states, 
$2^{2\cN - 1}$ bosonic and as many fermionic. However, if some of 
the central charges coincide with $M$ the multiplet shortens, 
since some of the creation operators annihilate the ground state. 
 
In the case of $\cN = 2$, this happens when $M = |Z|$. The 
corresponding multiplet is said to be 1/2 BPS since half the 
creation operators (4 out of 8) act trivially. As a result the 
supermultiplet contains only half as many states, \ie 8 (4 bosons 
and 4 fermions) instead of $16= 2^4$. 
 
In the case of the $\cN = 4$ superalgebra, one has two sub-cases $M = 
|Z_1| = |Z_2|$ (1/2 BPS) and $M = |Z_1| \neq |Z_2|$ (1/4 BPS). The 
corresponding multiplets are respectively 1/2 and 3/4 the length of 
ordinary $\calN=4$ multiplets. 
 
As discussed in Sect.~\ref{sec:N4sym}, in the conformal phase the 
$\calN=4$ SYM theory has a larger group of symmetries, the $\calN=4$ 
superconformal group, $PSU(2,2|4)$, see Appendix~\ref{sec:psu224}. 
 
In this situation the fundamental degrees of freedom of the theory are 
gauge-invariant composite operators which are organised into 
multiplets forming unitary irreducible representations (UIR's) of 
$PSU(2,2|4)$. Each composite operator in a multiplet can be labelled 
by the quantum numbers associated with the maximal bosonic sub-group 
of $PSU(2,2|4)$, $SO(2,4)\times SO(6)$, \ie two spins, $j_1$ and 
$j_2$, the scaling dimension, $\D$, and three $SO(6)$ Dynkin labels, 
$[k,l,m]$. 
 
A further generalisation of the concept of BPS multiplet arises in
this context. The UIR's of $PSU(2,2|4)$ have been classified
in~\cite{dobpet}.  For a review and applications to the AdS/CFT
correspondence see~\cite{MBrev,N4multi}. Ordinary long
representations contain a number of states proportional to $2^{16}$,
with the proportionality constant related to the dimension of the
representation of the bosonic sub-group under which the lowest
component transforms. Shorter representations arise when specific
relations occur among the $SO(2,4)\times SO(6)$ quantum numbers of the
lowest component of the multiplet. The correlation functions
considered in Sect.~\ref{sec:N4-1inst} involve operators belonging  to
multiplets classified as 1/2 BPS. These are characterised by the  fact
that their lowest component is a Lorentz scalar operator of  dimension
$\D=\ell$, with $\ell\ge2$, transforming in the $[0,\ell,0]$
representation of the $SO(6)$ R-symmetry group. Generic multiplets of
this type have $\fr{12}\ell^2(\ell^2-1)\,2^8$ components. The cases
$\ell=2,3$ are special in that they are characterised by a further
accidental shortening and are sometimes referred to as
ultra-short. Many other shortening conditions can be identified. For
instance 1/4 BPS multiplets arise when the lowest component is a
Lorentz scalar, double trace operator with $\D=2k+l$ transforming in
the representation $[k,l,k]$ of $SO(6)$. In the case of 1/2 and 1/4
BPS multiplets the range of spin is respectively 4 and 6 units,
whereas long multiplets have a spin range of 8 units.

\section{-- The $\calN=4$ superconformal group} 
\label{sec:psu224} 
 
In this Appendix we summarise some basic properties of the
four-dimensional $\calN=4$ superconformal group, $PSU(2,2|4)$. More
details and references can be found in the
reviews~\cite{adscftrevs}. The maximal bosonic subgroup of
$PSU(2,2|4)$ is the direct product of the four-dimensional conformal
group, $SO(2,4)\sim SU(2,2)$, and of the R-symmetry group of the
$\calN=4$ superalgebra, $SO(6)\sim SU(4)$. The conformal group is the
group of transformations which preserve the form of the metric up to a
(position dependent) scale factor. In four-dimensional Minkowski
space, with metric $\eta_{\mu\nu}={\rm diag}(-,+,+,+)$, it is
generated by translations, Lorentz transformations, dilations and
special conformal transformations. We denote the corresponding
generators respectively by $P_\mu$, $L_{\mu\nu}$, $D$ and $K_\mu$,
$\mu,\nu=0,1,2,3$. For the generators of the $SU(4)$ R-symmetry we use
$T^a$, $a=1,2,\ldots,15$. 
 
The action of infinitesimal conformal transformations on the 
coordinates, $x_\mu\to x^\pp_\mu(x)=x_\mu+\d x_\mu(x)$, is the 
following 
\ba 
&& \d x_\mu(x) = a_\mu \quad ({\rm translations}) \nn \\ 
&& \d x_\mu(x) = \Lambda_\mu{}^\nu x_\nu \quad ({\rm Lorentz 
~ transformations}) \nn \\ 
&& \d x_\mu(x) = \lambda x_\mu \quad ({\rm dilations}) 
\label{conftrans} \\ 
&& \d x_\mu(x) = 2b_\nu x^\nu x_\mu - x_\nu x^\nu b_\mu \quad 
({\rm special ~ conformal ~ transformations})\, , \nn 
\ea 
where $a_\mu$ and $b_\mu$ are constant vectors, $\lambda\in\R^+$ and 
the constant matrix $\Lambda_\mu{}^\nu$ satisfies $\eta^{\rho\sigma} 
\Lambda_\rho{}^\mu\Lambda_\sigma{}^\nu = \eta^{\mu\nu}$. 
 
The fermionic symmetries in $PSU(2,2|4)$ comprise sixteen Poincar\'e 
supersymmetries, generated by the supercharges $Q^A_\a$ and $\bar 
Q_A^\adot$, with $A=1,2,3,4$ and $\a,\adot=1,2$, and sixteen special 
(or conformal) supersymmetries, generated by the supercharges $S_A^\a$ 
and $\bar S^A_\adot$. 
 
As explained in Appendix~\ref{sec:BPSconf} the superconformal algebra 
also admits six complex scalar central charges as well as additional 
generalised central charges which carry Lorentz indices. 
 
Neglecting the central extensions the superconformal algebra reads 
\ba 
&& [L_{\mu\nu},P_\rho] = -i(\eta_{\mu\rho}P_\nu- 
\eta_{\nu\rho}P_\mu) \, , \quad [L_{\mu\nu},K_\rho] = 
-i(\eta_{\mu\rho}K_\nu-\eta_{\nu\rho}K_\mu) \, , \nn \\ 
&& [L_{\mu\nu},L_{\rho\sigma}] = -i\eta_{\mu\rho}L_{\nu\sigma} 
+{\rm \, permutations} \, , \quad [P_\mu,K_\nu] = 2iL_{\mu\nu} 
-2i\eta_{\mu\nu}D \, , \nn \\ 
&& [D,L_{\mu\nu}] = 0 \, , \quad [D,P_\mu] = -iP_\mu \, , \quad 
[D,K_\mu] = iK_\mu \, , \nn \\ 
&& \{Q^A_\a,Q^B_\b\} = \{S^\a_A,S^\b_B\} = 
\{Q^A_\a,\bar S^B_\adot\} = \{\bar Q^\adot_A,S^\b_B \} = 0 
\, , \nn \\ 
&& \{Q^A_\a,\bar Q_{\adot B}\} = 2\sigma^\mu_{\a\adot} 
P_\mu\d^A{}_B \, , \quad \{S_{\a A},\bar S_\adot^B\} 
= 2\sigma^\mu_{\a\adot}P_\mu\d_A{}^B \, , \nn \\ 
&& \{Q^A_\a,S_{\b B}\} = \veps_{\a\b}(\d^A{}_B D + T^A{}_B) 
+\half \d^A{}_B L_{\mu\nu}\veps_{\b\g}\s_{\,\a}^{\mu\nu\,\g} \, . 
\label{psu224alg} 
\ea 
The R-symmetry group of automorphisms of the generic $\calN$-extended 
supersymmetry algebra is $U(\calN)$. The $\calN=4$ case under 
consideration is special in that the $U(1)$ factor in the 
decomposition $U(\calN)=SU(\calN)\times U(1)$ of the R-symmetry 
becomes an outer automorphism: none of the other generators in the 
algebra is charged under this $U(1)$ symmetry~\cite{dobpet,intri} and 
all the fields and composite operators in $\calN=4$ SYM are neutral 
under this central $U(1)$. The absence of the abelian factor in the 
R-symmetry is reflected in the notation $PSU(2,2|4)$ as opposed to 
$SU(2,2|4)$. 
 
As has been discussed in Sect.~\ref{sec:N4sym}, the observables in 
the $\calN=4$ SYM theory are correlation functions of local 
gauge-invariant composite operators. Such operators are labelled by 
the quantum numbers characterising their transformation under the 
bosonic subgroup $SO(2,4)\times SO(6)$. A class of operators playing a 
special r\^ole in a conformal field theory such as $\calN=4$ SYM are 
the {\em conformal primary operators}. These are defined by the 
condition of being annihilated by special conformal transformations 
acting at the origin, 
\be 
[K_\mu, \calO(x)]\big|_{x=0} = 0 \, . 
\label{primary} 
\ee 
The existence of such operators is associated with the presence of a 
lower bound on the dimension of fields and operators in a unitary 
conformal field theory. Since the action of $K_\mu$ lowers the 
dimension of an operator, the existence of the unitarity bound implies 
that in every representation of the conformal group there must be an 
operator satisfying~(\ref{primary}). The action of the generators of 
the conformal group on primary operators is as follows, 
\ba 
&& [P_\mu, \calO(x)] = i\del_\mu \calO(x) \nn \\ 
&& [L_{\mu\nu},\calO(x)] = [i(x_\mu\del_\nu-x_\nu\del_\mu) 
+M_{\mu\nu}]\calO(x) \nn \\ 
&& [D,\calO(x)] = -i(\D-x^\mu\del_\mu)\calO(x) \label{confonprim} \\ 
&& [K_\mu,\calO(x)] = [i(x^2\del_\mu-2x_\mu x^\nu\del_\nu+ 
2x_\mu\D)-2x^\nu M_{\mu\nu}]\calO(x) \, . \nn 
\ea 
The functional form of two- and three-point functions of primary 
operators is fixed by conformal invariance. In the case of Lorentz 
scalars, for instance, the two-point function vanishes unless the two 
operators have the same scaling dimension, in which case it takes the 
form 
\be 
\la \calO_i(x_1)\,\calO_j(x_2) \ra = 
\frac{c_{ij}}{|x_{12}|^{2\D}} \, , 
\label{2ptprim} 
\ee 
where $x_{12} = x_1-x_2$, $c_{ij}$ are constants and $\D$ is the 
common dimension of $\calO_i$ and $\calO_j$. For three-point functions 
conformal invariance implies 
\be 
\hsp{-0.2}\la \calO_i(x_1)\calO_j(x_2)\calO_k(x_3) \ra \!=\! 
\frac{c_{ijk}}{|x_{12}|^{\D_i+\D_j-\D_k}|x_{13}|^{\D_i+\D_k-\D_j} 
|x_{23}|^{\D_j+\D_k-\D_i}} \, , 
\label{3ptprim} 
\ee 
where $c_{ijk}$ are numerical coefficients. The form of four- and
higher-point functions is not completely determined by the conformal
symmetry. Four-point functions, for instance, are determined up a
function of two conformally invariant cross-ratios constructed from
the four insertion points, \eg $r=x_{12}^2x_{34}^2/x_{13}^2x_{24}^2$
and $s=x_{14}^2x_{23}^2/x_{13}^2x_{24}^2$. The scaling dimensions and
the coefficients, $c_{ij}$ and $c_{ijk}$, in~(\ref{2ptprim})
and~(\ref{3ptprim})  are in general functions of the Yang--Mills
coupling  constant, $g$, and the $\vartheta$-angle. 
 
Local composite operators in $\calN=4$ SYM are organised in multiplets 
of the superconformal group. The bottom component of any such 
multiplet, \ie the operator of lowest dimension, is referred to as a 
{\em superconformal primary operator}. Superconformal primary 
operators are annihilated by the special supersymmetry generators 
acting at the origin, 
\be 
\{S^A_\a,\calO(x)]\big|_{x=0} = 0 \, , 
\label{suconfprimary} 
\ee 
where the symbol $\{S,\calO]$ indicates a commutator if $\calO$ is 
bosonic and an anti-commutator if $\calO$ is fermionic. Notice that 
superconformal primary operators are always also conformal primaries, 
but the opposite is not true. 
 
As discussed in Appendix~\ref{sec:BPSconf} there are special UIR's of 
the superconformal group corresponding to short BPS multiplets. 
Operators in such multiplets are protected and their two- and 
three-point functions do not receive quantum corrections. This implies 
that their scaling dimensions and three-point couplings are not 
renormalised.

\section{-- Compendium of differential geometry} 
\label{sec:diffgeo} 
 
An $n$ dimensional topological manifold is a set of points such that
the neighborhood of a point $P$ (any open set containing the point
$P$) looks like $\R^n$. In order to describe a manifold one needs an
atlas made of many patches that are related to one another by
transition functions. If the transition functions are continuous, then
the manifold is continuous. If the transition functions are
differentiable, then the manifold is differentiable. If the transition
functions are complex analytic, then the manifold is complex. 
 
One can add further structures. On a differentiable manifold one can 
define a metric which is a symmetric bilinear form on the vector 
fields such that $g(U, V) = g(V, U) = g_{ij} U^i V^j$ in a local 
coordinate patch. Parallel transport is achieved by means of a 
connection $\Gamma^i_{jk}$ that can be fixed to be the Christoffel 
connection imposing that the metric be covariantly constant, \ie $0= 
D_i g_{jk} \equiv \partial_i g_{jk} - \Gamma^l_{ik} g_{jl} - 
\Gamma^l_{ji} g_{lk}$. One can then construct the Riemann curvature 
tensor $R_{ij}^k{}_l$ and its contractions, the Ricci curvature tensor 
$R_{il}$ and the scalar curvature $R$. After parallel transport a 
vector gets transformed by means of a $SO(n)$ 
rotation. Transformations along closed paths form the holonomy group 
of the manifold, which is necessarily a subgroup of $SO(n)$.

On differentiable manifolds one can define $p$-forms with $p<n$. A 
0-form is a function, a 1-form is combination of the differentials 
of the local coordinates $A = A_i(x) \dr x^i$. In a local 
coordinate patch 
\be 
A_p = {1\over p!} \sum_{i_1,...,i_p} 
A_{i_1,...,i_p} \dr x^{i_1}\wedge ... \wedge \dr x^{i_p} \, . 
\ee 
where the (anti-)symmetric wedge product satisfies $A_p \wedge B_q = 
(-)^{p q}B_q \wedge A_p $. On forms one can define an exterior 
differential $\dr A_p = B_{p+1}$, that satisfies the (graded) Leibniz 
rule. On a Riemannian manifold, one can also define a Hodge star 
operator $*A_p = \tilde{A}_{n-p}$. Combining $\dr$ and $*$ one can 
define a differential operator $\delta$ such that $ \delta A_p \equiv 
* \,\dr\! * A_p = C_{p-1}$, that generalises the divergence. The Lie 
derivative of a $p$-form along a vector field $V$ is defined by 
\be 
\cL_V A_p = \iota_V \dr A_p + \dr (\iota_V A) \, , 
\ee 
where $\iota_V$ denotes contraction with $V$. In a local coordinate 
patch one has 
\be 
\cL_V A_{i_1,...,i_p} = V^i \de_i 
A_{i_1,...,i_p} - \sum_{k=1}^p A_{i_1,...,i,...,i_p}\de_{i_k} V^i 
\, . 
\ee 
Both $\dr$ and $\delta$ are nilpotent, \ie $\dr^2=0$ and 
$\delta^2=0$. The generalised Laplacian is given by $\Delta A_p = (\dr 
\delta + \delta \dr) A_p$. It coincides with the standard Laplacian 
$\Delta = {||g||}^{-1/2} \partial_i ({||g||}^{1/2} 
g^{ij}\partial_j)$ on 0-forms (scalars). A form is closed if $\dr A=0$ 
and exact if $A=\dr C$. A form is co-closed if $\delta A=0$ and 
co-exact if $A=\delta C$. A form which is closed and co-closed is 
harmonic, \ie $\Delta A = 0$. The equivalence classes of closed forms 
$\cC^p$ that differ by exact forms $\cE^p$ define the cohomology 
groups $\cH^p = \cC^p/\cE^p$. De Rham has shown that one can always 
find a harmonic representative in each cohomology 
class~\footnote{Given a closed $p$-form such that $\dr A = 0$ but 
$\delta A\neq 0$, one can always find a cohomologous form $A' = A + 
\dr\Lambda$ such that $\dr A' = 0$, by construction, and $\delta A'=0$ 
by requiring that $\delta \dr \Lambda = - \delta A$, \ie inverting the 
elliptic operator $\Lambda = - (\delta \dr)^{-1} \delta A$.}. 
 
A symplectic manifold is an even dimensional manifold that admits a 
closed 2-form, known as symplectic form, \eg for the phase space of a 
point in $\R^n$ one has $\Omega = \sum_i \dr p_i \wedge \dr x^i$. 
 
On complex manifolds one can decompose $\dr$ as $\dr = \partial + 
\bar\partial$, where both $\partial$ and $\bar\partial$ are 
nilpotent. By a complex coordinate change one can always put the 
metric in Hermitean form $\dr s^2 = g_{i\bar{j}} \dr z^i 
\dr\bar{z}^{\bar{j}}$ at least locally. 
 
The K\"ahler form on a complex Riemannian manifold is $\omega = 
g_{i\bar{j}} \dr z^i \wedge \dr\bar{z}^{\bar{j}}$. If $\omega$ is 
closed, $\dr\omega = 0$, which implies $\partial \omega = 0 = 
\bar\partial\omega$, the manifold is K\"ahler. Locally $\omega = 
\partial \bar\partial K$, where $K(z, \bar{z})$ is the K\"ahler 
potential. If the manifold has real dimension $4n$ and admits three 
closed K\"ahler forms, $\dr\omega^I = 0$, $I=1,2,3$, whose components 
satisfy the algebra of quaternions the manifold is said to be 
hyper-K\"ahler. If the three K\"ahler forms are not closed but rather 
$\dr\omega^I = c_n \epsilon^{IJK} \omega_J \wedge \omega_K$, the 
manifold is said to be quaternionic. 
 
An isometry of the metric is a coordinate transformation that leaves 
the metric invariant and is thus generated by a vector field $V$ that 
satisfies 
\be 
0 = \cL_V g_{ij} = V^k 
\partial_k g_{ij} - g_{ik} 
\partial_j V^k - g_{ik} \partial_j V^k \equiv - \nabla_i V_j - 
\nabla_j V_i \, , 
\ee 
where indices are raised and lowered with the metric. 
 
A holomorphic isometry is such that $\cL_V \omega = 0$. Thanks to the 
closure of $\omega$, $V$ admits a prepotential because $\dr(\iota_V 
\omega) = 0$ implies $\iota_V \omega = \dr\mu_V$ locally. The 
prepotential is known also as the holomorphic K\"ahler map. A 
tri-holomorphic isometry is such that $\cL_V \omega^I = 0$. Thanks to 
the closure of $\omega^I$, $V$ admits three prepotentials because 
$\dr(\iota_V \omega^I) = 0$ implies $\iota_V \omega^I = \dr\mu^I_V$ 
locally. The prepotentials are known also as the tri-holomorphic 
K\"ahler maps.



\begin{thebibliography}{99.} 
 
 
\bibitem{BEL} 
A.A.~Belavin, A.M.~Polyakov, A.M.~Schwartz and Yu.S.~Tyupkin, 
Phys. Lett. B \textbf{59}, 85 (1975). 
 
\bibitem{GTH} 
G.~'t~Hooft, Phys. Rev. Lett. \textbf{37}, 8 (1976); 
Phys. Rev. D \textbf{14}, 3432 (1976); {\it ibidem} \textbf{18}, 
2199 (1978) and Phys. Rept. \textbf{142}, 357 (1986). 
 
\bibitem{COL} 
S.R.~Coleman, ``{\it The uses of Instantons}'', 
Lecture delivered at 1977 Int. School of Subnuclear Physics, 
Erice, Italy, Jul 23-Aug 10, 1977, Plenum Press (New York, 1978). 
 
\bibitem{AKMRV} 
D.~Amati, K.~Konishi, Y.~Meurice, G.C.~Rossi and G.~Veneziano, 
Phys. Rept. \textbf{162}, 169 (1988). 
 
\bibitem{REV} 
A.I.~Vainshtein, V.I.~Zakharov, V.A.~Novikov, M.A.~Shifman, 
Sov. Phys. Usp. \textbf{25} 195 (1982) [Usp. Fiz. Nauk \textbf{136} 
553 (1982)], revised and updated version published in 
``{\it ITEP Lectures on Particle Physics and Field theory}'', 
World Scientific (Singapore, 1999), Volume \textbf{1} pp. 201-299;\\ 
M.A.~Shifman, Lectures given at the International School 
of Physics ``Enrico Fermi'', Varenna, Italy, July 3-6 1995, 
\hepth{9704114};\\ 
M.A.~Shifman and A.I.~Vainhstein, \hepth{9902016}. 
 
\bibitem{KHOREV} 
N.~Dorey, T.J.~Hollowood, V.V.~Khoze and M.P.~Mattis, Phys. Rept. 
\textbf{371}, 231 (2002) [\hepth{0206063}] and references therein. 
 
\bibitem{BOOK} 
M.~Nakahara, ``{\it Geometry, Topology and Physics}'', Graduate 
Student Series in Physics, Gen. Ed. D.F.~Brewer, Institute of Physics 
Publishing (Bristol and Philadelphia, 1990). 
 
\bibitem{GS} 
J.L.~Gervais and B.~Sakita, Phys. Rev. B \textbf{11}, 2943 (1975);\\ 
E.~Tomboulis, Phys. Rev. B \textbf{12}, 1678 (1975). 
 
\bibitem{BER} 
C.~Bernard, Phys. Rev. D \textbf{19}, 3013 (1979). 
 
\bibitem{YAF} 
L.G.~Yaffe, Nucl. Phys. B \textbf{151}, 247 (1979). 
 
\bibitem{TUN1} 
R.~Jackiw and C.~Rebbi, Phys. Rev. Lett. \textbf{37}, 172 (1976) and 
Phys. Rev. D \textbf{14}, 517 (1976). 
 
\bibitem{TUN2} 
C.~Callan, R.~Dashen and D.~Gross, Phys. Lett. B \textbf{63}, 334 
(1976) and Phys. Lett. B\textbf{66}, 375 (1977). 
 
\bibitem{SCIU} 
D.I.~Olive, R.J.~Crewther and S.~Sciuto, Riv. Nuovo Cim. \textbf{2N8}, 
1 (1979). 
 
\bibitem{BERE} 
F.A.~Berezin, ``{\it The Method of Second Quantization}'', Academic 
Press (New York, 1966);\\ 
L.D.~Faddeev, ``{\it Introduction to Functional Methods}'' in ``{\it 
Methods in Field Theory}'', Les Houches 1975, Eds. R. Balian and 
J.~Zinn-Justin, North Holland (Amsterdam, 1976);\\ 
P.~Ramond, ``{\it Field Theory: a Modern Primer}'', Benjamin-Cumming 
(Reading, Mass., 1981). 

\bibitem{MEU} 
Y.~Meurice, Phys. Lett. B \textbf{164}, 141 (1985). 
 
\bibitem{MRT}
L.~Maiani, G.C.~Rossi and M.~Testa, Phys. Lett. B \textbf{292}, 
397 (1992).

\bibitem{AS} 
M.F.~Atiyah and I.M.~Singer, Bull. Amer. Math. Soc. \textbf{69}, 
422 (1963); Ann. Math. \textbf{87}, 485 and 546 (1968);\\ 
M.F.~Atiyah and G.B.~Segal, Ann. Math. \textbf{87}, 531 (1968);\\ 
L.~Alvarez-Gaum\'e, Comm. Math. Phys. \textbf{90}, 161 (1983);\\ 
D.~Friedan and P.~Windey, Nucl. Phys. B \textbf{235}, 395 (1984). 
 
\bibitem{BCGW} 
C.W.~Bernard, N.H.~Christ, A.H.~Guth and E.J.~Weinberg, Phys. Rev. D 
\textbf{16}, 2967 (1977). 
 
\bibitem{ADHM} 
M.F.~Atiyah, V.~Drinfeld, N.~Hitchin and Y.~Manin, Phys. Lett. A 
\textbf{65}, 185 (1978). 
 
\bibitem{TEM} 
E.~Corrigan, D.~Fairlie, S.~Templeton and P.~Goddard, Nucl. Phys. B 
\textbf{140}, 31 (1978);\\ 
N.H.~Christ, E.G.~Weinberg and N.K.~Stanton, Phys. Rev. D \textbf{18}, 
2013 (1978);\\ 
E.~Corrigan, P.~Goddard and S.~Templeton, Nucl. Phys. B \textbf{151}, 
63 (1979). 
 
\bibitem{ADSALL1} 
I.~Affleck, M.~Dine and N.~Seiberg, Phys. Rev. Lett. \textbf{51}, 
1026 (1983); Nucl. Phys. B\textbf{241}, 493 (1984) and Nucl. Phys. 
B \textbf{256}, 557 (1985). 
 
\bibitem{RUS4} 
V.A.~Novikov, M.A.~Shifman, A.I.~Vainshtein and V.I.~Zakharov, 
JETP Lett. \textbf{39} 601 (1984). 
 
\bibitem{COR} 
S.F.~Cordes, Nucl. Phys. B \textbf{273}, 629 (1986). 
 
\bibitem{DADI} 
A.~D'Adda and P.~Di Vecchia, Phys. Lett. B \textbf{73}, 162 (1978). 
 
\bibitem{SWTI1} 
G.~Veneziano, Phys. Lett. B \textbf{124}, 357 (1983). 
 
\bibitem{ARV} 
D.~Amati, G.C.~Rossi and G.~Veneziano, Nucl. Phys. B \textbf{249}, 
1 (1985). 
 
\bibitem{WEIN} 
S.~Coleman, J.~Wess and B.~Zumino, Phys. Rev. \textbf{177}, 2239 
(1969);\\ 
C.G.~Callan, S.~Coleman, J.~Wess and B.~Zumino, Phys. Rev. 
\textbf{177}, 2247 (1969);\\ 
S.~Weinberg, Physica A \textbf{96}, 327 (1979). 
 
\bibitem{GL} 
J.~Gasser and H.~Leutwyler, Phys. Rept. \textbf{87}, 77 (1982); 
Annals Phys. \textbf{158}, 142 (1984) and Nucl. Phys. B \textbf{250}, 
465 (1985). 
 
\bibitem{VY} 
G.~Veneziano and S.~Yankielowicz, Phys. Lett. B \textbf{113}, 321 (1982). 
 
\bibitem{TVY} 
T.~Taylor, G.~Veneziano and S.~Yankielowicz, Nucl. Phys.B 
\textbf{218}, 493 (1983). 
 
\bibitem{ADSALL2} 
I.~Affleck, M.~Dine and N.~Seiberg, Phys. Lett. B \textbf{137}, 187 
(1983); Phys. Rev. Lett. \textbf{52}, 1677 (1984) and Phys. Lett. B 
\textbf{140}, 59 (1984). 
 
\bibitem{RUS1} 
V.A.~Novikov, M.A.~Shifman, A.I.~Vainshtein and V.I.~Zakharov, 
Nucl. Phys. B \textbf{223}, 445 (1983) and \textbf{229}, 407 (1983). 
 
\bibitem{RV} 
G.C.~Rossi and G.~Veneziano, Phys. Lett. B \textbf{138}, 195 (1984). 
 
\bibitem{SABJ} 
J.~Schwinger, Phys. Rev. \textbf{82}, 664 (1951);\\ 
S.~Adler, Phys. Rev. \textbf{177}, 2426 (1969);\\ 
J.S.~Bell and R.~Jackiw, Nuovo Cimento A \textbf{60}, 47 (1969). 
 
\bibitem{KON} 
K.~Konishi, Phys. Lett. B \textbf{135}, 439 (1984);\\ 
K.~Konishi and K.~Shizuya, Nuovo Cimento A \textbf{90}, 111 (1985);\\ 
T.E.~Clark, O.~Piguet and K.~Sibold, Nucl. Phys. B \textbf{143}, 
445 (1978); {\it ibidem} \textbf{159}, 1 (1979) and {\it ibidem} 
\textbf{169}, 77 (1980);\\ 
S.~Gates, Jr., M.T.~Grisaru, M.~Ro\v{c}ek and W.~Siegel, ``{\it Superspace}'' 
Benjamin/Cummings (New York, 1983) [\hepth{0108200}]. 
 
\bibitem{KP} 
K.~Konishi and H.~Panagopoulos, Phys. Lett. B \textbf{191}, 290 (1987). 
 
\bibitem{FIPU} 
D.~Finnell and P.~Pouliot, Nucl. Phys. B \textbf{453}, 227 (1995) 
[\hepth{9503115}]. 
 
\bibitem{KHO1} 
T.J.~Hollowood, V.V.~Khoze, W.J.~Lee and M.P.~Mattis, Nucl. Phys. B 
\textbf{570}, 241 (2000) [\hepth{9904116}]. 
 
\bibitem{RUS3} 
V.A.~Novikov, M.A.~Shifman, A.I.~Vainshtein and V.I.~Zakharov, 
Nucl. Phys. B \textbf{229}, 381 (1983) and Phys. Lett. B \textbf{166} 
329 (1986] [Sov. J. Nucl. Phys. \textbf{43}, 294 (1986); Yad. Fiz. 
\textbf{43}, 459 (1986)];\\ 
T.~Morris, D.~Ross and C.~Sachrajda, Phys. Lett. B \textbf{172}, 40 (1986). 
 
\bibitem{WITIND} 
E.~Witten, Nucl. Phys. B \textbf{202}, 253 (1982). 
 
\bibitem{SEI} 
N.~Seiberg, Phys. Rev. D \textbf{49}, 6857 (1994) [\hepth{9402044}] 
and Nucl. Phys. B \textbf{431} 484 (1995) [\hepth{9408099}]. 
 
\bibitem{APP} 
T.~Appelquist and J.~Carazzone, Phys. Rev. D \textbf{11}, 2856 (1975). 
 
\bibitem{AMRV} 
D.~Amati, Y.~Meurice, G.C.~Rossi and G.~Veneziano, Nucl. Phys. B 
\textbf{263}, 591 (1986). 
 
\bibitem{THOA} 
G.~'t~Hooft, Proceedings of the 1979 Carg\'ese Summer School, 
Eds. G.~'t~Hooft et al., Plenum Press (New York, 1980). 
 
\bibitem{RUS2} 
V.A.~Novikov, M.A.~Shifman, A.I.~Vainshtein and V.I.~Zakharov, 
JETP Lett. \textbf{39}, 601 (1984) and Nucl. Phys. B \textbf{260}, 
157 (1985);\\ 
M.A.~Shifman, A.I.~Vainshtein and V.I.~Zakharov, Usp. Fiz. Nauk 
\textbf{146}, 683 (1985) [Sov. Phys. Usp. \textbf{28}, 709 (1985)]. 
 
\bibitem{FS} 
J.~Fuchs and M.G.~Schmidt, Z. Phys. C \textbf{30}, 161 (1986);\\ 
J.~Fuchs, Nucl. Phys. B \textbf{272}, 677 (1986) and {\it ibidem} 
\textbf{282}, 437 (1987). 
 
\bibitem{AFF} 
I.~Affleck, Nucl. Phys. B \textbf{191}, 429 (1981). 
 
\bibitem{CV} 
G.~Curci and G.~Veneziano, Nucl. Phys. B \textbf{292}, 555 (1987). 
 
\bibitem{MONT} 
I.~Montvay, Int. J. Mod. Phys. A \textbf{17}, 2377 (2002) [{\tt 
hep-lat/0112007}]. 
 
\bibitem{BD} 
F.~Buccella, J.P.~Derendiger, S.~Ferrara and C.~Savoy, Phys. Lett. B 
\textbf{115}, 375 (1982). 
 
\bibitem{SW} 
N.~Seiberg and E.~Witten, Nucl.Phys. B \textbf{426}, 19 (1994); 
Erratum {\it ibidem} \textbf{430}, 485 (1994) [\hepth{9407087}]. 
 
\bibitem{NOREN} 
J.~Wess and B.~Zumino, Phys. Lett. B \textbf{49}, 52 (1974);\\ 
J.~Iliopoulos and B.~Zumino, Nucl. Phys. B \textbf{76}, 310 (1974);\\ 
S.~Ferrara, J.~Iliopoulos and B.~Zumino, Nucl. Phys. B \textbf{77}, 
413 (1974);\\ 
S.~Weinberg, Phys. Lett. B \textbf{62}, 111 (1976);\\ 
M.T.~Grisaru, M.~Ro\v{c}ek and W.~Siegel, Nucl. Phys. B \textbf{159}, 
429 (1979). 
 
\bibitem{GEGL} 
H.~Georgi and S.~Glashow, Phys. Rev. Lett. \textbf{32}, 438 (1974). 
 
\bibitem{MV} 
Y.~Meurice and G.~Veneziano, Phys. Lett. B \textbf{141}, 69 (1984). 
 
\bibitem{BGK1} 
E.~Guadagnini and K.~Konishi, Nuovo Cimento A \textbf{90}, 400 (1985);\\ 
A.~Bicci and K.~Konishi, Europhys. Lett. \textbf{1}, 275 (1986). 
 
\bibitem{BGK2} 
K.~Konishi, Nucl. Phys. B \textbf{289}, 253 (1987). 
 
\bibitem{RDWN} 
C.~Rosenzweig, J.~Schechter and G.~Trahern, Phys. Rev. D \textbf{21}, 
3388 (1980);\\ 
P.~Di~Vecchia and G.~Veneziano, Nucl. Phys. B \textbf{171}, 253 
(1980);\\ 
E.~Witten, Ann. Phys. \textbf{128}, 363 (1980);\\ 
K.~Kawarabayashi and N.~Ohta, Nucl. Phys. B \textbf{175} 477 (1980);\\
K.~Kawarabayashi and N.~Ohta, Prog. Theor. Phys. \textbf{66} 1789 (1981):\\ 
P.~Nath and A.~Arnowitt, Phys. Rev. D \textbf{23}, 473 (1981). 
 
\bibitem{SYM} 
K.~Symanzik, in ``{\it New Developments in Gauge Theories}'', 
page 313, Eds. G. 't~Hooft et al., Plenum (New York, 1980);\\ 
K.~Symanzik, ``Some topics in quantum field theory'' in 
``{\it Mathematical Problems in Theoretical Physics}'', Eds. R. 
Schrader et al., Lectures Notes in Physics, Vol. 153, Springer 
(New York, 1982);\\ 
K.~Symanzik, Nucl. Phys. B {\bf 226}, 187 and 205 (1983). 
 
\bibitem{REVET} 
G.~Shore and G.~Veneziano, Int. J. Mod. Phis. A \textbf{1}, 499 (1986);\\ 
M.~Peskin, Proc. of the 1996 Theoretical Advanced Study Institute, 
``{\it Fields, String and Duality}'', Boulder, Colorado, 2-28 June 1996, 
\hepth{9702094}. 
 
\bibitem{WP} 
K.G.~Wilson, Phys. Rev. B \textbf{4}, 3174 (1971);\\ 
J.~Polchinski, Nucl. Phys. B \textbf{231}, 269 (1984);\\ 
G.~Gallavotti, Rev. Mod. Phys. \textbf{57}, 471 (1985). 
 
\bibitem{KEN} 
S.~Arnone, C.~Fusi and K.~Yoshida, JHEP \textbf{9902}, 022 (1999) 
[\hepth{9812022}];\\ 
S.~Arnone, S.~Chiantese and K.~Yoshida, 
Int.\ J.\ Mod.\ Phys.\ A {\bf 16}, 1811 (2001) 
[\hepth{0012111}].\\ 
S.~Arnone, D.~Francia and K.~Yoshida, Mod. Phys. Lett. A \textbf{17}, 
1191 (2002) [\hepth{0110104}];\\ 
J.~Ambj\"orn and R.A.~Janik, Phys. Lett. B \textbf{569}, 81 (2003) 
[\hepth{0306242}];\\ 
S.~Arnone and K.~Yoshida, Int. J. Mod. Phys. B \textbf{18}, 469 (2004);\\ 
S.~Arnone, F.~Guerrieri and K.~Yoshida, JHEP \textbf{0405}, 031 (2004) 
[\hepth{0402035}]. 
 
\bibitem{ARKA} 
N.~Arkani-Hamed and H.~Murayama, Phys. Rev. D \textbf{57}, 6638 (1998) 
[\hepth{9705189}] and JHEP \textbf{0006}, 030 (2000) [\hepth{9707133}]. 
 
\bibitem{VW} 
R.~Dijkgraaf and C.~Vafa, \hepth{0208048};\\ 
R.~Dijkgraaf, M.T.~Grisaru, C.S.~Lam, C.~Vafa, and D.~Zanon, 
Phys. Lett. B \textbf{573}, 138 (2003) [\hepth{0211017}]. 
 
\bibitem{GEO} 
G.~Hailu and H.~Georgi, JHEP \textbf{0402}, 038 (2004) [\hepth{0401101}]. 
 
\bibitem{KY} 
H.~Kawai, T.~Kuroki, T.~Morita and K.~Yoshida, Phys. Lett. B 
\textbf{611}, 269 (2005) [\hepth{0412216}]. 
 
\bibitem{ILS} 
K.A.~Intriligator, R.G.~Leigh, N.~Seiberg, Phys. Rev. D \textbf{50}, 
1092 (1994) [\hepth{9403198}]. 
 
\bibitem{ASV} 
A.~Armoni, M.A.~Shifman and G.~Veneziano, Nucl. Phys. B 
\textbf{667}, 170 (2003) [\hepth{0302163}]; 
Phys. Rev. Lett. \textbf{91}, 191601 (2003) [\hepth{0307097}] 
and Phys. Lett. B \textbf{579}, 384 (2004) [\hepth{0309013}];\\ 
A.~Armoni, G.~Shore and G.~Veneziano, Nucl. Phys. B \textbf{740}, 23 
(2006) [\hepth{0511143}]. 
 
\bibitem{KV} 
K.~Konishi and G.~Veneziano, Phys. Lett. B \textbf{187}, 106 (1987). 
 
 
\bibitem{N2susy} 
S.~Ferrara and B.~Zumino, 
Nucl.\ Phys.\ B {\bf 79}, 413 (1974). \\ 
M.~Sohnius, K.S.~Stelle and P.C.~West, 
Phys.\ Lett.\ B {\bf 92}, 123 (1980). \\ 
W.~Lerche, ``Lecture on N = 2 supersymmetric gauge theory'', 
{\it Given at NATO Advanced Study Institute: Les Houches Summer 
School on Theoretical Physics, Session 64: Quantum Symmetries, Les 
Houches, France, 1 Aug - 8 Sep 1995.} \\ 
A.~Bilal, \hepth{9601007}. 
 
\bibitem{N4susy} 
L.~Brink, J.H.~Schwarz and J.~Scherk, 
Nucl.\ Phys.\ B {\bf 121}, 77 (1977). \\ 
F. Gliozzi, J. Scherk and D. Olive, Nucl. Phys. B 
{\bf 122}, 256 (1977). 
 
\bibitem{Zeq0hypers} 
P.S.~Howe, K.S.~Stelle and P.C.~West, 
Phys.\ Lett.\ B {\bf 124}, 55 (1983). 
 
\bibitem{N2frerev} 
L.~Andrianopoli, M.~Bertolini, A.~Ceresole, R.~D'Auria, S.~Ferrara, 
P.~Fre and T.~Magri, J.\ Geom.\ Phys.\ {\bf 23}, 111 (1997) 
[\hepth{9605032}]. 
 
\bibitem{emdual} 
C.~Montonen and D.I.~Olive, 
Phys.\ Lett.\ B {\bf 72} (1977) 117. 
 
\bibitem{seibwitt2} 
N.~Seiberg and E.~Witten, 
Nucl.\ Phys.\ B {\bf 431}, 484 (1994) 
[\hepth{9408099}]. 
 
\bibitem{matone} 
M.~Matone, 
Phys.\ Lett.\ B {\bf 357}, 342 (1995) 
[\hepth{9506102}]. \\ 
M.~Matone, 
JHEP {\bf 0104}, 041 (2001) 
[\hepth{0103246}]. 
 
 \bibitem{fuctrav} 
F.~Fucito and G.~Travaglini, 
Phys.\ Rev.\ D {\bf 55}, 1099 (1997) 
[\hepth{9605215}]. 
 
\bibitem{N2twist} 
E.~Witten, 
Commun.\ Math.\ Phys.\ {\bf 117}, 353 (1988). 
 
\bibitem{seibwitt3} 
N.~Seiberg and E.~Witten, 
JHEP {\bf 9909}, 032 (1999) 
[\hepth{9908142}]. 
 
\bibitem{nekra1} 
N.~Nekrasov and A.S.~Schwarz, 
Commun.\ Math.\ Phys.\ {\bf 198}, 689 (1998) 
[\hepth{9802068}]. 
 
\bibitem{mns} 
G.~W.~Moore, N.~Nekrasov and S.~Shatashvili, 
Commun.\ Math.\ Phys.\ {\bf 209}, 77 (2000) 
[\hepth{9803265}]. 
 
\bibitem{nekra2} 
N.A.~Nekrasov, 
Commun.\ Math.\ Phys.\ {\bf 241}, 143 (2003) 
[\hepth{0010017}]. 
 
\bibitem{nekra3} 
N.A.~Nekrasov, 
Adv.\ Theor.\ Math.\ Phys.\ {\bf 7}, 831 (2004) 
[\hepth{0206161}]. 
 
\bibitem{douglas} 
M.R.~Douglas, 
\hepth{9512077}. 
 
\bibitem{witten} 
E.~Witten, 
Nucl.\ Phys.\ B {\bf 460}, 335 (1996) 
[\hepth{9510135}]. \\ 
E.~Witten, 
JHEP {\bf 0204}, 012 (2002) 
[\hepth{0012054}]. 
 
\bibitem{billo1} 
M.~Billo, M.~Frau, I.~Pesando, F.~Fucito, A.~Lerda and A.~Liccardo, 
JHEP {\bf 0302}, 045 (2003) 
[\hepth{0211250}]. 
 
\bibitem{billo2} 
M.~Billo, M.~Frau, F.~Fucito and A.~Lerda, 
\hepth{0606013}. 
 
\bibitem{branetogaugerev} 
A.~Giveon and D.~Kutasov, 
Rev.\ Mod.\ Phys.\ {\bf 71}, 983 (1999) 
[\hepth{9802067}]. 
 
\bibitem{malda} 
J.M.~Maldacena, 
Adv.\ Theor.\ Math.\ Phys.\ {\bf 2}, 231 (1998) 
[Int.\ J.\ Theor.\ Phys.\ {\bf 38}, 1113 (1999)] 
[\hepth{9711200}]. 
 
\bibitem{GKP} 
S.S.~Gubser, I.R.~Klebanov and A.M.~Polyakov, 
Phys.\ Lett.\ B {\bf 428}, 105 (1998) 
[\hepth{9802109}]. 
 
\bibitem{Witthol} 
E.~Witten, 
Adv.\ Theor.\ Math.\ Phys.\ {\bf 2}, 253 (1998) 
[\hepth{9802150}]. 
 
\bibitem{adscftrevs}{O.~Aharony, S.S.~Gubser, J.M.~Maldacena, 
H.~Ooguri and Y.~Oz, Phys.\ Rept.\ {\bf 323}, 183 (2000) 
[\hepth{9905111}]. \\ 
E.~D'Hoker and D.Z.~Freedman, Lectures given at Theoretical Advanced 
Study Institute in Elementary Particle Physics: Strings, 
Branes and Extra Dimensions, Boulder, Colorado, 3-29 Jun 2001, 
\hepth{0201253}.} 
 
\bibitem{MBrev} 
M.~Bianchi, 
Nucl.\ Phys.\ Proc.\ Suppl.\ {\bf 102}, 56 (2001) 
[\hepth{0103112}]. \\ 
M.~Bianchi, 
Fortsch.\ Phys.\ {\bf 53}, 665 (2005) 
[\hepth{0409304}]. 
 
\bibitem{BFMR} 
M.~Bianchi, F.~Fucito, G.C.~Rossi and M.~Martellini, 
Nucl.\ Phys.\ B {\bf 440}, 129 (1995) 
[\hepth{9409037}]. \\ 
M.~Bianchi, F.~Fucito, G.C.~Rossi and M.~Martellini, 
Nucl.\ Phys.\ B {\bf 473}, 367 (1996) 
[\hepth{9601162}]. 
 
\bibitem{CW}{S.R.~Coleman and E.~Weinberg, 
Phys.\ Rev.\ D {\bf 7}, 1888 (1973).} 
 
\bibitem{thooftmonop} 
G.~'t Hooft, 
Nucl.\ Phys.\ B {\bf 79}, 276 (1974). 
 
\bibitem{polyamonop}{A.M.~Polyakov, 
{\it JETP Lett.} {\bf 20} (1974) 194 
[{\it Pisma Zh. Eksp. Teor. Fiz.} {\bf 20} (1974) 430].} 
 
\bibitem{dyons} 
B.~Julia and A.~Zee, 
Phys.\ Rev.\ D {\bf 11}, 2227 (1975). 
 
\bibitem{PraSomm} 
M.K.~Prasad and C.M.~Sommerfield, 
Phys.\ Rev.\ Lett.\ {\bf 35}, 760 (1975). 
 
\bibitem{Nahm} 
W.~Nahm, 
Phys.\ Lett.\ B {\bf 90}, 413 (1980). 
 
\bibitem{OliWitt} 
E.~Witten and D.I.~Olive, 
Phys.\ Lett.\ B {\bf 78}, 97 (1978). 
 
\bibitem{Osb} 
H.~Osborn, 
Phys.\ Lett.\ B {\bf 83}, 321 (1979). 
 
\bibitem{gno}{P.~Goddard, J.~Nuyts and D.I.~Olive, Nucl. Phys. 
B {\bf 125}, 1 (1977).} 
 
\bibitem{dualMeissconf} 
G.~'t Hooft, 
Nucl.\ Phys.\ B {\bf 153}, 141 (1979). 
 
\bibitem{gravFterms} 
R.~Dijkgraaf, M.T.~Grisaru, H.~Ooguri, C.~Vafa and D.~Zanon, 
JHEP {\bf 0404}, 028 (2004) 
[\hepth{0310061}]. 
 
\bibitem{th1}{T.J.~Hollowood, JHEP {\bf 0203}, 038 (2002) 
[\hepth{0201075}] and Nucl. Phys. B {\bf 639}, 66 (2002) 
[\hepth{0202197}].} 

\bibitem{hyperinstmonoWitFre} 
D.~Anselmi and P.~Fr\'e, Nucl. Phys. B {\bf 404}, 288 (1993) 
[\hepth{9211121}] and Phys. Lett. B {\bf 347}, 247 (1995) 
[\hepth{9411205}]. \\ 
E.~Witten, Math. Res. Lett. {\bf 1}, 769 (1994) 
[\hepth{9411102}]. 
 
\bibitem{SWcurveNf} 
A.~Klemm, W.~Lerche and S.~Theisen, 
Int.\ J.\ Mod.\ Phys.\ A {\bf 11}, 1929 (1996) 
[\hepth{9505150}]. 
 
\bibitem{DonKron} S.K.~Donaldson and P.B.~Kronheimer, 
{\it ``The Geometry of Four-Manifolds''}, Oxford Mathematical 
Monographs, Oxford University Press (Oxford, New York, 1997). 
 
\bibitem{veneziano} 
G.~Veneziano, 
Nuovo Cim.\ A {\bf 57}, 190 (1968). 
 
\bibitem{Dbranes} 
J.~Dai, R.G.~Leigh and J.~Polchinski, 
Mod.\ Phys.\ Lett.\ A {\bf 4}, 2073 (1989).\\ 
R.G.~Leigh, 
Mod.\ Phys.\ Lett.\ A {\bf 4}, 2767 (1989). \\ 
J.~Polchinski, 
Phys.\ Rev.\ Lett.\ {\bf 75}, 4724 (1995) 
[\hepth{9510017}]. \\ 
C.~Angelantonj and A.~Sagnotti, 
Phys.\ Rept.\ {\bf 371}, 1 (2002) 
[Erratum-ibid.\ {\bf 376}, 339 (2003)] 
[\hepth{0204089}]. 
 
\bibitem{wittbranes} 
E.~Witten, 
JHEP {\bf 9807}, 006 (1998) 
[\hepth{9805112}]. 
 
\bibitem{DM} 
M.R.~Douglas and G.W.~Moore, 
\hepth{9603167}. 
 
\bibitem{angarm} 
C.~Angelantonj and A.~Armoni, 
Phys.\ Lett.\ B {\bf 482}, 329 (2000) 
[\hepth{0003050}]. 
 
\bibitem{mbjfm} 
M.~Bianchi and J.F.~Morales, 
JHEP {\bf 0008}, 035 (2000) 
[\hepth{0006176}]. 
 
\bibitem{N1stringinst} 
{F.~Fucito, J.F.~Morales, R.~Poghossian 
and A.~Tanzini, JHEP {\bf 0601}, 031 (2006) 
[\hepth{0510173}]. \\ 
R.~Blumenhagen, M.~Cvetic and T.~Weigand, 
\hepth{0609191}. \\ 
M.~Haack, D.~Krefl, D.~Lust, A.~Van Proeyen and M.~Zagermann, 
\hepth{0609211}. \\ 
L.E.~Ibanez and A.M.~Uranga, \hepth{0609213}. \\ 
B.~Florea, S.~Kachru, J.~McGreevy and N.~Saulina, 
\hepth{0610003}. \\ 
N.~Akerblom, R.~Blumenhagen, D.~Lust, E.~Plauschinn and 
M.~Schmidt-Sommerfeld, \hepth{0612132}. \\ 
M.~Bianchi and E.~Kiritsis, \hepth{0702015}.} 
 
 
 
\bibitem{finN4}{L.V.~Avdeev, O.V.~Tarasov and A.A.~Vladimirov, 
Phys. Lett. B {\bf 96}, 94 (1980); \\ 
M.T.~Grisaru, M.~Ro\v{c}ek and W.~Siegel, Phys. Rev. Lett. 
{\bf 45}, 1063 (1980);\\ 
M.F.~Sohnius and P.C.~West, Phys. Lett. B {\bf 100}, 245 
(1981); \\ 
W.E.~Caswell and D.~Zanon, Nucl. Phys. B {\bf 182}, 125 (1981); \\ 
S.~Mandelstam, Nucl. Phys. B {\bf 213}, 149 (1983); \\ 
L.~Brink, O.~Lindgren and B.E.W.~Nilsson, 
Phys. Lett. B {\bf 123}, 323(1983); \\ 
P.S.~Howe, K.S.~Stelle and P.K.~Townsend, Nucl. Phys. 
B {\bf 236}, 125 (1984).} 
 
\bibitem{sen}{A.~Sen, Phys. Lett. B {\bf 329}, 217 (1994) 
[\hepth{9402032}].} 
 
\bibitem{N4multi}{M.~Bianchi, F.A.~Dolan, P.J.~Heslop and H.~Osborn, 
\hepth{0609179}.} 
 
 
\bibitem{VaWi}{C.~Vafa and E.~Witten, Nucl. Phys. B {\bf 431}, 3 (1994) 
[\hepth{9408074}].} 
 
\bibitem{langlands}{A.~Kapustin and E.~Witten, \hepth{0604151}; \\ 
S.~Gukov and E.~Witten, \hepth{0612073}; \\ 
A.~Kapustin, \hepth{0612119}.} 
 
\bibitem{KHETAL}{N.~Dorey, T.J.~Hollowood, V.V.~Khoze, M.P.~Mattis 
and S.~Vandoren, Nucl. Phys. \textbf{B552}, 88 (1999) 
[\hepth{9901128}].} 
 
\bibitem{bvnv}{A.V.~Belitsky, S.~Vandoren and P.~van Nieuwenhuizen, 
Class. Quant. Grav. {\bf 17}, 3521 (2000) [\hepth{0004186}].} 
 
\bibitem{gk}{M.B.~Green and S.~Kovacs, JHEP 
{\bf 0304}, 058 (2003) [\hepth{0212332}].} 
 
\bibitem{bgkr}{M.~Bianchi, M.B.~Green, S.~Kovacs and G.C.~Rossi, 
JHEP {\bf 9808}, 013 (1998) [\hepth{9807033}].} 
 
\bibitem{dkmv}{N.~Dorey, V.V.~Khoze, M.P.~Mattis and S.~Vandoren, 
Phys. Lett. B {\bf 442}, 145 (1998) [\hepth{9808157}].} 
 
\bibitem{bkrs1}{M.~Bianchi, S.~Kovacs, G.C.~Rossi and Ya.S.~Stanev, 
JHEP {\bf 9908}, 020 (1999) [\hepth{9906188}].} 
 
\bibitem{bccl}{L.S.~Brown, R.D.~Carlitz, D.B.~Creamer, C.~Lee, 
Phys. Rev. D {\bf 17}, 1583 (1978).} 
 
\bibitem{extr}{E.~D'Hoker, D.Z.~Freedman, S.D.~Mathur, 
A.~Matusis and L.~Rastelli, \hepth{9908160}.} 
 
\bibitem{bk}{M.~Bianchi and S.~Kovacs, 
Phys. Lett. B {\bf 468}, 102 (1999) 
[\hepth{9910016}].} 
 
\bibitem{ehssw}{B.~Eden, P.S.~Howe, C.~Schubert, E.~Sokatchev 
and P.C.~West, Phys. Lett. B {\bf 472}, 323 (2000) 
[\hepth{9910150}].} 
 
\bibitem{nextextr}{J.~Erdmenger and M.~Perez-Victoria, 
Phys. Rev. D {\bf 62}, 045008 (2000) 
[\hepth{9912250}]; \\ 
B.U.~Eden, P.S.~Howe, E.~Sokatchev and P.C.~West, 
Phys. Lett. B {\bf 494}, 141 (2000) [\hepth{0004102}].} 
 
\bibitem{nearextr}{E.~D'Hoker, J.~Erdmenger, D.Z.~Freedman 
and M.~Perez-Victoria, Nucl. Phys. B {\bf 589}, 3 (2000) 
[\hepth{0003218}].} 
 
\bibitem{wilson}{S.J.~Rey and J.T.~Yee, 
Eur. Phys. J. C {\bf 22}, 379 (2001) 
[\hepth{9803001}]. \\ 
J.M.~Maldacena, Phys. Rev. Lett. {\bf 80}, 4859 (1998) 
[\hepth{9803002}].} 
 
\bibitem{bgk}{M.~Bianchi, M.B.~Green and S.~Kovacs, 
JHEP {\bf 0204}, 040 (2002) 
[\hepth{0202003}]; \hepth{0107028}.} 
 
\bibitem{instadim}{S.~Kovacs, Nucl. Phys. B {\bf 684}, 3 (2004) 
[\hepth{0310193}].} 
 
\bibitem{hklr}{N.J.~Hitchin, A.~Karlhede, U.~Lindstrom and 
M.~Ro\v{c}ek, Commun. Math. Phys. {\bf 108}, 535 (1987).} 
 
\bibitem{kns}{W.~Krauth, H.~Nicolai and M.~Staudacher, 
Phys. Lett. B {\bf 431}, 31 (1998) [\hepth{9803117}]; \\ 
W.~Krauth and M.~Staudacher, Phys. Lett. B {\bf 435}, 
350 (1998) [\hepth{9804199}].} 
 
\bibitem{bg}{T.~Banks and M.B.~Green, 
JHEP {\bf 9805}, 002 (1998) [\hepth{9804170}].} 
 
\bibitem{gg}{M.B.~Green and M.~Gutperle, Nucl. Phys. B {\bf 498}, 195 
(1997) [\hepth{9701093}].} 
 
\bibitem{SIIB}{J.H.~Schwarz, Nucl. Phys. B {\bf 226}, 269 (1983).} 
 
\bibitem{krv}{H.J.~Kim, L.J.~Romans and P.~van Nieuwenhuizen, 
Phys. Rev. D {\bf 32}, 389 (1985).} 
 
\bibitem{gogr}{R.~Gopakumar and M.B.~Green, JHEP {\bf 9912}, 015 
(1999) [\hepth{9908020}].} 
 
\bibitem{bmn}{D.~Berenstein, J.M.~Maldacena and H.~Nastase, 
JHEP {\bf 0204}, 013 (2002) 
[\hepth{0202021}].} 
 
\bibitem{HigherSpinLimit}{B.~Sundborg, 
Nucl.\ Phys.\ B {\bf 573}, 349 (2000) [\hepth{9908001}] \\ 
S.E.~Konstein, M.A.~Vasiliev and V.N.~Zaikin, 
JHEP {\bf 0012}, 018 (2000) [\hepth{0010239}]; \\ 
E.~Witten, \textit{``Spacetime Reconstruction''}, Talk at JHS 60 
Conference, California Institute of Technology, Nov. 3-4, 2001 
{\tt http://quark.caltech.edu/jhs60/witten/1.html}; \\ 
E.~Sezgin and P.~Sundell, JHEP {\bf 0109}, 036 (2001) 
[\hepth{0105001}]; \\ 
E.~Sezgin and P.~Sundell, JHEP {\bf 0109}, 025 (2001) 
[\hepth{0107186}]; \\ 
A.M.~Polyakov, Int. J. Mod. Phys. A {\bf17S1}, 119 (2002) 
[\hepth{0110196}].} 
 
\bibitem{GrandeBouffe}{M.~Bianchi, J.F.~Morales and H.~Samtleben, 
JHEP {\bf 0307}, 062 (2003) [\hepth{0305052}]; \\ 
N.~Beisert, M.~Bianchi, J.F.~Morales and H.~Samtleben, 
JHEP {\bf 0402}, 001 (2004) [\hepth{0310292}]; \\ 
N.~Beisert, M.~Bianchi, J.F.~Morales and H.~Samtleben, 
JHEP {\bf 0407}, 058 (2004) 
[\hepth{0405057}]. } 
 
\bibitem{penlim}{R.~Penrose, in {\it Differential Geometry and 
Relativity}, eds. M.~Cahen and M.~Flato (Reidel, Dordrecht, 
Netherlands 1976); \\ 
R.~Gueven, Phys. Lett. B {\bf 482}, 255 (2000) [\hepth{0005061}].} 
 
\bibitem{bfhp}{M.~Blau, J.~Figueroa-O'Farrill, C.~Hull and 
G.~Papadopoulos, Class. Quant. Grav. {\bf 19}, L87 (2002) 
[\hepth{0201081}]and JHEP {\bf 0201}, 047 (2002) 
[\hepth{0110242}].} 
 
\bibitem{met}{R.R.~Metsaev, Nucl. Phys. B {\bf 625}, 70 (2002) 
[\hepth{0112044}].} 
 
\bibitem{mt}{R.R.~Metsaev and A.A.~Tseytlin, 
Phys. Rev. D {\bf 65}, 126004 (2002) [\hepth{0202109}].} 
 
\bibitem{bmnrevs}{A.~Pankiewicz, Fortsch. Phys. {\bf 51}, 
1139 (2003) [\hepth{0307027}]. \\ 
J.C.~Plefka, Lectures given at RTN Winter School 
on Strings, Supergravity and Gauge Theory, Turin, Italy, 7-11 
January 2003, Fortsch. Phys. {\bf 52}, 264 (2004) 
[\hepth{0307101}]. \\ 
J.~M.~Maldacena, Lectures given at the Theoretical Advanced Study 
Institute in Elementary Particle Physics (TASI 2003): Recent Trends 
in String Theory, Boulder, Colorado, 1-27 Jun 2003, 
\hepth{0309246}. \\ 
M.~Spradlin and A.~Volovich, Lectures given at ICTP Spring School 
on Superstring Theory and Related Topics, Trieste, Italy, 
31 Mar - 8 Apr 2003, \hepth{0310033}. \\ 
D.~Sadri and M.M.~Sheikh-Jabbari, 
Rev. Mod. Phys. {\bf 76}, 853 (2004) [\hepth{0310119}]. \\ 
R.~Russo and A.~Tanzini, Class. Quant. Grav. {\bf 21}, S1265 (2004) 
[\hepth{0401155}].} 
 
\bibitem{gks1}{M.B.~Green, S.~Kovacs and A.~Sinha, 
JHEP {\bf 0505}, 055 (2005) [\hepth{0503077}].} 
 
\bibitem{gks2}{M.B.~Green, S.~Kovacs and A.~Sinha, 
JHEP {\bf 0512}, 038 (2005) [\hepth{0506200}].} 
 
\bibitem{gks3}{M.B.~Green, S.~Kovacs and A.~Sinha, 
 Phys. Rev. D {\bf 73}, 066004 (2006) 
[\hepth{0512198}].} 
 
\bibitem{saletan}{E.J.~Saletan, J. Math. Phys. {\bf 7}, 53 
(1961).} 
 
\bibitem{kpss}{C.~Kristjansen, J.~Plefka, G.W.~Semenoff and 
M.~Staudacher, Nucl. Phys. B {\bf 643}, 3 (2002) [\hepth{0205033}].} 
 
\bibitem{mit1}{N.R.~Constable, D.Z.~Freedman, M.~Headrick, 
S.~Minwalla, L.~Motl, A.~Postnikov and W.~Skiba, 
JHEP {\bf 0207}, 017 (2002) [\hepth{0205089}].} 
 
\bibitem{gabgre}{M.R.~Gaberdiel and M.B.~Green, Annals Phys. 
{\bf 307}, 147 (2003) [\hepth{0211122}].} 
 
\bibitem{gg2}{M.B.~Green and M.~Gutperle, Nucl. Phys. 
B {\bf 476}, 484 (1996) [\hepth{9604091}].} 
 
 
 
 
\bibitem{WI} 
E.~Witten, Nucl. Phys. B \textbf{223} 422 (1983). 
 
\bibitem{DVNPV} 
P.~Di~Vecchia, F.~Nicodemi, R.~Pettorino and G.~Veneziano, Nucl. Phys. B 
\textbf{181} 318 (1981). 
 
\bibitem{WIVE} 
E.~Witten, Nucl. Phys. B \textbf{156}, 269 (1979);\\ 
G.~Veneziano, Nucl. Phys. B \textbf{159}, 213 (1979). 
 
\bibitem{GIU} 
For a recent review see, L.~Del Debbio, L.~Giusti, C.~Pica, 
Nucl. Phys. (Proc. Suppl) B \textbf{140} 
603 (2005) [{\tt hep-lat/0409100}] and references therein. 
 
\bibitem{VAST} 
C.~Vafa and A.~Strominger, Phys. Lett. \textbf{B379} 99 (1996) 
[\hepth{9601029}]. 
 
 
 
\bibitem{RSYM} 
J.~Wess and B.~Zumino, Nucl. Phys. B \textbf{70}, 39 (1974);\\ 
P.~Fayet and S.~Ferrara, Phys. Rept. \textbf{32}, 250 (1977);\\ 
J.~Wess and J.~Bagger, ``{\it Supersymmetry and Supergravity}'', 
2nd Edition, Princeton Univ. Press (Princeton, 1992). 
 
\bibitem{PEQU} 
R.~Peccei and H.~Quinn, Phys. Rev. Lett. \textbf{38}, 1440 (1977) 
and Phys. Rev. D \textbf{16}, 1791 (1977). 
 
\bibitem{FZPS} 
S.~Ferrara and B.~Zumino, Nucl. Phys. B \textbf{87}, 207 (1975);\\ 
O.~Piguet and K.~Sibold, Nucl. Phys. B \textbf{196}, 428 (1982); 
{\it ibidem} 447 (1982) and Helv. Phys. Acta \textbf{63}, 71 (1990). 
 
\bibitem{FP} 
L.~Faddeev and V.~Popov, Phys. Lett. B \textbf{25}, 29 (1967). 
 
\bibitem{TRAV} 
G.~Travaglini, Ph.D. lectures, University of Rome ``Tor Vergata'', 
unpublished. 
 
\bibitem{FHSCHR} 
R.P.~Feynman and A.R.~Hibbs, ``{\it Quantum Mechanics and Path 
Integrals}'', Mc Graw Hill (New York, 1977). 
 
\bibitem{PREV} 
G.C.~Rossi and M.~Testa, Nucl. Phys. B {\bf 163}, 109 (1980); {\it 
ibidem} \textbf{B176}, 477 (1980) and {\it ibidem} \textbf{B237}, 442 
(1984); \\ 
K.~Symanzik, Nucl. Phys. B \textbf{190}, 1 (1981);\\ J.P. Leroy, 
J.~Micheli, G.C.~Rossi and K.~Yoshida, Z. Phys. C \textbf{48}, 653 
(1990). 
 
\bibitem{LSCH} 
M.~L\"uscher, Comm. Math. Phys. \textbf{54}, 283 (1977);\\ 
G.~Marchesini and E.~Onofri, Nuovo Cim. A \textbf{65}, 298 (1981);\\ 
M.~L\"uscher, R.~Narayanan, P.~Weisz and U.~Wolff, Nucl. Phys. B 
\textbf{384}, 168 (1992);\\ 
M.~L\"uscher, R.~Sommer, P.~Weisz and U.~Wolff, Nucl. Phys. B 
\textbf{389}, 247 (1993) and {\it ibidem} \textbf{413}, 481 (1994);\\ 
S.~Sint, Nucl. Phys. B \textbf{421}, 135 (1994) and Nucl. Phys. B 
\textbf{451}, 416 (1995). 
 
\bibitem{OWLK} 
B.A.~Ovrut and J.~Wess, Phys. Rev. D \textbf{25}, 409 (1982);\\ 
W.E.~Lerche, Nucl. Phys. B \textbf{238}, 582 (1984);\\ 
T.~Kugo, I.~Ojima and T.~Yanagida, Phys. Lett. B \textbf{135}, 402 (1984). 
 
\bibitem{GeoGla}{H.~Georgi and S.~L.~Glashow, 
Phys.\ Rev.\ D {\bf 6}, 2977 (1972).} 
 
\bibitem{tong}{D.~Tong, Lectures given at Theoretical Advanced Study 
Institute in Elementary Particle Physics: Many Dimensions 
of String Theory, Boulder, Colorado, 5 Jun - 1 Jul 2005, 
\hepth{0509216}.} 
 
\bibitem{dobpet}{V.K.~Dobrev and V.B.~Petkova, 
Phys. Lett. B {\bf 162}, 127 (1985).} 
 
\bibitem{intri}{K.~Intriligator, Nucl. Phys. B {\bf 551}, 575 (1999) 
[\hepth{9811047}].} 
 
\end{thebibliography}
\end{document}